\documentclass[11pt,twoside,subeqns]{book}
\usepackage[ansinew]{inputenc}
\usepackage{ngerman,a4}
\usepackage{amssymb}   
\usepackage{amsmath}
\usepackage{amsthm,mathrsfs}
\usepackage{amscd}  

\usepackage{graphicx}
\usepackage{psfrag}
\usepackage{epsfig}
\usepackage{setspace}
\usepackage{fancyheadings}
\usepackage[english]{babel} 
\usepackage[setpagesize=false]{hyperref}
\usepackage{url}

\newtheorem{theorem}{Theorem}[section]
\newtheorem{proposition}[theorem]{Proposition}
\newtheorem{definition}[theorem]{Definition}
\newtheorem{lemma}[theorem]{Lemma}
\newtheorem{app-lemma}{Lemma}[subsection]
\newtheorem{corollary}[theorem]{Corollary}

\numberwithin{equation}{section}
\usepackage[ansinew]{inputenc}
\usepackage{ngerman,a4}
\usepackage{amssymb}   
\usepackage{amsmath}
\usepackage{amsthm,mathrsfs}
\usepackage{amscd}  
\usepackage{longtable}

\usepackage{graphicx}
\usepackage{psfrag}
\usepackage{epsfig}
\usepackage{setspace}
\usepackage{fancyheadings}
\usepackage[english]{babel} 
\usepackage{oldgerm}

\font\fa=bbm11
\newcommand{\R}{\mbox{\fa R}}   
\newcommand{\N}{\mbox{\fa N}}   

\newcommand{\Ibb}[1]{ {\rm I\ifmmode\mkern -3.6mu\else\kern -.2em\fi#1}}
\newcommand{\ibb}[1]{\leavevmode\hbox{\kern.3em\vrule
     height 1.2ex depth -.3ex width .2pt\kern-.3em\rm#1}}
\newcommand{\Cl}{{\ibb C}}           
\newcommand{\Rl}{{\Ibb R}}           
\newcommand{\Zl}{\mathbb{Z}}

\newcommand{\bref}[1]{(\ref{#1})}

\newcommand{\Om}{\Omega}
\newcommand{\Uhat}{\widehat{U}}

\newcommand{\Hil}{\mathcal{H}}
\newcommand{\VV}{\mathcal{V}}

\newcommand{\Q}{\mathcal{Q}}
\newcommand{\F}{\mathcal{F}}

\newcommand{\DD}{\mathcal{D}} 
 
\newcommand{\W}{\mathcal{W}}    
\newcommand{\Z}{\mathcal{Z}}    
\newcommand{\ZZZ}{\mathscr{Z}}  
\newcommand{\Ss}{\mathscr{S}}   
\newcommand{\OO}{\mathcal{O}}   
\newcommand{\PGpo}{\mathcal{P}_+^\uparrow}   
\newcommand{\PGpro}{\mathcal{P}_+} 
\newcommand{\PG}{\mathcal{P}} 
 
\newcommand{\OOO}{\mathscr{O}}

\newcommand{\NN}{\mathcal{N}}  
  
\newcommand{\X}{\mathcal{X}}  
\newcommand{\Y}{\mathcal{Y}}  

\newcommand{\frS}{\textfrak{S}} 

\newcommand{\LL}{\mathscr{L}} 
\newcommand{\Ds}{\mathscr{D}} 
\newcommand{\Ps}{\mathscr{P}} 
\newcommand{\DT}{\mathscr{D}}
\newcommand{\CC}{\mathscr{C}}

\newcommand{\pol}{\mathscr{P}}
\newcommand{\SF}{\mathcal{S}}

\newcommand{\lmto}{\longmapsto}

\def\bte{{\mbox{\boldmath{$\te$}}}}
\def\Bose{{\mbox{\boldmath{$S_2=+1$}}}}
\def\Fermi{{\mbox{\boldmath{$S_2=-1$}}}}
\def\bla{{\mbox{\boldmath{$\lambda$}}}}
\def\bze{{\mbox{\boldmath{$\zeta$}}}}
\def\bxi{{\mbox{\boldmath{$\xi$}}}}
\def\beps{{\mbox{\boldmath{$\varepsilon$}}}}

\def\bof{{\mbox{\boldmath{$f$}}}}
\def\bk{{\mbox{\boldmath{$k$}}}}
\def\bl{{\mbox{\boldmath{$l$}}}}

\newcommand{\bno}[1]{|\!|\!|#1|\!|\!|}
\def\bs{{\mbox{\boldmath{$s$}}}}

\def\br{{\mbox{\boldmath{$r$}}}}

\newcommand{\sbte}{{\mbox{\footnotesize \boldmath $\theta$}}}

\newcommand{\sbk}{{\mbox{\footnotesize \boldmath $k$}}}
\newcommand{\sbl}{{\mbox{\footnotesize \boldmath $l$}}}
\newcommand{\sbze}{{\mbox{\footnotesize \boldmath $\zeta$}}}
\newcommand{\sbla}{{\mbox{\footnotesize \boldmath $\lambda$}}}

\newcommand{\sbeps}{{\mbox{\footnotesize \boldmath $\varepsilon$}}}

\newcommand{\sbr}{{\mbox{\footnotesize \boldmath $r$}}}

\newcommand{\spp}{{\mbox{\footnotesize $-(\pi,\pi)$}}}
\newcommand{\spn}{{\mbox{\footnotesize $(-\pi,0)$}}}
\newcommand{\minpi}{{\mbox{\footnotesize $-\pi$}}}
\newcommand{\snp}{{\mbox{\footnotesize $(0,-\pi)$}}}
\newcommand{\snull}{{\mbox{\footnotesize $0$}}}
\newcommand{\skappa}{{\mbox{\footnotesize $\kappa(S_2)$}}}
\newcommand{\minskappa}{{\mbox{\footnotesize $-\kappa(S_2)$}}}
\newcommand{\fone}{{\mbox{\footnotesize $\langle\te_1,\te_2|\,A\Om\rangle$}}}
\newcommand{\ftwo}{{\mbox{\footnotesize $\langle\Om\,|\,[z(\te_2),A]\,|\,\te_1\rangle$}}}
\newcommand{\fthree}{{\mbox{\footnotesize $\langle A^*\Om\,|\te_2,\te_1\rangle$}}}
\newcommand{\fxi}{{\mbox{\footnotesize $(\Delta^{1/4}A\Om)_2(\te_1,\te_2)$}}}
\newcommand{\lone}{{\mbox{\footnotesize $-\La_2$}}}
\newcommand{\ltwo}{{\mbox{\footnotesize $-\La_2^{\tau_1}$}}}
\newcommand{\bnstrich}{{\mbox{\footnotesize $\B_2'(\kappa(S_2))$}}}
\newcommand{\bbase}{{\mbox{\footnotesize $\B_2(\kappa(S_2))$}}}
\newcommand{\sms}{{\mbox{\footnotesize $s$}}}
\newcommand{\ms}{{\mbox{\footnotesize $4m^2$}}}
\newcommand{\mss}{{\mbox{\footnotesize $4m^2-s$}}}

\newcommand{\supp}{\mathrm{supp}\,}

\newcommand{\te}{\theta}
\newcommand{\la}{\lambda}
\newcommand{\La}{\Lambda}
\newcommand{\eps}{\varepsilon}

\newcommand{\fti}{\widetilde{f}}
\newcommand{\ghat}{\hat{g}}
\newcommand{\fhat}{\hat{f}}

\newcommand{\gti}{\widetilde{g}}
\newcommand{\fbar}{\overline{f}}

\newcommand{\Zd}{Z^{\dagger}}
\newcommand{\Cd}{C^{\dagger}}
\newcommand{\zd}{z^{\dagger}}

\newcommand{\ad}{a^{\dagger}}

\newcommand{\lto}{\longrightarrow}

\newcommand{\Tu}{\mathcal{T}}
\newcommand{\Ba}{\mathcal{B}}
\newcommand{\Cu}{\mathcal{C}}  

\newcommand{\iin}{_\mathrm{in}}

\newcommand{\oout}{_\mathrm{out}}
\newcommand{\tp}[1]{^{\otimes #1}}    

\newcommand{\dom}{\mathrm{dom}}

\newcommand{\A}{\mathcal{A}}
\newcommand{\B}{\mathcal{B}}
\newcommand{\M}{\mathcal{M}}
\newcommand{\K}{\mathcal{K}}
\renewcommand{\LL}{\mathcal{L}}

\newcommand{\vol}{\mathrm{vol}}

\begin{document}

\thispagestyle{empty}

\begin{center}

\vspace*{3em}
\vfill

{\Huge \bf 
On the Construction of \\\vspace*{4mm} Quantum Field Theories \\\vspace*{6.5mm} with Factorizing S-Matrices
}
\vspace*{2cm}
\vfill
{\Large 
Dissertation \\
zur Erlangung des Doktorgrades \\
der Mathematisch-Naturwissenschaftlichen Fakult\"aten\\
der Georg-August-Universit\"at zu G\"ottingen

\vfill
  vorgelegt von \\
Gandalf Lechner\\
aus Hamburg

\vfill
 
G\"ottingen, 2006

\vfill

}
\end{center}

\clearpage  

\thispagestyle{empty}
\mbox{ } \vfill \mbox{ } \\
\noindent
D 7  \\
Referent: Prof.\ Dr.\ D.\ Buchholz \\
Korreferent: Prof.\ Dr.\ K. Fredenhagen \\
Tag der m\"undlichen Pr\"ufung: 24.05.2006

%
%
\tableofcontents

\chapter{Introduction}\label{chapter:introduction}

\section{The Construction of Models in Relativistic Quantum Field Theory}

The observed phenomena of high energy physics can be successfully described by relativistic quantum field theories. Such theories are the constituents of the standard model, and have led to our current understanding of the physics of elementary particles. But although the predictions of quantum field theories agree with the experimental results to a very high degree of accuracy in several cases, the construction of models within this framework is a delicate issue, which is only partly understood today.

The most common approach to construct quantum field theories starts from a quantized version of a classical Lagrangian and uses a formal perturbative expansion in the coupling constant around an interaction-free model theory. However, the perturbation series is believed to diverge in many cases, and can therefore not be used for a proper definition of models. These problems have attracted the attention of mathematical physicists for several decades, and have stimulated different approaches to constructive quantum field theory.

The first theories which were constructed without having to rely on cutoffs were defined by a Lagrangian with a polynomial interaction term on two-dimensional Minkowski spacetime. These models were established using the Hamiltonian strategy, and could be shown to satisfy the assumptions of axiomatic quantum field theory \cite{GJ1}.

With the advent of the Euclidian approach \cite{symanzik}, powerful new methods and strategies became available. Until the beginning of the 1980s, several interacting models on two- and three-dimensional spacetime had been constructed with the help of functional integral techniques, see \cite{glimmjaffe} and the references cited therein. These models have been thoroughly investigated, and their relation to perturbation theory is well understood.

However, after this period of successes, little progress has been made in constructive aspects of quantum field theory\footnote{See, however, \cite{tobias-phd,tobias} for interesting recent results concerning the local S-matrices in the $P(\varphi)_2$ models.}. In particular, comparable results in four dimensions are missing up to now.


A completely different approach to the construction of two-dimensional quantum field theories is taken up in the so-called bootstrap form factor program \cite{Babujian:2003sc,smirnov}. Here models are not defined in terms of Lagrangians, but rather with the help of a factorizing S-matrix. Such scattering operators have first been found in the context of integrable models like the Sine-Gordon theory or the scaling Ising model \cite{abdalla}, and have a simple structure in comparison with an S-matrix in higher dimensions. In particular, it is possible to explicitly specify non-trivial factorizing S-matrices, and use them as a sufficiently well understood description of a possible interaction. 

In the form factor program, one investigates local field operators associated with a given S-matrix by analyzing their matrix elements in scattering states. Due to the special form of the S-matrix, there exist numerous constraints on these matrix elements (called form factors in this context), which render their explicit calculation possible. In fact, the form factors of many models are known today.

After the computation of the form factors, the crucial step in the bootstrap program is the derivation of formulae for $n$-point Wightman functions of local fields associated with the considered S-matrix, which are given by certain infinite sums of integrals over form factors. These sums are very difficult to control as a consequence of the involved structure of local quantum fields, reminiscent of the perturbation series. Due to these problems, the construction envisaged in the form factor program can be presently carried out only for two special models \cite{Babujian:2003sc}.
\\
\\
In the present work, we present a new approach to the construction of quantum field theories with factorizing S-matrices. Our approach uses the insights of the structural analysis of relativistic quantum physics carried out in the algebraic framework of quantum field theory \cite{haag} and can be summarized as follows.

As in the form factor program, we consider the inverse scattering problem, i.e. the construction starts from a given factorizing S-matrix. But instead of aiming at the construction of local quantum fields in the first place, we begin by considering objects with weaker localization properties, which are easier to construct. These are certain fields describing asymptotic particle states connected to the given S-matrix. On the vacuum, they act by creating single particle states, without admixture of vacuum polarization clouds. Such operators have been introduced by B. Schroer \cite{Schroer:1997cx,Schroer:1997cq}, who coined the name polarization-free generators for them. He discovered that polarization-free generators are localized in causal closures of half lines, i.e. wedge-shaped spacelike regions of Minkowski space ({\em wedges}, for short). Later on, it was shown by Borchers, Buchholz and Schroer that such generators exist in any local theory, although they may in general have delicate domain properties \cite{BBS}.

In this work, we will show that polarization-free generators can be constructed in a non-perturbative manner for a large class of S-matrices. Using concepts of algebraic quantum field theory, these generators are then used to build an infinite family of models. Here the crucial insight is that for the construction of a model theory, it is not necessary to derive explicit formulae for field operators or $n$-point functions. Rather, it is sufficient to control the structure of the algebras generated by local observables.

It has been discovered by D. Buchholz that the rich algebraic structure of the observables in quantum field theory can be used to study observables localized in bounded regions of spacetime in an indirect manner, in terms of the algebras generated by the polarization-free generators \cite{BuLe}. In particular, it is possible to decide whether to a given factorizing S-matrix $S$ there exists a model of quantum field theory whose scattering is governed by $S$. The main technical tool underlying this observation is a condition on the modular operators of the wedge algebras, due to Bucholz, D'Antoni and Longo \cite{nuclearmaps1,nuclearmaps2}.

Using these methods, the long-standing question of the form factor program regarding the existence of models with a prescribed factorizing S-matrix is solved here. For a certain infinite class of S-matrices, corresponding model theories which comply with all basic assumptions of relativistic quantum physics are constructed in the framework of algebraic quantum field theory.



Among these models are the Sinh-Gordon theory and the scaling Ising model, which are usually formulated in terms of a Lagrangian with hyperbolic interaction potential and the scaling limit of a model of statistical mechanics, respectively. These systems are well-studied from different points of view \cite{Sklyanin:1989cf,FMS,ising2d,BKW}, but for most of the theories found here, an alternative description is not known. Thus they may be considered as new models.

After the existence of the models is established, it is necessary to analyze their scattering states and to compute their S-matrices in order to verify that the construction really yields the solution of the inverse scattering problem. In contrast to all other approaches, this is possible here. It will be shown how multi-particle scattering states can be explicitly calculated, and we find that the S-matrices used as an input in the construction can be recovered from the collision states of the finished models. Moreover, we prove that these models have a complete interpretation in terms of asymptotic particle states, and thus provide the very first interacting theories in which the property of asymptotic completeness can be established.

\section{Overview of this Thesis}

This thesis is organized as follows. In chapter \ref{chapter:introduction}, we briefly summarize the framework of algebraic quantum field theory in order to establish our notation and indicate which conventions are used. Furthermore, some basic geometrical facts about two-dimensional Minkowski space are collected here for later reference.

A model-independent construction procedure for two-dimensional quantum field theories in terms of wedge-localized objects is presented in chapter \ref{chapter:netsin2d}. We employ the framework of local quantum physics \cite{haag}, and address the question how a relativistic quantum theory can be constructed in terms of its observables which are localized in a particular wedge. The main result of this chapter is the derivation of a set of clear-cut conditions on the algebra generated by such wedge-localized observables which ensure that the corresponding model theory complies with all principles of relativistic quantum physics.

This general construction is then made concrete in a family of models which are characterized in terms of their S-matrices. These S-matrices are taken to be of the factorizing type, and the properties of such scattering operators are recalled in chapter \ref{chapter:factS}. In chapter \ref{chapter:wedgenet}, the construction is carried out by an inverse scattering approach. We fix a factorizing S-matrix from a certain infinite class, and construct a corresponding model theory. At the basis of this construction lies the Zamolodchikov-Faddeev algebra, which is given a spacetime interpretation in terms of a pair of associated wedge-local quantum fields. These fields constitute the basic objects of our models, and we study their properties in chapter \ref{chapter:wedgenet}. It is shown there that they determine a local quantum theory as in the general setup of chapter \ref{chapter:netsin2d}. Furthermore, we compute the two-particle scattering states of this theory, which are found to reproduce the initially given two-particle S-matrix.

Chapter \ref{chapter:nuclearity}, devoted to an analysis of the local observable content of these models, is a crucial step in the construction. For a large class of S-matrices, we prove the existence of local observables in arbitrarily small spacetime regions by verifying the conditions on the underlying wedge algebra proposed in chapter \ref{chapter:netsin2d}. For a still larger class, we obtain observables localized in regions above a minimal size.

Two aspects of the interaction in the constructed models are investigated in chapter \ref{chapter:reconstructS}. Our most important result in this context is the proof that the S-matrices, which were used as an input for the construction, can be recovered from the collision states of the model by application of the Haag-Ruelle scattering theory. This implies that the program carried out here provides the solution to the inverse scattering problem for the considered class of S-matrices. Moreover, we prove that the constructed models have a complete interpretation in terms of asymptotic particle states, providing the first examples of interacting quantum field theories for which this property has been established. In addition to the results related to scattering theory, we derive bounds on the thermodynamical partition function of the theory.

The main text of the thesis is completed in chapter \ref{chapter:conclusions} with a presentation of our conclusions, and an account of open problems and perspectives.
\\
\\
The three appendices cover the following subjects. In appendix \ref{chapter:mnc+-1}, the analysis of chapter \ref{chapter:nuclearity} is reconsidered for two special models, namely the interaction-free theory and the scaling Ising model. Due to the simpler algebraic structure of these theories, it is possible to treat them with different methods, leading to somewhat stronger results than in chapter \ref{chapter:nuclearity}. Appendix \ref{app:zamo-calcs} contains the technical proofs of certain statements needed in the main text. Finally, appendix \ref{chapter:math} provides some mathematical background material.
\\
\\
Parts of the content of chapter \ref{chapter:netsin2d} have been published in the joint paper \cite{BuLe} with D. Buchholz, and most of the material covered in chapter \ref{chapter:wedgenet} can be found in \cite{GL1}. The proceedings contribution \cite{GL-bros} contains a preliminary version of the results of chapter \ref{chapter:nuclearity}, whereas a treatment similar to the one presented here is the content of \cite{GL-nuci}. The analysis of the scaling Ising model in the context of our inverse scattering construction has been published in \cite{GL-1}.


\section{Geometrical Preliminaries on Two-Dimensional Minkowski Space}\label{Sec:Geo}

We consider Minkowski space as the two-dimensional real plane $\Rl^2$ endowed with proper coordinates $x=(x_0,x_1)$ and the inner product $x\cdot y = x_0y_0-x_1y_1$. Using units in which the speed of light is $c=1$, the subregions of points $x$ with $x\cdot x >0$, $x\cdot x< 0$ and $x\cdot x=0$ are called timelike, spacelike and lightlike, respectively. (Also Planck's constant $\hbar$ will be set to unity here.)

The invariance group of this product is the Poincar\'e group $\PG$, which is generated by the translations $t_a: x\mapsto x+a$, $a\in \Rl^2$, the proper boost transformations
\begin{align}\label{def:boost}
	\Lambda(\la) : x \longmapsto
	\left(
		\begin{array}{ll}
			\cosh\la & \sinh\la\\
			\sinh\la & \cosh\la
		\end{array}
	\right)
	\;x
	\,,\qquad
	\la\in\Rl,
\end{align}
and the two reflections $T:x\mapsto(-x_0,x_1)$ and $-T:x\mapsto(x_0,-x_1)$.

The {\em proper} Poincar\'e group $\PG_+$ is generated by the translations, boosts and the total reflection $-1 : x \longmapsto -x$. 

Finally, the {\em proper orthochronous} Poincar\'e group, $\PGpo$, is generated by the translations and boosts only.
 \\
 \\
For the localization of physical observables, different regions in $\Rl^2$ will become important in this thesis. We adopt the convention to work with open regions only, and thus define the spacelike complement $\OO'$ of a set $\OO\subset\Rl^2$ as the interior of $\{x\in\Rl^2\,:\, (x-y)^2<0\;\; \forall y\in\overline{\OO}\}$.

\begin{figure}[here]
    \noindent
    \psfrag{wr}{$W_R$}
    \psfrag{wl}{$W_L$}
    \psfrag{0}{$\,$}
    \psfrag{x0}{$x_0$}
    \psfrag{x1}{$x_1$}
    \centering\epsfig{file=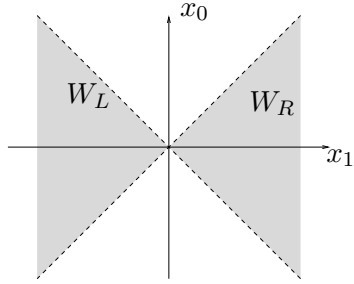,width=4.4cm}
    \caption{The left and the right wedge.}
\label{fig:wr-wl}
\end{figure}

Of particular significance for us is the family of {\it wedges}, which is defined as follows. The so-called {\it right wedge} is the set
\begin{align}\label{def:WR}
	W_R &:= \{x\in\Rl^2 \,:\, x_1 > |x_0|\}
\end{align}
and the {\it left wedge} is $W_L:=W_R'=-W_R$ (see figure \ref{fig:wr-wl}). An arbitrary wedge is defined to be a set of the form $gW_R$, where $g\in{\mathcal P}$ is a Poincar\'e transformation. The set of all wedges will be denoted $\W$.
\\
\\
Note that the sets $W_R$ and $W_L$ are invariant under the action of the boost transformations \bref{def:boost}, since the eigenvectors of $\La(\la)$ \bref{def:boost} are lightlike. Hence $\W$ has the form
\begin{align}\label{eq:wedgeset}
	\W &= \{W_L+x\,:\,x\in\Rl^2\} \cup \{W_R+x\,:\,x\in\Rl^2\}\,.
\end{align}
Wedges of the form $W_L+x$ and $W_R+x$ will be referred to as left and right wedges, respectively.

Besides these unbounded regions, we also consider {\it double cones}, usually denoted by the letter $\OO$. A double cone is defined to be a non-empty intersection of a forward and a backward lightcone, which in two dimensions is equivalent to saying that it is the intersection of a left and a right wedge. The family of all double cones will be denoted by $\OOO$.

Each double cone determines four associated wedges, as depicted in the following figure.
\begin{figure}[htbpc]
    \noindent
    \psfrag{O}{$\OO$}
    \psfrag{WRR}{$W_R^\OO$}
    \psfrag{WLL}{$W_L^\OO$}
    \psfrag{x0}{$x_0$}
    \psfrag{x1}{$x_1$}
    \psfrag{WR}{$\left(W_L^\OO\right)'$}
    \psfrag{WL}{$\left(W_R^\OO\right)'$}
    \centering\epsfig{file=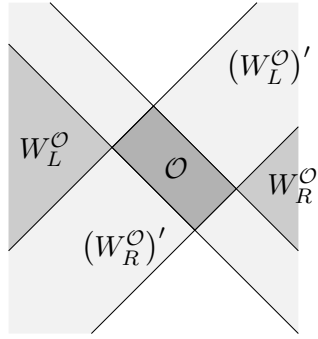,width=3.9cm}
     \caption{A double cone $\OO$ and its associated wedges}
	\label{fig:doublecone}
\end{figure}
\\
With the notation introduced here, there holds for any double cone $\OO$
\begin{align}
	\OO' = W_L^\OO \cup W_R^\OO\,,
	\qquad
	\OO = \left(W_L^\OO\right)' \cap \left(W_R^\OO\right)'\,.
\end{align}
The fact that two double cones $\OO_1, \OO_2$ are contained in each other or lie spacelike to each other can be expressed in terms of their associated wedges as
\begin{align}
	\OO_1\subset\OO_2	&\Longleftrightarrow W_R^{\OO_1} \supset W_R^{\OO_2} \text{ and } W_L^{\OO_1} \supset W_L^{\OO_2}\,,\\
	\OO_1\subset\OO_2'	&\Longleftrightarrow \OO_1  \subset W_L^{\OO_2} \text{ or } \OO_1\subset W_R^{\OO_2}\,.
\end{align}
Finally, we agree to write $\OO_1 \Subset \OO_2$ for an inclusion of two (arbitrary) regions $\OO_1,\OO_2\subset\Rl^2$ if the closure of $\OO_1$ is contained in the interior of $\OO_2$.\label{inclusionsymbol}

\section{The Algebraic Formulation of Quantum Field Theory}\label{Sec:Axioms}

In this section we briefly recall the framework of algebraic quantum field theory, mainly to indicate our notations and conventions. An introduction to the subject can be found in either of the books \cite{haag, araki, Horuzhy, baumwoll}.
\\
\\
We consider a relativistic quantum theory on Minkowski space $\Rl^d$ $(d\geq 2)$ in its vacuum representation. In the algebraic approach to quantum field theory, a model is characterized in terms of its algebra $\A$ of local observables, which are given by selfadjoint operators on a fixed Hilbert space $\Hil$ (the vacuum Hilbert space of the theory). The algebra $\A$ has a very rich structure, which we outline in the following.

To begin with, $\A$ contains all local algebras $\A(\OO)$, generated by the observables localized in a spacetime region $\OO\subset\Rl^d$. Usually, $\A(\OO)\subset\B(\Hil)$ is taken to be a von Neumann algebra, i.e. a $*$-subalgebra of $\B(\Hil)$ which is closed in the weak operator topology. The assignment
\begin{align}\label{def:aonet}
\Rl^d \supset \OO \lmto \A(\OO) \subset\B(\Hil)
\end{align}
contains all physical information of the theory, and therefore constitutes the main object of interest in algebraic quantum field theory. For \bref{def:aonet} to model the observables of a relativistic quantum system, the algebras $\A(\OO)$ must have a number of properties.

To begin with, the interpretation of the selfadjoint elements of $\A(\OO_1)$ as observables localized in $\OO_1$ implies that to a larger region $\OO_2$, there must correspond a larger algebra $\A(\OO_2)$, i.e.
\begin{align}
\A(\OO_1)	\subset	\A(\OO_2)		\qquad {\rm for}\qquad \OO_1\subset\OO_2\,.
\end{align}
This property will be referred to as {\rm isotony}, it gives the map $\OO\lmto\A(\OO)$ the mathematical structure of a net.

In relativistic theories, no effect can propagate faster than the speed of light. As a consequence, observables localized in regions of spacetime which cannot be connected by a sequence of light rays are commensurable and do not interfere with each other. In view of Heisenberg's uncertainty relation, this implies in particular that the corresponding operators must commute.  Causality can therefore be implemented in the mathematical framework by requiring that algebras of observables localized in spacelike separated regions must commute, 
\begin{align}
\A(\OO_1)	\subset \A(\OO_2)'		\qquad {\rm for}\qquad	\OO_1\subset\OO_2'\,.
\end{align}
As usual, we adopt the convention to write the causal (spacelike) complement of a region $\OO\subset\Rl^d$ as $\OO'$, and use the prime on an algebra of operators acting on a Hilbert space $\Hil$ to denote its commutant in $\B(\Hil)$.

The condition of locality can be strengthened to {\em Haag duality} by requiring that any operator commuting with all elements of $\A(\OO)$ is localized in $\OO'$, 
\begin{align}\label{def:Haagduality}
\A(\OO')=\A(\OO)'\,.
\end{align}
Whereas Haag-duality is known to hold if the region $\OO$ in \bref{def:Haagduality} is a wedge and the net is generated by finite-component Wightman fields \cite{BiWi2}, it is not valid for bounded regions in general.

The relativistic symmetries are assumed to act on the net by a representation of the identity component of the Poincar\'e group, i.e. one postulates that there exists a strongly continuous, unitary representation $U:\PGpo\lto\B(\Hil)$ of the proper orthochronous Poincar\'e group on $\Hil$. The covariance of the theory then demands
\begin{align}
U(g)\A(\OO)\,U(g)^{-1}	&=	\A(g\,\OO)\,,\qquad g\in\PGpo\,,
\end{align}
where $g\,\OO=\{g\,x\,:\,x\in\OO\}$ denotes the transformed region.

The translation group $(\Rl^d,+)$ is contained in $\PGpo$ as a subgroup, and as usual, the generators $P^\mu$ of $U(x)=e^{i x_\mu P^\mu}$ are interpreted as energy and momentum operators. In a vacuum representation, the stability of matter requires to have a positive energy spectrum in all Lorentz frames. This spectrum condition amounts to the joint spectrum of the $P^\mu$ being contained in the closure of the forward light cone $V^+:=\{(p_0,\vec{p}\,)\in\Rl^d\,:\,p_0 > |\vec{p}\,|\}$.

Finally, there is a vector $\Om\in\Hil$ which models the physical vacuum state. It is required to have energy and momentum zero, i.e. $\Om$ must be invariant under the action of $U$. Moreover, one usually requires that $\Om$ is unique in the sense that all $U$-invariant vectors in $\Hil$ are scalar multiples of $\Om$. 

In quantum field theory, the vacuum state has the Reeh-Schlieder property. In the algebraic framework, this property can be formulated by requiring $\Om$ to be cyclic for the local algebras, i.e.
\begin{align}
\overline{\A(\OO)\Om}		=	\Hil\,,	\qquad \OO\subset\Rl^d\;\;{\rm open}\,.
\end{align}
(The bar denotes closure in the norm topology of $\Hil$.) For regions $\OO$ with non-empty causal complement, it then follows that $\Om$ must also be separating for $\OO$, i.e. $A\Om=0$, $A\in\A(\OO)$, implies $A=0$.
\\
\\
In this work, we will refer to a net $\A:\OO\lmto\A(\OO)$ \bref{def:aonet} which has all the above described properties as a {\em local net on $\Rl^d$}. If the stronger condition \bref{def:Haagduality} is assumed, we speak of a {\em Haag-dual net}. In both cases, the algebras $\A(\OO)$ are called {\em local algebras} if $\OO$ is bounded, and their elements are called {\em local observables}.

\chapter{Construction of Two-Dimensional Local Nets from Wedge Algebras}
\label{chapter:netsin2d}

A large part of this thesis is devoted to the explicit construction of two-dimensional quantum field theory models with prescribed factorizing S-matrices. As mentioned in the Introduction, the methods we use differ significantly from those usually employed in constructive of quantum field theory. Instead of perturbatively studying a quantized classical field theory with functional integral techniques, the approach followed here focuses on the structure of the algebra of quantum observables.

The main insight is that the construction of models can be based on quantities with quite weak localization properties, and strictly local observables are obtained in a second step. The weak localization to be used is localization in wedge regions (section \ref{Sec:Geo}), and in the present chapter, we study the model-independent aspects of constructions of local nets in terms of wedge-local quantities.

Wedges are on the one hand large enough to allow a comparatively easy construction of observables localized within them. On the other hand, a double cone $\OO$ is the intersection of all wedges containing $\OO$. Hence it is possible to study observables localized in finite regions indirectly in terms of wedge-localized objects.
\\
\\
The framework best suited for our analysis is that of algebraic quantum field theory \cite{haag, araki, Horuzhy, baumwoll}, and accordingly, the most natural input for the construction is an algebra of observables localized in a wedge (a {\em wedge algebra}, for short).

As we shall see later, such a wedge algebra can be defined in terms of a factorizing S-matrix on two-dimensional Minkowski space, constituting the starting point of our construction of models by means of inverse scattering theory. In the present chapter, devoted to a model-independent discussion of the concept, we define a wedge algebra as an abstract algebra $\M$ satisfying an appropriate set of conditions which ensure that $\M$ can be consistently interpreted as being generated by quantum observables localized in a wedge region. The results of this chapter are therefore not restricted to the realm of models with a factorizing S-matrix, but hold for a much larger class of theories.

In algebraic quantum field theory, wedge algebras are well-studied objects because of the distinguished geometric action of their modular operators with respect to the vacuum \cite{BiWi1, BiWi2, borchers-2d, 4maenner, MundBiWi}. Exploiting this important property, there have been several constructive proposals making use of wedge algebras. The work which is most closely related to our discussion is that of Borchers, who also considers the two-dimensional case in \cite{borchers-2d}, and constructs a theory in terms of a single wedge algebra. Wiesbrock proposes constructions on two- \cite{Wiesbrock1+1}, three- \cite{Wiesbrock2+1}, and four-dimensional \cite{Wiesbrock3+1} Minkowski space by considering a set of wedge algebras in appropriate modular positions relative to each other. Longo and Rehren also use a form of wedge algebras in their analysis of algebraic boundary conformal field theory \cite{ABC-QFT}.
\\
\\
In combination with a representation $U$ of the translations, a wedge algebra $\M$ can be used to define a local net in a straightforward manner. This is done in section \ref{sec:ch2los}, where we define a {\em standard right wedge algebra} as a von Neumann algebra which transforms appropriately under the representation $U$. However, due to the infinite extension of the wedges, the characterization of {\em local} observables in this theory becomes a nontrivial issue. For example in the works of Borchers \cite{borchers-2d} and of Wiesbrock \cite{Wiesbrock1+1, Wiesbrock2+1, Wiesbrock3+1}, the existence of local observables had to be postulated as an additional assumption. This unsatisfactory situation is here improved by identifying suitable conditions on the underlying wedge algebra $\M$ which imply the existence of local observables.

It was noticed in \cite{BuLe} that such a condition is the so-called {\em split property for wedges}, which in combination with Haag duality fixes the structure of the local algebras completely. The split property is discussed in section \ref{sec:spw}. Strengthening the results of \cite{BuLe}, we prove that it implies that the associated local net complies with all principles of algebraic quantum field theory. In particular, the Reeh-Schlieder property is shown to be a consequence of the split property for wedges. 

Closely related to the split property is the {\em modular nuclearity condition}, a condition on the modular objects of the wedge algebra $\M$. This condition, its relation to thermodynamical properties, and its advantages over the split property regarding applications in models are explained in section \ref{sec:mnc}. In Theorem \ref{thm:Ch2summary}, we summarize the main results of this chapter: Given a standard right wedge algebra which satisfies the modular nuclearity condition, an associated local net can be defined which matches all requirements of algebraic quantum field theory, and moreover has some additional properties.

In the last section, we address the question how standard right wedge algebras can be constructed explicitly, as a prerequisite and motivation for the discussion of models in the later chapters.
\\
\\
Parts of the results shown here have been published in the joint paper \cite{BuLe} with D. Buchholz.

\section{Nets of Wedge Algebras and Double Cone Algebras}\label{sec:ch2los}
In this section, we start our construction of quantum field theory models from an abstract algebra $\M$ which is interpreted as a ``wedge algebra'' with respect to a given representation $U$ of the translation group. To begin with, we collect and motivate our assumptions.

\subsection{Assumptions}

Motivated by the idea of inverse scattering theory, we begin by fixing the particle spectrum of the model to be constructed, which is a prerequisite for the formulation of an S-matrix. In relativistic quantum field theory, the choice of the particle spectrum amounts to the choice of an appropriate positive energy representation of the proper orthochronous Poincar\'e group $\PGpo$. According to the classical analysis of Wigner \cite{Wigner}, the decomposition of the one particle subrepresentation into irreducible representations of $\PGpo$ yields the mass and spin quantum numbers of the particles present in the theory.

For our construction, however, it suffices to consider a representation $U$ of the two-dimensional translation subgroup $(\Rl^2,+)\subset\PGpo$. We will see later that this representation can be extended to the proper Poincar\'e group with the help of the modular data of wedge algebras.

To be precise, we consider a strongly continuous, unitary representation $U$ of the two-dimensional translation group on a Hilbert space $\Hil$ as the starting point of our construction. As usual, the stability requirement of positive energy in all Lorentz frames is made, i.e. the joint spectrum of the generators $P=(P_0,P_1)$ of $U(x)=e^{iP\cdot x}$ is supposed to be contained in the closed forward light cone $V^+=\{p\in\Rl^2\,:\,p_0\geq |p_1|\}$. Whereas the choice of a particular representation space $\Hil$ is largely a matter of convenience, the choice of the (unitary equivalence class of the) representation amounts to fixing the mass spectrum of the theory as the eigenvalues of the associated mass operator $M=(P_0^2-P_1^2)^{1/2}$.

In the first steps of the construction, we allow for arbitrary mass spectra, but later on, we will have to restrict to the case of finitely many species of massive particles. In any case, we require the existence of a unique vacuum by postulating that $U$ contains the trivial representation as a one-dimensional subrepresentation. This corresponds to the existence of a vacuum vector $\Om\in\Hil$ which is invariant under the action of $U$, and which is characterized by this condition uniquely up to scalar multiples. Clearly, our Hilbert space should be large enough to accommodate also other physical states besides the vacuum, and so we also assume $\dim\Hil>1$.

Having set the stage for the construction of a model theory by introducing the Hilbert space $\Hil$ and the representation $U$, we now need an input for the formulation of the net of observable algebras. As explained in the previous section, we will generate this net from wedge-local operators, and so we consider a von Neumann algebra $\M\subset\B(\Hil)$, the selfadjoint elements of which shall be interpreted as observables measurable in a wedge. Fixing once and for all the right wedge $W_R$ \bref{def:WR} as our reference wedge region, we now collect assumptions on $\M$ which guarantee that $\M$ can be consistently viewed as an algebra of observables localized in $W_R$.

A typical and important feature in quantum field theory is the Reeh-Schlieder property of the vacuum. In the context of a local net $\A$ it can be formulated by requiring that the sets $\A(\OO)\Om$ are dense in $\Hil$ for any open region $\OO$, i.e. the vacuum vector is cyclic for the local algebras. So if the algebra $\M$ shall represent the observables localized in $W_R$, it must have $\Om$ as a cyclic vector. Moreover, thinking of $\M=\A(W_R)$ as the algebra of the right wedge in a local net $\A$, the Reeh-Schlieder property for the observables localized in the left wedge, $\A(W_L)\subset\A(W_R)'=\M'$, implies that $\Om$ must also be cyclic for the commutant $\M'$ of $\M$. This is equivalent to $\Om$ being separating for the algebra itself \cite{BraRob1}, and we thus assume that the vacuum vector is cyclic and separating for $\M$.

To characterize $\M$ not only as an observable algebra which is localized in {\em some} region of spacetime, but more precisely in the right wedge $W_R$, note that $W_R$ has the geometric property that it is mapped into itself by translations $t_x$ with $x\in\overline{W_R}$, i.e. $W_R+x\subset W_R$ for $x\in\overline{W_R}$. In view of the isotony and covariance properties a relativistic quantum field theory is bound to have, we are therefore lead to postulate $U(x)\M\,U(x)^{-1}\subset\M$ for any $x\in\overline{W_R}$.

This requirement completes our list of assumptions on $\Hil$, $U$ and $\M$, and we summarize them in the definition of a {\em standard right wedge algebra} $(\M,U,\Hil)$, which is the basic input for the construction carried out in this chapter.

\begin{definition}\label{def:srwa}
A standard right wedge algebra is a triple $(\M,U,\Hil)$ consisting of a Hilbert space $\Hil$ with $\dim\Hil>1$, a representation $U$ of the translation group $(\Rl^2,+)$ on $\Hil$, and a von Neumann algebra $\M\subset\B(\Hil)$ such that the following conditions are satisfied.
\begin{itemize}
\item[a)] $U$ is strongly continuous and unitary. The joint spectrum of the generators $P_0$, $P_1$ of $U(\Rl^2)$ is contained in the forward light cone $\{(p_0,p_1)\in\Rl^2\,:\,p_0\geq|p_1|\}$.\\
There is an up to a phase unique unit vector  $\Om\in\Hil$ which is invariant under the action of $U$.
\item[b)] $\Om$ is cyclic and separating for $\M$.
\item[c)] For each $x\in \overline{W_R}$, the adjoint action of $U(x)$ induces endomorphisms on $\M$,
\begin{align}\label{def:Mx}
	\M(x) := U(x)\M U(x)^{-1} \subset \M,\qquad x\in \overline{W_R}\,.
\end{align}
\end{itemize}
\end{definition}

\noindent Thinking about explicit realizations of this structure, note that the representation $U$ can be easily constructed as follows: Choose a spectrum of masses $m_1,...,m_N\geq 0$ and corresponding translation representations $U_{1,k}$ with mass $m_k$, $k=1,...,N$. For example, $U_{1,k}$ can be represented on the space $L^2(\Rl,d\mu_k)$ of momentum wavefunctions which are square integrable with respect to the standard measure $d\mu_k=(p^2+m_k^2)^{-1/2}\,dp$ as
\begin{align}
(U_{1,k}(x)f)(p)		&:=	e^{i((p^2+m_k^2)^{1/2}x_0-p\,x_1)}\cdot f(p)\,.
\end{align}
By a standard procedure, one then obtains the Bose Fock space $\Hil$ over $\bigoplus_{k=1}^N L^2(\Rl,d\mu_k)$, which is acted upon by the second quantization $U$ of $\bigoplus_{k=1}^N U_{1,k}$ and contains a unique invariant vacuum vector $\Om$.

The fact that this Hilbert space and this representation are shared by the corresponding free field theory does not exclude the possibility to have other nets on $\Hil$, which also transform covariantly under $U$ and exhibit nontrivial interaction. In fact, in an asymptotically complete theory one can use either one of the unitary M{\o}ller operators to represent the interacting net on the Hilbert space of incoming or outgoing collision states, which is just the Fock space over the one particle space of the theory \cite{araki}. So the assumption that $\Hil$ and $U$ are of the above described forms is no essential restriction.

In contrast to the construction of the representation $U$, the construction of an associated wedge algebra $\M$ is a much more difficult problem. This is only to be expected, since, as we shall see in the following, the choice of $\M$ essentially fixes the complete theory. In particular, the interaction is encoded in the algebra $\M$. In chapter \ref{chapter:wedgenet}, we will take an inverse scattering approach to construct a wedge algebra, and define $\M$ in terms of a prescribed factorizing S-matrix $S$. In this case, the interaction is fixed by the choice of $S$. The important question how wedge algebras can be constructed without relying on an explicitly known S-matrix is presently open and a subject of current research\footnote{To satisfy assumption c) of Definition \ref{def:srwa}, one could take an operator $X\in\B(\Hil)$ and consider the von Neumann algebra $\M_X:=\{U(x)XU(x)^{-1}\,:\,x\in\overline{W_R}\}''$, which clearly satisfies $U(x)\M_X U(x)^{-1}\subset\M_X$, $x\in\overline{W_R}$ \cite{buchholz-pc}. But it is not clear how to choose $X$ in order to ensure that $\Om$ is cyclic and separating for $\M_X$. 
}.

In the present chapter, we take the point of view that $\Hil$, $U$ and $\M$ are given such that all the assumptions of Definition \ref{def:srwa} are satisfied, and construct a local net $\OO\mapsto \A(\OO)$ with the help of the standard right wedge algebra $(\M,U,\Hil)$. The explicit realization of examples of standard right wedge algebras, and hence of associated model theories, will then be carried out in chapter \ref{chapter:wedgenet} by the above mentioned methods.
\\
\\
The construction we want to present proceeds in two steps by first defining a net $W\longmapsto\A(W)$ of wedge algebras and then constructing the local observables within this net. The second step requires to impose one more condition on the standard right wedge algebra $(\M,U,\Hil)$, but for the beginning, the assumptions summarized in Definition \ref{def:srwa} are sufficient.

\subsection{Definition of the Net of Observable Algebras}\label{sec:constructAWO}

The construction of a net of wedge algebras using a standard right wedge algebra is rather immediate \cite{borchers-2d, BuLe}. In fact, the structure is uniquely fixed if we require Haag duality for wedges, a property which is known to hold in any net generated by Wightman fields \cite{BiWi2}.

 The following definitions are indispensable if we want to end up with a Haag-dual net which transforms covariantly under the adjoint action of $U$: We put
\begin{subequations}\label{def:wedgenet}
\begin{align}
\A(W_R)   &:= \M\,,\qquad  &\A(W_R+x) &:= U(x)\M\,U(x)^{-1}\,,\\
\A(W_L)   &:= \M'\,,\qquad &\A(W_L+x) &:= U(x)\M' U(x)^{-1}\,.
\end{align}
\end{subequations}
In view of the simple structure of the set of wedges $\W$ in two dimensions \bref{eq:wedgeset}, these assignments completely determine a net of wedge algebras.
\begin{lemma}\label{Lem:reconstructA(W)}
Consider a standard right wedge algebra $(\M,U,\Hil)$, and define $\A(W)$, $W\in\W$, as in \bref{def:wedgenet}.
\\
Then $\W\ni W \longmapsto \A(W)$ is a Haag-dual net of von Neumann algebras which transforms covariantly under the adjoint action of $U$. Moreover, $\Om$ is cyclic and separating for each $\A(W)$, $W\in\W$.
\end{lemma}
\begin{proof}
The translation covariance of $W\longmapsto\A(W)$ follows directly from the definition \bref{def:wedgenet}, and since $(U(x)\M U(x)^{-1})'=U(x)\M' U(x)^{-1}$, the duality $\A(W_R')=\A(W_R)'$ transports to all pairs $W, W'$ of wedges.

To show that $W\longmapsto\A(W)$ is isotonous, we consider an inclusion of wedges of the form $W_R+x\subset W_R$, $x\in \overline{W_R}$. In this situation, the corresponding algebras are included in each other,
\begin{align}\label{eq:inclusion-of-wedges-L}
\A(W_R+x)&=\M(x)\subset\M=\A(W_R),
\end{align}
in view of the assumption in Definition \ref{def:srwa} c). Taking commutants gives $\A(W_L+x)\supset\A(W_L)$, $x\in \overline{W_R}$. But any inclusion $W_1\subset W_2$ of wedges arises from one of these two cases by a translation, and therefore the isotony of $W\lmto\A(W)$ follows by covariance.

As a consequence of Definition \ref{def:srwa} b), $\Om$ is cyclic and separating for $\A(W_R)=\M$ and $\A(W_L)=\M'$. Taking into account the translation invariance of $\Om$, we see that this vector is cyclic and separating for each wedge algebra $\A(W)$, $W\in\W$, and the proof is finished.
\end{proof}

\noindent It has been discovered by Borchers \cite{borchers-2d} that in the present situation, the representation $U$ can be extended to an (anti-) unitary representation of the proper Poincar\'e group $\PG_+$. This can be achieved as follows: As $\Om$ is cyclic and separating for $\M$, the modular theory of Tomita and Takesaki \cite{KadRin2} is applicable and we may define the modular unitaries $\Delta^{it}$, $t\in\Rl$, and the modular conjugation $J$ of the pair $(\M,\Om)$. It has been shown in \cite{borchers-2d} (see \cite{florig} for a simplified proof) that as a consequence of the spectrum condition for $U$, the following commutation relations hold, $x\in\Rl^2$.
\begin{subequations}\label{borchers-CR}
\begin{align}
J\,U(x)J &= U(-x)\\
\Delta^{it}U(x)\Delta^{-it} &= U(\Lambda(-2\pi t)x)
\end{align}
\end{subequations}
Here $\Lambda(-2\pi t)$ is the boost transformation with rapidity parameter $-2\pi t$, defined in \bref{def:boost}. The equations \bref{borchers-CR} imply that a proper Poincar\'e transformation consisting of a boost $\Lambda(\la)$, a space-time reflection $(-1)^\eps$, $\eps=\pm 1$, and a subsequent translation $t_x$ along $x\in\Rl^2$, can be represented by the operator
\begin{align}\label{def:Uext}
	U(t_x(-1)^\eps\Lambda(\la)) &:= U(x)J^\eps \Delta^{-i\la/2\pi}\,.
\end{align}
This definition gives rise to an (anti-) unitary, strongly continuous representation of the proper Poincar\'e group $\PG_+$ on $\Hil$ under which $\Om$ is invariant, and under which the net $W\lmto\A(W)$ defined above transforms covariantly (see \cite{borchers-2d} or Proposition \ref{Lem:A(O)} below).

We adopt the convention to denote this representation by the same symbol $U$ as the representation of the translation subgroup, and introduce the notation
\begin{align}
U(x,\la)	:= 	U(t_x\La(\la))
\end{align}
for the transformations in the identity component $\PGpo$ of $\PG_+$. For simplicity, the notation $U(x)=U(x,0)$ for pure translations is also maintained.
\\
\\
According to Lemma \ref{Lem:reconstructA(W)} and Borchers' commutation relations, a net of wedge-local observable algebras which has the Reeh-Schlieder property and is moreover covariant under a representation of the proper Poincar\'e group, can be readily constructed from a standard right wedge algebra. It is not surprising that the construction of such a wedge net poses no difficulties since all relevant structures, like the algebra $\M$ and the translations $U(x)$, were part of our "input" $(\M,U,\Hil)$. However, the most important constituents of a relativistic theory are its {\it local} observables, which do not appear in our assumptions and therefore have to be constructed.

To fix ideas, we consider a double cone $\OO$ and look for observables $A$ localized in $\OO$. As pointed out in section \ref{Sec:Geo}, the causal complement of $\OO$ has two disconnected components, consisting of the right wedge $W_R^\OO$ and the left wedge $W_L^\OO$ (cf. figure \ref{fig:doublecone}, page \pageref{fig:doublecone}).

If $A\in\B(\Hil)$ is an operator corresponding to a measurement in $\OO$, then the principle of locality demands that $A$ is not influenced by operations in $\OO'=W_R^\OO \cup W_L^\OO$, and hence $A$ has to commute with all operators contained in $\A(W_R^\OO)$ or $\A(W_L^\OO)$. Thus it must be an element of the von Neumann algebra
\begin{align}\label{def:A(O)}
\A(\OO)	:= \left( \A(W_R^\OO) \vee \A(W_L^\OO) \right)'
		= \A(W_R^\OO)' \cap \A(W_L^\OO)'\,,
\end{align}
where the symbol $\A(W_R^\OO) \vee \A(W_L^\OO)$ denotes the von Neumann algebra generated by $\A(W_R^\OO)$ and $\A(W_L^\OO)$.

The assignment \bref{def:A(O)} is the maximal possible choice of $\A(\OO)$ compatible with locality, and will here be used as the definition of the observable algebras associated to double cones. By additivity, this definition extends to arbitrary bounded open regions $\Q\subset\Rl^2$, i.e. we put
\begin{align}\label{def:AQ}
	\A(\Q)	&:=	\bigvee_{\OO\in\OOO \atop \OO\subset\Q} \A(\OO)\;,
\end{align}
where $\OOO$ denotes the set of all double cones in $\Rl^2$ (see section \ref{Sec:Geo}).
\\
\\
The basic properties of these local algebras are specified in the following Proposition (see also \cite{borchers-2d}).
\begin{proposition}\label{Lem:A(O)} 
Let $(\M,U,\Hil)$ be as before and consider the algebras $\A(\OO)$ defined in (\ref{def:A(O)}, \ref{def:AQ}).
\\
The assignment $\OO\longmapsto\A(\OO)$ is a local net of von Neumann algebras transforming covariantly under the adjoint action of the representation $U$ \bref{def:Uext} of the proper Poincar\'e group.
\end{proposition}
\begin{proof}
We use the notation introduced in section \ref{Sec:Geo}. Considering an inclusion $\OO_1\subset\OO_2$ of double cones, we have $W_R^{\OO_1} \supset W_R^{\OO_2}$ and $W_L^{\OO_1} \supset W_L^{\OO_2}$ and hence
\begin{align*}
	\A(\OO_1) 	&= \A(W_R^{\OO_1})' \cap \A(W_L^{\OO_1})' 
			\subset \A(W_R^{\OO_2})' \cap \A(W_L^{\OO_2})'
			= \A(\OO_2)\,.
\end{align*}
So the definition \bref{def:AQ} implies that isotony holds, i.e. $\OO\longmapsto\A(\OO)$ is a {\em net} of von Neumann algebras.

To show locality, consider two spacelike separated double cones $\OO_1\subset\OO_2'$. In this situation, either $\OO_1\subset W_R^{\OO_2}$ and $\A(\OO_1)\subset \A(W_R^{\OO_2})$, or $\OO_1\subset W_L^{\OO_2}$ and $\A(\OO_1)\subset \A(W_L^{\OO_2})$. As $\A(\OO_2)$ is contained in the commutants $\A(W_R^{\OO_2})'$ and $\A(W_L^{\OO_2})'$, the algebras $\A(\OO_1)$ and $\A(\OO_2)$ commute with each other in both cases. For two arbitrary spacelike separated bounded regions $\Q_1\subset\Q_2'$, the same argument can be applied to the double cone algebras generating $\A(\Q_1), \A(\Q_2)$, leading also to the conclusion $\A(\Q_1)\subset\A(\Q_2)'$.

For the covariance of the net, we first consider the action of $U$ \bref{def:Uext} on wedge algebras. As the modular group induces automorphisms on $\A(W_R)$, the right wedge $W_R$ is invariant under the boosts $\La(t)$, and $J$ maps $\A(W_R)$ onto $\A(W_L)=\A(-W_R)$, we have
\begin{align*}
U(x,\la)\A(W_R) U(x,\la)^{-1}
&=
U(x)\Delta^{-i\la/2\pi}\A(W_R)\Delta^{i\la/2\pi}U(x)^{-1}
\\
&=
U(x)\A(W_R)U(x)^{-1}\\
&=
\A(W_R+x)
=
\A(t_x\La(\la)W_R)\,,\\
U(-1) \A(W_R) U(-1)^{-1}
&=
J \A(W_R) J
=
\A(W_R)'
=
\A(W_L)
=
\A(-W_R)
\,.
\end{align*}
Since each wedge is a Poincar\'e transform of $W_R$, this implies
\begin{align}
	U(g)\A(W)U(g)^{-1} &= \A(g\,W),\qquad g \in\PGpro,\;\,\;W\in\W\,.
\end{align}
The covariance of the local algebras is then an immediate consequence of their definition \bref{def:A(O)}. For arbitrary $g\in\PGpro$, $\OO\in\OOO$ we have
\begin{align*}
	U(g)\A(\OO)U(g)^{-1} 	&= U(g)\A(W_R^\OO)'U(g)^{-1} \cap U(g)\A(W_L^\OO)'U(g)^{-1}\\
					&= \A(g\,W_R^\OO)' \cap \A(g\,W_L^\OO)'\\
					&= \A(g\,\OO)\,,
\end{align*}
and by \bref{def:AQ}, this covariance property carries over to arbitrary regions $\Q\subset\Rl^2$.
\end{proof}

\noindent Lemma \ref{Lem:A(O)} shows that many important features of the net $\OO\longmapsto\A(\OO)$, in particular its locality, are inherited from the corresponding properties of the wedge algebras. Starting from a standard right wedge algebra, we thus obtain an associated local net, which is moreover uniquely fixed by $(\M,U,\Hil)$ if all observable algebras are chosen maximally.
\\
\\
However, the local algebras are defined in a rather indirect way as intersections of wedge algebras, and it is therefore a nontrivial issue to analyze their properties in more detail. For example, it is in general difficult to characterize the "size" of these algebras, and in particular, it is not clear if the Reeh-Schlieder property holds locally, i.e. if the vacuum vector $\Om$ is cyclic for $\A(\OO)$ if $\OO$ is bounded.

For the wedge algebras, the cyclicity of the vacuum follows from the assumption that $\Om$ is cyclic and separating for the initial algebra $\M$, and they must therefore be "big" in a certain sense. This notion of size can be made precise by studying the algebraic structure of $\M$, which turns out to be severely restricted if $\M$ is assumed to be part of a standard right wedge algebra $(\M,U,\Hil)$ in the sense of Definition \ref{def:srwa}.

The following theorem has first been shown by Driessler \cite{Driessler:1975cm} (see also the work of Longo, \cite[Thm. 3]{Lo79}).
\begin{theorem}\label{Thm:LongoType3} {\bf \cite{Driessler:1975cm,Lo79}}\\
Consider a standard right wedge algebra $(\M,U,\Hil)$. Then $\M$ is a type III$_1$ factor according to the classification of Connes.
\end{theorem}

\noindent Note that the trivial case $\M=\Cl\cdot 1$, which was found as a possibility in \cite{Lo79}, is here excluded by the assumption $\dim\Hil>1$ and the cyclicity of $\Om$ for $\M$. As all wedge algebras are (anti-) isomorphic to $\M$ by definition, Theorem \ref{Thm:LongoType3} implies that $\A(W)$ is a type III$_1$ factor for each wedge $W\in\W$, which is the typical situation in quantum field theory. In particular, $\M$ is purely infinite as a von Neumann algebra, and the underlying Hilbert space $\Hil$ must be infinite dimensional.

In comparison to this quite detailed information we have on the wedge algebras, little is known about the properties of their intersections \bref{def:A(O)}. In fact, even the question whether these algebras are nontrivial in the sense that they contain any operators apart from multiples of the identity has not been settled in general.
But the extreme case of a net with $\A(\OO) = \Cl\cdot 1$ for all bounded regions $\OO$ describes a theory without any local observables, which clearly has to be considered pathological from a physical point of view. We therefore need to impose additional conditions on the underlying standard right wedge algebra $(\M,U,\Hil)$ in order to obtain a physically reasonable net, the minimal demand being that the associated net contains local observables in regions of macroscopic extent.

\section{The Split Property for Wedges and its Consequences}\label{sec:spw}

In our setup, the algebra of observables localized in the double cone $\OO=(W_L^\OO)'\cap (W_R^\OO)'$ is defined as $\A(\OO)=\A(W_L^\OO)' \cap \A(W_R^\OO)'$, which is just the relative commutant of $\A(W_R^\OO)$ in $\A(W_L^\OO)'$. A general analysis of the structure of relative commutants of type III$_1$ factors would be favorable for the investigation of the local algebras, but does not yet exist in the literature. However, there does exist a distinguished case in which the algebraic structure of the relative commutant can be directly read off from the properties of the algebras $\A(W_L^\OO)$ and $\A(W_R^\OO)$. This is the case of a {\em split inclusion} $\A(W_R^\OO)\subset\A(W_L^\OO)'$, with which we will be concerned in the following. We begin by recalling the relevant definitions.
\begin{definition}\label{def:split}{\bf \cite{DoplicherLongo:StandardSplit}}\\
\vspace*{-11mm}
\\
	\begin{enumerate}
		\item An inclusion $\M_1\subset\M_2$ of two von Neumann algebras $\M_1$, $\M_2$ is called split if there exists a type I factor $\NN$ such that
		\begin{align}
			\M_1 \subset \NN \subset \M_2\,.
		\end{align}
 		\item An inclusion $\M_1\subset\M_2$ of two von Neumann algebras $\M_1,\M_2$
     		acting on a Hilbert space $\Hil$ is called standard if there exists a vector which is cyclic and separating for $\M_1$, $\M_2$, and the relative commutant $\M_1'\cap\M_2$.
		\item A local net $\A$ is said to have the split property for wedges if for each inclusion $W_1\Subset W_2$ of wedges $W_1,W_2\in\W$, the corresponding inclusion of wedge algebras is split \cite{Mue-SPW}.
	\end{enumerate}
\end{definition}

\noindent The split property of an inclusion $\M_1\subset\M_2$ of von Neumann algebras amounts to a form of statistical independence between $\M_1$ and $\M_2'$, and is sometimes also called "$W^*$-independence in the spatial product sense" \cite{Summers:1990tp}. This latter terminology is motivated by the fact that there exist a number of conditions that, when added to the split property of $\M_1\subset\M_2$, imply that $\M_1\vee\M_2'$ is naturally spatially isomorphic to the tensor product algebra $\M_1\otimes\M_2'$ (see, for example, the review article \cite[Thm. 3.9]{Summers:1990tp}). 
In the case of a standard inclusion, the following result holds.
\begin{lemma}\label{Lem:Split} {\bf \cite{D'AntoniLongo:Flip, DoplicherLongo:StandardSplit}}\\
  Let $\M_1\subset\M_2$ be a standard inclusion of von Neumann factors, acting on a Hilbert space $\Hil$. The following two
  statements are equivalent:
  \begin{enumerate}
  \item The inclusion $\M_1\subset\M_2$ is split.
  \item There exists a unitary $V$ mapping $\Hil$ onto $\Hil\otimes\Hil$
    such that
    \begin{eqnarray}\label{split-tensor}
      VM_1M_2'V^* &=& M_1\otimes M_2',\qquad\qquad M_1\in\M_1,\;\; M_2'\in\M_2'\;.
    \end{eqnarray}
  \end{enumerate}
\end{lemma}

\noindent The significance of the split property for our construction lies in the simplifying influence it has on the structure of the local algebras: Consider an inclusion $W_1\Subset W_2$ of wedges and the associated double cone algebra $\A(W_1\cap W_2)=\A(W_1)\cap\A(W_2)$. If the inclusion $\A(W_1)\subset\A(W_2)$ is split, the above result can be applied to realize $\A(W_1\cap W_2)$ as a tensor product of two wedge algebras on $\Hil\otimes\Hil$. In particular, the nontriviality $\A(W_1\cap W_2)\neq \Cl\cdot 1$ then follows.

This mechanism, to be explained in more detail in Proposition \ref{Prop:type31} below, was first observed by Schroer and Wiesbrock in \cite{Schroer-Wiesbrock}. However, the assumption of the split property for wedges as a tool to ensure the nontriviality of the local algebras was then discarded by these authors, probably because it was not clear how to establish the existence of interpolating type I factors for inclusions of wedge algebras. This issue will be discussed subsequently in section \ref{sec:mnc}. Irrespectively of the question how to check the split property in concrete applications, we now discuss whether it can possibly hold for inclusions of wedge algebras.

As mentioned above, the split property of $\M_1\subset\M_2$ amounts to a form of statistical independence between the subsystems described by the algebras $\M_1$ and $\M_2'$ of the larger system identified with $\M_1\vee\M_2'$. Namely, it implies that for any pair of normal states $\varphi_1$ on $\M_1$ and $\varphi_2$ on $\M_2'$, there exists a normal state $\varphi$ on $\M_1\vee\M_2'$ such that $\varphi|_{\M_1}=\varphi_1$, $\varphi|_{\M_2'}=\varphi_2$, expressing the fact that states in the subsystems $\M_1$ and $\M_2'$ can be prepared independently of each other. Moreover, $\varphi$ can be chosen in such a way that there are no correlations between "measurements" in $\M_1$ and $\M_2'$, i.e. as a product state
\begin{align*}
\varphi(M_1 M_2')	&=	\varphi_1(M_1)\cdot \varphi_2(M_2')\,,\qquad M_1\in\M_1\,,\;M_2'\in\M_2'\,.
\end{align*}
Taking $\M_1=\A(\OO_1)$ and $\M_2'=\A(\OO_2)$ as the observable algebras of two spacelike separated regions $\OO_1\subset\OO_2'$ in a quantum field theory given by a net $\A$, some form of statistical independence between $\M_1$ and $\M_2'$ can be expected on physical grounds. For the massive free field, the existence of normal product states for such pairs of local algebras was shown by Buchholz \cite{Buchholz:1973bk}. A corresponding analysis for algebras of free Fermi fields, and for the Yukawa$_2+P(\varphi)_2$ model has been carried by Summers \cite{Summers-split}.

Examples of theories violating the split property can be obtained by considering models with a non-compact global symmetry group or certain models with infinitely many different species of particles \cite{DoplicherLongo:StandardSplit}. Such theories have an immense number of local degrees of freedom, and according to the analysis in \cite{BuWi}, it is precisely this feature which is responsible for the breakdown of the split property.

So we may take the point of view that the split property is a reasonable assumption for inclusions of {\em local} algebras in theories which satisfy some rough bound on the number of their local degrees of freedom, such as theories of finitely many species of scalar particles. However, some care is needed when dealing with unbounded regions like wedges, even in such theories. In fact, there is an argument by Araki \cite[p. 292]{Buchholz:1973bk} to the effect that inclusions of wedge algebras cannot be split if the spacetime dimension is larger than two. Araki's argument exploits the translation invariance of wedges along their edges and does not apply in two dimensions, where these edges are zero-dimensional points. The split property for wedges is known to hold in the theory of a free, scalar, massive field \cite{Mue-SPW, BuLe}. It is, however, not fulfilled for arbitrary mass spectra. For example, the split property for wedges does not hold in massless theories, and is also violated in the model of a generalized free field with continuous mass spectrum \cite{DoplicherLongo:StandardSplit}. But for models describing finitely many species of massive particles, there is no a priori reason for the split property for wedges not to hold.
\\
\\
We will therefore take the split property for wedges as a tentative requirement on the net we constructed, and now investigate its many strong implications.

Its first consequence is stated in the following proposition, which is taken from \cite{BuLe}. 
\begin{proposition}\label{Prop:type31}{\bf \cite{BuLe}}\\
Consider a standard right wedge algebra $(\M,U,\Hil)$ which has the property that the inclusions $\M(x):=U(x)\M U(x)^{-1} \subset \M$ are split, $x\in W_R$.
\\
Then
\begin{enumerate}
	\item $\Hil$ is separable.
	\item $\M$ is isomorphic to the unique hyperfinite type III$_1$ factor.
	\item $\M(x)\subset\M$ is a standard inclusion, $x\in W_R$.
	\item The relative commutant $\M(x)'\cap\M$ is isomorphic to the unique hyperfinite type III$_1$ factor, $x\in W_R$. In particular, this algebra has cyclic vectors and is thus nontrivial.
\end{enumerate}
\end{proposition}
\begin{proof}
Let $\NN_x$ denote the type I factor interpolating between $\M(x)$ and $\M$. As $\Om$ is cyclic and separating for $\M$ and $\M(x)$, the same holds for $\NN_x$. It follows that $\NN_x$, being of type I, is separable in the ultraweak topology and consequently $\Hil$ is separable \cite[Prop. 1.2]{DoplicherLongo:StandardSplit}.

Now, as $U$ is continuous, $\M$ is continuous from the inside in the sense that $\M = \bigvee_{x\in W_R} \M(x)$. The split property thus implies that $\M$ can be approximated from the inside by the separable type I factors $\NN_x$ and is therefore hyperfinite \cite[Prop. 3.1]{Buchholz:1986bg}. As the type III$_1$ factor property also holds (Thm. \ref{Thm:LongoType3}), the claims a) and b) follow. It has been shown in \cite{Hag} that the hyperfinite type III$_1$ factor is unique.

To prove c), recall that on a separable Hilbert space, any factor of type III has cyclic and separating vectors \cite[Cor.2.9.28]{sakai}. Moreover, for any von Neumann algebra on $\Hil$ with a cyclic and separating vector, there exists a dense $G_\delta$ set of vectors which are both, cyclic and separating \cite{DixMar}. Now, taking into account that $\NN_x$ is isomorphic to $\B(\Hil)$ because it is a type I$_\infty$ factor, the relative commutant $\M(x)'\cap\NN_x$ of the type III factor $\M(x)$ in $\NN_x$ is (anti-) isomorphic to $\M(x)$ by modular theory. It is therefore of type III and has cyclic vectors in $\Hil$. Clearly, the latter property also holds for $\M(x)'\cap\M\supset\M(x)'\cap\NN_x$ and, as $\Om$ separates $\M$, it follows that there exists a dense $G_\delta$ set of cyclic and separating vectors for $\M(x)'\cap\M$. But the intersection of a finite number of dense $G_\delta$ sets is non-empty. So we conclude that the triple $\M$, $\M(x)$ and $\M(x)'\cap\M$ has a joint cyclic and separating vector, i.e. $\M(x)\subset\M$ is a standard split inclusion.

d) Having shown that $\M(x)\subset\M$ is standard, we can apply Lemma \ref{Lem:Split}. Hence there exists a unitary $V:\Hil\to\Hil\otimes\Hil$ implementing the isomorphism $\M(x)\vee\M' \cong \M(x)\otimes\M'$, and taking commutants yields
\begin{align}\label{tensor-iso}
 	\M(x)' \cap \M &\cong \M(x)' \otimes \M \,.
 \end{align}
To prove that $\M(x)'\cap\M$ is also isomorphic to the hyperfinite type III$_1$ factor, first note that this algebra is hyperfinite since $\M(x)'$ and $\M$ are. Secondly, the isomorphism $\M(x)'\cong\M$ implies $\M(x)'\cap\M\cong\M\otimes\M$.  But $(\M\otimes\M,U\otimes U,\Hil\otimes\Hil)$ is a standard right wedge algebra in the sense of Definition \ref{def:srwa}, with invariant vacuum vector $\Om\otimes\Om$. Thus we can apply Driessler's theorem (Thm. \ref{Thm:LongoType3}) once more, and conclude that $\M\otimes\M$, and hence $\M(x)'\cap\M$, is a factor of type III$_1$.
\end{proof}

Note that the assumption that $\M(x)\subset\M$ is split for all $x\in W_R$ implies that in the net $W\lmto\A(W)$ \bref{def:wedgenet}, all wedge inclusions $\A(W_1)\subset\A(W_2)$, $W_1\Subset W_2$, are standard split inclusions, since an arbitrary inclusion of wedge algebras can be transformed to a special inclusion of the form
\begin{align}
\A(W_R+x)=\M(x)\subset\M=\A(W_R)
\end{align}
by using the translation and TCP covariance of the net $\A$.

Proposition \ref{Prop:type31} implies that all double cone algebras \bref{def:A(O)} are isomorphic to the hyperfinite type III$_1$ factor, provided the split property for wedges holds. In particular, any local algebra $\A(\OO)$ associated to a non-empty open bounded region $\OO$ has cyclic vectors in this case and therefore contains many nontrivial observables. These results make the split property for wedges a very desirable feature for our construction, for it not only excludes the pathological case $\A(\OO)=\Cl\cdot 1$, corresponding to a theory without local observables, but also establishes the algebraic properties for $\A(\OO)$ which are expected to hold generally in quantum field theory \cite{Buchholz:1986bg}.
\\
\\
Having derived the existence of local observables from the assumption of the split property for wedges, we move on to the question whether the Reeh-Schlieder property holds locally, as the last axiom of algebraic quantum field theory to be checked for the constructed net.

The proof of this property amounts to showing that $\Om$ is a cyclic vector for any double cone algebra $\A(\OO)$. Although it is known that under the assumptions made above the local algebras have a dense set of vectors which are cyclic (and separating) \cite{DixMar}, it is not evident a priori that the vacuum vector $\Om$ belongs to this set. Nevertheless, it is possible to derive also the Reeh-Schlieder property in the present context, as will be done in the remainder of this section.

In order to transport the Reeh-Schlieder property from the wedges to the double cones, we need to establish a tight connection between the corresponding algebras $\A(W)$ \bref{def:wedgenet} and $\A(\OO)$ \bref{def:A(O)}. Such a connection is demonstrated in the following Lemma.
\begin{lemma}\label{lemma-loc}
  Assume that the net $\A$ \bref{def:wedgenet} has the split property for wedges. Then the double cone algebras generate the wedge algebras,
  \begin{align}\label{OW-additivity}
	\bigvee_{\OO\subset W \atop \OO\in\OOO}\A(\OO) = \A(W)\;,\qquad W\in\W.
  \end{align}
 \end{lemma}
\begin{proof}
By covariance, it suffices to consider the right wedge $W_R$. We choose an increasing sequence of double cones $\OO_1\subset\OO_2\subset\OO_3\subset...\subset W_R$ such that the left vertex of each $\OO_n$ lies in the origin of $\Rl^2$ and $W_R$ is exhausted by this sequence, i.e. for each double cone $\OO\subset W_R$ there exists an integer $n\in\N$ with $\OO\subset\OO_n$ (cf. figure \ref{fig:o-in-w}).
\begin{figure}[here]
    \noindent
    \psfrag{O0}{$\OO_1$}
    \psfrag{O1}{$\OO_2$}
    \psfrag{O2}{$\OO_3$}
    \psfrag{W}{$W_R$}
    \psfrag{w2}{$W_R^3$}
    \centering\epsfig{file=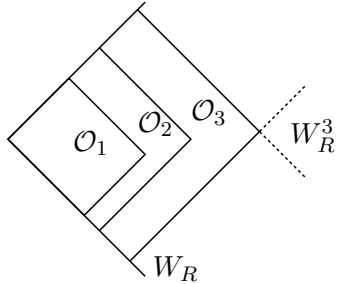,width=4cm}
    \caption{Generating the right wedge from a flag $\OO_1 \subset \OO_2
      \subset \OO_3 \subset \ldots \subset W_R$ of double cones with
      common left vertex.}
\label{fig:o-in-w}
\end{figure}
\\
As a consequence of isotony, we then have
\begin{align}
  \bigvee_{\OO\subset W_R \atop \OO\in\OOO}\A(\OO) = \bigvee_{n\in\mathbb{N}} \A(\OO_n)  =:  \A_{\rm loc}(W_R)
    \;.
\end{align}
We denote the right part of the causal complement of $\OO_n$ by $W_R^n$, i.e. $\OO_n'=W_L \cup W_R^n$. in view of the standard split property of $\A(W_R^1)\subset\A(W_R)$, there exists a unitary $V_1:\Hil\lto\Hil\otimes\Hil$ such that
\begin{align}\label{lem-iso}
    V_1 \A(\OO_1)' V_1^* &= V_1 \left(\A(W_L) \vee \A(W_R^1) \right) V_1^* =  \A(W_L) \otimes \A(W_R^1)\,, \\
    V_1 AB V_1^* &=  A \otimes B\,,
    \qquad        A\in\A(W_L),\;B\in\A(W_R^1)\,.
\end{align}
  As $\A(W_R^n) \subset \A(W_R^1)$, this isomorphism restricts to
  $V_1\A(\OO_n)'V_1^* = \A(W_L) \otimes \A(W_R^n)$, $n\in\N$. We thus obtain
\begin{align}
    \A_{\rm loc}(W_R)'
    &=
    \bigg(\bigvee_{n\in\mathbb{N}} \A(\OO_n)\bigg)'
    =
    \bigcap_{n\in\mathbb{N}} \A(\OO_n)'
    =
    V_1^* \bigcap_{n\in\mathbb{N}} \Big( \A(W_L) \otimes \A(W_R^n)\Big)\;V_1\nonumber
	\\
   &= V_1^* \Big( \A(W_L) \otimes \bigcap_{n\in\mathbb{N}} \A(W_R^n)\Big)\;V_1\,.\label{eq:tensorintersection}
\end{align}
In the last line, we used the commutation theorem for tensor products of von Neumann algebras (or rather, a consequence thereof, see \cite[Cor. IV.5.10]{Takesaki1}). Below we will show that the intersection appearing in \bref{eq:tensorintersection} is trivial, $\bigcap_n\A(W_R^n)=\Cl\cdot 1$. Anticipating this result for a moment, and noting that the isomorphism \bref{lem-iso} restricts to $V_1\A(W_L)V_1^*=\A(W_L)\otimes 1$, we have
\begin{align}
	\A_{\rm loc}(W_R)' &= V_1^* \big( \A(W_L) \otimes 1\big)\;V_1 = \A(W_L)\,.
\end{align}
Taking commutants then yields the claim $\A_{\rm loc}(W_R)=\A(W_L)'=\A(W_R)$.

So it remains to show that $\A_\infty := \bigcap_n\A(W_R^n)$ is trivial. To this end, we consider the commutant $\A_\infty'$ of this algebra and note that by construction, $\A_\infty'$ is stable under translations, $U(x)\A_\infty'U(x)^{-1} \subset \A_\infty'$, $x\in\Rl^2$. Furthermore, $\Om$ is cyclic for $\A_\infty'$ since $\A_\infty'\supset \A(W_L)$. Since the vacuum is (up to multiples) the only translation invariant vector and the spectrum condition holds, it follows by standard arguments (cf., for example, \cite{haag}) that $\A_\infty'=\B(\Hil)$, i.e. $\A_\infty=\Cl\cdot 1$. (see also \cite[Thm. 2.6.]{borchers-halfsided}.)

This completes the proof.
\end{proof}

In order to establish the Reeh-Schlieder property, we need another additivity result, which is due to M. M\"uger.

\begin{figure}[here]
  \noindent
  \psfrag{o1}{$\OO_1$}
  \psfrag{o2}{$\OO_2$}
  \psfrag{o}{$\hat{\OO}$}
  \psfrag{wr2}{$W_R^1$}
  \psfrag{wrr2}{$W_R^2$}
  \psfrag{wl1}{$W_L^2$}
  \psfrag{wll1}{$W_L^1$}
  \centering\epsfig{file=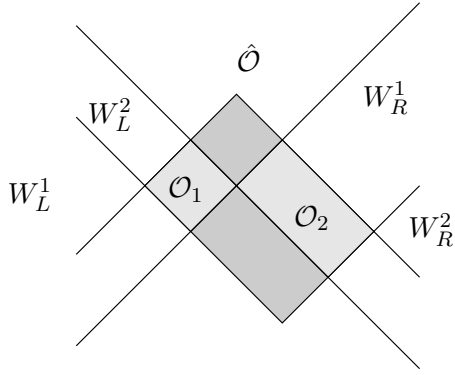,width=5.7cm}
  \caption{Adjacent double cones.}
\label{fig:adjDC}
\end{figure}

\begin{lemma}\label{lemma-mue}{\bf \cite{Mue-SPW}}\\
Let $\OO_1$, $\OO_2$ be two adjacent double cones and $\hat{\OO}=(\OO_1\cup\OO_2)''$ the double cone generated by
them (see figure \ref{fig:adjDC}).
\\
If the net $\A$ \bref{def:wedgenet} has the split property for wedges, there holds additivity for double cones in the form $\A(\OO_1)\vee\A(\OO_2)=\A(\hat{\OO})$.
\end{lemma}
\noindent As this Lemma is essential for our investigations, we will repeat M\"uger's proof here \cite{Mue-SPW, Mue-PhD}. We use the notation indicated in figure \ref{fig:adjDC}, namely we denote the left and right parts of the causal complement of the double cone $\OO_j$ by $W_L^j$ and $W_R^j$, $j=1,2$, respectively. With this notation, $\OO_j'=W_L^j \cup W_R^j$ and $(W_R^1)'=W_L^2$.
\begin{proof}
As all wedge inclusions are standard split inclusions, there exist unitaries $V_1$, $V_2:\Hil\lto\Hil\otimes\Hil$ such that
\begin{align}
  V_1\A(\OO_1)'V_1^* &= V_1 \Big( \A(W_L^1) \vee\A(W_R^1) \Big) V_1^* =  \A(W_L^1) \otimes \A(W_R^1)\,,
  \\
  V_2\A(\OO_2)'V_2^* &= V_2 \Big( \A(W_L^2) \vee\A(W_R^2) \Big) V_2^*= \A(W_L^2) \otimes \A(W_R^2)\,,
\end{align}
and the adjoint action of $V_1,V_2$ is natural in the sense of \bref{split-tensor}.
First note that by restriction we obtain
\begin{align*}
  V_2\A(W_L^2)V_2^*  = \A(W_L^2)\otimes 1,
  \qquad
  V_2\A(W_R^2)V_2^*  = 1\otimes\A(W_R^2)\,,
\end{align*}
and, by taking commutants,
\begin{align*}
V_2\A(\OO_2)V_2^*	&=	\A(W_L^2)'\otimes\A(W_R^2)'\,.
\end{align*}
Hence, using the commutation theorem for tensor products \cite{Takesaki1} and factoriality of the wedge algebras,
\begin{align*}
  V_2\left(\A(W_L^2)\vee\A(\OO_2)\right)V_2^*
  &=
  \big(\A(W_L^2)\otimes 1\big) \vee \big(\A(W_L^2)' \otimes \A(W_R^2)'\big)
  \\
  &=
  \B(\Hil) \otimes \A(W_R^2)'
  \\
  &=
  \left( 1 \otimes \A(W_R^2)\right)'
  \\
  &=
  V_2 \A(W_R^2)' V_2^*\,.
\end{align*}
Thus $\A(W_L^2)\vee\A(\OO_2)=\A(W_R^2)'$. On the other hand, \bref{split-tensor} implies that under the isomorphism implemented by $V_1$ we have
\begin{align*}
  V_1\A(\OO_2)V_1^*      = 1 \otimes \A(\OO_2), \qquad
  V_1\A(\hat{\OO})V_1^* = \A(W_L^1)' \otimes \A(W_R^2)'\,,
\end{align*}
because both, $\A(\OO_2)$ and $\A(W_R^2)$, are subalgebras of $\A(W_R^1)$. Together with the previously obtained equation $\A(W_L^2)\vee\A(\OO_2)=\A(W_R^2)'$ and the wedge duality $\A(W_R^1)'=\A(W_L^2)$, we arrive at
\begin{align*}
  V_1\big( \A(\OO_1) \vee \A(\OO_2) \big) V_1^*
  &=
  \big(\A(W_L^1)'\otimes\A(W_R^1)'\big) \vee \big(1\otimes\A(\OO_2)\big)
  \\
  &=
  \A(W_L^1)' \otimes \big(\A(W_L^2)\vee \A(\OO_2)\big)\\
  &=
  \A(W_L^1)' \otimes \A(W_R^2)'\\
  &=
  V_1 \A(\hat{\OO}) V_1^*\,.
\end{align*}
From this we conclude $\A(\OO_1) \vee \A(\OO_2)=\A(\hat{\OO})$, which is the claim of the Lemma.
\end{proof}

\noindent Next we show that Lemma \ref{lemma-loc} and Lemma \ref{lemma-mue} imply the Reeh-Schlieder property for double cones by a standard argument (cf., for example, \cite{BoBu-deSitter}).

\begin{proposition}\label{Prop:local-cyclicity}
Consider a standard right wedge algebra $(M,U,\Hil)$ and assume that $\M(x)\subset\M$ is split, $x\in W_R$.
\\
Then the vacuum vector $\Om$ is cyclic and separating for any algebra $\A(\OO)$ \bref{def:A(O)} associated to a non-empty open bounded region $\OO$.
\end{proposition}
\begin{proof}
In view of the definition of the local algebras for arbitrary regions \bref{def:AQ}, it suffices to consider double cones to prove the statement. The fact that $\Om$ is separating for the double cone algebras follows immediately because this vector even separates the larger wedge algebras.

So we have to show that $\Om$ is cyclic for $\A(\OO)$, where $\OO$ is an arbitrary double cone, which amounts to proving that the only vector orthogonal to  $\overline{\A(\OO)\Omega}$ is the zero vector. We thus consider a vector $\Psi\perp\overline{\A(\OO)\Om}$, and choose another double cone $\OO_1$ such that $\OO_1+\mathcal{U}\subset \OO$ for some open neighborhood $\mathcal{U}$ of the origin in $\Rl^2$. In this situation, we have
  \begin{align}\label{perp1}
    \langle \Psi, A(x)\Omega\rangle &= 0, \qquad A\in\A(\OO_1),\;x\in\mathcal{U}.
  \end{align}
But as a consequence of the spectrum condition, $x\longmapsto\langle \Psi, A(x)\Omega\rangle=\langle\Psi,e^{iP\cdot x}A\Omega\rangle$ is the boundary value of a function which is analytic in the forward tube $\Rl^2+i V^+$. As this function vanishes in an open region on the boundary of this tube, namely in ${\cal U}$ \bref{perp1}, it has to vanish for any $x\in\Rl^2$. Hence $U(x)\Psi$ lies orthogonal to $\overline{\A(\OO_1)\Omega}$ for all $x\in\Rl^2$. Since $\A(\OO_1)$ is a $*$-algebra, the same holds true for $A\,U(x)\Psi$, for arbitrary $A\in\A(\OO_1)$. Restricting to a slightly smaller double cone $\OO_2\subset\OO_1$, we obtain by repetition of the same argument $\langle \Psi,A_1(x_1)A_2(x_2)\Om\rangle=0$ for any $x_1,x_2\in\Rl^2$, $A_1,A_2\in\A(\OO_2)$. Thus we can proceed iteratively to conclude
  \begin{align}\label{rs-n}
    \langle \Psi, A_1(x_1)\cdots A_n(x_n)\Omega\rangle
	 &= 0,
	 \qquad
    A_1,...,A_n\in\A(\tilde{\OO}),\quad x_1,...,x_n\in\Rl^2\,,
  \end{align}
  where $\tilde{\OO}$ can be chosen as a fixed double cone $\tilde{\OO}\subset\OO_1$, independent of $n$.

Now consider some completely arbitrary double cone $\widehat{\OO}$. By appropriate choice of $n\in\N$ and $x_1,...,x_n\in\Rl^2$, we have $((\tilde{\OO}+x_1)\cup ... \cup (\tilde{\OO}+x_n))''\supset \widehat{\OO}$. So it follows from Lemma \ref{lemma-mue} and \bref{rs-n} that $\Psi$ lies in the orthogonal complement of $\overline{\A(\widehat{\OO})\Omega}$. As $\widehat{\OO}$ was arbitrary, this implies via Lemma \ref{lemma-loc} that $\Psi$ lies orthogonal to $\overline{\A(W)\Omega}$, where $W\supset\OO$ is a wedge. But $\Omega$ is cyclic for $\A(W)$, and so we conclude $\Psi=0$. This proves the claim.
\end{proof}

\noindent The above results show that the split property for wedges is a very powerful property, allowing for the derivation of many strong properties of the underlying net. In addition to the features discussed so far, even more can be said.

We first note that also the split property for inclusions of double cones $\OO_1\subset\OO_2$ follows \cite{Mue-SPW}. The split property for wedges implies the existence of a unitary $V_2$ which implements the isomorphisms
\begin{align}
\A(\OO_2)	&\cong	\A(W_L^{\OO_2})' \otimes \A(W_R^{\OO_2})' \,,\\
\A(\OO_1)	&\cong	\A(W_L^{\OO_1})' \otimes \A(W_R^{\OO_1})'\,,
\end{align}
and there are type I factors $\NN_L$ and $\NN_R$ such that $\A(W_L^{\OO_1})'\subset\NN_L\subset\A(W_L^{\OO_2})'$ and $\A(W_R^{\OO_1})'\subset\NN_R\subset\A(W_R^{\OO_2})'$. Hence $V_2^*(\NN_L\otimes\NN_R)V_2$ is a type I factor interpolating between $\A(\OO_1)$ and $\A(\OO_2)$.

Analogously to Lemma \ref{lemma-mue}, one can prove that the split property for wedges implies the {\em time slice property} \cite{Mue-SPW}. In the development of the algebraic approach quantum field theory, this property was suggested as a supplement for relativistic field equations \cite{Haag-Schroer}. It postulates that for any {\em time slice}, i.e. any "thickened Cauchy surface" of the form ${\cal C}=\{x\in\Rl^2\,:\,x\cdot \eta \in (a,b)\}$, where $\eta$ is timelike and $a<b$, the observables localized in ${\cal C}$ should generate all observables, $\A({\cal C})=\B(\Hil)$. Although this assumption is usually not regarded as a basic axiom of algebraic quantum field theory, it holds in the context considered here, as a consequence of the split property for wedges.

M\"uger also studied the superselection structure of theories having the split property for wedges, and found strong restrictions in comparison to the general case. Namely, he proved that any representation of the quasilocal algebra which is unitarily equivalent to the identical representation $\pi_0$ in restriction to (left and right) wedges, is in fact equivalent to a multiple of $\pi_0$ \cite[Thm. 3.1]{Mue-SPW}. This no-go theorem also applies to DHR representations, and shows that the split property for wedges is incompatible with the existence of DHR \cite{DHR1, DHR2} or Buchholz-Fredenhagen \cite{BF} sectors. Hence the only representations of such a theory besides the vacuum representation are soliton representations \cite{FRS1}, which interpolate between two different vacua on the left and right wedges \cite{Mue-PhD}. Our construction will therefore not lead to theories with a rich superselection structure, as charged representations and the occurrence of braid group statistics are ruled out by the split property for wedges. As suggested in \cite{Mue-PhD}, the construction of models with these characteristics should become possible if the Haag-duality of the net is relaxed to wedge duality, which amounts to a different (smaller) choice of the local algebras. However, we will not attempt to say anything concerning these questions here, and rather stick to the definition \bref{def:A(O)} of the local algebras introduced before.
\\
\\
To summarize, we have seen that a standard right wedge algebra $(\M,U,\Hil)$ which has the property that $\M(x)\subset\M$ is split for any $x\in W_R$, defines a local net $\OO\lmto\A(\OO)$ \bref{def:A(O)} which has all the indispensable features listed in section \ref{Sec:Axioms}. The split property for wedges amounts to an implicit restriction on the mass spectrum, and in combination with Haag duality, it also restricts the superselection structure of the theory.

From the point of view of applications in concrete examples, the split property is less convenient, however, because it seems to be difficult to verify the existence of interpolating type I factors for a given triple $(\M,U,\Hil)$. In the following section, we will therefore look for a more easily manageable condition on the underlying standard right wedge algebra, which implies the split property for wedges.

\section{The Modular Nuclearity Condition}\label{sec:mnc}

In the literature there exist several criteria which are known to imply the split property. Many of them can be formulated in terms of nuclear maps, and are therefore termed "nuclearity conditions". The definition and basic properties of nuclear maps are collected in appendix \ref{app:nuclearity} -- here we only recall that a linear map between two Banach spaces is nuclear if it can be decomposed into a series of operators of rank one whose norms are summable, and that a nuclear map is in particular compact.

In this section we discuss nuclearity conditions in quantum field theory and select the so-called {\it modular nuclearity condition} as the appropriate tool for our purposes. We then describe how this condition can be checked in concrete models, and summarize the main results of chapter \ref{chapter:netsin2d} in Theorem \ref{thm:Ch2summary}.
\\
\\
Nuclearity conditions were originally invented to amend the basic axioms of quantum field theory, like locality, covariance and positivity of the energy, by some additional requirement which ensures that the theory has a particle interpretation. In particular, a theory complying with such an additional requirement should exhibit the thermodynamical behavior expected from a theory of particles.

Improving on an earlier proposed compactness criterion by Haag and Swieca \cite{Haag-Swieca}, the first nuclearity condition was suggested by Buchholz and Wichmann \cite{BuWi}. This condition, nowadays commonly termed {\it energy nuclearity condition}, formalizes the insight that a quantum field theory must satisfy some restrictions on the number of its local degrees of freedom in order to exhibit regular thermodynamical properties.

The mathematical formulation of this condition can be given as follows \cite{Buchholz:1986bg}. Within the setting of a local net on $d$-dimensional Minkowski space ($d\geq 2$), one considers a region $\OO\subset\Rl^d$ and a parameter $\beta>0$, representing the inverse temperature. In analogy to the form of Gibbs equilibrium states in statistical mechanics, one defines the maps
\begin{align}\label{def:ThetaO}
	\Theta_{\beta,\OO} : \A(\OO) \lto \Hil\,,\qquad \Theta_{\beta,\OO}(A) &:= e^{-\beta H} A \Om\,,
\end{align}
where $H=P_0$ denotes the Hamiltonian with respect to the time direction $x_0$.

The energy nuclearity condition can then be formulated as follows:
\\
\\
{\bf Energy nuclearity condition}\\
{\it The maps $\Theta_{\beta,\OO}$ \bref{def:ThetaO} must be nuclear for any bounded region $\OO$ and any inverse temperature $\beta>0$, and there must exist constants $\beta_0,n>0$ (depending on $\OO$) such that the nuclear norm\footnote{The nuclear norm $\|\cdot\|_1$ of a nuclear map is defined in appendix \ref{app:nuclearity}.} of $\Theta_{\beta,\OO}$ is for $\beta\searrow 0$ bounded by
\begin{align}\label{theta-bound}
\|\Theta_{\beta,\OO}\|_1 \leq e^{(\beta_0/\beta)^n}\,.
\end{align}
}

\noindent The energy nuclearity condition and variants thereof have found several applications in relativistic quantum field theory. For example, it was shown that a theory complying with it has thermodynamical equilibrium states for all temperatures \cite{BuJu2}, which is not a consequence of the basic axioms alone \cite{BuJu1}. It thus indeed provides a criterion to test the thermodynamical properties of model theories, and the nuclear norm of $\Theta_{\beta,\OO}$ can be interpreted as the partition function of the system restricted to the "relativistic box" $\OO$ at temperature $\beta^{-1}$ \cite{BuWi}. More generally, energy nuclearity conditions turned out to be useful in considerations of phase space \cite{BuPo}.

The reason why we are interested in such a condition in the present context is apparent from the following theorem, which can be found in \cite[Thm. 2.1]{Buchholz:1986bg}.
\begin{theorem}\label{Thm:ENC->split}{\bf \cite{Buchholz:1986bg}}\\
Consider a net $\A$ on $d$-dimensional Minkowski space ($d\geq 2$) which satisfies the energy nuclearity condition.
\\
Then for any inclusion $\OO_1\Subset \OO_2$ of open, bounded regions $\OO_1, \OO_2 \subset \Rl^d$, the corresponding inclusion $\A(\OO_1) \subset \A(\OO_2)$ of observable algebras is split.
\end{theorem}

\noindent Theorem \ref{Thm:ENC->split} suggests the energy nuclearity condition as a sufficient condition for the split property, which is of crucial importance for our construction. In comparison to the search for interpolating type I factors, which is required  for the direct verification of the split property, the energy nuclearity condition is a much more concrete condition which is better manageable in applications. Its verification requires a spectral analysis of the operator $e^{-\beta H}$ on certain subspaces of $\Hil$, which essentially amounts to counting states of bounded extension in position and momentum space.

However, what is needed for our construction is the split property {\em for wedges}. But the above theorem was proven for inclusions of {\em bounded} regions only, and also the thermodynamical reasoning motivating the assumption of energy nuclearity \cite{BuWi} leads only to the conclusion that $\Theta_{\beta,\OO}$ should be nuclear for bounded regions $\OO$, and does not apply to wedges. In fact, given the infinite extension of wedges, it seems doubtful that the energy nuclearity condition is valid for these regions. Intuitively speaking, a wedge $W$ allows for too many degrees of freedom to be localizable within $W$, suggesting the breakdown of the nuclearity of the associated maps $\Theta_{\beta,W}$ \bref{def:ThetaO}.

More precisely, we find the following Lemma, which applies to nets on $d$-dimensional Minkowski space, $d\geq 2$. Note that the region $W$ appearing in its formulation can be chosen to be a wedge.

\begin{lemma}\label{Lem:NoEnNuc}
Consider a net $\A$ of local observables and let $W\subset\Rl^{d+1}$ be a region with the property that there exists a direction $x\in\Rl^d$ such that $W+x \subset W$.
\\
Then the maps $\Theta_{\beta,W}$ \bref{def:ThetaO} are compact if and only if $\A(W)=\Cl\cdot 1$.
\end{lemma}
\begin{proof}
The compactness of $\Theta_{\beta,W}$ in case $\A(W)=\Cl\cdot 1$ is trivial. So assume $\A(W)\neq \Cl\cdot 1$ and pick $A\in\A(W)$ with $\langle\Om,\,A\Om\rangle=0$ and $A\neq 0$. By the geometrical assumption on the region $W$ and the isotony of $\A$, the sequence $A_n := U(n\cdot x)A\,U(n\cdot x)^{-1}$, $n\in\N$, is contained in $\A(W)$. To prove that $\Theta_{\beta,W}$ is not compact, it suffices to show that the image of this sequence, $\Psi_n := \Theta_{\beta,W}(A_n)$, has no subsequence which converges in the norm topology of $\Hil$ \cite{SimonReed1}.

As the translations $U(n\cdot x)$ commute with $e^{-\beta H}$ and the vacuum is translation invariant, all vectors $\Psi_n$ have the same norm $\|\Psi_n\|=\|e^{-\beta H}A\Omega\| \neq 0$. Taking into account the cluster property of $\Om$ \cite{maison} yields
\begin{align}
    \langle \Psi_n,\Psi_m\rangle = \langle A\Omega,U((m-n)\cdot x)e^{-2\beta H}A\Omega\rangle \longrightarrow \langle \Om,\,A\Om\rangle\,\langle\Om,\,e^{-2\beta H}A\Om\rangle = 0
\end{align}
as $|m-n|\to\infty$. Hence for arbitrary small $\eps > 0$ there exists an integer $N_\eps$ such that $|\langle\Psi_n,\Psi_m\rangle| < \eps\,\|e^{-\beta H}A\Omega\|^2$ for $|m-n|\geq N_\eps$, and consequently we get
\begin{align}\label{est:cmpt}
    \|\Psi_n-\Psi_m\| \geq \left(2\,(1-\varepsilon)\right)^{1/2}\, \|e^{-\beta H}A\Omega\|\,,\qquad |m-n|\geq N_\eps\,.
\end{align}
Clearly, this inequality contradicts $\{\Psi_n\}_n$ having a convergent subsequence, and hence $\Theta_{\beta,W}$ is not compact.
\end{proof}

\noindent As compactness is a weaker property than nuclearity, the energy nuclearity condition cannot hold in the context of wedge algebras, and does therefore not provide the desired condition on the standard right wedge algebra underlying our construction to guarantee the split property for wedges.
\\
\\
There exists, however, a reformulation of the energy nuclearity condition in terms of the modular operators of local observable algebras instead of the Hamiltonian, and this so-called {\em modular nuclearity condition} is better suited in the present context. 


The modular nuclearity condition has been found by Buchholz, d'Antoni and Longo in \cite{nuclearmaps1}, and can be formulated in a rather abstract setting for an inclusion $\M_1\subset\M_2$ of von Neumann factors on a Hilbert space $\Hil$: Assume that there exists a cyclic and separating vector $\Omega$ for $\M_2$, and denote the modular operator of $(\M_2, \Omega)$ by $\Delta_2$. As the analogue of the maps $\Theta_{\beta,\OO}$ \bref{def:ThetaO} appearing in the energy nuclearity condition, one here considers
\begin{align}\label{def:XiAbstract}
	\Xi_{\M_1,\M_2} : \M_1\lto\Hil\,,\qquad \Xi_{\M_1,\M_2}(M) &:= \Delta_2^{1/4} M \Omega\,.
\end{align}
Roughly speaking, the modular operator $\Delta_2$, or rather its logarithm, takes the role of the Hamiltonian in this setting. In analogy to the energy nuclearity condition, the requirement that the maps \bref{def:XiAbstract} must be nuclear will be referred to as the  modular nuclearity condition for the inclusion $\M_1\subset\M_2$.

It has been shown in \cite{nuclearmaps2} that in application to the local algebras $\A(\OO)$ of a quantum field theory, the energy and modular nuclearity conditions are essentially equivalent. More precisely, the order of the map $\Xi_{\A(\OO),\A(\hat{\OO})}$ \bref{def:XiAbstract} can be estimated by the order of $\Theta_{\beta,\OO}$ for appropriate inverse temperature $\beta$. (The order of a bounded linear map is an alternative measure of the "size" of the image of this map.) On the other hand, under the assumption of the Bisognano-Wichmann property, the order of $\Theta_{\beta,\OO}$ can be estimated by the order of $\Xi_{\A(\OO),\A(\hat{\OO})}$ times a factor which depends on the inverse temperature and the geometry of the considered regions $\OO\subset\hat{\OO}$.  This latter estimate is not valid if $\OO$ is taken to be a wedge.

So the nuclearity properties of the maps \bref{def:ThetaO} and \bref{def:XiAbstract} are closely related in application to the local observable algebras of a quantum field theory. Therefore also the modular nuclearity condition implies the split property.
\begin{theorem}\label{MNC->split}{\bf \cite{nuclearmaps1}}\\
	Consider an inclusion $\M_1\subset\M_2\subset\B(\Hil)$ of von Neumann factors and the maps \bref{def:XiAbstract} as above. Then the following holds:
	\begin{enumerate}
		\item	If $\Xi_{\M_1,\M_2}$ is nuclear, the inclusion $\M_1\subset\M_2$ is split.
		\item If the inclusion $\M_1\subset\M_2$ is split, the map $\Xi_{\M_1,\M_2}$ is compact.
	\end{enumerate}
\end{theorem}
\noindent For possible generalizations to the non-factor case, see \cite{fid}.
\\\\
Although the two nuclearity conditions introduced here are closely related in applications local algebras, they are not equivalent as such. Following the arguments in \cite{nuclearmaps2} closely, one notes that the equivalence of the two concepts breaks down when applied to observable algebras which are localized in unbounded regions, like wedges. This opens up the possibility for the modular nuclearity condition to hold for inclusions of wedge algebras. The following argument shows that the maps $\Xi_{\M_1,\M_2}$ can in fact be nuclear if $\M_1\subset\M_2$ is an inclusion of wedge algebras in a quantum field theory on two-dimensional Minkowski space.

For the sake of concreteness, we consider an inclusion of right wedges
\begin{align}
	W_R +x \subset W_R,\qquad x\in W_R\,.
\end{align}
As the corresponding observable algebras are related by $\A(W_R+x)=U(x)\A(W_R)U(x)^{-1}$, and $A\lmto A(x)=U(x)A\,U(x)^{-1}$ is an invertible map of norm one between the Banach spaces\footnote{All operator algebras appearing here are considered as Banach spaces equipped with the norm $\|\cdot\|_{\B(\Hil)}$.} $\A(W_R)$ and $\A(W_R+x)$, the map $\Xi_{\A(W_R+x),\A(W_R)}$ is nuclear if and only if
\begin{align}\label{def:XiW}
	\Xi(x) : \A(W_R) \lto \Hil\,,\qquad \Xi(x)(A) &:= \Delta^{1/4}U(x)A\Omega\,,
\end{align}
is, with the same nuclear norm. (Here $\Delta$ denotes the modular operator of $\A(W_R)$ with respect to the vacuum vector $\Omega$.)

We now mimic the proof of Lemma \ref{Lem:NoEnNuc} in order to understand why the obstruction to the nuclearity of $\Theta_{\beta,W}$ found there does not show up in the analysis of the map $\Xi(x)$. So we pick a direction $y\in W_R$ and consider the sequence
\begin{align}\label{psi-mod}
	\Psi_n 	&:= \Xi(x) \left(U(n\cdot y)A\, U(n\cdot y)^{-1} \right)
			 = \Delta^{1/4} U(x+n\cdot y)A\Omega \,.
\end{align}
In contrast to $e^{-\beta H}$, the modular operator $\Delta^{1/4}$ does not commute with the translations $U(n\cdot y)$. In fact, one rather has Borchers' commutation relations between the translations and the modular unitaries $\Delta^{it}$ \bref{borchers-CR}, which after analytic continuation to $t=-\frac{i}{4}$ read (on an appropriate domain)
\begin{align}
	\Delta^{1/4} U(y) &= U(\La(\tfrac{i\pi}{2})y) \Delta^{1/4}\,.
\end{align}
The boost $\La(\tfrac{i\pi}{2})$ with imaginary rapidity parameter $\frac{i\pi}{2}$ appearing here is explicitly given by the matrix \bref{def:boost}
\begin{align}
 	\Lambda(\tfrac{i\pi}{2}) = \left( \begin{array}{cc}
 							\cosh\tfrac{i\pi}{2} & \sinh \tfrac{i\pi}{2} \\
 							\sinh\tfrac{i\pi}{2}  & \cosh\tfrac{i\pi}{2}
 	                                   		\end{array}
 						\right)
 					=  	\left(\begin{array}{cc}
 	                                   		0 & i \\
 							i & 0
 							\end{array}
 						\right)\,,
\end{align}
and hence we find
\begin{align}
U(\La(\tfrac{i\pi}{2})y)	&= e^{iP\cdot \La(\frac{i\pi}{2})y}	=	e^{-P\cdot y'},\qquad y':=(y_1,y_0)\,.
\end{align}
As $y\in W_R$, the flipped vector $y'$ lies in the forward light cone, which implies $P\cdot y' \geq 0$ because of the spectrum condition. Thus the vectors $\Psi_n$, given by
\begin{align}
	\Psi_n=\Xi(x)\left(U(y)AU(y)^{-1}\right) 	&= e^{-P\cdot y'} \Xi(x)A
\end{align}
have norms $\|\Psi_n\|$ converging to zero for $n\to\infty$, and the argument of Lemma \ref{Lem:NoEnNuc} cannot be applied.

The above calculation shows that observables $A\in\A(W_R)$ which are localized in the remote right part of $W_R$ are exponentially suppressed by the operator $e^{-P\cdot y'}$. Disregarding these damped observables, we are thus effectively left with the map $\Xi(x)$ restricted to the algebra of observables localized in a "tip" region of $W_R+x$, which is bounded in the case of two-dimensional spacetime. Taking into account the relations between the energy and modular nuclearity conditions mentioned before, and the thermodynamical significance of the former, we therefore expect the validity of the modular nuclearity condition in a two-dimensional quantum field theory which has the properties typical of a particle theory.

We mention as an aside that in higher-dimensional Minkowski space, the sequence \bref{psi-mod}, with $y$ chosen in the edge of $W_R$, can be used to prove that the maps $\Xi(x)$ are not compact. In view of Theorem \ref{MNC->split} b), this implies the breakdown of the split property for wedges in more than two spacetime dimensions, as mentioned before.
\\
\\
But in two dimensions, there is no reason for the modular nuclearity condition to fail if the theory does not exhibit to many wedge-local degrees of freedom, and we thus use the postulate that the maps $\Xi(x)$ \bref{def:XiW} are nuclear, $x\in W_R$, as the desired sufficient condition for the split property for wedges. Note that this is a condition on the input from which we started our construction, since the map $\Xi(x)$ \bref{def:XiW} in question is defined solely in terms of the objects $\M=\A(W_R)$, $U$ and $\Hil$ constituting the standard right wedge algebra $(\M,U,\Hil)$.

If the maps $\Xi(x)$, corresponding to the inclusions $\A(W_R+x)\subset\A(W_R)$, are nuclear for any $x\in W_R$, it follows that the net $\A$ \bref{def:wedgenet} has the split property for wedges. For any inclusion of wedges can be transformed to $W_R+x\subset W_R$ by translation and reflection, and the corresponding (anti-) automorphisms $A \lmto JAJ$, $A\lmto U(x)A\,U(x)^{-1}$ are, considered as (anti-) linear maps between the respective Banach spaces, invertible and bounded (cf. Theorem \ref{Lem:nuclearmaps--app} in appendix \ref{app:nuclearity}). In particular, all the results obtained in the previous section, which relied on the split property for wedges, are valid if $\Xi(x)$ is nuclear.

 In the following theorem, we summarize the results of our construction.
\begin{theorem}\label{thm:Ch2summary}
	Consider a standard right wedge algebra $(\M,U,\Hil)$ which has the property that the maps
	\begin{align}
		\Xi(x) : \M\lto\Hil\,,\qquad \Xi(x)A &:= \Delta^{1/4}U(x)A\Omega\,,\qquad x\in W_R\,,
	\end{align}
	are nuclear, where $\Delta$ denotes the modular operator of $(\M,\Om)$.
	\\
	Then
	\begin{itemize}
		\item[a)] The correspondence $\OO\longmapsto\A(\OO)$ defined by (\ref{def:wedgenet}, \ref{def:A(O)}, \ref{def:AQ}) is a local net which transforms covariantly under the representation $U$ \bref{def:Uext} of the proper Poincar\'e group.
		\item[b)] The net $\A$ has the split property for inclusions of wedges (and hence, also for inclusions of double cones).
		\item[c)] The wedge algebras \bref{def:wedgenet} and the double cone algebras \bref{def:A(O)} are all isomorphic to the hyperfinite type III$_1$ factor.
		\item[d)] Haag duality holds, i.e. $\A(\OO)'=\A(\OO')$ for any double cone $\OO\subset\Rl^2$.
		\item[e)] Strong additivity as expressed by Lemma \ref{lemma-loc} and Lemma \ref{lemma-mue} holds.
		\item[f)] The Reeh-Schlieder property holds, i.e. the vacuum vector $\Omega$ is cyclic and separating for $\A(\OO)$ if $\OO$ is a non-empty region with non-empty causal complement.
	\end{itemize}
	{\hfill $\square$}
\end{theorem}

\noindent It follows from the above mentioned estimates on the order of $\Theta_{\beta,\OO}$ that the nuclearity of $\Xi(x)$ also implies nuclearity properties of $\Theta_{\beta,\OO}$ for bounded regions $\OO$. To obtain the bound $\|\Theta_{\beta,\OO}\|_1\leq \exp(\beta_0/\beta)^n$ required in the energy nuclearity condition, however, one needs more detailed knowledge about the maps $\Xi(x)$. This mechanism will be used in chapter \ref{chapter:reconstructS}, where we compute a bound on the nuclear norm of $\Theta_{\beta,\OO}$ in terms of the nuclear norm of $\Xi(x)$ in concrete models.

To conclude this section, we briefly indicate how the modular nuclearity condition will be verified in application to the models with factorizing S-matrix discussed in chapters \ref{chapter:wedgenet} and \ref{chapter:nuclearity}. First note that the remarks made for the energy nuclearity condition apply also here: Whereas in the direct verification of the split property, interpolating type I factors need to be constructed, the verification of the modular nuclearity condition amounts to a kind of spectral analysis of the maps $\Xi(x)$.

These maps have the advantage to be given in a very concrete form, since the modular operator $\Delta$ acts as a boost transformation, with imaginary rapidity parameter $\frac{i\pi}{2}$. In the models we have in mind, the Hilbert space $\Hil$ will have a form similar to the Bose Fock space over $L^2(\Rl,d\te)$. In particular, its elements are sequences of $n$-particle wavefunctions $(\te_1,...,\te_n)\lmto \Psi_n(\te_1,...,\te_n)$. These functions can be most conveniently formulated in rapidity space, where the modular operator $\Delta^{1/4}$, acts according to 
\begin{align}
(\Delta^{1/4}\Psi)_n(\te_1,...,\te_n)	&=	\Psi_n(\te-\tfrac{i\pi}{2},...,\te_n-\tfrac{i\pi}{2})\,,
\end{align}
i.e. $\Delta^{1/4}$ maps the wavefunctions $\Psi_n$ onto certain analytic continuations of themselves. Thus the check of the nuclearity of the maps $\Xi(x)$ can be carried out with methods of complex analysis. As we shall see, this task can be accomplished in a large class of theories.

\section{The Construction of Wedge Algebras}

Up to now, our approach consisted in constructing a local net on a Hilbert space $\Hil$ in terms of a von Neumann algebra $\M$ and a representation $U$ of the translations, and in identifying appropriate conditions on these objects to ensure that the resulting theory is physically meaningful. But for the construction of concrete models we also have to specify how such a standard right wedge algebra is to be defined explicitly. This is the subject of chapters \ref{chapter:wedgenet} and \ref{chapter:nuclearity} of this thesis, where a family of models fitting into the framework developed here is constructed.  Nonetheless, we want to make some remarks about the problem of constructing wedge algebras already in the present more abstract discussion, partly to motivate the approach to the construction which is used in chapter \ref{chapter:wedgenet}, and partly to indicate the historical development of the subject.

\subsection{Modular Wedge-Localization and the Construction of Interaction-Free Theories}


To begin with, we will show how the definition of interaction-free theories can be adapted to our construction program, thereby providing first examples of standard right wedge algebras $(\M,U,\Hil)$ satisfying all assumptions. In doing so, we will refrain from using the usual field-theoretic formulation of free nets, but rather follow the work of Brunetti, Guido and Longo \cite{BGL}, who constructed interaction-free nets in a purely algebraic manner.

Deviating slightly from our setup in section \ref{sec:ch2los}, the authors of \cite{BGL} consider a Hilbert space $\Hil_1$, to be interpreted as the single particle space of the theory, and a unitary, strongly continuous representation $U_1$ of the proper Poincar\'e group $\PGpo$ acting on it. To come up with a definition of a wedge algebra, some localization concept is needed. As the essential point of their construction, Brunetti, Guido and Longo invent such a concept on the single particle space $\Hil_1$, which can be motivated as follows.

To recall the relevant notions, we consider the localization of states in the context of a Haag-dual net $\A$. In this setting, a vector $\Psi\neq 0$ can be defined to be localized in a spacetime region $\OO$ if the associated vector state $\|\Psi\|^{-2}\langle\Psi,\cdot\;\Psi\rangle$ coincides with the vacuum state $\langle\Om,\,\cdot\;\Om\rangle$ on the algebra $\A(\OO')=\A(\OO)'$.

It has been shown by Licht \cite{Licht1, Licht2} that there is a correspondence between the localization of operators in $\B(\Hil)$ and vector states (see also \cite{Knight}). Namely, to any vector $\Psi$ localized in $\OO$ there exists an operator $A\in\A(\OO)$ such that $\Psi=A\Omega$. Hence these vectors lie in the domain of the Tomita operator $S_\OO$ of $(\A(\OO),\Om)$ \cite{KadRin2}. Since $S_\OO$ is an antilinear involution, $S_\OO^2\subset 1$, its domain can be conveniently described in terms of its eigenspace $\K(\OO)$ to eigenvalue one as dom$S_\OO = \K(\OO)+i\K(\OO)$ \cite{BGL}. The observable algebra $\A(\OO)$ is then generated by the operators $A$ corresponding to vectors in $\K(\OO)$ via the above mentioned relation.

These observations also apply to the single particle space, and can be used for the definition of wedge localization on $\Hil_1$ by invoking the Bisognano-Wichmann theorem \cite{BiWi1, BiWi2}.
This theorem connects the algebraic structure of the net of observables in a quantum field theory with the geometric structure of the underlying Minkowski spacetime in the following way. It asserts that the modular unitaries $\Delta^{it}$ and modular involution $J$ of the couple $(\A(W_R),\Om)$ are related to the representation $U$ of the proper Poincar\'e group by\footnote{If the dimension of spacetime is larger than two, the formula \bref{eq:BiWiJ} has to be altered by a rotation.}
\begin{align}
	\Delta^{it} &= U(\La(-2\pi t)),\qquad t\in\Rl\,,\label{eq:BiWiD}\\
	J		&= U(-1)\,, \label{eq:BiWiJ}
\end{align}
if the net is generated from finite-component Wightman fields. In algebraic quantum field theory, Borchers' theorem \cite{borchers-2d} constitutes an important partial version of this result, which under additional assumptions has been used by J. Mund to derive the above equations also in this setting \cite{MundBiWi}.

So the formulas (\ref{eq:BiWiD}, \ref{eq:BiWiJ}) are valid in a wide range of theories, and were therefore used as an input in \cite{BGL}. More precisely, in the construction starting from the representation $U_1$ on $\Hil_1$, the operators $\Delta$ and $J$ can be {\em defined} by (\ref{eq:BiWiD}, \ref{eq:BiWiJ}), with $U$ replaced by $U_1$, and turn out to have all the algebraic properties familiar from modular theory. In particular, the "geometric Tomita operator of the right wedge" $S:=J\Delta^{1/2}$ exists as an unbounded, antilinear involution on $\Hil_1$, and one can define its eigenspace
\begin{align}\label{def:KWR}
	\K(W_R) &:= \big\{\Psi\in{\rm dom}\Delta^{1/2} \;:\; J\Delta^{1/2}\Psi=\Psi\big\}\,.
\end{align}
Vectors in this space are considered as single particle states localized in the right wedge $W_R$. Because of its relation to modular theory, this idea is commonly referred to as {\it modular localization}, see \cite{Schroer:1997cq,Mund:2002yc,Mund:2005cv} for applications of it.

Having defined a notion of wedge localization on $\Hil_1$, the construction of the wedge algebra $\M$ in the free theory can be accomplished easily. Considering the Fock space $\Hil$ over $\Hil_1$, we find a representation $U$ of $\PG_+$ and an invariant vacuum vector by second quantization. The unitary Weyl operators $V(\Psi)$, $\Psi\in\Hil_1$, act on the Fock space $\Hil$ and lead to the definition of a right wedge algebra as
\begin{align}\label{def:M-free}
	\M := \big\{V(\Psi) \;:\; \Psi \in \K(W_R) \big\}''\,.
\end{align}
It can be shown by standard arguments making use of the algebraic structure of the CCR algebra that the triple $(\M,U,\Hil)$ constructed in this way is a standard right wedge algebra in the sense of our Definition \ref{def:srwa}.

Also the modular nuclearity condition holds if there are not too many different kinds of particles in the mass spectrum of $U_1$ (\cite{BuLe}, see chapter \ref{chapter:nuclearity} of this thesis). The local net generated from the triple $(\M,U,\Hil)$ along the lines described in section \ref{sec:constructAWO} is (an abstract version of) the theory of free fields.
\\
\\
The general idea of defining wedge-localized observables with the help of wedge-localized vectors which are characterized by the modular localization concept is applicable to the construction of models with nontrivial interaction as well. But the Weyl operators $V(\Psi)$ used in \bref{def:M-free} are typical of free theories and therefore have to be substituted by operators with different commutation relations. This observation leads us to the concept of polarization-free generators.

\subsection{Polarization-Free Generators}

A simplifying feature of the free theory which was used in the above construction is the fact that the localization of a Weyl operator can be expressed on the one particle space. This property is closely related to the form of the free field operator $\phi_0$ in the standard formulation of this model, which is a solution of the Klein-Gordon equation. Put differently, $\phi_0$ generates only single particle states from the vacuum, a property an interacting Wightman field cannot have, as the Jost-Schroer theorem states \cite[Thm. 4-15]{streater}.

Intuitively, this fact can be understood by appealing to Heisenberg's uncertainty relation: The sharp localization of an operator $A$ in configuration space implies large fluctuations in momentum space. One would therefore expect that a vector of the form $A\Om$, where $A$ is localized in a bounded region, describes a "vacuum polarization cloud" having contributions with arbitrarily high particle numbers, and this is indeed the case in the presence of interactions.

But when the localization in configuration space is weakened from strict localization to localization in a wedge, the Jost-Schroer theorem does not apply, i.e. there is no general argument ruling out the existence of operators which are localized in wedges and generate single particle states from the vacuum. This was realized by Schroer, who coined the name {\it polarization-free generators} for such objects and studied their properties in models \cite{Schroer:1997cq,Schroer:1997cx,Schroer-Wiesbrock,Schroer:1999xi,Schroer:2000dg}. In \cite{BBS}, the concept was formalized in a model-independent way. Within the setting of a local net $\A$ on $d$-dimensional Minkowski space ($d\geq 2$),  a polarization-free generator is defined in the following way.
\begin{definition}\label{def:PFG}
A polarization-free generator $G$ is a closed operator satisfying the following conditions:
\begin{enumerate}
	\item $G$ is affiliated with a wedge algebra $\A(W)$, $W\in\W$.
	\item The vacuum vector $\Omega$ is contained in the domains of $G$ and $G^*$.
	\item The vectors $G\Omega$ and $G^*\Omega$ lie in the one-particle space.
\end{enumerate}
\end{definition}

\noindent It was shown in \cite{BBS} that polarization-free generators exist in fact in {\em any} theory. Hence these objects seemed to be the appropriate substitute for the free field operator to transport the modular localization concept from the level of one particle states to the level of operators.

However, it was also discovered that polarization-free generators have unwieldy domain properties, in general. The following definition, requiring some mild temperateness properties which are fulfilled by the free field, has drastic consequences.
\begin{definition}\label{def:tPFG}
A polarization-free generator is said to be temperate if there exists a dense, translation invariant subspace $\DD$ of its domain such that the functions $x\mapsto GU(x)\Psi$ are strongly continuous and polynomially bounded for $\Psi\in\DD$, and the same holds true for its adjoint $G^*$.
\end{definition}
\noindent Note that the requirements made in this definition are indispensable if one wants to do Fourier analysis with polarization-free generators, as is for example necessary to utilize them in scattering theory.

It was shown in \cite{BBS} that if the spacetime dimension is larger than two, the temperateness assumption implies that the underlying theory has a trivial S-matrix, thus limiting the use of polarization-free generators to the realm of two-dimensional theories. Also on two-dimensional Minkowski space, the assumption that the polarization-free generators of a theory are temperate puts strong constraints on the form of the S-matrix, albeit weaker ones than in the higher-dimensional case. For example, it follows that there can be no particle production \cite[Thm. 3.5]{BBS}. This observation indicates that the range of application for temperate polarization-free generators might be the family of theories with a factorizing S-matrix, in which the particle number is a conserved quantity -- despite these models being fully relativistic\footnote{The family of factorizing S-matrices is discussed in chapter \ref{chapter:factS}.}.

This conjecture is further supported by the relation between the temperate polarization-free generators and the S-matrix: In view of the absence of polarization clouds and the Reeh-Schlieder property, $x\longmapsto U(x)GU(x)^{-1}$ is a weak solution of the Klein-Gordon equation if $G$ is a temperate polarization-free generator \cite{BBS}. Thus $G$ can be split into a creation and an annihilation part by taking restrictions of its Fourier transform to the upper and lower mass shell, respectively. Under additional regularity assumptions on $G$, the analysis in \cite{BBS} indicates that the associated creation operators fulfill certain quadratic exchange relations, involving the two-particle S-matrix.
\\
\\
In the framework of field-theoretic models with a factorizing S-matrix, on the other hand, such relations are well-known as the Zamolodchikov-Faddeev algebra \cite{ZZ}. This algebra is usually described as a $*$-algebra of non-commuting distributions $Z(\te)$, $Z(\te)^*$ which are parametrized by the rapidity $\te$ and satisfy the relations, $\te_1,\te_2\in\Rl$,
\begin{subequations}\label{zrel:chap2}
\begin{align}
	Z(\te_1)Z(\te_2)		&=		S_2(\te_1-\te_2)\ Z(\te_2)Z(\te_1),\\
	Z^*(\te_1)Z^*(\te_2)	&=		S_2(\te_1-\te_2)\ Z^*(\te_2)Z^*(\te_1),\\
	Z(\te_1)Z^*(\te_2) 	&=		S_2(\te_2-\te_1)\,Z^*(\te_2) Z(\te_1) + \delta(\te_1-\te_2)\cdot 1 \,.
\end{align}
\end{subequations}
Here $S_2$ is the {\em scattering function} of the model, which is in one-to-one correspondence with its two-particle S-matrix, see chapter \ref{chapter:factS} for a discussion of this matter. The relations \bref{zrel:chap2} can be heuristically motivated, and are taken as an input in the bootstrap form factor program (section \ref{sec:ffp}). The conceptual relevance of the objects $Z(\te)$, $Z^*(\te)$ seems to be that of the creation and annihilation parts of temperate polarization-free generators.

In fact, it was precisely in the family of models based on such a Zamolodchikov algebra, where the concept of polarization-free generators was first introduced by Schroer by taking appropriate combinations of the operators $Z$ and $Z^*$ \cite{Schroer:1997cq}.
Utilizing these operators, wedge algebras can be defined in a manner similar to the free net construction by Brunetti, Guido and Longo, replacing the free field operator by the polarization-free generators.
\\
\\
In chapter \ref{chapter:wedgenet}, we will study a representation of the Zamolodchikov-Faddeev algebra and define certain semi-local quantum fields in terms of the representing operators. These fields are then used to generate a standard right wedge algebra and the associated local net, with respect to which they become temperate polarization-free generators.

\chapter{Factorizing S-Matrices and the Form Factor Program}\label{chapter:factS}

The common feature of the models to be constructed in chapter \ref{chapter:wedgenet} is that their S-matrices are of the factorizing type. As a prerequisite, we recall the definition and properties of S-matrices on two-dimensional Minkowski space in general, and of factorizing S-matrices in particular, in the first sections of this chapter. In section \ref{sec:ffp}, we describe the so-called bootstrap form factor program, which is the usual framework for investigations of models with a factorizing S-matrix, and discuss its relation to our work.

\section{S-Matrices in Two Dimensions}\label{sec:SM}
To introduce the notations and conventions used here, we recall some properties of the S-matrix in relativistic quantum physics  (cf., for example, \cite{iago, ovovov, araki}).

For simplicity, we consider a theory describing a single species of neutral scalar particles of mass $m>0$ which do not have any internal degrees of freedom. In this case\footnote{It is not strictly necessary to assume such a mass gap in the energy-momentum spectrum, as the Haag-Ruelle scattering theory also works for more general mass spectra \cite{buchholz-massless-fermis, buchholz-massless-bosons,herbst,dybalski}; but this simplest case is sufficient for our purposes.}, one can apply the Haag-Ruelle scattering theory \cite{bs-scatter, araki, haag-scatter, ruelle-scatter} to compute multiparticle collision states. The incoming and outgoing scattering states span the symmetric Fock space $\Hil^+$ over the one particle space $\Hil_1$ of the theory, and one obtains two isometries $V\iin$ and $V\oout$, mapping $\Hil^+$ onto certain subspaces $\Hil\iin\subset\Hil$ and $\Hil\oout\subset\Hil$ of the full Hilbert space $\Hil$ of the theory. The S-matrix $S$ is defined as the product of these generalized M{\o}ller operators, $S:={V\iin}{V\oout}^*:\Hil\oout\to\Hil\iin$, and maps outgoing onto incoming scattering states.

Unfortunately, there is no general agreement about the definition of the S-matrix; many authors use $V\oout{V\iin}^*$ as the definition of $S$ instead. Moreover, it is sometimes advantageous to consider the S-matrix as an operator on $\Hil^+$, i.e. to use 
\begin{align}
\hat{S}	&:=	{V\oout}^* V\iin	=	{V\iin}^* S V\iin 	= {V\oout}^* S {V\oout}
\,.
\end{align}
In this work, we define the {\em S-matrix on $\Hil$} as $S:={V\iin}{V\oout}^*$, and {\em the S-matrix on $\Hil^+$} as $\hat{S} := {V\oout}^* V\iin$. When speaking about the S-matrix without any further specification, either of the two operators is meant.

If every vector in $\Hil$ has an interpretation in terms of asymptotic particle states, i.e. $\Hil\iin=\Hil\oout=\Hil$, the theory is said to be {\em asymptotically complete}, and the S-matrix $S$ becomes a unitary on $\Hil$ (and accordingly, $\hat{S}$ is a unitary on $\Hil^+$).

It has to mentioned, however, that up to now, the property of asymptotic completeness has not been fully established for any relativistic theory with non-trivial interaction yet \cite{bs-scatter} -- the models constructed here are therefore the very first theories of this type where asymptotic completeness is known to hold.

In case of a particle spectrum as described above, the single particle space $\Hil_1$ of the theory can be identified with the space $L^2(\Rl^s,d\mu({\bf p}))$ of square integrable momentum wavefunctions on the upper mass shell, where $d\mu({\bf p})=({\bf p}^2+m^2)^{-1/2}d^s{\bf p}$ is the usual Lorentz invariant measure and $s\geq 1$ the spatial dimension. On two-dimensional Minkowski space, however, it is more convenient to use as a variable the rapidity instead of the momentum. The rapidity $\te$ can be regarded as a particular parametrization of the (one-dimensional) upper mass shell $H_m^+ = \{((p^2+m^2)^{1/2},p)\,:\,p\in\Rl\}$\label{uppermassshell}, related to the on-shell two-momentum by
\begin{align}\label{def:pte}
p(\te)		&:=
m
\left(
\begin{array}{c}
\cosh\te\\\sinh\te
\end{array}
\right)\,,
\qquad\te\in\Rl\,.
\end{align}
Note that $\te\lmto p(\te)$ is a bijection between $\Rl$ and $H_m^+$, and that the invariant measure $d\mu(p)$ takes after reparametrization the simple form of Lebesgue measure in $\te$.
We may therefore identify the one-particle space with $L^2(\Rl,d\te)$, and the symmetric Fock space over it is $\Hil^+=\bigoplus_{n=0}^\infty L^2(\Rl^n,d^n\bte)_+$, where $d^n\bte=d\te_1\cdots d\te_n$ and the subscript "+" denotes total symmetrization in all variables. 

On $\Hil^+$ we have a second quantization representation $U$ of the proper orthochronous Poincar\'e group with mass $m$ and spin zero. Recall that a boost transformation is a translation in rapidity space, described by a parameter $\la\in\Rl\,$. The transformation $(x,\la)$ consisting of such a boost and a subsequent spacetime translation along $x\in\Rl^2$ is thus represented on $\Hil^+$ according to
\begin{align}
(U(x,\la)\Psi)_n(\te_1,...,\te_n)	&=	\prod_{k=1}^n e^{ip(\te_k)\cdot x}\cdot \Psi_n(\te_1-\la,...,\te_n-\la)\,.
\end{align}
In this setup, incoming and outgoing $n$-particle states are described by square integrable, totally symmetric functions of $n$ rapidity variables $\te_1,...,\te_n$, and S-matrix elements take the form
\begin{align*}
\langle \Psi,\,\hat{S}\,\Phi\rangle
&=
\sum_{n,m=0}^\infty \int d^n\bte \int d^m\bte'\,\overline{\Psi_n(\te_1,...\te_n)}\,S_{n,m}(\te_1,...,\te_n;\te_1',...,\te_m')\,\Phi_m(\te_1',...,\te_m')\,,
\end{align*}
where $\Psi_n\in\Hil_n^+$, $\Phi_m\in\Hil_m^+$ are the totally symmetric wavefunctions of the respective asymptotic states. The kernels $S_{n,m}$ are tempered distributions on $\Ss(\Rl^{n+m})$. Formally they are given by scalar products of improper rapidity states,
\begin{align}\label{Snmelement}
S_{n,m}(\te_1,...,\te_n;\te_1',...,\te_m')
&=\;
n!\cdot 
\oout\langle \te_1,...,\te_n\,|\,\te_1',...,\te_m'\rangle\iin\,.
\end{align}
The normalization factor $n!$ is a matter of convention. 

The form of the distributions $S_{n,m}$ is restricted by several constraints. For example, energy momentum conservation demands that $S_{n,m}(\te_1,...,\te_n;\te_1',...,\te_m')$ contains the factor $\delta(\sum_{k=1}^n p(\te_k)-\sum_{l=1}^m p(\te_l'))$, and covariance under proper Lorentz transformations and the TCP operator implies that in the sense of distributions \cite{araki},
\begin{align*}
S_{n,m}(\te_1-\la,...,\te_n-\la;\te_1'-\la,...,\te_m'-\la)
&=
S_{n,m}(\te_1,...,\te_n;\te_1',...,\te_m')\,,\qquad\la\in\Rl\,,
\\
S_{n,m}(\te_1,...,\te_n;\te_1',...,\te_m')
&=
S_{m,n}(\te_1',...,\te_m';\te_1,...,\te_n)\,.
\end{align*}
In two dimensions, the energy momentum conservation law for collision processes with two incoming and two outgoing particles forces $S_{2,2}(\te_1,\te_2;\te_1',\te_2')$ to vanish unless $\te_1=\te_1'$, $\te_2=\te_2'$ or $\te_1=\te_2'$, $\te_2=\te_1'$. Separating the corresponding delta distributions from $S_{2,2}$, the remaining part becomes a function of the usual Lorentz invariant Mandelstam variable $s$, the square of the total energy of the collision process.
Hence the two-particle S-matrix elements can be written as
\begin{align}
S_{2,2}(\te_1,\te_2;\te_1',\te_2')
&=
\frac{1}{2}\left(\delta(\te_1-\te_1')\delta(\te_2-\te_2') + \delta(\te_1-\te_2')\delta(\te_2-\te_1')\right)\cdot F(s(\te_1,\te_2))\,,
\nonumber
\\
s(\te_1,\te_2)
&:=
(p(\te_1)+p(\te_2))^2
=
2m^2(1+\cosh(\te_1-\te_2))\,.
\end{align}
In contrast to the situation in higher-dimensional Minkowski space, where the two-particle S-matrix element depends on two Mandelstam variables, $s$ and one of the momentum transfers $t=(p(\te_1)-p(\te_1'))^2$ and $u=(p(\te_1)-p(\te_2'))^2$, in two dimensions, $F$ is a function of the single variable $s$ only. Parametrized as above, $S_{2,2}$ is manifestly invariant under Poincar\'e transformations, including the TCP symmetry.

The analytic structure of $s\lmto F(s)$ has been studied extensively in the past \cite{crossing,Bros:1974ad,Bros:1985gy,Bros:1983vf}. It is known that $F$ is the boundary value of a function which is analytic outside a large enough disc in the cut $s$-plane, with branch cuts (arising from two-particle thresholds) along $[4m^2,\infty)$ and the negative real line \cite{eden, abdalla}. The original function $F$ is recovered from its analytic continuation as the boundary value at $[4m^2,\infty)$, taken from Im$\,s>0$ \cite{eden}. A priori, $F$ could have poles corresponding to other stable particles (bound states) in the theory. But as we are considering a model of only a single kind of particles, these poles can be excluded. We are thus dealing with S-matrices for which $F$ is analytic in the complete cut $s$-plane. Moreover, the analytic continuation of $F$ has the properties of {\em crossing symmetry} and {\em hermitian analyticity}, which we recall now.

\begin{figure}[here]
    \noindent
    \psfrag{crossing}{crossing}
    \psfrag{Pol}{pole}
    \psfrag{Im(s)}{${\rm Im}\,s$}
    \psfrag{Re(s)}{${\rm Re}\,s$}
    \psfrag{Schnitt}{branch cut}
    \psfrag{0}{$\snull$}
    \psfrag{s}{$\sms$}
    \psfrag{4m2-s}{$\mss$}
    \psfrag{F(s)}{$F(s)$}
    \psfrag{F(s)c}{$\overline{F(s)}$}
    \psfrag{4m2}{$\ms$}
    \centering\epsfig{file=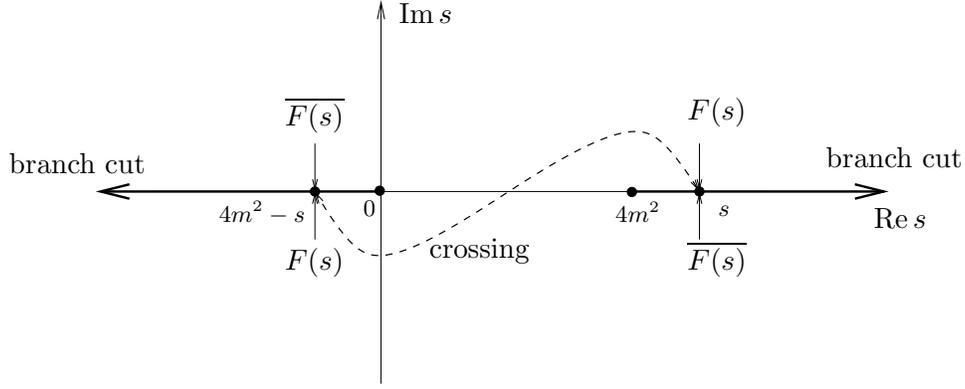,width=12cm}
   \caption{The relations between the different boundary values of $F$ as a consequence of the crossing symmetry and hermitian analyticity.}
\label{fig:crossing}
\end{figure}

The phenomenon of crossing symmetry provides a relation between a scattering process with incoming momenta $p_1$, $p_2$, and outgoing momenta $p_1'$, $p_2'$, with the "crossed" process\footnote{ The exchange of particles with their antiparticles can be ignored here since the particles under consideration carry no charges.} with incoming momenta $p_1$, $-p_2'$ and outgoing momenta $p_1',-p_2$. Namely, it asserts that the functions $F$ corresponding to these processes are boundary values of the same analytic function. As the mentioned exchange of momenta amounts to the change $s\to 4m^2-s$ in the square of the total energy, this implies that the boundary value of $F$ on the cut Re$\,s<0$, approached from Im$\, s<0$, coincides with the physical boundary value of $F$ \cite{iago,araki,eden} (see figure \ref{fig:crossing}). For two-particle amplitudes, the crossing property has been proven in the framework of the LSZ formalism \cite{crossing}.

Hermitian analyticity \cite{olive, Miramontes} states that the boundary value of $F$ on the right cut, approached from Im$\,s<0$, is given by the complex conjugate of the physical boundary value on the other side of the cut. This property is believed to hold quite generally \cite{iago}, but there seems to exist no rigorous proof of it. It is definitely valid in the family of S-matrices we will be studying later. In combination with crossing symmetry, hermitian analyticity implies that the boundary values of $F$ at the lower side of the right cut and at the upper side of the left cut coincide, see figure \ref{fig:crossing}.
\\
\\
In the rapidity picture, we consider instead of $F$ the function
\begin{align}
S_2(\te)
&:=
F(s(\te,0))
=
F(2m^2(1+\cosh\te))\,,\qquad \te\geq 0\,,
\end{align}
which contains all information about scattering processes with two incoming and two outgoing particles.

\begin{figure}[here]
    \noindent
    \psfrag{imt}{${\rm Im}\,\te$}
    \psfrag{ret}{${\rm Re}\,\te$}
    \psfrag{0}{$0$}
    \psfrag{ipi}{$i\,\pi$}
    \psfrag{t}{$\te$}
    \psfrag{-t}{$-\te$}
    \psfrag{ipi+t}{$i\pi+\te$}
    \psfrag{ipi-t}{$i\pi-\te$}
    \psfrag{s2t}{$S_2(\te)$}
    \psfrag{s2-t}{$\overline{S_2(\te)}$}
    \psfrag{s2-tp}{$S_2(\te)$}
    \psfrag{s2tp}{$\overline{S_2(\te)}$}
    \centering\epsfig{file=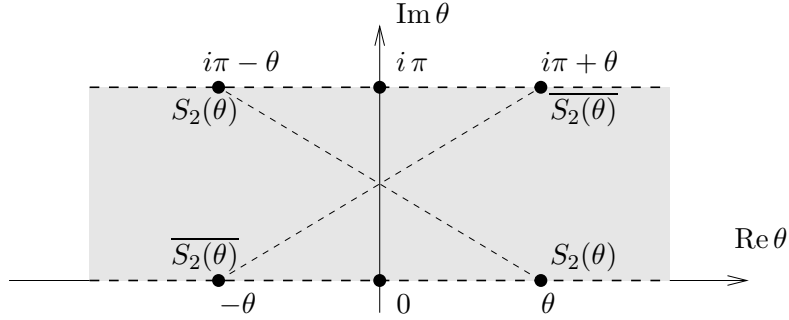,width=10cm}
   \caption{The relations between the different boundary values of $S_2$ as a consequence of the crossing symmetry and hermitian analyticity.}
\label{fig:s2strip}
\end{figure}

We now translate the properties of $F$ to properties of $S_2$. To begin with, note that the function $\te\lmto s(\te):=2m^2(1+\cosh\te)$ is a biholomorphic map from the strip
\begin{align}
S(0,\pi)
&:=
\{\zeta\in\Cl\,:\,0<{\rm Im}\,\zeta<\pi\}
\end{align}
to the cut plane in which $F$ is analytic. Hence $S_2=F\circ s$ is analytic in $S(0,\pi)$, and since the positive real half line, forming part of the boundary of $S(0,\pi)$, is mapped onto the upper boundary of the cut along $[4m^2,\infty)$ by $s$, the physical values of $S_2$ are obtained for real, positive rapidities. The lower boundary of this cut is the image of the negative real half line in the rapidity picture, and therefore the property of hermitian analyticity reads for $S_2$
\begin{align}\label{s2-hermitiananalyticity}
S_2(-\te)
&=
\overline{S_2(\te)}
\,,\qquad \te\in\Rl\,.
\end{align}
The change of variables $s\to 4m^2-s$ inherent in the crossing symmetry is in the rapidity parametrization given by $\te\to i\pi-\te$, and we thus have
\begin{align}\label{s2-crossing}
S_2(\te)
&=
S_2(i\pi-\te)
\,,\qquad\te\in\Rl\,.
\end{align}
The relations between the boundary values of $S_2$ are illustrated in figure \ref{fig:s2strip}.

\section{Factorizing S-Matrices and their Scattering Functions}\label{sec:factorizingS}

Whereas the structure of the two-particle S-matrix has been studied extensively, much less is known about the higher S-matrix elements $S_{n,m}$, $n,m>2$. The full S-matrix of an interacting quantum field theory is in general a very complicated object, as can be inferred from the existence of several no-go theorems for "simple" S-matrices. For example, it is known that if there is no particle production in a theory (i.e. $S_{n,m}=0$ for $n\neq m$), there can be no interaction at all \cite{aks}. Also if there exist conserved quantities in collision processes which transform like higher Lorentz tensors, the S-matrix has to be trivial, as the Coleman-Mandula theorem (under some additional assumptions) states \cite{CM}. Both of these examples apply if the spacetime dimension is higher than two \cite{Shankar-Witten}, and in fact, not a single non-trivial S-matrix is known in this case.

In two spacetime dimensions, however, the situation is quite different, the interesting point being that there do exist S-matrices which admit higher spin conserved charges despite describing non-trivial interaction. These special scattering operators have first been found in the context of quantized versions of completely integrable classical field theories, for example in the Sine-Gordon theory \cite{Zamo-SG}. In such models, there exists an infinite number of conservation laws which severely restrict the dynamics. In particular, the particle number is a conserved quantity in collision processes, despite the dynamics being fully relativistic. Moreover, the $n$-particle S-matrix factorizes into a product of several two-particle S-matrices, which motivated the name "factorizing S-matrices" for these objects.

A model-independent treatment of factorizing S-matrices was given in \cite{Iago-fact}. In the case of a single species of massive particles without any internal degrees of freedom an S-matrix is defined to be factorizing if its kernels are of the form
\begin{align}\label{factsnm}
S_{n,n}(\te_1,...,\te_n;\te_1',...,\te_n')
&=
S^0_{n,n}(\te_1,...,\te_n;\te_1',...,\te_n')\cdot\prod_{1\leq l<k\leq n} S_2(|\te_k-\te_l|)\,,\\
S_{n,m}&=0\;,\qquad n\neq m\,,
\end{align}
where
\begin{align}\label{id-kernels}
S^0_{n,m}(\te_1,...,\te_n;\te_1',...,\te_m')
&=
\frac{\delta_{nm}}{n!}\sum_{\pi\in\frS_n}\prod_{k=1}^n\delta(\te_k-\te_{\pi(k)}')
\end{align}
are the kernels of the trivial S-matrix $S^0={\rm id}$ and $\frS_n$ denotes the group of permutations of $n$ objects.

A factorizing S-matrix shares two important properties with the free S-matrix $S^0$: Firstly, $S_{n,m}$ vanishes for $n\neq m$, expressing the fact that transition amplitudes from $n$-particle states to $m$-particle states vanish, i.e. there is no particle production in the theory. Secondly, $S_{n,n}(\te_1,...,\te_n;\te_1',...,\te_n')$ vanishes if $\{\te_1,...,\te_n\}\neq\{\te_1',...,\te_n'\}$, that is, the sets of incoming and outgoing rapidities coincide.

Physically, the factorization of the multiparticle S-matrix elements into two-particle S-matrix elements is expected to hold in theories where collision processes of $n$ particles can be treated as a succession of two-particle scattering processes, the movement between the two particle collisions being free. Despite these strong constraints on the interaction, there are observable effects in a collision process governed by a factorizing S-matrix which clearly show that there is non-trivial interaction. For example, due to the non-constant phase shift of $S_2$, time delays appear, and in theories describing particles with additional quantum numbers, these are dynamic quantities\footnote{If there are several species of particles present in the theory, or the particles carry internal quantum numbers, the two-particle S-matrix elements become (finite-dimensional) matrices. In this case, the order of multiplication in the product \bref{factsnm} has to be specified, which leads to postulating the Yang-Baxter equations as an additional requirement.}.

Factorizing S-matrices have been found in many two-dimensional quantum field theory models. As mentioned before, these are usually quantized versions of integrable classical field theories with an infinite number of conservation laws, prominent examples being the Sine-Gordon and Sinh-Gordon models, the Thirring model, the Ising model, and the nonlinear $\sigma$-model \cite{abdalla}. However, it is by no means clear that all factorizing S-matrices are realized by Lagrangian field theories, and in fact, we will construct an infinity of models to which a Lagrangian formulation or a classical counterpart is not known.
\\
\\
The assumption that an S-matrix factorizes simplifies its mathematical structure drastically, as now the full S-matrix can be described in terms of a single function. This is the two-particle S-matrix element $S_2$, which will be called the {\em scattering function} in the context of a factorizing S-matrix. In particular, the so-called unitarity constraint, namely the requirement that the operator $S$ belonging to a family of kernels $S_{n,m}$ must be a unitary, translates into a simple condition on $S_2$, much in contrast to the general situation with particle production. In view of \bref{factsnm}, $S$ acts on each $n$-particle space of the Fock space of asymptotic wavefunctions as a multiplication operator\footnote{In the study of S-matrices in their own right, the operator multiplying with $\prod_{l<k}S_2(\te_k-\te_l)$, which differs from \bref{snfacts} by the missing absolute value in the scattering function, is often called the $n$-particle S-matrix, the identification with the physical S-matrix being understood.} , 
\begin{align}\label{snfacts}
(S\Psi)_n(\te_1,...,\te_n)
&=
\prod_{1\leq l < k \leq n}S_2(|\te_k-\te_l|)\cdot\Psi_n(\te_1,...,\te_n)
\,.
\end{align}
Hence the unitarity of $S$ is equivalent to $S_2$ being a phase, {\em i.e.}
 \begin{align}\label{s2-unitarity}
\overline{S_2(\te)}=S_2(\te)^{-1}
\,,\qquad
\te\in\Rl\,.
\end{align}
To summarize, a factorizing S-matrix is uniquely determined by its scattering function $S_2$, which satisfies the equations (\ref{s2-hermitiananalyticity}, \ref{s2-crossing}, \ref{s2-unitarity}). As we intend to do inverse scattering theory, we give a formal definition of the term "scattering function", which will be used as the starting point for the construction of models.
 \begin{definition}\label{def:s2}
  A scattering function is an analytic function $S_2: S(0,\pi) \lto\Cl$ which is bounded and continuous on the closure of this strip and satisfies the equations
  \begin{align}\label{s2-rel}
    \overline{S_2(\te)}
    \;=\;
    S_2(\te)^{-1}
    \;=\;
    S_2(-\te)
    \;=\;
    S_2(\te+i\pi)
    \,,\qquad
    \te\in\Rl\,.
  \end{align}
  The set of all scattering functions is denoted by $\SF$.
\end{definition}
Note that we have slightly strengthened the properties of $S_2$ in this definition by requiring continuous boundary values and boundedness on the strip. However, as $S_2$ must have modulus one on the boundary of $S(0,\pi)$ as a consequence of unitarity and crossing symmetry, this strengthening is very slight, since already a bound of the form $|S_2(\te+i\la)|\leq \exp(e^{c|\te|})$, $c<1$, for $\te+i\la\in S(0,\pi)$ implies that $S_2$ is in fact uniformly bounded \cite[Thm. 12.9]{rudin}.

From a mathematical point of view, it is also interesting to note that $\SF$ has the structure of a semi group under pointwise multiplication, i.e. the constant function $S_2(\te)=1$ is contained in $\SF$ as a neutral element, and $\SF$ is stable under taking products. As mentioned before, $S_2=1$ represents the trivial S-matrix $S^0={\rm id}$ and hence the interaction-free theory. A typical non-trivial example for a scattering function in $\SF$ is provided by the Sinh-Gordon model, i.e. the integrable model defined by the Lagrangian
\begin{align}
\LL_{\rm ShG}
&=
\frac{1}{2}\partial_\mu\phi(x)\,\partial^\mu\phi(x) - \frac{m^2}{g^2}\cosh(g\phi(x))
\,,
\end{align}
where $g$ is the (real) coupling constant. Extrapolating results obtained in perturbation theory, the scattering function of the Sinh-Gordon model is expected to be \cite{AFZ, BraSa}
\begin{align}
S_2(\te)
&=
\frac{\sinh\te-i\sin(\pi B)}{\sinh\te+i\sin(\pi B)}
\,,\qquad B:=\frac{g^2}{4\pi+g^2}\,.
\end{align}
Taking into account $0\leq B<1$, this function is easily seen to belong to the class $\SF$.
\\
\\
The constraints summarized in Definition \ref{def:s2} are so strong that they actually allow to calculate the most general form of $S_2$ by methods of complex analysis. More precisely, each scattering function $S_2\in\SF$ is uniquely fixed by its zeros and two more parameters, as stated in the following Proposition.
\begin{proposition}\label{prop:s2}
The set $\SF$ of scattering functions is
\begin{align}\label{s2-rep}
      \SF &=
      \left\{
        \zeta\lmto  \eps
      \cdot
      e^{ia\sinh\zeta}
      \cdot
      \prod_{k}
      \frac{\sinh\beta_k-\sinh\zeta}{\sinh\beta_k+\sinh\zeta}
      \;:\;
      \eps=\pm1,\,a\geq 0,\;\{\beta_k\}\in\ZZZ
      \right\}\,,
    \end{align}
    where the family $\ZZZ$ consists of the finite or infinite sequences $\{\beta_k\}\subset\Cl$ satisfying the following conditions:
\begin{itemize}
\item[i)] $0<\mathrm{Im}\beta_k\leq\frac{\pi}{2}$,
\item[ii)] $\beta_k$ and $-\overline{\beta_k}$ appear the same (finite) number of times in the sequence $\{\beta_k\}$,
\item[iii)] $\{\beta_k\}$ has no finite limit point,
\item[iv)] $\sum_k \mathrm{Im}\frac{1}{\sinh\beta_k}<\infty$.
\end{itemize}
The product in \bref{s2-rep} converges absolutely and uniformly in $\zeta$ on compact subsets of the strip $S(0,\pi)$.
\end{proposition}
\noindent A similar result has been obtained by Mitra \cite{mitra}. The proof of Proposition \ref{prop:s2} can be found in appendix \ref{sec:formofs2}.
\\
\\
The only {\em constant} scattering functions are $S_2=\pm 1$. Whereas $S_2=+1$ corresponds to the free theory, $S_2=-1$ is realized in the scaling limit of the Ising model \cite{BKW}. As mentioned above, the scattering function of the Sinh-Gordon model is an element of $\SF$ for any value $g>0$ of the coupling constant. But for the more general functions \bref{s2-rep} in $\SF$, no corresponding quantum field theoretic model is known.
\\
\\
It is interesting to notice, and will turn out to be of some relevance later, that the structure of $S_2$ implies stronger analyticity properties than just analyticity in the physical sheet $S(0,\pi)$. Denoting open strips in $\Cl$ by
\begin{align}\label{strip-not}
S(a,b)	:= \big\{\zeta\in\Cl\,:\,a<{\rm Im}\,<b\big\}
\,,\qquad
a<b\,,
\end{align}
we observe that as a consequence of $S_2(-\te)=S_2(\te)^{-1}$ and the continuity of $S_2$ on the real line, this function extends to a meromorphic function on $S(-\pi,\pi)$. The zeros $\beta_k$ of $S_2$ correspond to poles at $-\beta_k\in S(-\pi,0)$, which cannot accumulate to an essential singularity since $\{\beta_k\}$ has no finite limit point. Moreover, the crossing symmetry implies
\begin{align}
S_2(\te-i\pi)	=	S_2(-\te+i\pi)^{-1}	=	S_2(\te)^{-1}	=	S_2(\te+i\pi)\,,
\end{align}
and hence each $S_2$ continues to a $(2\pi i)$-periodic, meromorphic function on all of $\Cl$. The connection between zeros in $S(0,\pi)$ and poles in $S(-\pi,0)$ (and $S(\pi,2\pi)$) also shows that $S_2$ is actually analytic in the strip $S(-\kappa(S_2),\pi+\kappa(S_2))$, where
\begin{align}\label{def:kappa}
\kappa(S_2)
:=
\inf\big\{{\rm Im}\,\zeta\,:\,\zeta\in S(0,\tfrac{\pi}{2})\,,\quad S_2(\zeta)=0\big\}\,.
\end{align}
There exist scattering functions which have an infinite sequence of zeros approaching the real line, resulting in $\kappa(S_2)=0$. But for a large subfamily of $\SF$, in particular for all scattering functions having finitely many zeros in the physical strip, $\kappa(S_2)$ is strictly larger than zero, and hence $S_2$ can be analytically continued to a strip broader than $S(0,\pi)$. For later reference, we state the following Lemma.
\begin{lemma}
Consider a scattering function $S_2\in\SF$ with finitely many zeros in $S(0,\pi)$, and with parameter $a=0$ in \bref{s2-rep}. Then $S_2$ can be analytically continued to $S(-\kappa(S_2),\pi+\kappa(S_2))$, where $\kappa(S_2)>0$ \bref{def:kappa}. Moreover, $S_2$ is uniformly bounded on each strip $S(-\kappa,\pi+\kappa)$, $\kappa<\kappa(S_2)$.
\end{lemma}
\begin{proof}
By Proposition \ref{prop:s2}, the assumptions imply that $S_2$ is of the form
\begin{align}
S_2(\zeta)
=
\pm \prod_{k=1}^N  \frac{\sinh\beta_k-\sinh\zeta}{\sinh\beta_k+\sinh\zeta}
\;,\qquad 0< {\rm Im}\,\beta_1,...,{\rm Im}\,\beta_N \leq \tfrac{\pi}{2}\,,\;\;N<\infty\,.
\end{align}
The continuation to $S(-\kappa(S_2),\pi+\kappa(S_2))$, with $\kappa(S_2)=\min_{k=1,...,N}{\rm Im}\,\beta_k$, is clear from this formula. For the boundedness, note that $|S_2(\te+i\la)|\to 1$ for $\te\to\pm\infty$, $\la\in[0,\pi]$ fixed, and this convergence is uniform in $\la$. As $S_2$ has no zeros in $S(0,\kappa(S_2))$, this implies via the minimum principle that we have the lower bound  
\begin{align*}
|S_2(\zeta)|\geq\min\{|S_2(\te+i\kappa)|\,:\,\te\in\Rl\}
=:
\|S_2\|_\kappa^{-1}
>0
\,,\qquad
\zeta\in S(0,\kappa)\,,\;\kappa<\kappa(S_2)\,.
\end{align*}
The relations $S_2(-\zeta)=S_2(\zeta)^{-1} = S_2(i\pi+\zeta)$ then yield $|S_2(\zeta)|\leq \|S_2\|_\kappa<\infty$ for each $\zeta\in S(0,\kappa) \cup S(\pi,\pi+\kappa)$. Together with the bound $|S_2(\zeta)|\leq 1$, $\zeta\in S(0,\pi)$, this gives the desired result.
\end{proof}
We close this section by pointing out that some restriction on the distribution of the zeros of the scattering function which goes beyond the properties summarized in Proposition \ref{prop:s2} can be expected for physical reasons. As was shown above, a zero of $S_2$ in $S(0,\pi)$ corresponds to a pole in $S(-\pi,0)$ and vice versa. Whereas poles in the physical sheet  $S(0,\pi)$ are related to stable bound states, and are excluded in the present discussion, poles in the "unphysical sheet" $S(-\pi,0)$ are interpreted as evidence for unstable particles. Heuristically, an unstable particle is modelled by a complex mass $m_R$ with negative imaginary part, $m_R=m_R^{\rm phys}-i\Gamma/2$, where $\Gamma^{-1}>0$ is taken to be the lifetime of this unstable particle \cite{weinberg,castro}. Hence in the rapidity picture, a pole in $S(-\pi,0)$ corresponds to such a resonance, and by expressing the Mandelstam variable $s= (m_R^{\rm phys}-i\Gamma/2)^2$ in terms of the rapidity one sees that the lifetime of the resonance is the longer the closer the pole lies to the real line.

If a scattering function exhibits a sequence of infinitely many zeros approaching the real line, there are also infinitely many resonances present in the theory, the lifetimes of which can become arbitrarily long. In fact, it is possible to choose a distribution of zeros complying with the conditions listed in Proposition \ref{prop:s2} such that the associated unstable particles have unbounded lifetimes and ``masses'' $m_k$ so that $\sum_k e^{-m_k/T}$ diverges for all temperatures $T>0$. But a model with these characteristics
cannot be expected to have a regular thermodynamical behavior or only a finite partition function \cite{BuWi, BuJu2}.

Later on, we will therefore restrict to scattering functions which do not exhibit this behavior.

\section{The Form Factor Program}\label{sec:ffp}

The form factor program is a constructive approach to quantum field theory which has been developed in the late seventies, with the aim to construct model theories in the Wightman framework from the input of a factorizing S-matrix. As the goals of the form factor program and our constructions are very similar, we give here a brief introduction to the basic ideas of the form factor program. For a more complete treatment of the subject, see \cite{karo, BKW,Babujian:2001wp,smirnov,FMS,Babujian:2003sc}, and the references cited therein.
\\
\\
As we are interested in the essential concepts, we describe the program in the setting of theories with the simple particle spectrum specified before, although it is not restricted to this case. Fixing a factorizing S-matrix by means of its scattering function $S_2$, the task is to construct the $n$-point functions of a Wightman quantum field theory \cite{streater} which has the S-matrix corresponding to $S_2$.

The construction is carried out on the space of scattering states, and at its basis lies a certain algebraic structure of these states, which is named the {\em Zamolodchikov algebra} or {\em Zamolodchikov's algebra} after its inventors \cite{ZZ}. It can be motivated as follows: Consider $n$ idealized particles on the one-dimensional line with sharp rapidities $\te_1,...,\te_n$. Each particle is represented by a symbol $Z^*(\te)$, where $\te=\te_k$ is its rapidity. States of more than one particle are symbolized by expressions like $Z^*(\te_1) \cdots Z^*(\te_n)$, where the ordering of the symbols $Z^*(\te_k)$ in this product is given by the spatial ordering of the corresponding particles on the line. If the particles are arranged from the left to the right in the order of increasing rapidities, the symbol $Z^*(\te_1)\cdots Z^*(\te_n)$, $\te_1<...<\te_n$, is taken to represent an outgoing $n$-particle scattering state, since the faster particles are always located to the right of the slower ones and hence there is no interaction between them. (The interaction range is assumed to be zero.) Analogously, symbols arranged in order of decreasing rapidities, $Z^*(\te_1)\cdots Z^*(\te_n)$, $\te_1>...>\te_n$, stand for incoming collision states which have not interacted in the past. As outgoing and incoming scattering states should be related by the factorizing S-matrix based on $S_2$ \bref{snfacts}, one imposes the relation
\begin{align}\label{zfalg1}
Z^*(\te_n)\cdots Z^*(\te_1)
=
\prod_{1\leq l<k\leq n}S_2(\te_k-\te_l)
\cdot
Z^*(\te_1)\cdots Z^*(\te_n)
\,,\qquad 
\te_1<...<\te_n\,.
\end{align}
The symbols $Z^*(\te)$ are elements of an abstract non-commutative algebra, and their physical interpretation is that they create a single particle state with rapidity $\te$ from the vacuum state. Multiplying \bref{zfalg1} from the left with the inverse $n$-particle S-matrix, and taking into account $S(\te)^{-1}=S(-\te)$, the Zamolodchikov brothers obtained the exchange relation \cite{ZZ}
\begin{align}\label{zalgffp1}
Z^*(\te_1)Z^*(\te_2)
=
S_2(\te_1-\te_2)
\,Z^*(\te_2)Z^*(\te_1)
\,,\qquad \te_1,\te_2\in\Rl\,.
\end{align}
This structure was completed by Faddeev \cite{Faddeev} by adding a corresponding annihilation operator $Z(\te)$ and postulating the algebraic relations
\begin{align}\label{zalgffp2}
Z(\te_1)Z(\te_2) 		&= 		S_2(\te_1-\te_2) \,Z(\te_2)Z(\te_1)\,,\\
Z(\te_1)Z^*(\te_2) 	&=		S_2(\te_2-\te_1) \,Z^*(\te_2) Z(\te_1) + \delta(\te_1-\te_2)\cdot 1 \,,
\end{align}
between $Z$ and $Z^*$. (Here $1$ denotes the identity in the algebra.) The algebra given by the relations \bref{zalgffp1} and \bref{zalgffp2} is usually referred to as the Zamolodchikov algebra or the Zamolodchikov-Faddeev algebra. Setting $Z(\te)^*=Z^*(\te)$, it acquires a $*$-structure, and can be represented on a Hilbert space by defining a vacuum state on it which is annihilated by $Z(\te)$. (This  representation will be discussed in detail in the next chapter.)
\\
\\
Taking the Zamolodchikov-Faddeev algebra as represented on a Hilbert space with vacuum vector $\Om$, one considers in the form factor program expressions of the form
\begin{align}
F^{A(x)}_n(\te_1,...,\te_n;\te_1',...,\te_m')
&:=
\langle Z^*(\te_1)\cdots Z^*(\te_n)\Om,\,A(x)\,Z^*(\te_1')\cdots Z^*(\te_m')\Om\rangle\,,
\end{align}
where $A(x)$ is a (still to be constructed) local operator, taken to be localized at the spacetime point $x$. The $F^{A(x)}_n$ are called (generalized) {\em form factors} and constitute the main objects of interest in this approach. The relations of Zamolodchikov's algebra, the interpretation of $Z^*(\te_1)\cdots Z^*(\te_n)\Om$ as asymptotic scattering states for certain orderings of the rapidities, and the assumed locality of $A(x)$ give rise to many relations between different form factors, for example
\begin{align}
F^{A(x)}_n(\te_1,...,\te_n;\te_1',...,\te_m')
&=
F^{A(x)}_n(\te_1,...,\te_n;\te_2',\te_1',...,\te_m')\cdot S_2(\te_1'-\te_2')
\end{align}
as a consequence of \bref{zalgffp1}. Other relations are given by the crossing symmetry of $S_2$, or follow from the requirement of Lorentz covariance for the field $A(x)$, which is attributed some spin; but we do not write down all these equations here as we are  interested primarily in the general strategy of the program. 
The emerging set of equations is then promoted to a system of axioms \cite{smirnov}, usually amended by additional assumptions about maximal analyticity domains for the form factors.
\\
\\
Starting from these axioms, the construction of a model theory in the form factor program consists of the following three steps \cite{Babujian:2003sc}.

To begin with, the scattering function, respectively the two-particle S-matrix in the case of a richer particle spectrum, is calculated. For the simplest case, the possible scattering functions are given by Proposition \ref{prop:s2}, but in general, also the more complicated Yang-Baxter equations have to be solved. In a second step, the form factors are calculated by solving the system of axiom equations for certain local objects $A(x)$. Finally, the $n$-point Wightman functions of the theory are expressed in terms of the form factors by inserting a complete set of intermediate scattering states in the vacuum expectation values of the field operators. For example, the two-point function of a hermitian local operator $A$ is given (up to some constant factor) by \cite{Babujian:2001wp}
\begin{align}\label{ffp-npf}
\langle\Om,\,A(x)A(0)\Om\rangle
=
\sum_{n=0}^\infty
\frac{1}{n!}
\int d\te_1\cdots\int d\te_n\,e^{-ix\cdot\sum_{k=1}^n p(\te_k)}
\left|\langle\Om\,|\, A(0)\,|\te_1,...,\te_n\rangle\iin\right|^2
\,,
\end{align}
where $|\te_1,...,\te_n\rangle\iin$ denotes the incoming $n$-particle state corresponding to $Z^*(\te_1)\cdots Z^*(\te_n)\Om$.
The distributions obtained in this way are then interpreted as the Wightman functions of the constructed theory.
\\
\\
The first two steps of this program, the determination of appropriate two-particle S-matrices and the calculation of their associated form factors, have been carried out in many models, for example in the Sinh-Gordon \cite{FMS} and Sine-Gordon models \cite{Babujian:1998uw} and their generalizations \cite{castro}, the Ising model \cite{BKW}, and many more. The knowledge of form factors already allows to extract physical information, which has even led to concrete applications \cite{ffp-applications}.

The crucial step of completing the construction of these models, however, requires controlling series of the type \bref{ffp-npf}, and is a long-standing open problem. In fact, the converge properties of such sums have been thoroughly investigated only in two special models, namely the scaling limit of the Ising model, and the Yang-Lee model \cite{Babujian:2003sc, fringpr}. Even for these two best understood cases, no proof of the Wightman axioms for the resulting family of $n$-point functions is known to us\footnote{However, the convergence of the form factor series for the two-point function can be shown in the case of the Yang-Lee model \cite{smirnov-pc}.}. Furthermore, it has not been shown that the collision states of the so-defined theories (if they exist) reproduce the initially taken S-matrix, and also the property of asymptotic completeness, which is used as an assumption in the construction, seems to be hard to establish in this manner.
\\
\\
These problems point at the complicated structure local quantum fields have in the presence of non-trivial interactions, despite the relatively simple form of the S-matrix.
The problem of establishing convergence of series as \bref{ffp-npf} is reminiscent of the problems one faces in the usual perturbative approach to quantum field theory, and seems to be very difficult to overcome.

From a conceptual point of view, one might expect serious problems in trying to find the $n$-point functions of a quantum field from its S-matrix, as it is well-known that the field is not uniquely fixed by the scattering operator \cite{borchers:class}. Moreover, the explicit construction of field operators in the presence of interactions is a very ambitious task, and much more than what is needed to construct a quantum field theory with a given S-matrix, which requires only control over asymptotic properties.
\\
\\
Schroer suggested a complementary approach \cite{Schroer:1997cx,Schroer:1997cq,Schroer:1999xi,Schroer-Wiesbrock} to the problem of constructing quantum field theories from a factorizing S-matrix, which is the basis for our construction in the following chapter. In contrast to the form factor program, the starting point is an appropriate spacetime interpretation of (the vacuum representation of) Zamolodchikov's algebra. It turns out that field operators can be formed in terms of Zamolodchikov annihilation and creation operators in a simple way, and that these fields can be consistently interpreted as being localized in wedge regions. This observation links the concrete problem of finding quantum field theory models with a given factorizing S-matrix with the more abstract problem of defining quantum field theories using wedge algebras, as discussed in chapter \ref{chapter:netsin2d}.

As we shall see, by generating wedge algebras with Schroer's wedge-local fields, one can obtain a local quantum field theory along the same lines as in chapter \ref{chapter:netsin2d}, which for an infinite family of scattering functions can be shown to have all physically important properties. Moreover, it is possible to explicitly compute the multiparticle scattering states of this theory, and prove that it is asymptotically complete. The S-matrix is found to coincide with the initially given one, thus establishing the construction as the solution to the inverse scattering problem for this family of S-matrices.

\chapter{A Family of Models with Factorizing S-Matrices}
\label{chapter:wedgenet}

This chapter is devoted to the construction of a family of two-dimensional quantum field theory models with prescribed factorizing S-matrices. We consider a single species of massive, scalar, neutral particles and a factorizing S-matrix $S$ which is fixed by its scattering function $S_2$ (Def. \ref{def:s2}). As in the form factor program, the construction is based on the Zamolodchikov-Faddeev algebra $\Z(S_2)$ with scattering function $S_2$, which is rigorously defined in section \ref{sec:zf-algebra}. We then define the vacuum Hilbert space for the models to be constructed as a representation space of the Zamolodchikov algebra.

Following the proposals of Schroer \cite{Schroer:1997cq,Schroer:1997cx}, a quantum field $\phi$ is then defined explicitly as an unbounded operator on $\Hil$. This field is a non-local auxiliary object in the construction, not to be confused with a local physical field defining the model in the sense of Wightman theory. The localization properties of $\phi$ are discussed in section \ref{sec:phi}. It is a crucial point for the constructive program followed here that $\phi$ is not completely delocalized, but can be consistently interpreted as being localized in a wedge region. This opens up the possibility to generate a Poincar\'e-covariant net of wedge-local observable algebras in terms of $\phi$. As in the abstract construction in chapter \ref{chapter:netsin2d}, these wedge algebras determine a net of {\em local} algebras, which constitute the definition of the model constructed from the scattering function $S_2$.

The investigation of these local observables is postponed to chapter \ref{chapter:nuclearity}. In the present chapter, we compute the two-particle scattering states of the constructed models and prove that they reproduce the initially given scattering function $S_2$ (section \ref{sect:wedgescattering}). In the last section, we study the models in the framework of algebraic quantum field theory and make contact with the formalism of chapter \ref{chapter:netsin2d}.
\\
\\
The basic idea underlying the construction presented here is due to B. Schroer \cite{Schroer:1997cq,Schroer:1997cx,Schroer:1999xi} and Schroer and Wiesbrock \cite{Schroer-Wiesbrock}, who observed that the field $\phi$ might be localizable in a wedge, and investigated its properties as a polarization-free generator. The proof of the wedge-locality of $\phi$ and the computation of the two-particle scattering states have been found in \cite{GL-1}, and the analysis of the modular structure of the wedge net was carried out in the joint paper \cite{BuLe} with D. Buchholz.

\section{The Zamolodchikov-Faddeev Algebra}\label{sec:zf-algebra}

The algebraic structure lying at the root of our construction is the so-called Zamolodchikov-Faddeev algebra \cite{ZZ, Faddeev},  an algebra of creation and annihilation operators which satisfy quadratic exchange relations involving a scattering function. The Zamolodchikov-Faddeev algebra, mostly called Zamolodchikov algebra for brevity, is a common tool in the framework of the form factor program \cite{Babujian:2003sc,castro,FMS}. Fixing a scattering function $S_2\in\SF$ (Def. \ref{def:s2}), it is usually described as a $*$-algebra generated by non-commuting distributions $Z(\te)$, $Z^*(\te)$, which are parametrized by the rapidity $\te$ and satisfy the commutation relations, $\te_1,\te_2\in\Rl$,
\begin{subequations}\label{ZFdis}
\begin{align}
	Z(\te_1)Z(\te_2)		&=		S_2(\te_1-\te_2)\ Z(\te_2)Z(\te_1),\\
	Z^*(\te_1)Z^*(\te_2)	&=		S_2(\te_1-\te_2)\ Z^*(\te_2)Z^*(\te_1),\\
	Z(\te_1)Z^*(\te_2) 	&=		S_2(\te_2-\te_1)\,Z^*(\te_2) Z(\te_1) + \delta(\te_1-\te_2)\cdot 1 \,.
\end{align}
\end{subequations}
Here $1$ denotes the unit in the algebra, and all equations have to be understood in the sense of distributions. For the physical motivation of the relations \bref{ZFdis} in the context of scattering theory with the factorizing S-matrix given by $S_2$, see section \ref{sec:ffp}.

In order to give a rigorous definition of Zamolodchikov's algebra, we work with "smeared" quantities, formally given by $Z(\psi)=\int d\te\,Z(\te)\psi(\te)$, $\Zd(\psi)=\int d\te\,Z^*(\te)\psi(\te)$, where $\psi$ is a test function. Abstractly speaking, we consider the symbols $Z(\Psi_1),\Zd(\Psi_1)$, $\Psi_1\in L^2(\Rl,d\te)$, and more generally
\begin{align}
(Z^{\#_1}\times ... \times Z^{\#_n})(\Psi_n)\,,\qquad \Psi_n\in L^2(\Rl^n)\,,
\end{align}
where $Z^{\#_l}$ stands for $Z$ or $\Zd$, independently in each entry. These symbols are assumed to depend complex linearly on $\Psi_n$. They generate a linear space, and a product is defined by
\begin{align*}
	(Z^{\#_1} \times \ldots \times Z^{\#_n})(\Psi_n)\cdot (Z^{\#_1'} \times &\ldots \times Z^{\#'_m})(\Psi_m)
	\\
	&=
	(Z^{\#_1} \times \ldots \times Z^{\#_n} \times Z^{\#_1'} \times \ldots \times Z^{\#'_m})(\Psi_n \otimes \Psi_m)
	\,.
\end{align*}
We adjoin a unit $1$ as the neutral element. An involutive, antilinear star operation is fixed by setting $1^*=1$ and (with $Z^{\dagger\dagger}:=Z$)
\begin{align}
\big((Z^{\#_1}\times \ldots\times Z^{\#_n})(\Psi_n)\big)^*\label{ZFstar2}
 		&=
 	(Z^{\#_n\dagger} \times\ldots\times Z^{\#_1\dagger})(\Psi_n^*)
 	\,,\qquad
 	\Psi_n\in L^2(\Rl^n),\\
\Psi_n^*(\te_1,...,\te_n)&=\overline{\Psi_n(\te_n,...,\te_1)}\,.
\end{align}
{\em The Zamolodchikov-Faddeev algebra $\Z(S_2)$ with scattering function $S_2\in\SF$} is obtained by imposing the commutation relations \bref{ZFdis} on this free $*$-algebra. To formulate them, we adopt the convention to regard the scattering function $S_2$ also as a multiplication operator, acting on functions $\Psi_2\in L^2(\Rl^2)$ according to 
\begin{align}
	(S_2 \Psi_2)(\te_1,\te_2)	&:=	S_2(\te_1-\te_2)\cdot \Psi_2(\te_1,\te_2)\,.
\end{align}
With this notation, the definition of the algebraic structure of $\Z(S_2)$ is completed by requiring the relations
\begin{subequations}\label{ZFrelations}
\begin{align}
	Z(\psi)Z(\varphi)		&=	(Z\times Z)(S_2^*(\varphi \otimes \psi)), \label{ZFrelation1}
	\\
	Z(\psi)\Zd(\varphi)		&=	(\Zd\times Z)(S_2(\varphi\otimes \psi)) + \langle \overline{\psi},\,\varphi\rangle\cdot 1\,.\label{ZFrelation2}
\end{align}
\end{subequations}
Here the brackets $\langle\,.\,,\,.\,\rangle$ denote the scalar product on $L^2(\Rl,d\te)$, given as usual by the integral $\langle \psi,\varphi\rangle = \int d\te\,\overline{\psi(\te)}\varphi(\te)$. Note that the equations \bref{ZFrelations} arise from \bref{ZFdis} by formal integration against $\psi(\te_1)\varphi(\te_2)$ over $\te_1,\te_2\in\Rl$.

Recall that as a scattering function, $S_2$ satisfies the relations (Def. \ref{def:s2})
\begin{align}\label{eq:s2}
\overline{S_2(\te)}
    \;=\;
    S_2(\te)^{-1}
    \;=\;
    S_2(-\te)
    \;=\;
    S_2(\te+i\pi)
    \,,\qquad
    \te\in\Rl\,.
\end{align}
These imply in particular that the second equation in \bref{ZFrelations} is consistent with the involution \bref{ZFstar2}. Applying the star operation to the first commutation relation \bref{ZFrelation1}, we also obtain
\begin{align}
	\Zd(\psi)\Zd(\varphi)	&=	(\Zd\times \Zd)(S_2^*(\varphi \otimes \psi))\,.
\end{align}
It is worth noting that in case the scattering function is constant, $S_2=\pm 1$, the commutation relations \bref{ZFrelations} read
\begin{align}
	Z(\psi)Z(\varphi)		&=	\pm \,Z(\varphi)Z(\psi),\\
	Z(\psi)\Zd(\varphi)		&=	\pm \,\Zd(\varphi)Z(\psi) + \langle\overline{\psi},\,\varphi\rangle\cdot 1\,.
\end{align}
Disregarding the difference to the common convention according to which the "annihilation operator" $Z(\psi)$ depends {\em anti}\,linearely on $\psi$, the Zamolodchikov algebras $\Z(+1)$ and $\Z(-1)$ are thus isomorphic to the CCR and CAR algebras over $L^2(\Rl)$, respectively \cite{BraRob2}.
\\  
\\
We now turn to the description of the vacuum Hilbert space $\Hil$ of the model theory to be constructed, which carries a representation of $\Z(S_2)$. In the special cases $S_2=\pm 1$, the space $\Hil$ coincides with the Bose and Fermi Fock space over $L^2(\Rl,d\te)$, respectively. Also for a generic scattering function $S_2\in\SF$, it is a proper subspace of  the unsymmetrized Fock space $\F_{\Hil_1}:=\bigoplus_{n=0}^\infty \Hil_1\tp{n}$\label{defFH1} over $\Hil_1 := L^2(\Rl,d\te)$. As we shall see below, $\Hil$ carries a grading with respect to the "particle number", i.e. it has the structure $\Hil=\bigoplus_{n=0}^\infty\Hil_n$, $\Hil_n\subset\Hil_1\tp{n}=L^2(\Rl^n)$.
\\
\\
Proceeding analogously to the construction of the Bose Fock space as a subspace of $\F_{\Hil_1}$, we distinguish the functions in $\Hil_n$ by requiring invariance under a representation of the symmetric group $\frS_n$ on $\Hil_1\tp{n}$, the only discrepancy lying in a different representation of the permutations.

Let $\frS_n$ denote the group of permutations of $n$ elements, and $\tau_k\in\frS_n$, $k=1,...,n-1$, the transposition which exchanges $k$ and $(k+1)$. Instead of only permuting the arguments of functions in $L^2(\Rl^n)$, our symmetrization prescription uses the $S_2$-dependent operators
\begin{align}\label{def:Dn}
	(D_n(\tau_k) f_n)(\te_1,...,\te_n) 
	&:=
	S_2(\te_{k+1}-\te_k)\cdot f_n(\te_1,...,\te_{k+1},\te_k,...,\te_n) \,.
\end{align}
\begin{lemma}\label{lem:Dn}
Consider the map 
\begin{align}
	\frS_n \ni \tau_k \longmapsto D_n(\tau_k) \in \B(L^2(\Rl^n))\,.
\end{align}
\begin{enumerate}
\item $D_n$ defines a unitary representation of $\frS_n$ on $L^2(\Rl^n)$ which acts explicitly as
\begin{subequations}\label{dns-uebel}
\begin{align}
	\left(D_n(\pi) f_n\right)(\te_1,...,\te_n)
	&=
	S^\pi(\te_1,...,\te_n)\cdot f_n(\te_{\pi(1)},...,\te_{\pi(n)}), 
	\\
	S^\pi(\te_1,...,\te_n)
	&:=
	\prod_{1\leq l < k \leq n \atop \pi(l) > \pi(k)} S_2(\te_{\pi(l)}-\te_{\pi(k)})\,.
\end{align}
\end{subequations}
\item  The mean over $D_n$,
  \begin{align}\label{def:Pn}
    P_n := \frac{1}{n!}\sum_{\pi\in\frS_n}D_n(\pi)\,,
  \end{align}
is an orthogonal projection.
\end{enumerate}
\end{lemma}
\begin{proof}
{\em a)} In order to show that $D_n$ is a representation of $\frS_n$, one has to check the relations $D_n(\tau_k)^2=1$ for $k=1,...,n-1$ and $[D_n(\tau_j),D_n(\tau_k)]=0$ for $|j-k|>1$, as well as 
\begin{align}\label{frS-3}
	D_n(\tau_k)D_n(\tau_{k+1})D_n(\tau_k)
	&=
	D_n(\tau_{k+1})D_n(\tau_k)D_n(\tau_{k+1})\,,\;\qquad k=1,...,n-2\;.
\end{align}
The first relation follows from $S_2(-\te)=S_2(\te)^{-1}$ \bref{eq:s2}: Let $f_n\in L^2(\Rl^n)$, then
\begin{align*}
	\left(D_n(\tau_k)^2 f_n\right)(\te_1,...,\te_n)
	&=
	S_2(\te_{k+1}-\te_k)\cdot \left(D_n(\tau_k)f_n\right)(\te_1,...,\te_{k+1},\te_k,...,\te_n)
	\\
	&=
	S_2(\te_{k+1}-\te_k) \cdot S_2(\te_k-\te_{k+1}) \cdot f_n(\te_1,...,\te_n)
	\\
	&=
	f_n(\te_1,...,\te_n)\,.
\end{align*}
The second relation, $[D_n(\tau_j),D_n(\tau_k)]=0$ for $|j-k|>1$, holds because $D_n(\tau_k)$ acts only on the variables $\te_k$ and $\te_{k+1}$. Since $S_2$ is a multiplication operator, the different $S_2$-factors occurring in the computation of $D_n(\tau_k)D_n(\tau_{k+1})D_n(\tau_k)$ and $D_n(\tau_{k+1})D_n(\tau_k)D_n(\tau_{k+1})$ commute with each other, which implies the third relation \bref{frS-3}. Finally, $D_n(\tau_k)$ is unitary since $S_2$ has modulus unity, and this property carries over to arbitrary $D_n(\pi)$, $\pi\in\frS_n$, because the transpositions generate the symmetric group.

To verify the formula \bref{dns-uebel}, we first note that it agrees with the definition \bref{def:Dn} on the transpositions $\tau_j$. Now assume the formula is valid for some permutation $\pi$. Then there holds, $f_n\in L^2(\Rl^n)$, $j=1,...,n-1$, 
\begin{align}
\big(D_n(\pi\tau_j)f_n\big)(\te_1,..,\te_n)
&=
S^\pi(\te_1,..,\te_n)\cdot \big(D_n(\tau_j)f_n\big)(\te_{\pi(1)},..,\te_{\pi(n)})
\label{s2-factor}
\\
&=
S^\pi(\te_1,..,\te_n)\, S_2(\te_{\pi(j+1)}-\te_{\pi(j)})\, f_n(\te_{\pi(1)},..,\te_{\pi(j+1)},\te_{\pi(j)},..,\te_{\pi(n)}).
\nonumber
\end{align}
On the other hand, the function $S^{\pi\tau_j}$ is
\begin{align*}
	S^{\pi\tau_j}(\te_1,...,\te_n)
	&=
	\prod_{1\leq l<k\leq n\atop {\pi(l)>\pi(k) \atop (l,k)\neq (j,j+1)}} S_2(\te_{\pi(l)}- \te_{\pi(k)}) \cdot \prod_{\pi(j+1) > \pi(j)} S_2(\te_{\pi(j+1)}-\te_{\pi(j)})
	\\
	&=
	S^\pi(\te_1,...,\te_n)\cdot S_2(\te_{\pi(j+1)}-\te_{\pi(j)})\,,
\end{align*}
as can be seen by considering the two cases $\pi(j)<\pi(j+1)$ and $\pi(j)>\pi(j+1)$. Hence the factor of scattering functions appearing in \bref{s2-factor} coincides with $S^{\pi\tau_j}$, and so \bref{dns-uebel} holds for the permutations $\pi\tau_j$, provided it holds for $\pi$. Since the transpositions generate $\frS_n$, the validity of \bref{dns-uebel} follows by induction.

{\em b)} As $D_n$ is a unitary representation, we have $P_n=P_n^*$. The equation $P_n^2=P_n$ holds because $\frS_n$ is a group of $n!$ elements.
\end{proof}

\noindent{\em Remark:} A similar symmetrization procedure has also been used by Liguori and Mintchev in the context of Fock spaces with generalized statistics \cite{LiMi}.
\\\\
With the help of the projections $P_n$ we can symmetrize the unsymmetrized Fock space $\mathcal{F}_{\Hil_1}$ with respect to $D_n$, which yields the definition of our model Hilbert space $\Hil$:
\begin{align}\label{def:Hil}
  \Hil := \bigoplus_{n=0}^\infty \Hil_n,\qquad \Hil_n := P_n\Hil_1\tp{n},\qquad\Hil_0 := \Cl\,.
\end{align}
The projections $P_n$ will be considered as operators on $\mathcal{F}_{\Hil_1}$ by putting $P_n\Psi:=0$ for $\Psi\perp \Hil_n$, and we also write $P:=\bigoplus_{n=0}^\infty P_n \in\B(\F_{\Hil_1})$ for the orthogonal projection onto the subspace $\Hil$.

The interpretation of $\Hil$ is as follows. In a relativistic quantum field theory describing a single species of scalar, neutral particles of mass $m>0$, the single particle space can be realized in the momentum picture as the space of functions on the upper mass shell $H^+_m = \{((p_1^2+m^2)^{1/2},p_1)\,:\,p_1\in\Rl\}$, which are square integrable with respect to the measure $(p_1^2+m^2)^{-1/2}dp_1$. In the rapidity parametrization of $H^+_m$ \bref{def:pte}, this measure is simply the Lebesgue measure $d\te$. So $\Hil_1=L^2(\Rl,d\te)$ will be interpreted as the one particle space of our model, and $\Hil_n$ will be referred to as "$n$-particle spaces". Generalizing the totally symmetric functions known from the free bosonic field, its elements are those functions $\Psi_n\in L^2(\Rl^n)$ which are $S_2${\em-symmetric} in the sense that
\begin{align}\label{eq:S2sym}
	\Psi_n(\te_1,...,\te_{k+1},\te_k,...\te_n) 
  		&=
	S_2(\te_k-\te_{k+1}) \cdot \Psi_n(\te_1,...,\te_n)\,,\quad k\in\{1,...,n-1\}\,.
\end{align}
In the zero-particle space $\Hil_0$ we fix a unit vector $\Omega$, representing the physical vacuum. Generic elements of $\Hil$ are denoted $\Psi=(\Psi_0,\Psi_1,...\,)$, $\Psi_n\in\Hil_n$, and have the norm $\|\Psi\|^2=\sum_{n=0}^\infty\int d^n\bte|\Psi_n(\bte)|^2<\infty$.

We introduce the particle number operator $N$ as $(N\Psi)_n:=n\cdot\Psi_n$ on the vectors $\Psi$ with $\sum_n n^2\|\Psi_n\|^2<\infty$. The dense subspace of $\Hil$ which consists of vectors $\Psi$ with finite particle number, i.e. $\Psi_n=0$ for sufficiently large $n$, is denoted $\DD$.\label{defineDD}
\\
\\
On the unsymmetrized Fock space $\F_{\Hil_1}$, the proper orthochronous Poincar\'e group $\PGpo$ (see section \ref{Sec:Geo}) acts via the representation $\Uhat$, defined as
\begin{align}\label{def:Uconc}
	\big(\Uhat(x,\la) \Psi\big)_n(\te_1,...,\te_n) 
		&:=
	\exp\bigg(i\sum_{k=1}^n p(\te_k) \cdot x \bigg) \cdot \Psi_n(\te_1-\la,...,\te_n-\la)\,,
\end{align}
where $p(\te)=m(\cosh\te,\sinh\te)^T$ \bref{def:pte} and $m>0$ is the mass of the particles in our model. This representation can be restricted to $\Hil$: The translations $\Uhat(x,0)$ preserve the symmetry structure \bref{eq:S2sym} because they act by multiplication with totally symmetric functions, and also the boosts $\Uhat(0,\la)$ map $\Hil$ onto $\Hil$ because the scattering function in \bref{eq:S2sym} depends only on differences of rapidities.

We denote the restriction of $\Uhat$ to $\Hil$ by $U$, and take it as the definition of the representation of the Poincar\'e symmetries in the model to be constructed.

Clearly, $U$ is a strongly continuous positive energy representation\label{U-discussion} of $\PGpo$, and $\Om\in\Hil_0$ is the (up to a phase) unique $U$-invariant unit vector, justifying its interpretation as representing the physical vacuum state. The energy and momentum operators\footnote{Using standard notation, we denote these operators by $P_0,P_1$, as no confusion with the projections $P_n$ \bref{def:Pn} is likely to arise.} $P_0$, $P_1$ are defined as the generators of the translation subgroups in $x_0$- and $x_1$-direction, as usual. The mass operator $M:= (P_0^2-P_1^2)^{1/2}$ has $m$ as its only positive eigenvalue, with eigenspace $\Hil_1$. So also the interpretation of vectors $\Psi_1\in\Hil_1$ as single particle states complies with the definition of $U$. The interpretation of vectors $\Psi_n\in\Hil_n$ as $n$-particle states will be justified later in scattering theory. Until then, the term "$n$-particle space" will be used just as a name for $\Hil_n$.

Note that as a representation on $\F_{\Hil_1}$, $\Uhat$ can be extended to the proper Poincar\'e group $\PG_+\supset\PGpo$, which also contains the spacetime reflection $-1$, mapping $x\in\Rl^2$ to $-x$. Its representative $\Uhat(-1)$ is defined as complex conjugation,
\begin{align}\label{uhat-1}
	(\Uhat(-1) \Psi)_n(\te_1,...,\te_n) &:= \overline{\Psi_n(\te_1,...,\te_n)}\,,\qquad\Psi\in\F_{\Hil_1}\,.
\end{align}
But complex conjugation does not preserve the $S_2$-symmetry \bref{eq:S2sym}, since the scattering function is not real, in general. (The only exceptions are the constant functions $S_2=\pm 1$.) Hence $\Uhat(-1)$ cannot be restricted to the subspace $\Hil$. Later on, we will find a unitary $U(-1)\in\B(\Hil)$, extending $U$ to $\PG_+$, as an important step in the construction of the wedge-local observables.

\subsection{Representation of Zamolodchikov's Algebra}
Guided by the Fock representation of the CCR algebra, we now introduce properly symmetrized creation and annihilation operators on $\Hil$. We start from their unsymmetrized counterparts $a(\psi),\ad(\psi)$, $\psi\in\Hil_1$, which are defined on the subspace $\F_{\Hil_1}^{(0)}\subset\F_{\Hil_1}$ of finite particle number by
\begin{align}
   a(\psi)\Omega &:= 0\,,
   	\qquad&
(a(\psi)\Phi)_n(\te_1,...,\te_n) &:= \sqrt{n+1}\,\int d\te \, \psi(\te)\Phi_{n+1}(\te,\te_1,...,\te_n) \label{def:a}\,,\\
 \ad(\psi)\Omega &:=  \psi\,,
\qquad&
(\ad(\psi)\Phi)_n  &:= \sqrt{n}\;\psi(\te_1) \cdot \Phi_{n-1}(\te_2,...,\te_n)\,. \label{def:ad} 
\end{align}
Note that these operators are related to each other according to $a(\psi)^* \supset \ad(\overline{\psi})$, as can be seen by taking adjoints on $\F_{\Hil_1}$.

The domain of the representation of the Zamolodchikov algebra will be the subspace $\DD=P\F_{\Hil_1}^{(0)}\subset\Hil$ of $S_2$-symmetric vectors of finite particle number.
\begin{lemma}\label{z-lemma}
{\quad}\\
\vspace*{-12mm}\\
\begin{enumerate}
\item The map 
  \begin{align}\label{def:zzd}
  Z(\psi) \longmapsto z(\psi):=P a(\psi)P,\qquad \Zd(\psi) \longmapsto \zd(\psi):=P\ad(\psi)P
  \end{align}
  extends to a representation of the Zamolodchikov-Faddeev algebra $\Z(S_2)$ on $\DD$ which has $\Omega$ as a cyclic vector.
\item On $\Phi\in\DD$ one has, $\psi\in L^2(\Rl)$,
  \begin{equation}
    \zd(\psi)\Phi = P\ad(\psi)\Phi,\quad\qquad z(\psi)\Phi = a(\psi)\Phi\,,
  \end{equation}
and explicitly
\begin{align}
	(z(\psi)\Phi)_n(\te_1,...,\te_n)		&=	\sqrt{n+1} \int d\te\, \psi(\te)\,\Phi_{n+1}(\te,\te_1,...,\te_n)
	\,,\label{act-z}
	\\
	(\zd(\psi)\Phi)_n(\te_1,...,\te_n)
		&=
	\frac{1}{\sqrt{n}} \sum_{k=1}^{n} \prod_{j=1}^{k-1}S_2(\te_k-\te_j)\, \psi(\te_k)\, \Phi_{n-1}(\te_1,\te_2,...,\widehat{\te}_k,...,\te_n)
	\,.\label{act-zd}
\end{align}
Here the hat on $\te_k$ indicates that this variable is omitted.
\end{enumerate}
\end{lemma}
\begin{proof}
We begin with the proof of {\em b)}. The annihilator $a(\psi)$ preserves the $S_2$-symmetry \bref{eq:S2sym} in the remaining $n-1$ variables when applied to a function $\Phi_n\in\Hil_n$ \bref{def:a}. Hence $a(\psi) \Hil_n \subset \Hil_{n-1}$, and thus the projections in the definition of $z(\psi)$ may be omitted, $z(\psi)\Phi=a(\psi)\Phi$ for all $\Phi\in\DD$. Equation \bref{act-z} is just a reformulation of this fact. Since $P=1$ on $\DD$, the equation $\zd(\psi)\Phi=P\ad(\psi)\Phi$ follows as well.
	
To compute the action of the creation operator $\zd(\psi)$, we introduce the special permutations $\sigma_k\in\frS_n$, $k=1,...,n$, defined by
\begin{align}\label{def:sk}
	\sigma_k 	&:= 	\tau_{k-1}\tau_{k-2}\cdots\tau_1\,,\qquad k=2,...,n\,,\qquad \sigma_1={\rm id}\,.
\end{align}
Any permutation $\pi\in\frS_n$ can be uniquely decomposed as $\pi=\sigma_k\rho$, where $\sigma_k\in\frS_n$ and $\rho\in\frS_{n-1}$ acts on the subset $\{2,...,n\}\subset\{1,...,n\}$. For the projection $P_n$ this entails
\begin{align}\label{eq:Pn}
	P_n	&=	\frac{1}{n!}\sum_{k=1}^{n}\sum_{\rho\in\frS_n} D_n(\sigma_k)\left(1\otimes D_{n-1}(\rho)\right)
		= 
	\frac{1}{n}\sum_{k=1}^{n} D_n(\sigma_k)\left(1\otimes P_{n-1}\right)\,.
\end{align}
As $\sigma_k(1,2,...,n)=(k,1,2,...,\widehat{k},...,n)$, it follows from \bref{dns-uebel} that the permutations $\sigma_k$ are represented as, $f_n\in L^2(\Rl^n)$,
\begin{align}\label{eq:sk}
	\left(D_n(\sigma_k)f_n\right)(\te_1,...,\te_n)	&=	\prod_{j=1}^{k-1}S_2(\te_k-\te_j) \cdot f_n(\te_k,\te_1,\te_2,...,\widehat{\te}_k,...,\te_n)\,.
\end{align}
The hat on $\te_k$ indicates that this variable is omitted. With these formulae, the action of the creation operator, $(\zd(\psi)\Phi)_n = \sqrt{n}\,P_n(\psi\otimes\Phi_{n-1})$ \bref{def:ad}, can be explicitly evaluated as
\begin{align*}
	(\zd(\psi)\Phi)_n(\te_1,...,\te_n)
	&=
	\frac{1}{\sqrt{n}} \sum_{k=1}^{n} \prod_{j=1}^{k-1}S_2(\te_k-\te_j) \psi(\te_k) \Phi_{n-1}(\te_1,\te_2,...,\widehat{\te}_k,...,\te_n)\,,
\end{align*}
in agreement with \bref{act-zd}.

{\em a)} The linearity of $\psi\lmto z^\#(\psi)$ follows directly from \bref{def:zzd}. As $a(\psi)^* \supset a(\overline{\psi})$, the $*$-operation acts like
\begin{align}
	z(\psi)^*	&= (Pa(\psi)P)^*	\supset P\ad(\overline{\psi})P = \zd(\overline{\psi})\,,
\end{align}
as in the abstract Zamolodchikov algebra \bref{ZFstar2}. Using \bref{act-z} and the $S_2$-symmetry of $\Phi\in\DD$, we obtain, $\psi,\varphi\in\Hil_1$, 
\begin{align}
	(z(\psi)z(\varphi)\Phi)_n&(\te_1,...,\te_n)
	=
	\sqrt{(n+1)(n+2)}\int d\te \, \psi(\te)\int d\te'\,\varphi(\te')\,\Phi_{n+2}(\te',\te,\te_1,\te_2,...,\te_n)
	\nonumber
	\\
	&=
	\sqrt{(n+1)(n+2)}\int d\te \int d\te' S_2(\te-\te')  \psi(\te) \varphi(\te') \,\Phi_{n+2}(\te,\te',\te_1,\te_2,...,\te_n)
	\,.\label{zz-exp}
\end{align}
Linear and continuous extension in $\psi\otimes\varphi\in L^2(\Rl^2)$ yields the definition of $(z\times z)$ as
\begin{align*}
\big((z\times z)(\Psi_2)\Phi\big)_n(\te_1,...,\te_n)
&=
\sqrt{(n+1)(n+2)}\int d\te \int d\te' \Psi_2(\te,\te') \,\Phi_{n+2}(\te',\te,\te_1,\te_2,...,\te_n).
\end{align*}
Since $S_2^*(\varphi\otimes\psi)(\te',\te)=S_2(\te-\te')\varphi(\te')\psi(\te)$, \bref{zz-exp} implies the first commutation relation of Zamolodchikov's algebra, $z(\psi)z(\varphi)=(z\times z)(S_2^*(\varphi\otimes\psi))$ \bref{ZFrelation1}. To derive the second relation \bref{ZFrelation2}, we compute
\begin{align}
	( z(\psi)\zd(\varphi) \Phi)_n(\te_1,..,\te_n)
	&=			
	\int \!\!d\te_0\, \psi(\te_0) \sum_{k=1}^{n+1} \prod_{j=1}^{k-1} S_2(\te_{k-1}-\te_{j-1}) \varphi(\te_{k-1}) \Phi_n(\te_0,..,\widehat{\te}_{k-1},..,\te_n)
	\nonumber
	\\
	&=
	\int\!\! d\te_0\, \psi(\te_0) \sum_{k=0}^{n} \prod_{j=0}^{k-1} S_2(\te_k-\te_j) \varphi(\te_k) \Phi_n(\te_0,\te_1,..,\widehat{\te}_k,..,\te_n)
	.\label{inlem:zref}
\end{align}
As above, linear and continuous extension in $\psi\otimes\varphi\in L^2(\Rl^2)$ yields the operators $(z\times \zd)$ ($(\zd\times \zd)$ and higher products $(z^{\#_1}\times ... \times z^{\#_n})(\Psi_n)$, $\Psi_n\in L^2(\Rl^n)$, are defined accordingly). 

The term corresponding to $k=0$ in \bref{inlem:zref} gives $\langle\overline{\psi},\varphi\rangle\cdot \Phi_n(\te_1,...,\te_n)$, as required in \bref{ZFrelation2}. The flipped operator $\zd(\varphi)z(\psi)$ acts as
\begin{align*}
	(\zd(\varphi)z(\psi)\Phi)_n(\te_1,...,\te_n)
	&=
	\sum_{k=1}^n \prod_{j=1}^{k-1} S_2(\te_k-\te_j)\, \int d\te_0\,\varphi(\te_k)  \psi(\te_0)\, \Phi	_n(\te_0,\te_1,...,\widehat{\te}_k,...,\te_n)
	\,.
\end{align*}
If $\varphi(\te_k)\psi(\te_0)$ is replaced by $S_2(\te_k-\te_0)\varphi(\te_k)\psi(\te_0)=S_2(\varphi\otimes \psi)(\te_k,\te_0)$, this expression coincides with the partial sum \bref{inlem:zref}, running over $k=1,...,n$. Hence we arrive at
\begin{align}
	(z(\psi)\zd(\varphi))\Phi
	&=
	\langle\overline{\psi},\varphi\rangle\cdot \Phi + ((\zd\times z)(S_2(\varphi\otimes \psi))\,\Phi,
\end{align}
reproducing \bref{ZFrelation2}. To finish the proof, note that the cyclicity of $\Om$ follows from
\begin{align}
	\zd(\psi_1)\cdots\zd(\psi_n)\Om	&= 	\sqrt{n!} \, P_n(\psi_1\otimes ...\otimes \psi_n)
\end{align}
since $P_n$ is linear and maps $\F_{\Hil_1}$ onto $\Hil$.
\end{proof}

If $S_2=+1$, the operators $z(\psi)$, $\zd(\psi)$ form a representation of the CCR algebra and are unbounded. In the case $S_2=-1$, however, it is well known that the CAR algebra $\Z(-1)$ is represented by bounded operators \cite{BraRob2}. For general scattering functions, the operators $z(\te_1)$, $\zd(\te_2)$ interpolate between these two extremal cases. Considering for example the typical scattering function
\begin{align}
S_2(\te)=\frac{\sinh\te-ib}{\sinh\te+ib}\,,\qquad 0<b<\pi\,,
\end{align}
the commutation relations between $z(\te_1), \zd(\te_2)$ resemble the relations of the CAR algebra if $\te_1$ and $\te_2$ are close to each other, since $S_2(0)=-1$. For far separated rapidities $\te_1-\te_2\to\infty$, however, $z(\te_1), \zd(\te_2)$ approximately obey the relations of the CCR algebra since $S_2(\te)\to+1$ for $\te\to\pm\infty$. So in general, the operators $z(\psi)$, $\zd(\psi)$ are unbounded. 
\\
\\
The bounds with respect to the particle number which are familiar from the CCR {\em and} CAR algebras, hold for arbitrary scattering functions, as the following simple Lemma shows.

\begin{lemma}\label{lemma-z-bounded}
For arbitrary scattering functions $S_2\in\SF$, the following bounds hold with respect to the particle number operator $N$:
\begin{align}\label{Z-bounds}
	\|z(\psi)\Phi\| 	\leq	\|\psi\|\cdot\|N^{1/2}\Phi\|
	\,,\quad
	\|\zd(\psi)\Phi\| 	\leq	\|\psi\|\cdot\|(N+1)^{1/2}\Phi\|
	\,,\quad \Phi\in\DD\,.
\end{align}
\end{lemma}
\begin{proof}
The unsymmetrized creation and annihilation operators are related to the particle number operator $N$ by  $a(\psi_1)\ad(\psi_2) = \langle\overline{\psi_1},\psi_2\rangle\cdot (N+1)$. Hence
\begin{align*}
	\|\ad(\psi)\Phi\|^2	&=		\|\psi\|^2\|(N+1)^{1/2}\Phi\|^2\,, \qquad \Phi\in\DD\,,
\end{align*}
which implies $\|\ad(\psi)(N+1)^{-1/2}\|=\|\psi\|$. By taking adjoints, we see that the operator  \,$(N+1)^{-1/2}a(\psi) = a(\psi)N^{-1/2}$ has also norm $\|\psi\|$. The claimed bounds for the Zamolodchikov operators $z^\#(\psi)=Pa^\#(\psi)P$ now follow from $\|P\|=1$.
\end{proof}

\section{Wedge-Local Quantum Fields}\label{sec:phi}

With the $S_2$-symmetric Hilbert space, the corresponding creation and annihilation operators, and the representation of the Poincar\'e symmetries on $\Hil$, we have introduced all objects necessary for the discussion of physical observables and their localization properties. These observables will be constructed with the help of two quantum fields $\phi$, $\phi'$ on $\Hil$, which we define and analyze in this section.

As mentioned before, in the special case of the scattering function $S_2=1$, the Zamolodchikov algebra coincides with the CCR algebra, and the Hilbert space $\Hil$ is the usual totally symmetric Fock space over $\Hil_1$. So in this case, the free, scalar field $\phi_0$ of mass $m$ can be defined as a sum of a creation and an annihilation operator $z,\zd$. Mimicking this construction, we now introduce a quantum field $\phi$ on two-dimensional Minkowski space in terms of the Zamolodchikov operators, but allow for an arbitrary scattering function $S_2\in\SF$.

In the following, $\Ss$ denotes the space of Schwartz test functions.
\begin{definition}\label{def:phi}
  Let $f\in\Ss(\R^2)$ 
 and set
  \begin{align}\label{phi}
    f^\pm(\te) := \frac{1}{2\pi}\int d^2x\,f(\pm x)e^{ip(\te)\cdot x},\qquad 
p(\te)=
m\left(
      \begin{array}{c}
        \cosh\te\\
        \sinh\te
      \end{array}
\right)\;.
  \end{align}
  The field operator $\phi(f)$ is defined as 
  \begin{align}\label{deff:phi}
    \phi(f) := \zd(f^+)+z(f^-)\;.
  \end{align}
\end{definition}
\noindent Since the functions $\te\lmto f^\pm(\te)=\fti(\mp m\cosh\te,\pm m\sinh\te)$ are defined as restrictions of the Fourier transform $\fti$ of $f\in\Ss(\Rl^2)$, we may consider $f^+$ and $f^-$ as vectors in $\Hil_1=L^2(\Rl)$.

The field $\phi$ was invented by B. Schroer \cite{Schroer:1997cx}. Note that it reproduces the free field $\phi_0$ in the case $S_2=1$. In the following Proposition we show that also for generic $S_2$, the field operator $\phi$ shares many properties with $\phi_0$, except locality. 
\begin{proposition}\label{prop:phi}
The field operator $\phi(f)$ has the following properties:
\begin{enumerate}
\item For any $S_2\in\SF$, $\phi(f)$ is defined on $\DD$ and leaves this space invariant. There holds the bound, $\Psi\in\DD$,
\begin{align}\label{bnd:phi}
	\|\phi(f)\Psi\|	&\leq		\big(\|f^+\|+\|f^-\|\big) \cdot \|(N+1)^{1/2}\Psi\|\,.
\end{align}
\item For $\Psi\in\DD$ one has
\begin{align}
	\phi(f)^*\Psi = \phi(\fbar)\Psi.
\end{align}
All vectors in $\DD$ are entire analytic for $\phi(f)$. If $f\in\Ss(\R^2)$ is real, $\phi(f)$ is
essentially selfadjoint on $\DD$.
\item $\phi$ is a solution of the Klein-Gordon equation of mass $m$: For every $f\in\Ss(\Rl^2)$, $\Psi\in\DD$ one has 
\begin{align}
	\phi((\Box+m^2)f)\Psi = 0\,.
\end{align}
\item $\phi(f)$ transforms covariantly under the representation $U$ of $\PGpo$ \bref{def:Uconc}:
\begin{align}\label{phi_cov}
	U(x,\la)\phi(f)U(x,\la)^{-1} 	&=	\phi(f_{(x,\la)})\,,\\
      	f_{(x,\la)}(y)				&=	f(\La(\la)^{-1}(y-x)),\quad x\in\Rl^2,\;\la\in\Rl\,.
\end{align}
Here $\La(\la)$ denotes the boost with rapidity $\la$ \bref{def:boost}.
\item The vacuum vector $\Omega$ is cyclic for the field $\phi$: Given any open set $\OO\subset\Rl^2$, the subspace 
\begin{align}
	\Ds_\OO &:= \mathrm{span}\big\{\phi(f_1)\cdots\phi(f_n)\Omega \,:\, f_1,...,f_n \in \Ss(\OO),\,n\in\N_0\big\}
\end{align}
   is dense in $\Hil$.
\item $\phi$ is local if and only if $S_2 = 1$. 
\end{enumerate}
\end{proposition}
\begin{proof}
{\em a)} These statements follow directly from the definition of $\phi(f)$, and the bounds given in Lemma \ref{lemma-z-bounded}.

To establish {\em b)}, one calculates $(\overline{f})^\pm = \overline{f^{\mp}}$, which implies
\begin{align*}
	\phi(f)^*\Psi = \big(\zd(f^+)^* + z(f^-)^*\big)\Psi = \big(z(\overline{f^+}) + \zd(\overline{f^-})\big)\Psi = \phi(\fbar)\Psi
\end{align*}
for $\Psi\in\DD$. In particular, $\phi(f)$ is hermitian for real $f$.
\\
Now let $\Psi_n\in\Hil_n$ and $c_f := \|f^+\|+\|f^-\|$. In view of the bound in $a)$, we have the estimates $\|\phi(f)\Psi_n\| \leq \sqrt{n+1}\,c_f\|\Psi_n\|$ and 
\begin{align*}
  \|\phi(f)^k\Psi_n\|
  \leq
  \sqrt{n+k}\,c_f\,\|\phi(f)^{k-1}\Psi_n\|
  \leq
  \sqrt{n+k}\cdots\sqrt{n+1}\, c_f^k\|\Psi_n\|\,,\qquad k\in\N.
\end{align*}
Thus, for arbitrary $\zeta\in\Cl$ there holds
\begin{equation*}
  \sum_{k=0}^\infty	\frac{|\zeta|^k}{k!}	\|\phi(f)^k \Psi_n\|
  \leq
  \|\Psi_n\|	\sum_{k=0}^\infty \sqrt{\frac{(n+k)!}{n!}}	\frac{1}{k!}(|\zeta|\,c_f)^k < \infty\,,
\end{equation*}
which shows that every $\Psi\in\DD$ is an entirely analytic vector for $\phi(f)$. Since $\DD$ is dense in $\Hil$, we can use Nelson's theorem \cite[Thm. X.39]{SimonReed2} to conclude that $\phi(f)$ is essentially selfadjoint on $\DD$ if $f$ is real. In the following we use the same symbol $\phi(f)$ for the selfadjoint closure of this operator.

{\em c)} is an immediate consequence of $((\Box+m^2)f)^\pm=0$. 

\noindent To prove {\em d)}, we choose $\varphi\in\Hil_1,\Psi\in\DD$,
$(x,\la)\in\PGpo$. Using the fact that $U$ commutes with the symmetrization $P_n$, and the second quantization structure of this representation \bref{def:Uconc}, we calculate 
\begin{align*}
	\big(U(x,\la)\zd(\varphi)U(x,\la)^*\Psi\big)_n
  	&=
	\sqrt{n}\;U(x,\la)P_n\big(\varphi\otimes U(x,\la)^*\Psi_{n-1}\big)
	\\
  	&=
	\sqrt{n}\;P_n\big(U(x,\la)\varphi\otimes\Psi_{n-1}\big)
	=
	\big(\zd(U(x,\la)\varphi)\Psi\big)_n\;.
\end{align*}
This implies (cf. \bref{def:Uconc})
\begin{align*}
	U(x,\la)z(\varphi)U(x,\la)^*
	&=
	\big(U(x,\la)\zd(\overline{\varphi})U(x,\la)^*\big)^*
	=
	z(\overline{U(x,\la)\overline{\varphi}})
	=
	z(U(-x,\la)\varphi) \,.
\end{align*}
One readily verifies $U(\pm x,\la)f^\pm = {f_{(x,\la)}}^\pm$, which yields the covariance of $\phi(f)$:
\begin{align}
	U(x,\la)\phi(f)U(x,\la)^{-1}
	&=
	\zd(U(x,\la)f^+) + z(U(-x,\la)f^-)
	=
	\phi(f_{(x,\la)})
	\,.
\end{align}
{\em e)} Let $\Ps(\OO)$ denote the algebra generated by all polynomials in the field $\phi(f)$ with test functions $f\in\Ss(\OO)$. By the standard Reeh-Schlieder argument making use of the spectrum condition \cite{streater} it follows that $\Ps(\OO)\Om$ is dense in $\Hil$ if and only if $\Ps(\Rl^2)\Om$ is. Choosing $f\in\Ss(\Rl^2)$ such that $f^-=0$, it follows that $\zd(f^+)\in\Ps(\Rl^2)$. Varying $f$ gives a dense set of $f^+$ in $\Hil_1$, implying that $\Om$ is cyclic for $\Ps(\Rl^2)$ and hence for $\Ps(\OO)$.

{\em f)} In the case $S_2=1$, the operator $\phi(f)$ is the free field, and well-known to be local. To show the non-local behavior of $\phi(f)$ for the other scattering functions, let $f,g\in\Ss(\Rl^2)$ be two test functions with spacelike separated supports, and consider
\begin{align}
  (P_2	[\phi(f),\phi(g)]	\Om)(\te_1,\te_2)
  &=
\frac{1}{\sqrt{2}}\,\left(f^+(\te_1)g^+(\te_2) -g^+(\te_1)f^+(\te_2)\right)\cdot(1-S_2(\te_2-\te_1))
\label{phi-not-local}
\end{align}
This expression vanishes for arbitrary (spacelike separated) test functions $f,g$ if and only if $S_2=1$.
\end{proof}

\noindent Proposition \ref{prop:phi} establishes most of the usual properties of Wightman fields for $\phi$: It is defined on a stable, Poincar\'e-invariant dense domain $\DD$, transforms covariantly under the representation $U$, and has the vacuum $\Om$ as a cyclic vector. Its matrix elements are tempered distributions on $\Ss(\Rl^2)$, and in particular, the two-point function of $\phi$ has the familiar form
\begin{align}\label{2point}
\langle\Om,\phi(x)\phi(y)\Om\rangle
&=
\frac{1}{2\pi^2}
\int d^2p\,\Theta(p_0)\delta(p^2-m^2)\,e^{-ip\cdot (x-y)}
\,,
\end{align}
independently of $S_2$. (The $S_2$-dependence of the $n$-point functions of $\phi$ shows up for even $n\geq 4$.) As the Jost-Schroer Theorem \cite[Thm. 4-15]{streater} states, this form of the two-point function implies that $\phi$ is either the free field or non-local. We have shown above that the former case is realized by the scattering function $S_2=1$, and the latter case for all other scattering functions. We also see from \bref{2point} that $\phi$ does not anticommute at spacelike distances, either -- independently of the underlying $S_2$.

In view of the lacking locality of $\phi$, this field might at first sight appear to be of little physical significance. In fact, we do not regard $\phi$ as a "physical" quantum field, but rather as an auxiliary object, which, however, will turn out to be very important for the construction of the theory. 

It was discovered by Schroer that $\phi$, despite not being local in the sense of Wightman theory, admits a weaker form of spacetime localization \cite{Schroer:1997cq,Schroer:1997cx}. Namely, he argued that $\phi$ can be interpreted as being localized in a wedge $W\subset\Rl^2$, cf. section \ref{Sec:Geo} for the definition of these regions. In the following, we will motivate the localization of $\phi$ and define what is meant by saying "$\phi$ can be interpreted as being localized in a wedge".

We first give a heuristic motivation by considering the time zero fields $\varphi(x_1)=\phi(0,x_1)$ and $\pi(x_1)=\dot{\phi}(0,x_1)$, $x_1\in\Rl$, which can also be expressed as linear combinations of Zamolodchikov creation and annihilation operators. According to the usual interpretation of Zamolodchikov's algebra (section \ref{sec:ffp}), the operator $\zd(\te)$ acts on the scattering states $\zd(\te_1)\cdots\zd(\te_n)\Om$, $\te_1>...>\te_n$, incoming from $x_1=-\infty$, by adding a particle with rapidity $\te$. Combining $\zd(\te)$ with a corresponding annihilation operator, we obtain the time zero fields $\varphi$, $\pi$. Because of this relation to incoming particles, one might conjecture that $\varphi(x_1)$ and $\pi(x_1)$ are localized on the half line $(-\infty,x_1)$. In analogy to free field theory, this suggests that $\phi(x_0,x_1)$ is localized in the causal closure of $\{x_0\}\times(-\infty,x_1)$, which is the wedge $W_L+x$. So $\phi$ differs significantly from a Wightman field $\phi_0$ in its localization properties: Whereas a Wightman field $\phi_0(x)$ is localized at the point $x$, $\phi(x)$ is localized in the infinitely extended wedge region $W_L+x$.

There is also a more mathematical motivation for this interpretation. Namely, the crossing symmetry of the scattering function, $S_2(\te+i\pi)=S_2(-\te)=S_2(\te-i\pi)$, is reminiscent of the KMS condition for wedge-local observables with respect to the boost group. The crossing symmetry is the reason for one of the form factor equations in the axiomatic system of Smirnov \cite{smirnov}, the so-called cyclic form factor equation. For a discussion of the relations between this equation and the KMS property, see \cite{Schroer:1997cq,schroer-crossing,niedermaier}. In fact, it is possible to show that polynomials in the field $\phi(f)$, with test functions having support in the left wedge $W_L$, satisfy the KMS property with respect to the boost group \cite{GLdipl}.

Given these motivations, we consider the field $\phi$ as being localized in a (left) wedge and now construct a quantum field theory out of it. In the end, this construction will lead to a strictly local theory, but as an intermediate step, it is important to analyze the wedge-localized quantities more closely.

In view of the covariance of $\phi$, $U(x)\phi(f)U(x)^{-1}$ is clearly localized in the wedge $W+x$ if $\phi(f)$ is localized in $W$, and in this way, we obtain quantum fields localized in every left wedge. The crucial issue is now to find a second field $\phi'$ which is localized in a {\em right} wedge, such that $\phi(x)$ and $\phi'(y)$ commute in an appropriate sense if $W_L+x$ and $W_R+y$ are spacelike separated. If such a field operator exists, and has the usual properties concerning covariance and cyclicity of the vacuum, the interpretation of $\phi$ as a Bose field localized in a wedge is consistent. In this case, quantum observables localized in arbitrary (left and right) wedge regions of Minkowski space can be constructed. In contrast, if $\phi$ does not admit such a second field $\phi'$, it might turn out that there are no corresponding observables localized in right wedges, and the interpretation of $\phi$ as a wedge-local field would be inconsistent. 
\\
\\
Before constructing $\phi'$, we make contact with the algebraic formalism employed in chapter \ref{chapter:netsin2d}. To this end, we define the algebra of quantum observables which are localized in the left wedge $W_L$ as
\begin{align}\label{def:AWL}
	\A(W_L)		&:=	\big\{ e^{i\phi(f)} \,:\, f\in \Ss(W_L)\,{\rm real} \big\}''\,.
\end{align}
Note that $e^{i\phi(f)}$ is unitary for real test functions $f$. In chapter \ref{chapter:netsin2d}, we considered a {\em standard right wedge algebra} $(\M,U,\Hil)$ consisting of a von Neumann algebra $\M$ and a representation $U$ of the translations, acting on a Hilbert space $\Hil$ (Def. \ref{def:srwa}), as the basic object of the construction. In the present more concrete setting, $\M$ is defined as
\begin{align}\label{def:M}
\M	&:=	\A(W_L)'\,,
\end{align}
and acts on the $S_2$-symmetric Fock space $\Hil$. The representation of the translations is obtained by restricting $U$ to the translation subgroup. (The restriction will be denoted $U$, too.)

In order to verify that the triple $(\M,U,\Hil)$ defines a standard right wedge algebra, we need to check the assumptions summarized in Definition \ref{def:srwa}. These are in detail:
\begin{itemize}
\item[a)] $U$ is strongly continuous and unitary. The joint spectrum of the generators $P_0$, $P_1$ of $U(\Rl^2)$ is contained in the forward light cone $\{(p_0,p_1)\in\Rl^2\,:\,p_0\geq|p_1|\}$.
\\
There is an up to a phase unique unit vector  $\Om\in\Hil$ which is invariant under the action of $U$.
\item[b)] $\Om$ is cyclic and separating for $\M$.
\item[c)] For each $x\in \overline{W_R}$, the adjoint action of $U(x)$ induces endomorphisms on $\M$,
\begin{align}\label{def:Mxx}
	\M(x) := U(x)\M U(x)^{-1} \subset \M,\qquad x\in \overline{W_R}\,.
\end{align}
\end{itemize}
Assumption a) is clearly satisfied in our construction, and c) can be deduced from the covariance of the field $\phi$. Regarding b), the cyclicity of $\Om$ for the field $\phi$ (Prop. \ref{prop:s2} {\em e)}) implies the cyclicity of $\Om$ for the algebra $\A(W_L)$ generated by $\phi$, and hence this vector separates the commutant $\M$ of $\A(W_L)$. (These statements are formally proven in section \ref{sec:modular} below.)

In the algebraic formulation, the crucial question whether there exists a second field $\phi'$ with the above described properties amounts to the question whether $\Om$ is also cyclic for $\M$. The cyclicity of $\Om$ for $\A(W_L)$ states that there are many observables localized in $W_L$, namely all bounded functions of the field $\phi(f)$, $\supp f\subset W_L$. But a priori, we do not have any observables localized in the right wedge $W_R$, which ensure the cyclicity of the vacuum for $\M$ -- these will be defined as bounded functions of $\phi'$.
\\
\\
We now turn to the construction of $\phi'$ and recall that $\phi'(x)$ is required to commute with $\phi(y)$ if $W_L+y$ and $W_R+x$ are spacelike separated. Thinking of a Wightman theory with TCP operator\footnote{We denote the TCP operator by $J$ instead of $\Theta$ since it coincides with the modular conjugation of the algebra of observables localized in the left wedge with respect to the vacuum, which is commonly denoted $J$. This connection has been established for the case of finite-component Wightman fields by Bisognano and Wichmann \cite{BiWi1, BiWi2}. In two dimensions, it is known to hold also in the more general framework of a theory of local observables, as a consequence of Borchers' theorem \cite{borchers-2d}.} $J$, fields localized in $W_L$ and $W_R=-W_L$ should be related by the action of $J$. This suggests the definition $\phi'(x):=J \phi(-x) J$ for the second field. The problem is, however, that we a priori have no TCP operator in our model.

To find an appropriate TCP operator nonetheless, we recall that $J$ should implement the total spacetime reflection $-1:x\lmto -x$. Hence it must extend the representation $U$ to the proper Poincar\'e group by $U(-1):=J$, which amounts to the commutation relations 
\begin{align}\label{commTheta}
J\,U(x,\la)J &= U(-x,\la)\,,\qquad x\in\Rl^2\,,\;\la\in\Rl\,.
\end{align}
The action of $J$ on the vacuum and on the one particle space is uniquely fixed by these relations, since $U$ acts irreducibly on $\Hil_1$. We find
\begin{align}\label{J-0-1}
 J\Om=\Om\,,\qquad (J\psi)(\te)=\overline{\psi(\te)}\,,\qquad \psi\in\Hil_1\,.
\end{align}

\noindent On the multiparticle spaces $\Hil_n$, $n>1$, however, $U$ acts reducibly and consequently $J$ is not fixed by the commutation relations \bref{commTheta}.

To motivate a particular choice for the implementation of the spacetime reflection, recall that it is our aim to construct a model which has the S-matrix $S$ \bref{snfacts} given by the scattering function $S_2$. So we assume for a moment that we had already constructed an asymptotically complete, local quantum field theory with S-matrix $S$ and TCP operator $J$, in which $\phi$ is a field localized in the left wedge.

It is well-known from scattering theory that the TCP operator $J$ is related to a corresponding "free" TCP operator $J_0$ by the M{\o}ller operators $V\iin$, $V\oout$. $J_0$ acts on the symmetric Fock space $\Hil^+$ over $\Hil_1$, consisting of the asymptotic collision states. More precisely, we define $J_0$ (as an antilinear operator on $\Hil^+$) as the second quantization of the restriction of $J$ to the single particle space, which in view of \bref{J-0-1} amounts to
\begin{align}
(J_0\Psi^+)_n(\te_1,...,\te_n)		&:=	\overline{\Psi^+_n(\te_1,...,\te_n)}\,.
\end{align}
Here $\Psi^+=(\Psi^+_0,\Psi^+_1,...\,)$ is a vector in the Bose Fock space $\Hil^+$. The relation between the TCP operator $J$ and $J_0$ is (cf., for example, \cite[Lemma 8]{MundBiWi})
\begin{align}
J 	=	V\iin J_0 V\oout^*
\,.
\end{align}
Taking into account the physical picture of factorized scattering which underlies Zamolodchikov's algebra (section \ref{sec:ffp}), the M{\o}ller operators are expected to act on idealized $n$-particle states with sharp rapidities as
\begin{subequations}\label{moellers}
\begin{align}
V\oout 	&: \ad(\te_1)\cdots \ad(\te_n)\Om^+	\lmto \zd(\te_1)\cdots\zd(\te_n)\Om\,,\qquad \te_1<...<\te_n\,,
\\
V\iin   	&: \ad(\te_1)\cdots \ad(\te_n)\Om^+	\lmto \zd(\te_1)\cdots\zd(\te_n)\Om\,,\qquad \te_1>...>\te_n\,.
\end{align}
\end{subequations}
Here $\ad(\te)$ denotes the asymptotic creation operators, representing the CCR algebra on the Bose Fock space $\Hil^+$, and $\Om^+\in\Hil^+$ is the Fock vacuum. These heuristic formulae lead to a definition of $J$ as follows. Let $\Psi_n(\te_1,...,\te_n):=\zd(\te_1)\cdots\zd(\te_n)\Om$, $\te_1<...<\te_n$. Then
\begin{align*}
(V\iin J_0V\oout^*\Psi_n)(\te_1,...,\te_n)
&=
V\iin J_0 \ad(\te_1)\cdots \ad(\te_n)\Om^+
=
V\iin \overline{\ad(\te_n)\cdots \ad(\te_1)\Om^+}
\\
&=
\overline{\zd(\te_n)\cdots\zd(\te_1)\Om}
=
\overline{\Psi_n(\te_n,...,\te_1)}
\,.
\end{align*}
Consequently, we {\em define}
\begin{align}\label{def:J}
	(J\Psi)_n(\te_1,...,\te_n)		&:=	\overline{\Psi_n(\te_n,...,\te_1)}\,.
\end{align}
Let us emphasize that none of the indicated properties which were used for the motivation of this definition, like asymptotic completeness of the theory, existence of M{\o}ller operators, or the formulae \bref{moellers}, are assumed in the following. Rather, we take \bref{def:J} as a definition.

Besides $J$, we also introduce the involution
\begin{align}\label{def:Gamma}
	(\Gamma\Psi)_n(\te_1,...,\te_n)		:=&\;	\overline{\Psi_n(-\te_1,...,-\te_n)}\,,
\end{align}
and study the properties of $J$ and $\Gamma$ in the following Lemma.
\begin{lemma}\label{lem:extendU}{\quad}\\
\vspace*{-10mm}\\
\begin{enumerate}
\item $J$ and $\Gamma$ are commuting antiunitary involutions which leave $\Hil$ invariant.
\item Denote by $T:\Rl^2\lto\Rl^2$ the time reflection, $T(x_0,x_1)=(-x_0,x_1)$, and put
\begin{align}
U(-1)		&:=J\,,\qquad U(T)	:=\Gamma\,,\qquad U(-T)	:= \Gamma J\,.
\end{align}
 These assignments extend $U$ to a representation of the full Poincar\'e group $\PG$ on $\Hil$.
\end{enumerate}
\end{lemma}
\begin{proof}
{\em a)} It is apparent from \bref{def:J} and \bref{def:Gamma} that $J$ and $\Gamma$ are commuting antiunitary involutions on the unsymmetrized Fock space $\F_{\Hil_1}=\bigoplus_{n=0}^\infty L^2(\Rl^n)$. To prove that they leave the subspace $\Hil\subset\F_{\Hil_1}$ invariant, we compute the commutation relations between these operators and the representations $D_n$ \bref{def:Dn} of the symmetric group. Let $\Psi_n\in L^2(\Rl^n)$.
\begin{align*}
	(JD_n(\tau_k)\Psi_n)(\te_1,...,\te_n)
	&=
	\overline{S_2(\te_{n-k}-\te_{n-k+1})}\,\overline{\Psi_n(\te_n,..,\te_{n-k},\te_{n-k+1},..,\te_1)}
	\\
	&=
	S_2(\te_{n-k+1} - \te_{n-k}) \, \overline{\Psi_n(\te_n,..,\te_{n-k},\te_{n-k+1},..,\te_1)}
	\\
	&=
	(D_n(\tau_{n-k})J\Psi_n)(\te_1,...,\te_n)
\end{align*}
So we have $JD_n(\tau_k)J=D_n(\tau_{n-k})$, and as the transpositions generate $\frS_n$, this shows that $J$ induces a group automorphism on $\frS_n$. The mean over the group, $P_n$, is therefore left invariant, $JP_n=P_nJ$, and hence $\Hil$ is stable under the action of $J$.

For $\Gamma$ we obtain
\begin{align*}
	(D_n(\tau_k)\Gamma \Psi_n)(\te_1,...,\te_n)
	&=
	S_2(\te_{k+1}-\te_k)\,\overline{\Psi_n(-\te_1,...,-\te_{k+1},-\te_k,...,-\te_n)}
	\\
	&=
	\overline{S_2(-\te_{k+1} - (-\te_k))}\cdot\overline{\Psi_n(-\te_1,...,-\te_{k+1},-\te_k,...,-\te_n)}
	\\
	&=
	(\Gamma D_n(\tau_k)\Psi_n)(\te_1,...,\te_n)\,.
\end{align*}
So $\Gamma$ commutes with $D_n(\tau_k)$ and hence with the projections $P_n$. This implies $\Gamma\Hil=\Hil$.

{\em b)} We need to check the commutation relations $J\, U(x,\la) J=U(-x,\la)$ and $\Gamma U(x,\la) \Gamma=U(Tx,-\la)$. So let $\Psi\in\Hil$, $(x,\la)\in\PGpo$ and consider
\begin{align*}
	(J\,U(x,\la)J\Psi)_n(\te_1,...,\te_n)
	&=
	\overline{(U(x,\la)J\Psi)_n(\te_n,...,\te_1)}
	\\
	&=
	\prod_{k=1}^n e^{-ip(\te_k)\cdot x} \cdot \Psi_n(\te_1-\la,...,\te_n-\la)
	\\
	&=
	(U(-x,\la)\Psi)_n(\te_1,...,\te_n)
\end{align*}
and, taking into account $T\,p(\te)=(-m\cosh\te,m\sinh\te)=-p(-\te)$ \bref{phi},
\begin{align*}
(\Gamma\, U(x,\la)\Gamma\Psi)_n(\te_1,...,\te_n)
&=
\prod_{k=1}^n e^{-ip(-\te_k)\cdot x} \overline{(\Gamma\Psi)_n(-\te_1-\la,...,-\te_n-\la)}
\\
&=
\prod_{k=1}^n e^{ip(\te_k)\cdot T x} \Psi_n(\te_1+\la,...,\te_n+\la)
\\
&=
(U(Tx,-\la)\Psi)_n(\te_1,...,\te_n)\,.
\end{align*}
As $\Psi$ and $(x,\la)$ were arbitrary, this proves that $J$ and $\Gamma$ implement the reflections $-1$ and $T$, respectively. The corresponding property of $J\Gamma$ follows since $J$ and $\Gamma$ commute.
\end{proof}
\noindent {\em Remark:} By a simple calculation which we omit here, one can show
\begin{align}
\Gamma z(\psi)\Gamma		&=	z(\Gamma\psi)
\,,\qquad 
\Gamma \zd(\psi)\Gamma		=	\zd(\Gamma\psi)
	\,,\qquad\psi\in\Hil_1\,,\\
\Gamma \phi(f) \Gamma		&=	\phi(f_T)			\,,\qquad f_T(x_0,x_1)= \overline{f(-x_0,x_1)}\,,\label{gamma-phi-covariance}
\end{align}
that is, the field $\phi$ also transforms covariantly under $\Gamma$.
\\\\
In view of Lemma \ref{lem:extendU}, we will consider $J$ as the TCP operator of the model. According to the strategy explained above, we now introduce the field $\phi'(x):=J\phi(-x)J$, properly defined by
\begin{align}\label{def:phi'}
	\phi'(f)		&:=		J\phi(f^*)J,\qquad f^*(x) := \overline{f(-x)}\,.
\end{align}
This field shares many properties with $\phi$.
\begin{lemma}
The field $\phi'$ also has the properties a)-f) listed in Proposition \ref{prop:phi} for $\phi$.
\end{lemma}
\begin{proof}
These properties follow in a straightforward manner from the definition of $\phi'$, and we can be brief about the proof.

As a consequence of the representation properties of $J$, $\phi'$ transforms covariantly under $U$, and in view of the antiunitary of $J$, the bound in {\em a)} and $\phi'(f)^*\Psi=\phi'(\fbar)\Psi$, $\Psi\in\DD$, follow. As $(f^*)^\pm = \overline{f^\pm}$, $\phi'$ is also a solution of the Klein-Gordon equation. The cyclicity of the vacuum follows from
\begin{align}
\phi'(f_1)\cdots\phi'(f_n)\Om
=
J\,\phi(f_1^*)\cdots\phi(f_n^*)\Om
\end{align}
since $J$ is an involution. Finally, $\phi'$ is local if and only if $\phi$ is, and thus we also have {\em f)}.
\end{proof}
\noindent Note that $\phi$ and $\phi'$ coincide if and only if the underlying scattering function is $S_2=1$, in which case they are given by the free field. In spite of these fields being different in general, $\phi$ and $\phi'$ still create the same single particle state from the vacuum, irrespectively of $S_2$:
\begin{align}
\phi'(f)\Om	=	J\,(f^*)^+	=	f^+	=\phi(f)\Om
\,,\qquad f\in\Ss(\Rl^2)\,.
\end{align}

\noindent But the most important property of the fields $\phi$, $\phi'$ is that they are relatively wedge-local to each other in the following sense: $\phi(f)$ and $\phi'(g)$ commute (on $\DD$) if $f$ has support in $W_L$ and $g$ in $W_R$. This result will enable us to complete the construction of the wedge-local observables of our model, and a posteriori justifies the interpretation of $\phi$ as being localized in a wedge.

To derive it, we consider the "reflected" Zamolodchikov operators
\begin{align}
	z(\psi)'	&:=		J z(\psi) J
	\,,\qquad
	\zd(\psi)'	:=		J \zd(\psi) J
	\,,
\end{align}
and calculate their commutation relations with $z(\psi)$, $\zd(\psi)$.
\begin{lemma}\label{lemma:zz'}
Let $\psi_1,\psi_2\in\Hil_1$. The following commutation relations hold on $\DD$:
\begin{align}
	[z(\psi_1)',\, z(\psi_2) ] = 0\,,
	\qquad
	[\zd(\psi_1)',\, \zd(\psi_2) ] = 0\,.
\end{align}
The mixed commutators act as multiplication operators on $\Hil_n$:
\begin{align}
\big[z(\overline{\psi_1})',\zd(\psi_2)\big]\Psi_n
&=
B_n^{\psi_1,\psi_2}\cdot\Psi_n
,\;\; B_n^{\psi_1,\psi_2}=+
\int d\te\,\,\psi_1(\te)\,\psi_2(\te)\prod_{j=1}^n
S_2(\te-\te_j)\;,
    \label{Bn}
\\
\big[\zd(\overline{\psi_1})',z(\psi_2)\big]\Psi_n
&=
C_n^{\psi_1,\psi_2}\cdot\Psi_n
,\;\; C_n^{\psi_1,\psi_2}=
    -\int d\te\,\psi_1(\te)\,\psi_2(\te)\prod_{j=1}^n
    S_2(\te_j-\te)\,.
    \label{Cn}
\end{align}
\end{lemma}
\begin{proof}
We first compute the explicit action of the reflected annihilation operator $z(\psi)'$ on $\Phi\in\DD$. On the basis of \bref{def:J} and \bref{act-z} we find
\begin{align*}
	(z(\psi)'\Phi)_n(\te_1,...,\te_n)
	&=
	\overline{(z(\psi)J\Phi)_n(\te_n,...,\te_1)}
	=
	\sqrt{n+1}\int d\te\, \overline{\psi(\te)}\,\Phi_{n+1}(\te_1,...,\te_n,\te)
	\,.
\end{align*}
The claimed commutation relations can be verified by straightforward calculation. Let $\psi_1,\psi_2\in\Hil_1$, $\Phi\in\DD$.
\begin{align*}
	\big([z(\psi_1)',\,z(\psi_2)]\Phi\big)_n&(\te_1,...,\te_n)
	=
	\sqrt{n+1}\int d\te'\, \overline{\psi_1(\te')} \big(z(\psi_2)\Phi\big)_{n+1}(\te_1,...,\te_n,\te')
	\\
	&-\sqrt{n+1}\int d\te\, \psi_2(\te) \big(z(\psi_1)'\Phi\big)_{n+1}(\te,\te_1,...,\te_n)
	\\
	=&\;
	\sqrt{(n+1)(n+2)}\int d\te'\, \overline{\psi_1(\te')}\int d\te\, \psi_2(\te) \Phi_{n+2}(\te,\te_1,...,\te_n,\te')
	\\
	&-\sqrt{(n+1)(n+2)}\int d\te\, \psi_2(\te) \int d\te'\, \overline{\psi_1}(\te') \Phi_{n+2}(\te,\te_1,...,\te_n,\te')
	\\
	=&\; 0\,.
\end{align*}
So $[z(\psi_1)',\,z(\psi_2)]=0$ on $\DD$, and by taking adjoints, we also obtain $[\zd(\psi_1)',\,\zd(\psi_2)]=0$. For the calculation of the mixed commutators, recall the formula \bref{act-zd} for the creation operator.
\begin{align*}
\big([z(\overline{\psi_1})',\zd(\psi_2)]\Phi\big)_n&(\te_1,..,\te_n)
\\
=&\;
\sqrt{n+1}\int d\te_{n+1}\,\psi_1(\te_{n+1})\,\big(\zd(\psi_2)\Phi\big)_{n+1}(\te_1,..,\te_n,\te_{n+1})
\\
& - \frac{1}{\sqrt{n}}\sum_{k=1}^n \prod_{j=1}^{k-1} S_2(\te_k-\te_j)\,\psi_2(\te_k)\, \big(z(\overline{\psi_1})'\Phi\big)_{n-1}(\te_1,..,\widehat{\te}_k,..,\te_n)
\\
=&
\sum_{k=1}^{n+1}\int d\te_{n+1}\psi_1(\te_{n+1})\prod_{j=1}^{k-1} S_2(\te_k-\te_j) \psi_2(\te_k)\Phi_n(\te_1,..,\widehat{\te}_k,..,\te_{n+1})
\\
&- \sum_{k=1}^n \prod_{j=1}^{k-1} S_2(\te_k-\te_j)\,\psi_2(\te_k)\int d\te_{n+1} \psi_1(\te_{n+1}) \Phi_n(\te_1,..,\widehat{\te}_k,..,\te_{n+1})
\\
=&
\int d\te_{n+1}\,\psi_1(\te_{n+1})\psi_2(\te_{n+1}) \prod_{j=1}^n S_2(\te_{n+1}-\te_j) \cdot \Phi_n(\te_1,..,\te_n)
\end{align*}
This calculation identifies the restriction of $[z(\overline{\psi_1})',\zd(\psi_2)]$ to the $n$-particle space as the operator multiplying with the function $B_n^{\psi_1,\psi_2}$ \bref{Bn}. By taking adjoints, we also have
\begin{align*}
	[\zd(\overline{\psi_1})',\,z(\psi_2)]\Phi_n
	&=
	-[z(\psi_1)',\,\zd(\overline{\psi_2})]^* \Phi_n
	=
	- \overline{ B_n^{ \overline{\psi_1},\overline{\psi_2}}}\cdot\Phi_n\,.
\end{align*}
The function $-\overline{ B_n^{ \overline{\psi_1},\overline{\psi_2}}}$ agrees with $C_n^{\psi_1,\psi_2}$ \bref{Cn} since $\overline{S_2(\te)}=S_2(-\te)$, and the proof is finished.
\end{proof}
\noindent After this preparation, we can prove the relative locality of the fields $\phi$ and $\phi'$.
\begin{proposition}\label{prop:phiphiprime}
The field operators $\phi$ \bref{phi} and $\phi'$ \bref{def:phi'} are relatively wedge-local to each other in the following sense: For $f\in \Ss(W_R)$, $g\in \Ss(W_L)$, there holds
\begin{align}\label{comm=0}
	[\phi'(f),\phi(g)]\Psi		&=	0,\qquad\Psi\in\DD\,.
\end{align}
\end{proposition}
\begin{proof}
We first note that it is sufficient to prove \bref{comm=0} for compactly supported test functions $f\in C_0^\infty(W_R)$, $g\in C_0^\infty(W_L)$: According to Proposition \ref{prop:phi} a), the map $(f,g)\longmapsto\langle\Phi,[\phi(f),\phi'(g)]\Psi\rangle$, $\Phi,\Psi\in\DD$, is a tempered distribution in $f$ and $g$. So if it vanishes on $C_0^\infty(W_R) \times C_0^\infty(W_L)$, it also vanishes on $\Ss(W_R)\times\Ss(W_L)$, which implies \bref{comm=0}.

Let $f\in C_0^\infty(W_R)$, $g\in C_0^\infty(W_L), \Psi_n\in\Hil_n$. In view of Lemma \ref{lemma:zz'} we have
\begin{align*}
	[\phi'(f),\phi(g)]\Psi_n
	&=
	[\zd({f^*}^+)'+z({f^*}^-)',\,\zd(g^+)+z(g^-)]\Psi_n\\
	&=
	[\zd(\overline{f^+})',\,z(g^-)]\Psi_n + [z(\overline{f^-})',\,\zd(g^+)]\Psi_n
	\\
	&=
	\big(C_n^{f^+\!,\,g^-} + B_n^{f^-\!,\,g^+} \big) \cdot \Psi_n\,,
\end{align*}
where the functions appearing in the last line are defined in \bref{Bn} and \bref{Cn}. So in order to establish the desired result, we need to show $C_n^{f^+,\,g^-} + B_n^{f^-,\,g^+}=0$. For this purpose, we recall some analytic properties of the functions involved.

Since $g$ has compact support, its Fourier transform is entire analytic, and hence also its restriction to the mass shell, $g^+(\te)=(2\pi)^{-1}\int d^2x \, g(x)\exp(i p(\te)\cdot x)$, is an entire analytic function. To estimate the exponential factor, note that for ${\rm Im}(p(\te)) \in W_R$, there holds ${\rm Im}(p(\te))\cdot x > 0 $ for all $x\in W_L$. Using the rapidity parametrization \bref{phi}, one calculates that 
\begin{align*}
  \mathrm{Im}(p(\te + i\mu)) = m\sin\mu\left(
    \begin{array}{c}
      \sinh\te\\\cosh\te
    \end{array}
\right) \in W_R\;\;\;\;\mathrm{for}\;\;\;\; 0<\mu<\pi\;.
\end{align*}
Making use of the fact that $\supp g$ is compact, it follows that $g^+$ is bounded on the strip\footnote{The notation for the strip regions appearing here is as in \bref{strip-not}.} $S(0,\pi)$, and $|g^+(\te+i\la)|$ converges rapidly to zero as $\te\to\pm\infty$ and $\la\in[0,\pi]$ is fixed. Since $p(\te+i\pi) = -p(\te)$, the value at the upper boundary of the strip is given by $g^+(\te+i\pi)=g^-(\te)$.

All these considerations apply to $f^*(x)=\overline{f(-x)}$ as well because $f^*\in C_0^\infty(W_L)$. In view of $(f^*)^+=\overline{f^-}$ we have analyticity for $f^-$ in $S(0,\pi)$, with $f^-(\te+i\pi)=f^+(\te)$, as well as exponential decay for $\mathrm{Re}(\te)\to\pm\infty$ in this strip.

Also recall that the scattering function $S_2$ is analytic in $S(0,\pi)$ and bounded and continuous on the closure of this region, with boundary values connected by the crossing symmetry relation $S_2(\te+i\pi)=S_2(-\te)$.

We now consider the integrand of $B_n^{f^-\!,\,g^+}$ \bref{Bn}, which is analytic on $S(0,\pi)$ and of fast decrease for $\mathrm{Re}(\te)\to\pm\infty$. This enables us to shift the integration from $\Rl$ to $\Rl+i\pi$:
\begin{align*}
  B_n^{f^-\!,\,g^+}(\te_1,...,\te_n)
  &= \int d\te\, f^-(\te)g^+(\te)\prod_{j=1}^n S_2(\te-\te_j)
  \\
  &=
   \int d\te\, f^-(\te+i\pi)g^+(\te+i\pi)\prod_{j=1}^n S_2(\te+i\pi -\te_j)\\
  &=
  \int d\te\, f^+(\te)g^-(\te)\prod_{j=1}^n S_2(\te_j-\te)
  \\
  &=
   -\, C_n^{f^+\,,g^-}(\te_1,...,\te_n)\,.
\end{align*}
So $B_n^{f^-\!,\,g^+}+C_n^{f^+\!,\,g^-}=0$ follows, and hence $[\phi(f)',\phi(g)]$ vanishes on $\DD$.
\end{proof}

\noindent In view of this result, we can consistently interpret $\phi'$ to be localized in the right wedge, i.e. the localization region of $\phi'(f)$ is the wedge $(W_R+\supp f)''$, whereas $\phi(f)$ is localized in $(W_L+\supp f)''$. By choosing the support of $f$ appropriately, we find for every wedge $W\subset\Rl^2$ quantum field operators which are localized in $W$, and have thus constructed a wedge-local quantum field theory as an important intermediate step in the definition of a {\em local} model.

Before we discuss the structure of the wedge algebras generated by $\phi$ and $\phi'$, we consider collision processes of two particles, which can be analyzed in terms of the field operators $\phi$, $\phi'$.

\section{Two-Particle Scattering States}\label{sect:wedgescattering}

The aim of our construction is to find a model theory which has the S-matrix $S$ corresponding to the scattering function $S_2$ we used in the definition of the Hilbert space and the fields, cf. \bref{snfacts}. In the end, the S-matrix of the constructed model has to be computed and compared to $S$ in order to check if the construction solves the inverse scattering problem.

The collision theory needed for the computation of the S-matrix relies on quasilocal one-particle generators to obtain multiparticle scattering states. But for the computation of two-particle scattering states, the wedge-localized fields $\phi$, $\phi'$ are sufficient, since two opposite wedges can be spacelike separated by translation. Moreover, these operators are especially convenient to use in collision theory since they generate single particle states from the vacuum.

In this section, we analyze scattering processes with two incoming and two outgoing particles. It will turn out that the incoming and outgoing two-particle scattering states are indeed connected by the two-particle S-matrix corresponding to $S_2$, as was expected from the motivation of our construction.
\\
\\
For the analysis of collision processes we employ the Haag-Ruelle scattering theory \cite{haag-scatter, ruelle-scatter, araki} in a form used in \cite{BBS} for wedge-localized fields. Let us recall how two-particle scattering states can be constructed in this case.

Instead of the usual quasi-local operators appearing in scattering theory, we here have to rely on quasi-wedgelocal fields of the form $\phi(f_t)$, $\phi'(f_t)$, where the functions $f_t$, $t\in \Rl$, are defined as usual by
\begin{align}
f_t(x)
&:=
\frac{1}{2\pi}
\int d^2p\,
\fti(p_0,p_1)\,e^{i(p_0-\omega_p)t}\,e^{-ip x}\;,\qquad \omega_p := \big(m^2+p_1^2\big)^{1/2}\,,
\end{align}
and the $\fti$ are smooth momentum space wavefunctions of compact support. For the construction of collision states, the asymptotic properties of $f_t$ as $t\to\pm\infty$ are important. We introduce the velocity support of a test function $f\in\Ss(\Rl^2)$ as
\begin{align}\label{velsup-4}
\VV(f)
&:=
\left\{(1,p_1/\omega_p)\,:\,(p_0,p_1)\in\supp\fti\,\right\}
\end{align}
and recall that the support of $f_t$ is essentially contained in $t\,\VV(f)$ for $t\to\pm\infty$ \cite{hepp}. More precisely, let $\chi$ be a smooth function which is equal to 1 on $\VV(f)$ and vanishes in the complement of a slightly larger region. Then $\fhat_t(x):=\chi(x/t)f_t(x)$ is the asymptotically dominant part of $f_t$, i.e. $f_t-\fhat_t\to 0$ for $t\to\pm\infty$, in the topology of $\Ss(\Rl^2)$ \cite{BBS}.

Furthermore, we adopt the notation to write $f\prec g$ if $\VV(g)-\VV(f)\subset\{0\}\times(0,\infty)$ \cite{BBS}.
\\
\\
To construct outgoing two-particle scattering states, we consider test functions $f_t$ and $g_t$ with $f\prec g$, such that the supports of $\fti$, $\gti$ are disjoint and concentrated around points on the upper mass shell $H^+_m$. Thus at very late or early times $t$ the operators $\phi(f_t)$ and $\phi'(g_t)$ are essentially localized in $W_L+t\,\VV(f)$ and $W_R+t\,\VV(g)$, respectively. Since $f\prec g$, these localization regions are spacelike separated and their distance increases linearly with $t$ as $t\to+\infty$. In view of $\phi(f)\Omega=f^+$, $\phi'(g)\Omega=g^+$, the outgoing two-particle state $(f^+\times g^+)\oout$ is given by the limit
\begin{align}
\lim_{t\to\infty}\phi(f_t)\phi'(g_t)\Omega = (f^+ \times g^+)\oout\,,\qquad f\prec g\,,
\end{align}
and similarly 
\begin{align}
\lim_{t\to\infty}\phi'(g_t)\phi(f_t)\Omega = (g^+\times f^+)\oout\,,\qquad f\prec g\,.
\end{align}
As the operators $\phi(f_t)$ and $\phi'(g_t)$ commute for $t\to\infty$, we have symmetric scattering states, $(g^+\times f^+)\oout= (f^+\times g^+)\oout$, as required for a Boson.

To construct incoming scattering states one has to exchange $f$ and $g$ because the regions $W_R+t\,\VV(f)$ and $W_L+t\,\VV(g)$ are far apart and spacelike separated in the limit $t\to-\infty$ if $f\prec g$. Thus
\begin{align}
\lim_{t\to-\infty}\phi(g_t)\phi'(f_t)\Omega = (g^+\times f^+)\iin
=
(f^+\times g^+)\iin\,,\qquad f\prec g\,.
\end{align}
In the models at hand, the limits needed for the computation of these collision states can be carried out easily. In view of the support properties of $\fti$ and $\gti$, there holds $f_t^+=f^+$, $g_t^+=g^+$ and $f_t^-=0$, $g_t^-=0$ . Hence all time-dependence drops out and we arrive at 
\begin{align}\label{2inout}
(f^+ \times g^+)\oout
&=
\lim_{t\to+\infty}\phi(f_t)\phi'(g_t)\Om
=
\sqrt{2}\,P_2(f^+\otimes g^+)\,,& \quad & f\prec g
\,,\\
(f^+ \times g^+)\iin
&=
\lim_{t\to-\infty}\phi(g_t)\phi'(f_t)\Om
=
\sqrt{2}\,P_2(g^+\otimes f^+)\,,&\quad & f\prec g\,.
\end{align}
Varying $f$ and $g$ within the above mentioned limitations, we obtain total sets of incoming and outgoing two-particle states in $\Hil_2$: Note that $f\prec g$ implies $\supp g^+-\supp f^+ \subset (0,\infty)$, as can be seen from the definition of the velocity support \bref{velsup-4}. On the other hand, for smooth one particle functions $\psi_1,\psi_2\in\Hil_1$ with compact support and $\supp \psi_2-\supp \psi_1 \subset (0,\infty)$, we can find $f_1,f_2\in\Ss(\Rl^2)$ such that $f_1^+=\psi_1$, $f_2^+=\psi_2$ and $f_1\prec f_2$. Therefore we also write $\psi_1\prec\psi_2$ in this situation. By continuity of \bref{2inout} in $f^+$ and $g^+$, we obtain
\begin{align}\label{2pout}
(\psi_1\times\psi_2)\oout	=	\sqrt{2}\,P_2(\psi_1\otimes\psi_2)\,,\qquad \psi_1\prec\psi_2\,,
\end{align}
and analogously
\begin{align}\label{2pin}
(\psi_1\times\psi_2)\iin	=	\sqrt{2}\,P_2(\psi_2\otimes\psi_1)\,,\qquad \psi_1\prec\psi_2\,,
\end{align}
Any smooth function with compact support in $R:=\{(\te_1,\te_2)\,:\,\te_1\leq\te_2\}$ can be approximated by linear combinations of functions of the form $\psi_1\otimes\psi_2$ with $\psi_1\prec\psi_2$. But the projection $P_2:L^2(R,d\te_1 d\te_2)\lto\Hil_2$ is linear, continuous and onto, implying that the above constructed scattering states \bref{2pout} and \bref{2pin} both form total sets in $\Hil_2$.

In particular, the interpretation of $\Hil_2$ as the "two-particle space" is justified by these results: Any $\Psi_2\in\Hil_2$ can be written as a superposition of incoming or outgoing two-particle collision states.
\\
\\
We now turn to the computation of the two-particle M{\o}ller operators $V\iin^{(2)}$, $V\oout^{(2)}:\Hil_2^+\lto\Hil_2$ and the S-matrix $S_{2,2}=V\oout^{(2)*} V\iin^{(2)}$ for $2\to 2$ processes. Here $\Hil_2^+$ denotes the symmetric two-particle subspace of the Bose Fock space $\Hil^+$ over $L^2(\Rl)$, cf. section \ref{sec:SM}.

The Bosonic scattering states \bref{2pout} and \bref{2pin} are represented in $\Hil_2^+$ by the vectors 
\begin{align}
V\oout^{(2)*}(\psi_1\times\psi_2)\oout	&=	\sqrt{2}P_2^+(\psi_1\otimes \psi_2)\,,\qquad \psi_1\prec\psi_2\,\\
V\iin^{(2)*}(\psi_1\times\psi_2)\iin		&=	\sqrt{2}P_2^+(\psi_1\otimes \psi_2)\,,\qquad \psi_2\prec\psi_1\,.
\end{align}
The symmetrization projection $P_2^+$ is given by $P_2$ with $S_2=1$. Hence ${V\oout^{(2)}}^*$ acts according to
\begin{align}
V\oout^{(2)*}(\psi_1\otimes\psi_2+S_2^*(\psi_2\otimes\psi_1))
=
\psi_1\otimes\psi_2+\psi_2\otimes\psi_1\,,\qquad
\psi_1\prec\psi_2\,.
\end{align}
Taking into account the support properties of $\psi_1\prec\psi_2$, it becomes apparent that $V\oout^{(2)*}$ is the multiplication operator multiplying with 
\begin{align}
V\oout^{(2)*}(\te_1,\te_2)
=
\left\{
\begin{array}{ccc}
1&;&\te_1\leq\te_2\\
S_2(\te_1-\te_2)&;&\te_1>\te_2
\end{array}
\right.\quad.
\end{align}
Analogously, $V\iin^{(2)}$ is seen to multiply with
\begin{align}
V\iin^{(2)}(\te_1,\te_2)
=
\left\{
\begin{array}{ccc}
S_2(\te_2-\te_1)&;&\te_1\leq\te_2\\
1&;&\te_1>\te_2
\end{array}
\right.\quad.
\end{align}
Hence the two-particle S-matrix $S_{2,2}=V\oout^{(2)*} V\iin^{(2)}$ is given by the operator multiplying with the symmetric function $(\te_1,\te_2)\lmto S_2(|\te_1-\te_2|)$, $\te_1,\te_2\in\Rl$, and thus coincides with the two-particle S-matrix given by the scattering function $S_2$ \bref{snfacts}.

We summarize these results in the following Proposition.
\begin{proposition}
Consider a scattering function $S_2\in\SF$ and the associated quantum field theory defined in terms of the wedge-local fields $\phi$ and $\phi'$.

This theory is asymptotically complete at the two-particle level, i.e. there exist total sets of incoming and outgoing two-particle scattering states in the subspace $\Hil_2\subset\Hil$. These states have the explicit forms
\begin{align}
(\psi_1 \times \psi_2)\oout	&= \sqrt{2}\,P_2(\psi_1\otimes\psi_2)\,,\qquad	\psi_1\prec\psi_2\,,
\\
(\psi_1 \times \psi_2)\iin	&= \sqrt{2}\,P_2(\psi_2\otimes\psi_1)\,,\qquad	\psi_1\prec\psi_2\,.
\end{align}
The two-particle S-matrix $S_{2,2}$ is given by the underlying scattering function,
\begin{align}
(S_{2,2}\Psi_2)(\te_1,\te_2)
&=
S_2(|\te_1-\te_2|)\cdot \Psi_2(\te_1,\te_2)\,,\qquad \Psi_2\in\Hil_2\,.
\end{align}
{\hfill  $\square$}
\end{proposition}
\noindent This Proposition shows that at least at the two-particle level, our construction is successful as a solution to the inverse scattering problem: The initial scattering function $S_2$, defining a particular model, also describes the two-particle collision processes in this theory.

To compute scattering states of more than two incoming or outgoing particles, compactly localized observables are needed. We will return to this question in chapter \ref{chapter:reconstructS}.

\section{Algebraic Formulation of the Models}\label{sec:modular}

Having verified the correct two-particle interaction, we now formulate the family of models defined by the fields $\phi$ and $\phi'$ in the language of algebraic quantum field theory, in order to make contact with the general construction of chapter \ref{chapter:netsin2d}.

Fixing a scattering function $S_2\in\SF$, the model we have constructed gives rise to the net $W\lmto\A(W)$ of wedge algebras defined by
\begin{subequations}\label{def:AWnet}
\begin{align}
\A(W_L+x)	&:=	\big\{e^{i\phi(f)} \,:\, f\in \Ss(W_L+x)\;{\rm real}\big\}''
\,,\\
\A(W_R+x)	&:=	\big\{e^{i\phi'(f)} \,:\, f\in \Ss(W_R+x)\;{\rm real}\big\}''
\,.
\end{align}
\end{subequations}
This net inherits its basic properties from the corresponding properties of the fields $\phi$, $\phi'$, as shown in the following Proposition.
\begin{proposition}\label{prop:AWnet}\quad\\
\vspace*{-12mm}\\
\begin{enumerate}
\item The map $W\lmto\A(W)$ defined in \bref{def:AWnet} is a local net of von Neumann algebras which transforms covariantly under the adjoint action of the extended representation $U$  (cf. Lemma \ref{lem:extendU}) of the full Poincar\'e group $\PG$.
\\
Moreover, the vacuum vector $\Om$ is cyclic and separating for each $\A(W)$, $W\in\W$.
\item Let $\tilde{U}$ denote the restriction of the representation $U$ to the translations. The triple $(\A(W_R),\tilde{U},\Hil)$ is a standard right wedge algebra in the sense of Definition \ref{def:srwa}.
\item With respect to the net $W\lmto\A(W)$, the fields $\phi(f)$ and $\phi'(f)$ are temperate polarization-free generators in the sense of Definition \ref{def:tPFG}, affiliated to $\A((W_L+\supp f)'')$ and $\A((W_R+\supp f)'')$, respectively.
\end{enumerate}
\end{proposition}
\begin{proof}
{\em a)} In view of the covariance properties of $\phi$, the exponentiated field operator satisfies $U(x,\la)e^{i\phi(f)}U(x,\la)^{-1}=e^{i\phi(f_{(x,\la)})}$ for any $(x,\la)\in\PGpo$. The three reflections $U(-1)=J$, $U(T)=\Gamma$ and $U(-T)=\Gamma J$ (Lemma \ref{lem:extendU}) satisfy for real test functions $f$
\begin{align}
J\phi(f)J					&=	\phi'(f_{-1})	\,,&		f_{-1}(x_0,x_1)&=f(-x_0,-x_1)\,,	\\
\Gamma\phi(f)\Gamma			&=	\phi(f_{T})	\,,&		f_{T}(x_0,x_1)&=f(-x_0,x_1)\,,	\\
\Gamma J\phi(f)J\Gamma		&=	\phi'(f_{-T})	\,,&		f_{-T}(x_0,x_1)&=f(x_0,-x_1)\,.
\end{align}
The first of these equations is an immediate consequence of the definition of $\phi'$,  the second states the covariance of $\phi$ under $\Gamma$ \bref{gamma-phi-covariance}, and the third one is a consequence of the other two. All three equations hold accordingly for the exponentiated fields operators $e^{i\phi(f)}$, and also for exponentials of $\phi'$. Since these operators generate the wedge algebras $\A(W)$ and since $\supp f_g = g\,\supp f$, $g\in\PG$, we conclude $U(g)\A(W)U(g)^{-1}=\A(g\, W)$, $g\in\PG$, $W\in\W$,  from the unitarity of $U$.

To prove the locality of the net, we consider the selfadjoint operators $\phi(f)$, $\phi'(g)$, $f\in \Ss(W_L)$, $g\in \Ss(W_R)$, which commute on $\DD$ in the sense of Prop. \bref{prop:phiphiprime}. In order to conclude that also the unitaries $e^{i\phi(f)}$ and $e^{i\phi'(g)}$ commute, we apply a theorem of Driessler and Fr\"ohlich \cite{DrFr}: These authors showed that this conclusion is valid if
\begin{align}
\|\phi(f)(1+H)^{-1}\|	<\infty,\qquad 
\|\phi'(g)(1+H)^{-1}\|	<\infty,
\end{align}
where $H$ is the Hamiltonian. But such $H$-bounds follow directly from the proof of Prop. \ref{prop:phi} b) and the fact $H\geq m\cdot N$. Hence $e^{i\phi(f)}$ and $e^{i\phi'(g)}$ commute for arbitrary real test functions $f\in \Ss(W_L)$, $g\in \Ss(W_R)$. This implies $\A(W_R)\subset\A(W_L)'$, and so locality of the net follows by covariance.

To show that the vacuum is cyclic and separating for the wedge algebras, we apply arguments of  \cite{BiWi1, BY90}. Let $f_1,...,f_n \in \Ss(W_L)$ be real, and denote by $E_k(t)$ the spectral projection of the selfadjoint operator $\phi(f_k)$, corresponding to spectral values in $[-t,t]$. Then $F_k(t):=E_k(t)\phi(f_k) \in \A(W_L)$ for all $t\in\Rl$, and $F_k(t)\to\phi(f_k)$ strongly on $\DD$ as $t\to\infty$. Hence $F_1(t)\cdots F_n(t)\Om$ converges to $\phi(f_1)\cdots\phi(f_n)\Om$ as $t\to\infty$, and we conclude the cyclicity of $\Om$ for $\A(W_R)$ from the cyclicity of $\Om$ for $\phi$ (Prop. \ref{prop:phi} {\em e)}). The identical argument can be applied to $\phi'$ as well, yielding the cyclicity of $\Om$ for $\A(W_R)$. But $\A(W_L)$ and $\A(W_R)$ commute, and so it follows that $\Om$ is cyclic and separating for these algebras. By covariance of $\A$ and the invariance of $\Om$ under $U$, this statement carries over to all wedge algebras.

{\em b)} The necessary properties of the triple $(\A(W_L),\tilde{U},\Hil)$ have been shown in {\em a)}.
\\
{\em c)} By virtue of the theorem of Driessler and Fröhlich, already used for the proof of {\em a)}, it follows that $\phi(f)$ commutes with $\A(W_R+x)$ if $W_R+x$ and $(W_L+\supp f)''$ are spacelike separated, i.e. $\phi(f)$ is affiliated with $\A((W_L+\supp f)'')$, and the same holds true for its adjoint, $\phi(f)^*=\phi(\fbar)$. Moreover, $\phi(f)\Om=f^+$ and $\phi(\fbar)\Om=\fbar^+$ are single particle states, which implies that $\phi(f)$ is a polarization-free generator. The temperateness assumptions made in Definition \ref{def:tPFG} are easily seen to be satisfied by taking $\DD$ as the domain of temperateness and using the bound from Proposition \ref{prop:phi} a).

The arguments for showing that $\phi'(f)$ is a temperate polarization-free generator are analogous.
\end{proof}

\noindent Proposition \ref{prop:AWnet} provides the link between the model constructions in this chapter and the more abstract analysis in chapter \ref{chapter:netsin2d}, as it shows that the models defined here are examples for constructions of quantum field theories in terms of a standard right wedge algebra. There is, however, a slight difference between the approach taken here and in chapter \ref{chapter:netsin2d}: In the latter context, the algebra of the left wedge was defined as the commutant of the standard right wedge algebra, resulting in a Haag-dual net. Here the algebras of the left and right wedge have been constructed in terms of the fields $\phi$ and $\phi'$. Below it is shown that both definitions are equivalent by computing the modular objects of the wedge algebras.

The following Proposition is due to D. Buchholz \cite{BuLe}.
\begin{proposition}\label{prop:BiWi}
Consider the net $W\lmto\A(W)$ of wedge algebras \bref{def:AWnet}.\\
 \vspace*{-11mm}\\
\begin{enumerate}
\item The Bisognano-Wichmann property holds, i.e. the modular unitaries $\Delta^{it}$ and modular conjugation $\tilde{J}$ affiliated with $(\A(W_R),\Om)$ are given by
\begin{align}
	\Delta^{it}	&=	U(0,-2\pi t),\qquad t\in\Rl,\label{modular-covariance}
	\\
	\tilde{J}		&=	J\,.
\end{align}
\item Haag-duality holds, i.e.
\begin{align}
	\A(W)'		&= 	\A(W')
	\,,\qquad W\in\W\,.
\end{align}
\end{enumerate}
\end{proposition}
\begin{proof}
{\em a)} It follows from modular theory that any boost $U(0,\la)$ commutes with $\Delta$ and $\tilde{J}$ since $\Om$ is invariant and $\A(W_R)$ is stable under its (adjoint) action. Hence the unitaries
\begin{align}
	 V(t) 	&:=	  U(0,2 \pi t) \Delta^{i t}
	 \,,\qquad t\in\Rl\,,
\end{align}
commute with any boost $U(0, \la)$, and as a consequence of Borchers' commutation relations \bref{borchers-CR}, also with all translations $U(x,0)$. 
Since $U$ acts irreducibly on $\Hil_1$, and $V$ is a representation of $(\Rl,+)$, this implies that $V(t) \upharpoonright \Hil_1 = e^{i t c}$ for fixed real $c$ and any $t \in \R$. 

Now, for real $f$ with $\text{supp} f \subset W_R$, $\phi'(f)$ is a selfadjoint operator which is affiliated with $\A(W_R)$, and the same holds for $\phi'_t(f) := V(t)\phi'(f) V(t)^{-1} $, $t \in \R$, because of the stability of $\A(W_R)$ under the adjoint action of $V(t)$. So both operators commute with all elements of  $\A(W_R)'$. Since $\Om$ is invariant under the action of $V(t)$ and since $\phi'(f) \Om\in \Hil_1$, the preceding result implies 
\begin{align}
\big(\phi'_t(f) - e^{i t c} \phi'(f) \big) A' \Omega = 0,
\quad A' \in \A(W_R)'.
\end{align}
It will be shown below that the dense set of vectors $\A(W_R)' \Om$ is a core for both, $\phi'(f)$ and $\phi'_t(f)$. Hence $\phi'_t(f) = e^{i tc} \phi'(f)$, which, in view of the selfadjointness of the field operators, is only possible if $c = 0$. This holds for any choice of $f$ within the above limitations, so $V(t)$ acts trivially on $\PG(W_R)\Om$, where $\PG(W_R)$ denotes the algebra consisting of polynomials in $\phi'(f)$, with test functions $f$ supported in $W_R$. As $\PG(W_R)\Om \subset \Hil$ is dense, one arrives at $V(t) = 1$, $t \in \R$, from which the claimed action of the modular unitaries follows.

Similarly, modular theory and the theorem of Borchers mentioned above imply that the unitary operator $I := \tilde{J}J$ commutes with all Poincar\'e transformations $U(x,\la)$. Furthermore, we have $Je^{i\phi'(f)}J=e^{i\phi(f_-)}$, $f_-(x)=f(-x)$, for real test functions $f$, and 
\begin{align*}
I\A(W_R) I	
		=	\tilde{J} \big\{e^{i\phi'(f)}\,:\,f\in \Ss(W_R)\,{\rm real}\big\}'' \tilde{J}
		=	\A(W_R)\,.
\end{align*}
Hence, putting $\phi'_I(f) := I \phi'(f) I^{-1}$, one finds by the same reasoning as in the preceding step that $\phi'_I(f) = \phi'(f)$. Thus $I=1$ and $\tilde{J}=J$.

The statement {\em b)} about Haag duality then follows from the equalities
\begin{equation}
 \A(W_R)'	=	\tilde{J} \A(W_R) \tilde{J}	=	J \A(W_R) J	=	\A(W_L)
\end{equation}
and covariance.

It remains to prove the assertion that $\A(W_R)'\Om$ is a core for the selfadjoint operators $\phi'(f)$, $\phi'_t(f)$ and $\phi'_I(f)$, respectively. To this end one makes use of the bound given in Proposition \ref{prop:phi} {\em a)}: For $\Psi\in\DD$ one has $\| \phi(f) \Psi \| \leq c_f \, \| (N + 1)^{1/2} \Psi \|$, where $N$ is the particle number operator and $c_f:=\|f^+\|+\|f^-\|$. Denoting the generator of the time translations by $P_0$ and recalling the structure of $U$ \bref{def:Uconc}, it is also clear that $m \, (N + 1) \leq  (P_0 + m1)$. So for $\Psi \in {\cal D} \cap {\cal D}_0$, where ${\cal D}_0$ is the domain of  $P_0$, one arrives at the inequalities
\begin{align}
\| \phi'(f) \Psi \|
\leq
c_f \, \| (N + 1)^{1/2} \Psi \|
\leq
m^{-1/2} c_f \, \| (P_0 + m1)^{1/2} \Psi \|
	\,.
\end{align}
It follows from this estimate by standard arguments that any core for $P_0$ is also a core for the field operators $\phi(f)$. Since the unitary operators $V(t)$ and $I$ in the preceding steps were shown to commute with the time translations, this domain property is also shared by the transformed fields $\phi'_t(f)$ and $\phi'_I (f)$, respectively.

In order to complete the proof, one has only to show that $\A(W_R)'\Om \cap \DD_0$ is a core for $P_0$. Now $\A(W_R)'\Om$ is mapped into itself by all translations $U(x)$, $x \in W_R$. Hence,  taking into account the invariance of $\Omega$ under translations, one finds that $\widetilde{f}(P) \A(W_R)'\Om \subset \A(W_R)'\Om \cap \DD_0$ for any test function $f$ with $\supp f \subset W_R$. But this space of functions contains elements $f$ such that $\widetilde{f}(P)$ is invertible. Hence $(P_0 \pm i 1) \widetilde{f}(P) \A(W_R)' \Om \subset (P_0 \pm i 1) (\A(W_R)'\Om\cap \DD_0)$ both are dense subspaces of $\Hil$, proving the statement.
\end{proof}

\noindent Having identified the modular objects as stated above, the net \bref{def:AWnet} coincides with the net generated from the standard right wedge algebra $(\A(W_R),\tilde{U},\Hil)$ as in chapter \ref{chapter:netsin2d}. Accordingly, we can define a net of {\em local} algebras by taking intersections of wedge algebras: The algebra of observables localized in a double cone $\OO = W_1 \cap W_2$ is defined as
\begin{align}\label{def:AOconc}
\A(W_1\cap W_2)		&:= \A(W_1)\cap\A(W_2)\,,
\end{align}
and for arbitrary open regions $\Q\subset\Rl^2$ we put\footnote{As before, $\OOO$ denotes the family of all double cones in $\Rl^2$.}
\begin{align}
\A(\Q)
&:=
\bigvee_{\OO\subset\Q\atop\OO\in\OOO}\A(\OO)
\,.
\end{align}
As was shown in chapter \ref{chapter:netsin2d}, it follows that $\OO\lmto\A(\OO)$ is a local net of von Neumann algebras which transforms covariantly under the representation $U$ of $\PG$, and we take this net as the definition of the model theory associated to a scattering function $S_2$.
\\
\\
In doing so, we avoid the subtle problem of finding explicit expressions for local quantum fields associated to $S_2$, which is the construction strategy in the form factor program. The formulation of such local fields is not uniquely fixed by $S_2$. Rather, it is long since known that a given S-matrix $S$ may have a multitude of different Wightman fields realizing $S$ as their scattering operator \cite{borchers:class}. Moreover, the construction of local interacting quantum fields can be expected to lead to complicated convergence questions, as witnessed by the bootstrap approach.

Using Smirnov's axiomatic formulation of the form factor equations, Schroer and Wiesbrock proposed a construction of elements of the local algebras \bref{def:AOconc} as certain infinite series of the Zamolodchikov operators \cite{Schroer-Wiesbrock}. But the convergence of such series can presently not be controlled, in close analogy to the problem the form factor program faces in the construction of $n$-point functions.
\\
\\
In contrast, we do not try to derive formulae for local observables or fields affiliated to the above defined net, but rather analyze the structure of the local algebras. Our strategy is the same as in chapter \ref{chapter:netsin2d}: For the net \bref{def:AOconc} to describe a physically meaningful theory, it must at least satisfy the basic assumptions of local quantum physics. Most of these assumptions, like locality and covariance, follow automatically from the corresponding properties of the wedge algebras. What remains to be proven in the models at hand is the Reeh-Schlieder property of the vacuum, and in particular, the {\em existence} of local observables. In the following chapter, we will investigate the modular nuclearity condition, introduced in chapter \ref{chapter:netsin2d}, as a sufficient condition for these properties.

\chapter[The Nuclearity Condition in Models with Factorizing S-Matrices]{The Modular Nuclearity Condition in Models with Factorizing S-Matrices}\label{chapter:nuclearity}

The models discussed in the previous chapter are defined in terms of the wedge-local fields $\phi$ and $\phi'$. Since these field operators are explicitly known, we have good control over all wedge-local quantities, as exemplified in the calculation of the two-particle scattering states and the modular data of the wedge algebras. In contrast, observables localized in bounded spacetime regions are not given by explicit formulae, but characterized in a less concrete way as elements of intersections of certain wedge algebras, and their properties are therefore more difficult to extract.

According to the general construction in chapter \ref{chapter:netsin2d}, it is possible to derive local properties, like the Reeh-Schlieder property for bounded regions, by a refined analysis of the wedge algebras. A sufficient condition for the models to comply with all principles of quantum field theory is the modular nuclearity condition (section \ref{sec:mnc}) for these algebras, because this condition implies the split property for wedges and all of its consequences (section \ref{sec:spw}).

In the present chapter, we consider the modular nuclearity condition in the family of the previously defined models. We begin by recalling some basic facts about nuclear maps between Banach spaces and outline our strategy for the verification of the nuclearity condition in section \ref{sec:mnc-strategy}. As a prerequisite for this proof, analytic properties of wedge-local wavefunctions are studied in section \ref{sec:analytic}, which depend on the structure of the underlying scattering function $S_2\in\SF$. Two subfamilies $\SF_0^- \subset \SF_0 \subset \SF$ of the family $\SF$ of all scattering functions are defined, and we establish the modular nuclearity condition for inclusions of sufficiently far separated wedges if $S_2\in\SF_0$ (Theorem \ref{thm:distalsplit}), and without restriction on the splitting distance if $S_2\in\SF_0^-$ (Theorem \ref{thm:-1nuclearity}). 

Thematically, also appendix \ref{chapter:mnc+-1} belongs to this chapter. There the two special models given by the constant scattering functions $S_2=\pm 1$ are considered, and due to the simpler structure of Zamolodchikov's algebra in these cases, the modular nuclearity condition can be proven by different methods than in the main text.

Most of the results of this chapter have been published in \cite{GL-nuci}, and a preliminary version can be found in \cite{GL-bros}.

\section{How to Prove the Modular Nuclearity Condition}\label{sec:mnc-strategy}

As in the previous chapter, we consider the family of models defined in terms of scattering functions $S_2\in\SF$. Within such a model, the algebra $\A(W_R)$ of observables localized in the right wedge $W_R$ is generated by the field $\phi'$,
\begin{align}
\A(W_R)=\{e^{i\phi'(f)}\,:\,f\in \Ss(W_R)\;{\rm real}\,\}''
\,.
\end{align}
Recall that this algebra, and hence the whole model theory, depends on the underlying scattering function $S_2\in\SF$, although this dependence is not reflected in our notation.

It has been shown that the vacuum vector $\Om$ is cyclic and separating for $\A(W_R)$ (Prop. \ref{prop:AWnet}), and that the associated modular operator $\Delta^{1/2}$ acts geometrically as a boost transformation with imaginary rapidity parameter $i\pi$ (Prop. \ref{prop:BiWi}). The modular nuclearity condition is the condition that the maps
\begin{align}\label{Xis}
\Xi(s) : \A(W_R)\lto\Hil\,,\qquad \Xi(s)A:=\Delta^{1/4}U(\bs)A\Om,\qquad \bs=(0,s),\quad s>0,
\end{align}
are nuclear for each $s>0$. In view of the covariance properties of the theory, the nuclearity of $\Xi(s)$ implies that the maps $\Xi_{\A(W_1),\A(W_2)}$ \bref{def:XiAbstract} corresponding to arbitrary inclusions $\A(W_1)\subset\A(W_2)$, $W_1\Subset W_2$, of wedge algebras are nuclear, as discussed in section \ref{sec:mnc}.

In \bref{Xis}, $\Hil$ is the $S_2$-symmetric Fock space over $L^2(\Rl,d\te)$, and the translations $U(\bs)$ are defined in \bref{def:Uconc}. The notation $\bs=(0,s)$, $s>0$, will be used throughout, and we refer to the parameter $s$ as the {\em splitting distance}. 
\\
\\
To begin with, let us recall the definition and some basic properties of nuclear maps between two Banach spaces $\X$ and $\Y$. The proofs of the statements made here can be found in \cite{jarchow, pietsch, nuclearmaps2}, and also in appendix \ref{app:nuclearity} of this thesis.
\begin{definition}
	Let $\X$ and $\Y$ be two Banach spaces. A linear map $T:\X\lto\Y$ is said to be nuclear if there exists a sequence of vectors $\{\Psi_k\}_k\subset\Y$ and a sequence of linear functionals $\{\rho_k\}_k\subset\X_*$ such that
	\begin{align}\label{nuc-decc}
		T(X) &= \sum_{k=1}^\infty \rho_k(X)\,\Psi_k\,,\qquad
		\sum_{k=1}^\infty \|\rho_k\|_{\X_*} \|\Psi_k\|_\Y < \infty \,.
	\end{align}
	The nuclear norm of such a mapping is defined as
	\begin{align}\label{def:nuclearnorm}
		\|T\|_1 &:= \inf_{\rho_k,\Psi_k} \sum_{k=1}^\infty \|\rho_k\|_{\X_*} \|\Psi_k\|_\Y\,,
	\end{align}
	where the infimum is taken over all sequences $\{\Psi_k\}_k\subset\Y$, $\{\rho_k\}_k\subset\X_*$  complying with the above conditions.
\end{definition}

There also exist more specific versions of nuclearity, known as "$p$-nuclearity" \cite{p-nuc}, but the above defined notion is sufficient for our purposes.

To interpret the maps in question \bref{Xis} as linear maps between two Banach spaces, we consider the von Neumann algebra $\A(W_R)$ as a Banach space with norm $\|\cdot\|_{\B(\Hil)}$, and the Hilbert space $\Hil$ is of course also a Banach space with its norm $\|\cdot\|_\Hil$. As no confusion is likely to arise, we keep on using the same symbol $\|\cdot\|$ for both, $\|\cdot\|_{\B(\Hil)}$ and $\|\cdot\|_\Hil$. Equipping $\A(W_R)$ and $\Hil$ with these norms, $\Xi(s)$ becomes a bounded operator, as can be seen with the help of modular theory as follows. (We denote by $J$ the modular conjugation of $(\A(W_R),\Om)$.)  In view of the (anti-) unitarity of $U(\bs)$ and $J$ and the selfadjointness of $\Delta$, we have
\begin{align*}
\|\Xi(s)A\|^2
&=
\langle U(\bs)A\Om,\,\Delta^{1/2}U(\bs)A\Om\rangle
=
\langle U(\bs)A\Om,\,J U(\bs)A^*\Om\rangle
\leq
\|A\|^2
\,,
\end{align*}
and hence the norm of $\Xi(s)$ as a linear operator between $(\A(W_R),\|\cdot\|_{\B(\Hil)})$ and $(\Hil,\|\cdot\|_\Hil)$ is not larger than 1. To show that $\Xi(s)$ is also nuclear, however, requires a much more detailed analysis.
\\
\\
Some properties of nuclear maps are collected in the following well-known Lemma and will be used throughout, mostly without any further mentioning. A proof of these statements can be found in appendix \ref{app:nuclearity}.

We denote by $\NN(\X,\Y)$ the set of nuclear maps between two Banach spaces $\X,\Y$, and by $\K(\X,\Y)$ the set of compact operators between $\X$ and $\Y$.
\begin{lemma}\label{Lem:nuclearmaps}
	Let $\X,\X_1,\Y,\Y_1$ be Banach spaces.
	\begin{enumerate}
		\item $(\NN(\X,\Y),\|\cdot\|_1)$ is a Banach space.
		\item $\NN(\X,\Y) \subset \K(\X,\Y)$, and $\|T\| \leq \|T\|_1$ for $T\in\NN(\X,\Y)$.
		\item Let $T\in \NN(\X,\Y), A_1\in \B(\Y,\Y_1), A_2 \in\B(\X_1,\X)$. Then $A_1 T A_2 \in \NN(\X_1,\Y_1)$, and 
			\begin{align}
				\|A_1 T A_2\|_1 &\leq \|A_1\| \cdot \|T\|_1 \cdot \|A_2\|\,.
			\end{align}
		\item Let $\Hil$ be a separable Hilbert space. Any trace class operator $T$ on $\Hil$ lies in $\NN(\Hil,\Hil)$ and satisfies $\|T\|_1\leq{\rm Tr}\,|T|$.
	\end{enumerate}
\end{lemma}
\noindent Having recalled these facts about nuclear maps, we now outline our strategy for the proof of the nuclearity of the maps $\Xi(s)$. To establish this property, we need to estimate the "size" of the image of $\A(W_R)$  under $\Xi(s)$ in $\Hil$ by exploiting the localization of $\A(W_R)$ in $W_R$ and the form of the operator $\Delta^{1/4}U(\bs)$ appearing in the definition of $\Xi(s)$. 
We will therefore study properties of state vectors $A\Omega$, $A\in\A(W_R)$, which reflect the localization of $A$ in the right wedge and the boundedness $\|A\|<\infty$.

Note that in view of the second quantization structure of the modular operator $\Delta^{1/4}$ and the translation $U(\bs)$, these operators can be restricted to the $n$-particle spaces $\Hil_n$, $n\in\N_0$. We may thus introduce the $n$-particle restrictions of $\Xi(s)$,
\begin{align}\label{def:Xin}
\Xi_n(s):\A(W_R)\lto\Hil_n
\,,\qquad
\Xi_n(s)A	:=	P_n\Xi(s)A	= \Delta^{1/4}U(\bs)(A\Om)_n\,,
\end{align}
where $(A\Om)_n$ denotes the $n$-particle rapidity wavefunction of $A$, 
\begin{align}\label{a-wave}
  (A\Omega)_n
  &:=
  P_n A \Omega
  \,\in\, \Hil_n \subset L^2(\Rl^n)
  \,,\qquad n\in\N_0\,.
\end{align}
The original map $\Xi(s)$ is the sum of its $n$-particle restrictions, 
\begin{align}\label{xi-series}
\Xi(s)
=
\sum_{n=0}^\infty \Xi_n(s)\,.
\end{align}
To show that $\Xi(s)$ is nuclear, we must prove that all the maps $\Xi_n(s)$ are nuclear, with summable nuclear norms, $\sum_{n=0}^\infty\|\Xi_n(s)\|_1<\infty$. For in this case, the series \bref{xi-series} converges in nuclear norm, and since the nuclear maps between $\A(W_R)$ and $\Hil$ form a Banach space with respect to the norm $\|\cdot\|_1$, this implies the desired result.

Recall that the translation $U(\bs)$, $\bs=(0,s)$, acts by multiplication with $\prod_{k=1}^n e^{-ims\sinh\te_k}$ \bref{def:Uconc}, and $\Delta^{1/4}$ acts by translation in the center of mass rapidity by $-\frac{i\pi}{2}$ \bref{modular-covariance}. As $\sinh(\te-\frac{i\pi}{2})=-i\cosh\te$, the functions $\Xi_n(s)A$ take the explicit form
\begin{align}\label{xi-n-ana}
(\Xi_n(s)A)(\te_1,...,\te_n)
=
\prod_{k=1}^n e^{-ms\cosh\te_k}\cdot 
(A\Om)_n(\te_1-\tfrac{i\pi}{2},...,\te_n-\tfrac{i\pi}{2})\,.
\end{align}
This equation has to be understood in terms of analytic continuation, and will be made more precise later. From \bref{xi-n-ana}, it is apparent that the map $\Xi(s)$ can be studied in terms of the wavefunctions $(A\Om)_n$, $A\in\A(W_R)$, in particular in terms of the analytic properties of these functions.

Intuitively speaking, the localization of $A$ in the right wedge corresponds to a kind of support restriction in position space, and by Fourier transformation, we therefore expect analytic properties of the wavefunctions $(A\Om)_n(\te_1,...,\te_n)$ and improper matrix elements $\langle\te_1,...,\te_k|\,A\,|\, \te_1',...,\te_l'\rangle$ in rapidity space. As analyticity is a very strong property, such features of $(A\Om)_n$ indicate that the image of $\Xi(s)$ is "small" in an appropriate sense, and will be useful for the proof of the nuclearity condition.

We therefore study analytic properties of the wavefunctions $(A\Om)_n$ in the following section. After some steps of successive analytic continuation, we will find that $(A\Omega)_n$ extends to a bounded analytic function in a tube domain in $\Cl^n$, the shape of which depends on the underlying scattering function (Prop. \ref{prop:ana}). This result will enable us to derive bounds on the nuclear norm of $\Xi_n(s)$ in the models at hand.

\section{Analytic Properties of Wedge-Local Wavefunctions}\label{sec:analytic}

One possibility of extracting information about the analytic structure of the functions $(A\Om)_n$ from the localization of $A\in\A(W_R)$ is to consider the commutators of $A$ with the time zero fields $\varphi$, $\pi$ of $\phi$,
\begin{align}\label{def-phi-pi-x}
\varphi(x_1):= \sqrt{2\pi}\,\phi(0,x_1)\,,\qquad \pi(x_1):=\sqrt{2\pi}\,(\partial_0\phi)(0,x_1)\,,\qquad x_1\in\Rl\,.
\end{align}
The prefactor $\sqrt{2\pi}$ appearing here is chosen solely for convenience.

As $\phi$ is localized in $W_L$, its time zero fields are localized on the left half line. To formulate this property precisely, we must take into account that the assignments \bref{def-phi-pi-x} have to be understood in the sense of (operator-valued) distributions. Evaluated on testfunctions $f\in\Ss(\Rl)$, we find
\begin{align}\label{def:phipi}
\varphi(f) &= \zd(\fhat) + z(\fhat_-),
    &\fhat(\te)&:=\fti(m\sinh\te),\\
    \pi(f)     &= i\big(\zd(\omega\fhat)-z(\omega\fhat_-)\big),
    &\fhat_-(\te)&:=\fhat(-\te)\,,
\end{align}
as can be easily inferred from the definition \bref{deff:phi} of $\phi$. The one particle Hamiltonian $\omega$ acts by multiplication with $\omega(\te)=m\cosh\te$ on its domain in $\Hil_1$.

The operators $\varphi(f)$, $\pi(f)$ are well-defined on the subspace $\DD\subset\Hil$ of finite particle number and satisfy
$\varphi(f)^*\supset \varphi(\fbar)$, $\pi(f)^*\supset \pi(\fbar)$. In particular, they are hermitian for real $f$. Along the same lines as in Proposition \ref{prop:phiphiprime}, one can prove ($A\in\A(W_R)$)
\begin{align*}
\langle\Psi_1,\,[\varphi(f),A]\,\Psi_2\rangle
=0
\,,\qquad
\langle\Psi_1,\,[\pi(f),A]\,\Psi_2\rangle
=0
\,,\qquad
\supp f\subset \Rl_-\,,\;\Psi_1,\Psi_2\in\DD\,.
\end{align*}
These commutators are studied in more detail in the following Lemma. Here and in the following, it turns out be convenient to formulate our results in terms of the Hardy spaces $H^2(\Tu)$, where $\Tu=\Rl^n+i\,\Cu\subset\Cl^n$ is a tube based on an open  convex domain $\Cu\subset\Rl^n$. Recall that the elements of $H^2(\Tu)$ are those analytic functions $F:\Tu\to\Cl$ for which $F_\sbla:\bte\lmto F(\bte+i\bla)$ is an element of $L^2(\Rl^n)$ for each $\bla\in\Cu$, and which have finite Hardy norm
\begin{align}
\bno{F}
&:=
\sup_{\sbla\in\Cu}\|F_\sbla\|_2
=
\sup_{\sbla\in\Cu}\left(\int_{\Rl^n} d^n\bte \,|F(\bte+i\bla)|^2\right)^{1/2}
<
\infty\,.
\end{align}
Some important properties of Hardy spaces on tube domains, such as their completeness with respect to $\bno{\cdot}$, or the $L^2$-convergence of Hardy functions to their boundary values, are collected in appendix \ref{app-hardy}.

We will use the symbols $\bte,\bla$ for real and $\bze$ for complex vectors in $n$-dimensional space, their components being written as $\bze=(\zeta_1,...,\zeta_n)$, $\bte=(\te_1,...,\te_n)$ etc. Also the notation $F_\sbla (\bte) := F(\bte+i\,\bla)$ for functions $F$ defined on a tube $\Tu=\Rl^n+i\,\Cu$, $\bte\in\Rl^n$, $\bla\in\Cu\subset\Rl^n$, will be used without further mentioning. Finally, $S(a,b)$ denotes the open strip region in $\Cl$ consisting of the complex numbers $\zeta$ with $a<{\rm Im}\,\zeta<b$, as before.
\begin{lemma}\label{lemma:Cphipi}
Let $A\in\A(W_R)$, $n_1,n_2\in\N_0$, $\Psi_1\in\Hil_{n_1}$, $\Psi_2 \in \Hil_{n_2}$ and consider the two functionals $C_\pm : \Ss(\Rl)\lto\Cl$,
  \begin{align}\label{cpm}
  C_-(f) := \langle\Psi_1,\,[\varphi(f),A]\,\Psi_2\rangle\,,
  \qquad
  C_+(f) := \langle\Psi_1,\,[\pi(f),A]\,\Psi_2\rangle\,,
\end{align}
  where $\fhat(\te):=\fti(m\sinh\te)$.
\\
There exist functions $\hat{C}_\pm\in H^2(S(-\pi,0))$ (depending linearly on $\Psi_2$ and $A$, and conjugate linearly on $\Psi_1$) which satisfy
\begin{align}
C_\pm(f)
=
\int_\Rl d\te\,\hat{C}_\pm(\te)\fhat(\te)
\,,\qquad f\in\Ss(\Rl)\,.
\end{align}
Their Hardy norms are bounded by
\begin{align}\label{Chardybound}
\bno{\hat{C}_\pm}\leq     c(n_1,n_2)\|\Psi_1\|\|\Psi_2\|\|A\|
\,,\qquad
c(n_1,n_2) := \sqrt{2n_1+1}+\sqrt{2n_2+1}\,,
\end{align}
and there holds
\begin{align}\label{pmsym}
  \hat{C}_\pm(\te-\tfrac{i\pi}{2}+i\mu)
  =
  \pm\,
  \hat{C}_\pm(-\te-\tfrac{i\pi}{2}-i\mu)
  \,,\qquad 
  -\tfrac{\pi}{2}\leq\mu\leq\tfrac{\pi}{2}\,.
\end{align}
\end{lemma}
\begin{proof}
Recall the form \bref{def:phipi} of the time zero field $\varphi$. To derive bounds on $|C_-(f)|$, we first note (with $\|f\|_2:=(\int dx|f(x)|^2)^{1/2}$)
\begin{align}
  \|\omega^{1/2}\fhat\|^2
  =
  \int d\te\,m\cosh\te\,|\fhat(\te)|^2
  =
  \int dp\,|\fti(p)|^2
  =
  \|f\|_2^{\;2}
  ,\quad
  f\in L^2(\Rl,dx)\,.
\end{align}
Combining this equation with the particle number bound \bref{Z-bounds} and taking into account the Fock structure of $\Hil$, we obtain, $j=1,2$,
\begin{align}
  \|\varphi(\omega^{1/2}f)\Psi_j\|^2
  =
  \|\zd(\omega^{1/2}\fhat)\Psi_j\|^2 +
  \|z(\omega^{1/2}\fhat_-)\Psi_j\|^2
  \leq (2n_j+1)\|\Psi_j\|^2\|f\|_2^{\,2}
  \,.
\end{align}
So the Schwarz inequality gives
\begin{align}\label{n-est-1}
  |C_-(\omega^{1/2}f)|
  &\leq
  \|\varphi(\omega^{1/2}\overline{f})\Psi_1\|\|A\Psi_2\|
  +
  \|A^*\Psi_1\|\|\varphi(\omega^{1/2}f)\Psi_2\|\nonumber\\
  &\leq
  \left(\sqrt{2n_1+1}+\sqrt{2n_2+1}\right)
  \|\Psi_1\|\|\Psi_2\|\|A\|\|f\|_2\nonumber
  \,.
\end{align}
Applying an analogous argument to $\pi$ yields exactly the same bound for $|C_+(\omega^{-1/2}f)|$,
\begin{align}
  \frac{|C_\pm(\omega^{\mp 1/2}f)|}{\|f\|_2}
  \leq
  c(n_1,n_2)\|\Psi_1\|\|\Psi_2\|\|A\|
  ,
  \quad
  c(n_1,n_2) := \sqrt{2n_1+1}+\sqrt{2n_2+1}
  \,.
\end{align}
These estimates imply that $C_+$ and $C_-$ are tempered distributions, and by application of Riesz' theorem, the functionals $f\lmto C_\pm(\omega^{\mp 1/2}f)$ are given by integration against functions in $L^2(\Rl)$ whose norm is bounded by $c(n_1,n_2)\|\Psi_1\|\|\Psi_2\|\|A\|$. In particular, the Fourier transforms $\tilde{C}_\pm$ exist as well-defined functions.

In view of the localization of $\varphi$, $\pi$ in the negative half line and of $A$ in the right wedge, we have $\supp C_\pm\subset \Rl_+$, and conclude that $\tilde{C}_\pm$ has an analytic continuation to the lower half plane, satisfying polynomial bounds at the boundary and at infinity \cite[Thm. IX.16]{SimonReed2}.

In the rapidity picture, we consider
\begin{align}\label{def:chat}
  \hat{C}_+(\te) := \tilde{C}_+(m\sinh\te)
  \,,\qquad
  \hat{C}_-(\te) := m\cosh\te\cdot\tilde{C}_-(m\sinh\te).
\end{align}
As $\sinh(.)$ maps $S(-\pi,0)$ to the lower half plane, $\hat{C}_\pm$ is analytic in this strip. In view of the estimate on $|C_\pm(\omega^{\mp 1/2}f)|$, the boundary values \bref{def:chat} are functions in $L^2(\Rl,d\te)$, with norm bounded by $(\omega_p=(p^2+m^2)^{1/2})$
\begin{align}
\int d\te\,|\hat{C}_\pm(\te)|^2		&=	\int dp\,\left(\omega_p^{\mp 1/2} |\tilde{C}_\pm(p)|\right)^2 \leq c(n_1,n_2)\|\Psi_1\|\|\Psi_2\|\|A\|\,.
\end{align}
The reflection symmetry \bref{pmsym} can be read off directly from \bref{def:chat}.

For the proof that $\hat{C}_\pm$ lies in the Hardy space $H^2(S(-\pi,0))$, we need an estimate on the $L^2$-norm of $\te\mapsto \hat{C}_\pm(\te-i\la)=\hat{C}_{\pm,-\la}(\te)$. To this end we consider the "shifted" functions $\hat{C}_\pm^{(s)}(\zeta):=e^{-ims\,\sinh\zeta}\cdot\hat{C}_\pm(\zeta)$, $s>0$, which decay rapidly for $|\te|\to\infty$, $0<\la<\pi$:
\begin{align}
  |\hat{C}_\pm^{(s)}(\te-i\la)|
  &=
  |e^{-ims\sinh(\te-i\la)}
  \,\hat{C}_\pm(\te-i\la)|
  =
  e^{-ms\sin\la\, \cosh\te}
  \,|\hat{C}_\pm(\te-i\la)|
  \;.\label{mono}
\end{align}
As $|\hat{C}_\pm(\te-i\la)|$ is bounded by a polynomial in $\cosh\te$ for $|\te|\to\infty$, we have $\hat{C}^{(s)}_{\pm,-\la}\in
L^2(\Rl)$ for all $\la\in[0,\pi]$ and $s>0$. In view of the previous estimates and \bref{pmsym}, $\|\hat{C}_\pm\|_2$ and $\|\hat{C}_{\pm,-\pi}\|_2$ are bounded by $c(n_1,n_2)\|\Psi_1\|\|\Psi_2\|\|A\|$. So the three lines theorem (see appendix \ref{app-hardy}) can be applied and we get
\begin{align}\label{l2bnd}
  \|\hat{C}^{(s)}_{\pm,-\la}\|_2
  \leq
   c(n_1,n_2)\|\Psi_1\|\|\Psi_2\|\|A\|
  \;,\qquad
  0\leq \la\leq\pi
  \;.
\end{align}
But as \bref{mono} is monotonically increasing as $s\to 0$, this uniform bound holds also for $\hat{C}_{\pm,-\la}$, $0\leq\la\leq\pi$. Hence we have proven
\begin{align}
\hat{C}_\pm \in H^2(S(-\pi,0))\,,\qquad \bno{\hat{C}_\pm} \leq c(n_1,n_2)\|\Psi_1\|\|\Psi_2\|\|A\|\,,
\end{align}
as claimed above.
\end{proof}

\noindent Instead of using the time zero fields, one could also study the wedge-localized field $\phi$ itself, and obtain analyticity results from the commutators $[\phi(f),A]$. In the context of free field theory on four-dimensional Minkowski space, such an analysis has been carried out by J. Bros \cite{bros-compact}. The support properties of $[\phi(f),A]$ in $\Rl^2$ lead to analytic properties in two-dimensional complex momentum space, whereas the support properties of the time zero commutators $[\varphi(f),A]$, $[\pi(f),A]$ in $\Rl$ lead to analytic properties on the one-dimensional complex mass shell, i.e. to analytic continuation in the rapidity. Since this is the feature we are interested in, we work in the time zero formalism.

The following corollary of Lemma \ref{lemma:Cphipi} is our basic tool for the derivation of analytic properties of the wavefunctions $(A\Om)_n$.

\begin{corollary}\label{lemma:chat}
With the notations introduced in Lemma \ref{lemma:Cphipi}, consider the two functionals $C,C^\dagger : \Ss(\Rl)\lto\Cl$,
  \begin{align}\label{c-cd}
    C(f) := \langle\Psi_1,\,[z(\fhat),A]\,\Psi_2\rangle,
    \qquad
    C^\dagger(f) := \langle\Psi_1,\,[\zd(\fhat),A]\,\Psi_2\rangle \,.
  \end{align}
There exists a function $\hat{C}\in H^2(S(-\pi,0))$ satisfying the bound \bref{Chardybound} and
  \begin{align}
    C(f)
    =
    \int_\Rl d\te\,\hat{C}(\te)\fhat(\te),
    \qquad\qquad
    C^\dagger(f)
    =
    -\int_\Rl d\te\,\hat{C}(\te-i\pi)\fhat(\te)\,.
  \end{align}
\end{corollary}
\begin{proof}
We define the function $\hat{C}$ in terms of $\hat{C}_\pm$ from Lemma \ref{lemma:Cphipi} as
\begin{align}
  \hat{C}(\zeta)
  &:=
  \frac{1}{2}(\hat{C}_-(\zeta)+i\,\hat{C}_+(\zeta))
  \,,\qquad
  \zeta\in\overline{S(-\pi,0)}\,.
\end{align}
Clearly $\hat{C}$ lies in $H^2(S(-\pi,0))$ and satisfies the bound \bref{Chardybound}. To show that its boundary values reproduce the functionals $C$, $C^\dagger$, we express the Zamolodchikov operators in terms of the time zero fields \bref{def:phipi}. The annihilation operator is  $z(\fhat)=\frac{1}{2}(\varphi(f_-)+i\pi(\omega^{-1}f_-))$, $f_-(x)=f(-x)$. Inserted in $C(f)$, this gives
\begin{align*}
  C(f)
  &=
  \frac{1}{2}(C_-(f_-)+i\,C_+(\omega^{-1}f_-))
  =
  \frac{1}{2}\int dp\,
  \bigg(\tilde{C}_-(p)+\frac{i\,\tilde{C}_+(p)}{\sqrt{p^2+m^2}}\bigg)\fti(p)
  \\
  &=
  \frac{1}{2}\int d\te\,
  \big(\hat{C}_-(\te)+i\,\hat{C}_+(\te)\big)\fhat(\te)
  =
  \int d\te\,
  \hat{C}(\te)\fhat(\te)\,.
\end{align*}
For the creation operator we have $\zd(\fhat)=\frac{1}{2}(\varphi(f)-i\pi(\omega^{-1}f))$ and, by taking into account the symmetry relation $\hat{C}_\pm(\te-i\pi)=\pm\hat{C}_\pm(-\te)$ \bref{pmsym},
\begin{align*}
  \Cd(f)
  &=
  \frac{1}{2}(C_-(f)-i\,C_+(\omega^{-1}f))
  =
  \frac{1}{2}\int d\te\,
  \big(\hat{C}_-(-\te)-i\,\hat{C}_+(-\te)\big)\fhat(\te)
  \\
  &=
  -\frac{1}{2}\int d\te\,
  \big(\hat{C}_-(\te-i\pi)+i\,\hat{C}_+(\te-i\pi)\big)\fhat(\te)
  =
  -\int d\te\,
  \hat{C}(\te-i\pi)\fhat(\te)
  \,.
\end{align*}
\end{proof}

\noindent Corollary \ref{lemma:chat} can be used to derive analytic properties of single particle
wavefunctions $(A\Omega)_1$ corresponding to operators $A\in\A(W_R)$. In fact, by putting $\Psi_1=\Psi_2=\Omega$ in \bref{c-cd} we obtain, $\fhat\in L^2(\Rl,d\te)$, 
\begin{align*}
  \int d\te\, \fhat(\te)\,\hat{C}(\te)
  =
  \langle \Omega, [z(\fhat),A]\,\Omega\rangle
  =
  \langle \overline{\fhat},\,A\Omega\rangle
  =
  \int d\te\, \fhat(\te)\,(A\Omega)_1(\te)
\,.
\end{align*}
Hence the single particle wavefunctions of operators localized in $W_R$ are boundary values of functions in $H^2(S(-\pi,0))$, with norm\footnote{It is not difficult to see that in this case, the bound can be improved to $\bno{(A\Om)_1}\leq\|A\|$.} $\bno{(A\Om)_1}\leq 2\|A\|$. This observation leads to a proof of the nuclearity of the single particle map $\Xi_1(s)$ \bref{def:Xin} as follows.

We consider $\Xi_1(s)$ as the concatenation of two maps, the first mapping $\A(W_R)$ into $H^2(S(-\pi,0))$ according to $A\lmto(A\Om)_1$, and the second mapping $H^2(S(-\pi,0))$ into $L^2(\Rl,d\te)$ according to $(A\Om)_1\lmto(A(\bs)\Om)_{1,-\pi/2}=\Xi_1(s)A$. We have shown above that the former map is bounded as a linear map between the Banach spaces $(\A(W_R),\|\cdot\|_{\B(\Hil)})$ and $(H^2(S(-\pi,0)),\bno{\cdot})$. The latter is explicitly given by (cf. \bref{xi-n-ana})
\begin{align}
\Delta_1(s) : H^2(S(-\pi,0)) \lto \Hil_1
\,,\qquad
(\Delta_1(s) F)(\te) &:= e^{-ms\cosh\te}\cdot F(\te-\tfrac{i\pi}{2})\,.
\end{align}
Making use of the analyticity and boundedness properties of $F\in H^2(S(-\pi,0))$, as well as of its convergence to its boundary values, we may express $F(\te-\frac{i\pi}{2})$ as a Cauchy integral over a closed curve $\gamma$ around $\te-\frac{i\pi}{2}$, and then deform $\gamma$ to the boundary of $S(-\pi,0)$. This yields
\begin{align*}
F(\te-\tfrac{i\pi}{2})
&=
\frac{1}{2\pi i}\oint_\gamma d\zeta'\,\frac{F(\zeta')}{\zeta'-\te+\frac{i\pi}{2}}
=
\frac{1}{2\pi i}\int_\Rl d\te'\,\left(\frac{F(\te'-i\pi)}{\te'-\te-\frac{i\pi}{2}} - \frac{F(\te')}{\te'-\te+\frac{i\pi}{2}}\right)
\,.
\end{align*}
Hence we find
\begin{align}
\Delta_1(s)F	=	\frac{1}{2}(T_s^- F_{-\pi} - T_s^+ F_0)\,,
\end{align}
where $T_s^\pm$ are integral operators on $L^2(\Rl)$, defined by the integral kernels
\begin{align}
T_s^\pm(\te,\te')
&:=
\frac{1}{\pi i}\;\frac{e^{-ms\cosh\te}}{\te'-\te\pm\frac{i\pi}{2}}\,.
\end{align}
It is shown in appendix \ref{app:intop} by a standard argument that $T_s^\pm$ are trace class operators on $L^2(\Rl)$ for any $s>0$. As $F\longmapsto F_0$ and $F\lmto F_{-\pi}$ are bounded maps from $H^2(S(-\pi,0))$ to $L^2(\Rl)$, we conclude that $\Xi_1(s)$ is nuclear (cf. Lemma \ref{Lem:nuclearmaps} c), d)).
\begin{corollary}\label{xi1nuc}
Consider a model theory with scattering function $S_2\in\SF$. The map $\Xi_1(s):\A(W_R)\lto\Hil_1$, $A\lmto\Delta^{1/4}U(\bs)(A\Om)_1$ \bref{def:Xin} is nuclear for arbitrary splitting distances $s>0$. {\hfill $\square$}
\end{corollary}

The property of $(A(s)\Om)_1$ which was used in the proof was that this function extends analytically to the one-dimensional tube $\Rl + i\,(-\pi,0)$ and defines a Hardy type function. The tube $\Rl + i\,(-\pi,0)$ contains the line $\Rl-\frac{i\pi}{2}$, and analytic continuation to this line yields $\Delta^{1/4}(A\Om)_1$ \bref{xi-n-ana}.

The same mechanism will also be used to establish the nuclearity of $\Xi_n(s)$ for $n>1$, namely, we will prove that $(A(s)\Om)_n$ extends to a function in the Hardy space $H^2(\Tu)$ over an $n$-dimensional tube $\Tu$ containing the subspace $\Rl^n-i(\frac{\pi}{2},...,\frac{\pi}{2})$. Expressing $(A(\bs)\Om)_n(\te_1-\frac{i\pi}{2},...,\te_n-\frac{i\pi}{2})$ as a Cauchy integral over the boundary of $\Tu$, we then arrive at the nuclearity of $\Xi_n(s)$. But in contrast to the situation on the single particle space, the properties of the underlying scattering function $S_2$ have an important influence on the analyticity domain of $(A\Om)_n$, and hence on the nuclearity properties of $\Xi_n(s)$, in the higher-dimensional case. We will have to restrict to a subclass $\SF_0\subset\SF$ of scattering functions, and now motivate the choice of $\SF_0$.
\\
\\
It is well known from the study of analytic properties of scattering amplitudes that in general, analyticity of the S-matrix in a domain larger than the physical region (which in the present setting is $S(0,\pi)$) cannot be expected \cite{martin}. However, if the theory is required to have decent thermodynamical properties, one is led to consider only models whose scattering functions can be analytically continued to a slightly larger region, as the following heuristic argument suggests. 

Poles of $S_2$ lying in the strip $S(-\pi,0)$ are usually interpreted as evidence for unstable particles with a finite lifetime \cite{eden}, cf. also section \ref{sec:factorizingS}. The lifetime of such a resonance becomes arbitrarily long if the corresponding pole lies sufficiently close to the real axis. In fact, there exist scattering functions having a sequence of zeros $\beta_k\in S(0,\pi)$ (which is accompanied by a sequence of poles at $-\beta_k$ because of $S_2(-\te)=S_2(\te)^{-1}$) giving rise to infinitely many resonances with arbitrarily long lifetimes and ``masses'' $m_k$ so that $\sum_k e^{-m_k/T}$ diverges for all temperatures $T>0$. But a model with these characteristics cannot be expected to have a regular thermodynamical behavior or only a finite partition function \cite{BuWi, BuJu2}.

On the other hand, if the modular nuclearity holds in a given model, also the thermodynamically significant map $\Theta_{\beta,\OO}:\A(\OO)\to\Hil$, $A\mapsto e^{-\beta H}A\Om$, which is at the basis of the energy nuclearity condition satisfies certain nuclearity properties, as discussed in section \ref{sec:mnc}. We therefore expect the maps $\Xi(s)$ \bref{Xis} {\it not} to be nuclear in a model with the previously described distribution of poles in its scattering function (although there might still exist local observables even in this situation). To exclude such models, we require \bref{def:kappa}
\begin{align}
\kappa(S_2)
:=
\inf\big\{{\rm Im}\,\zeta\,:\,\zeta\in S(0,\tfrac{\pi}{2})\,,\quad S_2(\zeta)=0\big\}\;>0
\,.
\end{align}
In this case all singularities of $S_2$ lie a finite distance off the real axis so that the lifetimes of all resonances are bounded from above. Hence regular thermodynamical properties can be expected, and the modular nuclearity condition might be satisfied.

In addition to this requirement we make a second, more technical restriction on the scattering function, namely we require that $S_2$ is bounded on the strip $S(-\kappa,\pi+\kappa)$ for $0\leq\kappa<\kappa(S_2)$. This is a condition on the phase shift we need in order to get bounds on the analytic continuations of the wavefunctions $(A\Om)_n$, see \cite{karo} for a similar assumption.

In the following, we will therefore restrict ourselves to scattering functions which are contained in the following subfamily of $\SF$.
\begin{definition}\label{def:S0family}
The subfamily $\SF_0\subset\SF$ consists of those scattering functions $S_2$ which satisfy $\kappa(S_2)>0$ and for which
\begin{align}\label{def:S2kappa}
\|S_2\|_\kappa
&:=
\sup\big\{|S_2(\zeta)|\,:\,\zeta\in \overline{S(-\kappa,\pi+\kappa)}\big\}<\infty
\,,\qquad \kappa\in(0,\kappa(S_2))\,.
\end{align}
\end{definition}
\noindent Note that the boundedness condition requires the parameter $a$ in the factorization formula \bref{s2-rep} to vanish. The family $\SF_0$ contains in particular all scattering functions with finitely many zeros in $S(0,\pi)$ and parameter $a=0$ \bref{s2-rep}, i.e. functions of the form
\begin{align}\label{rappo}
     S_2(\zeta) = \pm\prod_{k=1}^M\frac{\sinh\beta_k-\sinh\zeta}{\sinh\beta_k+\sinh\zeta}
\,,
\qquad
0<{\rm Im}\beta_1,...,{\rm Im}\beta_M<\frac{\pi}{2}\,.
\end{align}
A prominent example is given by the scattering function of the Sinh-Gordon model, which consists of a single factor in the product \bref{rappo}.

It should also be noted that scattering functions $S_2\in\SF_0$ have smooth boundary values on the real line. Moreover, by application of Cauchy's integral formula to the functions $S_2\in\SF_0$, it follows that all derivatives of $S_2$ are uniformly bounded on $\Rl$, i.e. there exist constants $c_m$ (depending on $S_2$) such that $|\partial^m_\te S_2(\te)|\leq c_m$ for all $\te\in\Rl$.
\\
\\
Having clarified which class of scattering functions will be considered in the following, we now turn to the study of the analyticity and boundedness properties of the wavefunctions $(A\Om)_n$, $n>1$. We begin by analyzing matrix elements of the form $\langle \zd(\te_{k+1})\cdots\zd(\te_n)\Om,\,A\,\zd(\te_k)\cdots\zd(\te_1)\Om\rangle$, $A\in\A(W_R)$, with certain contractions between the variables $\te_{k+1},...,\te_n$ on the left and $\te_1,...,\te_k$ on the right hand side. 

Some notations need to be introduced. Given two integers $0\leq k\leq n$, we define the {\em set $\CC_{n,k}$ of contractions} to be the power set of $\{k+1,...,n\}\times\{1,...,k\}$. We parametrize a contraction $C\in\CC_{n,k}$ by an ordered set of "right" indices $1\leq r_1<r_2<...<r_N\leq k$, an unordered set of pairwise different "left" indices $k+1\leq l_1,...,l_N\leq n$, and a permutation of $\{l_1,...,l_N\}$ to form the pairs $(l_i,r_i)\in C$. The number of such pairs will be referred to as the {\em length of $C$}, and denoted $|C|:=N\leq\min\{k,n-k\}$.

Writing $\bl_C:=(l_1,...,l_N)$, $\br_C:=(r_1,...,r_N)$ for the "left" and "right" indices of a given $C\in\CC_{n,k}$, we define for $A\in\B(\Hil)$
\begin{align}\label{def-arl}
\langle \bl_C |\,A\,|\br_C\rangle_{n,k}
&:=
\langle
\zd_{k+1}\cdots\widehat{\zd_{l_1}}\cdots\widehat{\zd_{l_N}}\cdots\zd_n\Om
\,,A\,
\zd_k\cdots\widehat{\zd_{r_N}}\cdots\widehat{\zd_{r_1}}\cdots\zd_1\Om
\rangle\,,
\end{align}
where $\zd_a := \zd(\te_a)$ is considered as an operator-valued distribution in $\te_a$ and the hats indicate omission of the corresponding creation operators. Note that in view of the particle number bounds \bref{Z-bounds} and the boundedness of $A$, these contracted matrix elements are well-defined tempered distributions on $\Ss(\Rl^{n-2|C|})$. For square-integrable functions $F_L\in L^2(\Rl^{n-k-|C|})$, $F_R\in L^2(\Rl^{k-|C|})$, there hold the bounds
\begin{align}\label{lr-bnd}
\left|\langle\bl_C|\,A\,|\br_C\rangle_{n,k}(F_L\otimes F_R)\right|
&\leq
\sqrt{(n-k-|C|)!}\sqrt{(k-|C|)!}\,\|F_L\| \|F_R\|\|A\|\,.
\end{align}
Employing the shorthand notations
\begin{align}
\delta_{l,r}		&:=	\delta(\te_l-\te_r),\quad
S_{a,b}		:=	S_2(\te_a-\te_b),\quad
S^{(k)}_{a,b}
			:=
			\left\{
			\begin{array}{lll}
			S_{b,a}	&;	& a\leq k<b\;\;{\rm or}\;\; b\leq k< a\\
			S_{a,b}	&;	& {\rm otherwise}
			\end{array}
			\right.
\,,
\end{align}
we associate with each contraction $C\in\CC_{n,k}$ the following distributions and functions:
\begin{align}\label{def-dS}
\delta_C		&:=	\prod_{j=1}^{|C|} \delta_{l_j,r_j}
\,,\qquad 
S_C^{(k)}		:=	\prod_{j=1}^{|C|} \prod_{m_j=r_j+1}^{l_j-1}
				S^{(k)}_{m_j,r_j}
				\cdot
				\prod_{r_i<r_j \atop l_i < l_j}
				S^{(k)}_{r_j,l_i}
				\,.
\end{align}
Here and in the following, the indices $l_i,r_i$ refer to the pairs in $C$, with the convention $r_1<...<r_{|C|}$.

The main objects of interest will be the completely contracted matrix elements of observables $A\in\A(W_R)$, defined as
\begin{align}\label{def-Acon}
\langle A\rangle^{\rm con}_{n,k}
&:=
\sum_{C\in\CC_{n,k}} (-1)^{|C|}\cdot\delta_C\cdot S_C^{(k)}\cdot\langle\bl_C|\,A\,|\br_C\rangle_{n,k}
\,.
\end{align}

The product $\delta_C\cdot S_C^{(k)}\cdot\langle\bl_C|\,A\,|\br_C\rangle_{n,k}$ is defined in the sense of distributions. Note that the product of $\delta_C$ and $\langle\bl_C|\,A\,|\br_C\rangle_{n,k}$ is well-defined because these distributions act on different variables. For the product of $\delta_C\cdot\langle\bl_C|\,A\,|\br_C\rangle_{n,k}$ and the function $S_C^{(k)}$ to be well-defined, too, we require $S_2\in\SF_0$. In this case, $S_2$ is smooth and has bounded derivatives on $\Rl$. Hence $\delta_C\cdot S_C^{(k)}\cdot\langle\bl_C|\,A\,|\br_C\rangle_{n,k}$ exists as a tempered distribution on $\Ss(\Rl^n)$. 

The relevant properties of the contracted matrix elements $\langle A\rangle^{\rm con}_{n,k}$ are explained in the following Lemma.
\begin{lemma}\label{lemma:recursion}
In a model with scattering function $S_2\in\SF_0$, consider $A\in\A(W_R)$. 
\begin{enumerate}
\item $\langle A\rangle_{n,k}^{\rm con}$ has an analytic continuation in the variable $\te_{k+1}$ to the strip $S(-\pi,0)$, $k\leq n-1$. The boundary value at Im$(\te_{k+1})=-\pi$ is given by
\begin{align}
\langle A\rangle_{n,k}^{\rm con}(\te_1,...,\te_{k+1}-i\pi,...,\te_n)
&=
\langle A\rangle_{n,k+1}^{\rm con}(\te_1,...,\te_{k+1},...,\te_n)
\,.
\end{align}
\item There holds the bound, $f_1,...,f_n\in\Ss(\Rl)$, $0\leq\la\leq\pi$,
\begin{align}\label{acon-bnd}
\left|\int d^n\bte \langle A\rangle_{n,k}^{\rm con}(\te_1,...,\te_{k+1}-i\la,...,\te_n) \prod_{j=1}^n f_j(\te_j)\right|
&\leq
2^n\sqrt{n!}\,\|A\|\prod_{j=1}^n\|f_j\|_2
\,.
\end{align}
\end{enumerate}
\end{lemma}
\begin{proof}
a) We will need to distinguish between those contractions $C\in\CC_{n,k}$ which do not contract $k+1$, i.e. fulfill $k+1\neq l_1,...,l_{|C|}$, and those contractions which satisfy $l_j=k+1$ for some $j\in\{1,...,|C|\}$. The former set will be denoted $\hat{\CC}_{n,k}$, and the latter $\check{\CC}_{n,k}$. The set of all contractions is the disjoint union $\CC_{n,k}=\hat{\CC}_{n,k}\sqcup \check{\CC}_{n,k}$.

Note that a contraction $C'\in\check{\CC}_{n,k}$ is always a union $C'=C\cup\{(k+1,r)\}$, where $C\in\hat{\CC}_{n,k}$ has length $|C|=|C'|-1$, and $r\notin \br_C$. In this situation, there holds
\begin{align}
(-1)^{|C'|}	&= -(-1)^{|C|}\,,\qquad \delta_{C'} = \delta_{k+1,r}\cdot\delta_C\,, 
\\
S_{C'}^{(k)}	&=
				\prod_{j=1}^{|C|} \prod_{m_j=r_j+1}^{l_j-1}
				S^{(k)}_{m_j,r_j}
				\cdot
				\prod_{m=r+1}^k S^{(k)}_{m,r}
				\cdot
				\prod_{r_i<r_j \atop l_i < l_j}
				S^{(k)}_{r_j,l_i}
				\cdot
				\prod_{r_i < r \atop l_i<k+1} S^{(k)}_{r,l_i}
				\cdot
				\prod_{r<r_j\atop k+1<l_j}S^{(k)}_{r_j,k+1}
				\nonumber\\
				&=
				S_C^{(k)} \cdot \prod_{m=r+1}^k S_{m,r} \cdot \prod_{r<r_j}S_{k+1,r_j}\,, 
\end{align}
since $l_1,...,l_{|C|}>k+1$. Taking into account $S_{a,b}={S_{b,a}}^{-1}$, we get
\begin{align}\label{congehampel1}
\delta_{C'}\cdot S_{C'}^{(k)}
&=
\delta_C\cdot S_C^{(k)}\cdot \delta_{k+1,r}\cdot\prod_{m=r+1\atop m\neq r_j\,{\rm for}\, r_j>r}^k S_{m,k+1}\,.
\end{align}

Similarly, contractions $C''\in\check{\CC}_{n,k+1}$ contracting $k+1$ (as a "right" index) are unions of the form $C''=\{(l,k+1)\}\cup C$, with $C\in\hat{\CC}_{n,k+1}$ and $l\notin\bl_C$. By a computation similar to the one above one finds in this situation
\begin{align}\label{congehampel2}
(-1)^{|C''|} 	&=	-(-1)^{|C|},\qquad
\delta_{C''}=\delta_C\cdot\delta_{l,k+1}
\,,\qquad
S_{C''}^{(k+1)}
=
S_C^{(k+1)}\cdot\!\!\!\!\prod_{m=k+2\atop m\neq l_i\,{\rm for}\, l_i<l}^{l-1} S_{k+1,m}
\,.
\end{align}

After these preparations, consider $C\in\hat{\CC}_{n,k}$ and $A\in\A(W_R)$. By repeated application of the relations of Zamolodchikov's algebra, we find
\begin{align}
\langle\bl_C|\,A\,|\br_C\rangle_{n,k}
=&\langle
\zd_{k+2}\cdots\widehat{\zd_{l_1}}\cdots\widehat{\zd_{l_{|C|}}}\cdots\zd_n\Om
\,,z_{k+1}A\,
\zd_k\cdots\widehat{\zd_{r_{|C|}}}\cdots\widehat{\zd_{r_1}}\cdots\zd_1\Om
\rangle\nonumber
\\
=&\;
\langle
\bl_C \cup \{k+1\}
|
\,[z_{k+1},A]\,
|
\br_C
\rangle_{n,k}\label{z-kalk-1}
\\
&+
\sum_{r=k\atop r\notin \sbr_C}^1 \delta_{k+1,r}
\!\!\!\!
\prod_{m=k\atop m\neq r_i\,{\rm for}\, r_j>r}^{r+1} 
\!\!\!\!
S_{m,k+1} 
\cdot
\langle
\bl_C \cup \{k+1\}
|\,A\,|
\br_C\cup\{r\}
\rangle_{n,k}
\,.\nonumber
\end{align}
Consider the second term, multiplied with $(-1)^{|C|}\delta_C S_C^{(k)}$ and summed over all $C\in\hat{\CC}_{n,k}$. Taking into account the remarks made at the beginning of the proof, there holds $\sum_{r=k,r\notin\sbr_C}^1\sum_{C\in\hat{\CC}_{n,k}} = \sum_{C'\in\check{\CC}_{n,k}}$, with $C$ and $C'$ being related by $C'=C\cup\{(k+1,r)\}$. Moreover, the delta distributions and scattering functions appearing in \bref{z-kalk-1} are the same as in \bref{congehampel1}, and $|C'|=|C|+1$. So we conclude
\begin{align*}
\sum_{C\in\hat{\CC}_{n,k}}
(-1)^{|C|}\delta_C S_C^{(k)}
\langle \bl_C|\,A\,|\br_C\rangle_{n,k}
=&
\sum_{C\in\hat{\CC}_{n,k}}
(-1)^{|C|}\delta_C S_C^{(k)}
\langle \bl_C\cup\{k+1\}|\,[z_{k+1},A]\,|\br_C\rangle_{n,k}
\\
&- \sum_{C'\in\check{\CC}_{n,k}}
(-1)^{|C'|}\delta_{C'} S_{C'}^{(k)}
\langle \bl_{C'}|\,A\,|\br_{C'}\rangle_{n,k}
\,,
\end{align*}
and since $\CC_{n,k}=\hat{\CC}_{n,k}\sqcup \check{\CC}_{n,k}$,
\begin{align}\label{acon-comm}
\langle A\rangle^{\rm con}_{n,k}
&=
\sum_{C\in\hat{\CC}_{n,k}}
(-1)^{|C|}\cdot\delta_C\cdot S_C^{(k)}\cdot
\langle \bl_C\cup\{k+1\}|\,[z_{k+1},A]\,|\br_C\rangle_{n,k}
\,.
\end{align}
This form of $\langle A\rangle^{\rm con}_{n,k}$ is most convenient to discuss its analytic properties. Consider a single term of \bref{acon-comm}, smeared with test functions $f_1(\te_1),...,f_k(\te_k),f_{k+2}(\te_{k+2}),...,f_n(\te_n)$ in all variables but $\te_{k+1}$.

As $k+1$ is not contracted in $C\in\hat{\CC}_{n,k}$, the delta distribution $\delta_C$ does not depend on $\te_{k+1}$. The function $S_C^{(k)}$ depends on $\te_{k+1}$ only via $m_j=k+1$ in $S_{m_j,r_j}^{(k)}$ in \bref{def-dS} since $l_i,r_j\neq k+1$. The factor $S_{k+1,r_j}^{(k)} = S_{r_j,k+1}$ has an analytic continuation in $\te_{k+1}$ to the strip $S(-\pi,0)$, with the crossing-symmetric boundary value $S_{k+1,r_j}=S_{k+1,r_j}^{(k+1)}$. All other factors in $S_C^{(k)}$ are of the form $S_{a,b}^{(k)}$, $a,b\neq k+1$, and therefore satisfy $S_{a,b}^{(k)}=S_{a,b}^{(k+1)}$. Thus $S_C^{(k)}$ can be analytically continued in $\te_{k+1}$ to the strip $S(-\pi,0)$, with boundary value at Im$(\te_{k+1})=-\pi$ given by $S_C^{(k+1)}$.

According to Corollary \ref{lemma:chat}, also the matrix element $\langle \bl_C\cup\{k+1\}|\,[z_{k+1},A]\,|\br_C\rangle_{n,k}$ has an analytic continuation in $\te_{k+1}\in S(-\pi,0)$, and its boundary value at Im$(\te_{k+1})=-\pi$ is obtained by exchanging $[z_{k+1},A]$ with $[A,\zd_{k+1}]$. So the statement about the analytic structure of $\langle A\rangle^{\rm con}_{n,k}$ follows, and it remains to show that the boundary value of this distribution at Im$(\te_{k+1})=-\pi$ is $\langle A\rangle^{\rm con}_{n,k+1}$. We have shown already
\begin{align*}
\langle A\rangle^{\rm con}_{n,k}(\te_1,...,\te_{k+1}-i\pi,...,\te_n)
&=
\sum_{C\in\hat{\CC}_{n,k}}
(-1)^{|C|}\delta_C S_C^{(k+1)}
\langle \bl_C\cup\{k+1\}|\,[A,\zd_{k+1}]\,|\br_C\rangle_{n,k}\,.
\end{align*}
Making use of the relations of Zamolodchikov's algebra once again, we obtain
\begin{align}
\langle \bl_C\cup\{k+1\}|\,[A,\zd_{k+1}]\,|\br_C\rangle_{n,k}
&=
\langle \bl_C|\,A\,|\br_C\rangle_{n,k+1}
\label{refforend}
\\
&\quad-
\sum_{l=k+2\atop l\notin\sbl_C}^n
\delta_{l,k+1}\!\!\!\!
\prod_{m=k+2\atop m\neq l_i\,{\rm for}\,l_i<l}^{l-1}
\!\!\!\!
S_{k+1,m}
\cdot \langle \bl_C|\,A\,|\{k+1\}\cup\br_C\rangle_{n,k+1}
\,.\nonumber
\end{align}
As $C\in\hat{\CC}_{n,k}$ does not contract $k+1$, we may consider $C$ also as an element of $\hat{\CC}_{n,k+1}$. According to the remarks made at the beginning of the proof, the contractions $C'':=C\cup\{(l,k+1)\}$, $l\notin\bl_C$, form all of $\check{\CC}_{n,k+1}$, i.e. $\sum_{l=k+2,l\notin\sbl_C}^n\sum_{C\in\hat{\CC}_{n,k+1}}=\sum_{C''\in\check{\CC}_{n,k+1}}$. Taking into account the relations \bref{congehampel2}, it follows that the second term on the right hand side in \bref{refforend}, multiplied with $(-1)^{|C|}\delta_C S_C^{(k+1)}$ and summed over $C\in\hat{\CC}_{n,k+1}$, gives $\sum_{C''\in\check{\CC}_{n,k+1}}(-1)^{|C''|}\delta_{C''}S_{C''}^{(k+1)}\langle\bl_{C''}|\,A\,|\br_{C''}\rangle_{n,k+1}$. Together with the first term in \bref{refforend}, we obtain
\begin{align*}
\langle A\rangle^{\rm con}_{n,k}(\te_1,...,\te_{k+1}-i\pi,...,\te_n)
&=
\sum_{C\in{\CC}_{n,k+1}}
(-1)^{|C|}\delta_C S_C^{(k+1)}
\langle \bl_C|\,A\,|\br_C\rangle_{n,k+1}
=
\langle A\rangle_{n,k+1}^{\rm con}(\bte)\,,
\end{align*}
completing the proof of part a) of the Lemma.
\\

b) Let $C\in\CC_{n,k}$ and put $\bte_\sbr := (\te_{r_1},...,\te_{r_{|C|}})$. Note that each factor in the product $S_C^{(k)}$ \bref{def-dS} depends either on one contracted variable and one uncontracted variable, or on two contracted variables $\te_{r_j}$, $\te_{l_i}$. Hence we may split the product $S_C^{(k)}$ into three factors  $S_C^{(k)}= S_{C,L}^{(k)}\cdot S_{C,M}^{(k)}\cdot S_{C,R}^{(k)}$, where the "left" factor $S_{C,L}^{(k)}$ depends on $\{\te_{k+1},...,\te_n\}\backslash\{\te_{l_1},...,\te_{l_{|C|}}\}$ and $\bte_\sbr$, the "middle" factor $S_{C,M}^{(k)}$ depends on $\te_{l_1},...,\te_{l_{|C|}},\te_{r_1},...,\te_{r_{|C|}}$, and the "right" factor $S_{C,R}^{(k)}$ depends on $\{\te_1,...,\te_k\}$.

For $f_1,...,f_n\in\Ss(\Rl)$, let
\begin{align}
F^L_{\sbte_\sbr}
&:=
S_{C,L}^{(k)} \cdot 
\big(
	f_{k+1}\otimes ... \otimes \widehat{f_{l_1}}\otimes ... \otimes \widehat{f_{l_{|C|}}}\otimes ... \otimes f_n
\big)\,,
\\
F^R_{\sbte_\sbr}
&:=
S_{C,R}^{(k)} \cdot 
\big(
	f_k\otimes ... \otimes \widehat{f_{r_{|C|}}}\otimes ... \otimes \widehat{f_{r_1}}\otimes ... \otimes f_1
\big)
\,,
\end{align}
where the hats indicate omission of the corresponding factors. The functions $F^{L/R}_{\sbte_\sbr}$ are considered as functions of $n-k-|C|$ and $k-|C|$ variables, respectively, which depend on the parameter $\bte_\sbr\in\Rl^{|C|}$. In view of the boundedness of the scattering function $S_2$, the $L^2$-norms of $F^{L/R}_{\sbte_\sbr}$ are bounded by
\begin{align}
\|F^L_{\sbte_\sbr}\| 
\leq
\prod_{j=k+1\atop j\notin \sbl_C}^n \|f_j\|_2
\,,\qquad
\|F^R_{\sbte_\sbr}\| 
\leq
\prod_{j=1\atop j\notin \sbr_C}^k \|f_j\|_2
\,,\qquad
\bte_\sbr\in\Rl^{|C|}\,.
\end{align}

With these notations, we consider the distribution $\langle A \rangle^{\rm con}_{n,k}$, smeared with a test function of the product form $f_1(\te_1)\cdots f_n(\te_n)$. After carrying out the integration over the delta distributions in \bref{def-Acon}, we find
\begin{align*}
&\langle A \rangle^{\rm con}_{n,k}(f_1\otimes ... \otimes f_n)
\\
&=
\sum_{C\in\CC_{n,k}}(-1)^{|C|}
\int d^{|C|}\bte_\sbr \,
S_{C,M}^{(k)}(\bte_{\sbr};\bte_{\sbr})
 \prod_{j=1}^{|C|} \left(f_{l_j}(\te_{r_j})f_{r_j}(\te_{r_j})\right)
\cdot
\langle\bl_C|\,A\,|\br_C\rangle_{n,k}
(F^L_{\sbte_\sbr} \otimes F^R_{\sbte_\sbr})
\,.
\end{align*}
Taking into account \bref{lr-bnd} and the bounds on $\|F^{L/R}_{\sbte_\sbr}\|$ and $|S_{C,M}^{(k)}(\bte_\sbr;\bte_\sbr)|=1$, Cauchy-Schwarz gives
\begin{align}\label{esoest1}
\left|\langle A \rangle^{\rm con}_{n,k}(f_1\otimes ... \otimes f_n)\right|
&\leq
\sum_{C\in\CC_{n,k}} \sqrt{(n-k-|C|)!(k-|C|)!} \cdot\|A\|\cdot \|f_1\|_2\cdots\|f_n\|_2\,.
\end{align}
In particular, it follows that $\te_{k+1}\mapsto\int \langle A\rangle^{\rm con}_{n,k}(\bte)\prod_{j=1\atop j\neq k+1} f_j(\te_j)d\te_j$ is an element of $L^2(\Rl)$. Making use of the fact that $S_2$ is bounded on $S(0,\pi)$, and the bounds of Cor. \ref{lemma:chat}, it follows that also $\te_{k+1}\mapsto\int \langle A\rangle^{\rm con}_{n,k}(\te_1,...,\te_{k+1}-i\la,...,\te_n)\prod_{j=1\atop j\neq k+1} f_j(\te_j)d\te_j$ is square-integrable, for any $0\leq\la\leq\pi$. By virtue of the three lines theorem, this implies that the bound \bref{esoest1} also holds for $|\int \langle A\rangle^{\rm con}_{n,k}(\te_1,...,\te_{k+1}-i\la,...,\te_n)\prod_{j=1} f_j(\te_j)d\te_j|$ instead of $|\langle A \rangle^{\rm con}_{n,k}(f_1\otimes ... \otimes f_n)|$.

It remains the combinatiorial problem to estimate the sum over all contractions. Note that the number of all contractions $C\in\CC_{n,k}$ with fixed length $|C|=N$ is $N!\left(k\atop N\right)\left(n-k\atop N\right)$, since each such contraction is given by two $N$-element subsets $\{r_1,...,r_N\}\subset\{1,...,k\}$ and $\{l_1,...,l_N\}\subset\{k+1,...,n\}$, and a permutation of $\{1,...,N\}$ to determine which element of $\{l_1,...,l_N\}$ is contracted with which element of $\{r_1,...,r_N\}$. Using $|C|\leq\min\{k,n-k\}$ and the simple inequality $a!b!\leq (a+b)!$, $a,b\in\N$, we find
\begin{align*}
\sum_{C\in\CC_{n,k}} \!\!\!\!\sqrt{(n-k-|C|)!(k-|C|)!}
\;&=
\sum_{N=0}^{\min\{k,n-k\}}\sqrt{(n-k-N)!(k-N)!}\,N!\left( k \atop N\right) \left( n-k \atop N \right)
\\
&\leq
\sqrt{n!}\sum_{N=0}^{\min\{k,n-k\}}\left( k \atop N\right) \left( n-k \atop N \right)
\\
&\leq
\sqrt{n!} \sum_{N=0}^k \sum_{M=0}^{n-k} \left( k \atop N\right) \left( n-k \atop M \right)
=
\sqrt{n!}\,2^k\cdot 2^{n-k}	
=
\sqrt{n!}\,2^n\,.
\end{align*}
In combination with \bref{esoest1}, this yields the desired bound \bref{acon-bnd}.
\end{proof}

The analytic properties of the contracted matrix elements $\langle A\rangle_{n,k}^{\rm con}$ imply analytic properties of the wavefunctions $(A\Om)_n$ as follows. Noting that $\CC_{n,0}=\CC_{n,n}=\emptyset$, we find
\begin{align}
\langle A\rangle_{n,0}^{\rm con}(\bte)
&=
\langle \zd(\te_1)\cdots\zd(\te_n)\Om,\,A\Om\rangle
=
\sqrt{n!}\cdot (A\Om)_n(\bte)\,,
\label{an0con}
\\
\langle A\rangle_{n,n}^{\rm con}(\bte)
&=
\langle\Om,\,A\zd(\te_n)\cdots\zd(\te_1)\Om\rangle
=
\sqrt{n!}\cdot\overline{(A^*\Om)_n(\te_n,...,\te_1)}
=
\sqrt{n!}\cdot (JA^*\Om)_n(\bte)
\,.
\nonumber
\end{align}
By successive application of Lemma \ref{lemma:recursion} a) to $\langle A\rangle_{n,0}^{\rm con}$, it follows that (along a certain path) $(A\Om)_n$ has an analytic continuation from $\Rl^n$ to $\Rl^n-i(\pi,...,\pi)$. The corresponding boundary value is given by $(JA^*\Om)_n(\bte) = (\Delta^{1/2}A\Om)_n(\bte)$, in agreement with modular theory (cf. \bref{modular-covariance}). But whereas the strong analyticity of $\zeta\lmto \Delta^{i\zeta} A\Omega$ in $S(-\frac{1}{2},0)$, following from modular theory, implies analyticity of $(A\Omega)_n$ only in the center of mass rapidity $n^{-1}(\te_1+...+\te_n)$ in the strip $S(-\pi,0)$, the results of Lemma \ref{lemma:recursion} lead to analyticity of $(A\Omega)_n$ considered as a function of $n$ complex variables, in a certain tube domain, which is formulated as a corollary below (Cor. \ref{cor}).

\begin{figure}[here]
    \noindent
    \psfrag{0}{\snull}
    \psfrag{im1}{\small{Im}$\,\zeta_1$}
    \psfrag{im2}{\small{Im}$\,\zeta_2$}
    \psfrag{k}{\skappa}
    \psfrag{p0}{\spn}
    \psfrag{0p}{\snp}
    \psfrag{-pp}{\spp}
    \psfrag{f1}{\fone}
    \psfrag{f2}{\ftwo}
    \psfrag{f3}{\fthree}
    \psfrag{fxi}{\fxi}
    \centering\epsfig{file=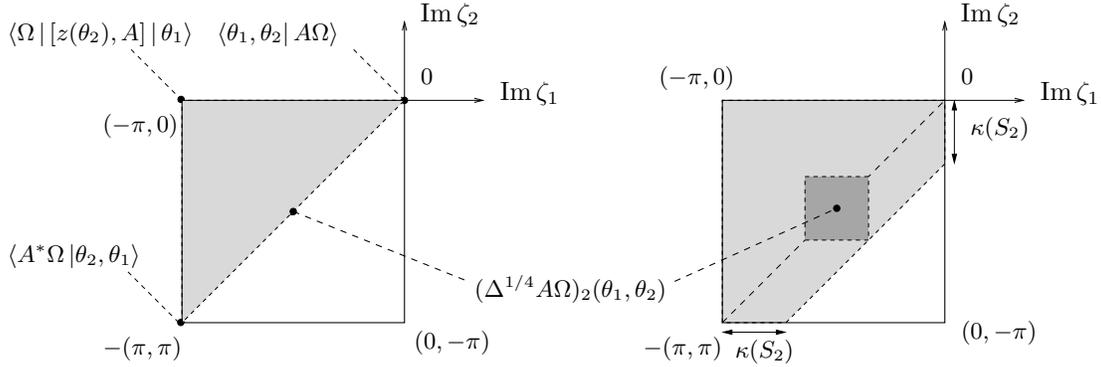,width=14cm}
     \caption{Base of the tube of analyticity of $(A\Om)_2$ (shaded region) with (right) and without (left) continuation of $S_2$ to the enlarged strip $S(-\kappa(S_2),\pi+\kappa(S_2))$}
    \label{fig:ana2}
\end{figure}

In the following, we use the same symbol $(A\Omega)_n$ also for the analytic continuation of the wavefunction \bref{a-wave} in order not to overburden our notation. The domain of holomorphy is the tube
  \begin{align}\label{def:Tn}
      \Tu_n:=\Rl^n-i\Lambda_n
      ,\qquad
      \Lambda_n
      &:=
      \left\{\bla\in\Rl^n \,:\, \pi>\la_1>\la_2>...>\la_n>0\right\}\,.
    \end{align}
\begin{corollary}\label{cor}
  Let $A\in\A(W_R)$.
  \begin{enumerate}
  \item $(A\Omega)_n$ has an analytic continuation to the tube
    $\Tu_n$. The wavefunctions \bref{a-wave} are recovered from
    $(A\Omega)_n(\,.-i\bla)$ as a limit in $\Ss(\Rl^n)'$ for $\bla\to
    0$ in $\Lambda_n$.
  \item Let 
\begin{align}\label{def-dla}
d(\bla)
&:=
\min\left\{\pi-\la_1,\tfrac{1}{2}(\la_1-\la_2),\tfrac{1}{2}(\la_2-\la_3),...,\tfrac{1}{2}(\la_{n-1}-\la_n),\la_n\right\}\,.
\end{align}
There holds the bound, $\bte\in\Rl^n$, $\bla\in\Lambda_n$,
    \begin{eqnarray}\label{bnd33}
      |(A\Omega)_n(\bte-i\bla)|
      &\leq&
	\left(\frac{4}{\pi \, d(\bla)}\right)^{n/2}\cdot\|A\|\,.
    \end{eqnarray}
  \end{enumerate}
\end{corollary}
\begin{proof}
{\em a)} Let $\bof\in\Ss(\Rl^n)$. We claim that the convolution $(A\Omega)_n*\bof$, considered as a function of $\te_1,...,\te_k$, with $\te_{k+1},...,\te_n\in\Rl$ fixed, is analytic in the tube $\Rl^k-i\Lambda_k$ and continuous on its closure. Our proof is based on induction in $k\in\{1,...,n\}$. In view of $\sqrt{n!}(A\Omega)_n=\langle A\rangle^{\rm con}_{n,0}$ \bref{an0con} and $\Rl^1-i\Lambda_1=S(-\pi,0)$, the claim for $k=1$ follows directly from Lemma \ref{lemma:recursion} a).

So assume analyticity of $(\te_1,...,\te_k)\lmto ((A\Omega)_n*\bof)(\te_1,...,\te_n)$ in $\Rl^k-i\Lambda_k$. According to Lemma \ref{lemma:recursion}, the boundary value at ${\rm Im}\,\te_1=...={\rm Im}\,\te_k=-\pi$ is given by $\langle A\rangle^{\rm con}_{n,k}/\sqrt{n!}$, which in turn has an analytic continuation in $\te_{k+1}\in S(-\pi,0)$. By application of the flat tube theorem (cf., for example, \cite{Epstein66}), we conclude that $(A\Omega)_n*\bof$, considered as a function of the first $k+1$ variables, has an analytic continuation to the convex closure of the set
\begin{align*}
  \Rl^{k+1}
  -i \big(
  \left\{(\la_1,...,\la_k,0)\,:\,(\la_1,...,\la_k)\in\Lambda_k\right\}
  \cup
  \left\{(\pi,...,\pi,\la_{k+1})\,:\, \pi > \la_{k+1} > 0 \right\}\!\big)
  \,.
\end{align*}
This convex closure is easily seen to coincide with $\Rl^{k+1}-i\Lambda_{k+1}$. So by induction, $(A\Omega)_n*\bof$ is analytic on $\Tu_n$ and continuous on $\overline{\Tu_n}$. As $\bof$ was arbitrary, statement {\em a)} follows \cite{ovovov2}.

To prove {\em b)}, let $f_1,...,f_n\in \Ss(\Rl)$ and put $\bof:= f_1\otimes ... \otimes f_n$. Lemma \ref{lemma:recursion} {\em c)} implies that at points $\bte-i\bla\in\overline{\Tu_n}$ with $\bla=(\pi,...,\pi,\la_{k+1},0,...,0)$,  $0\leq \la_{k+1}\leq\pi$, there holds the bound
\begin{align}\label{bnd34}
  |((A\Omega)_n*\bof)(\bte-i\bla)|
  \leq
  2^n\|A\|\,\prod_{j=1}^n\|f_j\|_2\,.
\end{align}
By a standard argument (cf., for example the proof of \cite[Lemma A.2]{GL-bros}), this bound can be seen to hold for arbitrary $\bla\in\overline{\Lambda}_n$. Moreover, it extends to $f_1,...,f_n\in L^2(\Rl)$ by continuity.

To establish \bref{bnd33}, we consider discs $D_r(\zeta_k)\subset\Cl$ of radius $r$ and center $\zeta_k\in\Cl$. In view of the mean value property for analytic functions, 
\begin{align}
(A\Om)_n(\bze)
&=
(\pi r^2)^{-n} \int\limits_{D_r(\zeta_1)}d\te'_1 d\la'_1\cdots \int\limits_{D_r(\zeta_n)} d\te_n' d\la_n' \; (A\Om)_n(\bte'+i\bla')
\,,
\end{align}
as long as the polydisc $D_r(\zeta_1)\times...\times D_r(\zeta_n)$ is contained in the analyticity domain $\Tu_n$. Denoting the imaginary part of $\bze$ by $-\bla$, this is the case if the conditions (cf. the definition \bref{def:Tn} of $\Tu_n$)
\begin{align*}
0<\la_n-r, \qquad \pi>\la_1+r\,\qquad \la_j- r>\la_{j+1}+r,\qquad j=1,...,n-1,
\end{align*}
hold. So we have to choose a radius $r<d(\bla)$, with 
\begin{align}
d(\bla)
&:=
\min\{\pi-\la_1,\tfrac{1}{2}(\la_1-\la_2),\tfrac{1}{2}(\la_2-\la_3),...,\tfrac{1}{2}(\la_{n-1}-\la_n),\la_n\}\,.
\end{align}
Let $r(\la_k'):=\sqrt{r^2-(\la_k')^2}$. Then
\begin{align*}
(A\Om)_n(\bte-i\bla)
 &=
 (\pi r^2)^{-n} \int\limits_{-r}^r d\la'_1\int\limits_{-r(\la_1')}^{r(\la_1')}d\te_1' \cdots \int\limits_{-r}^r d\la'_n\int\limits_{-r(\la_n')}^{r(\la_n')}d\te_n'
 \,(A\Om)_n(\bte+\bte'-i\bla+i\bla')
 \\
&=
(\pi r^2)^{-n} \int_{[-r,r]^{\times n}} d^n\bla'\, \left( (A\Om)_n * (\chi_{r(\la_1')}\otimes ... \otimes \chi_{r(\la_n')})\right) (\bte-i\bla+i\bla')
\,,
\end{align*}
where $\chi_{r(\la_k')}$ denotes the characteristic function of $[-r(\la_k'),r(\la_k')]$. Taking into account \bref{bnd34}, this leads us to the estimate
\begin{align}
\left|(A\Om)_n(\bte-i\bla)\right|
&\leq
(\pi r^2)^{-n}\cdot (2r)^n \cdot 2^n\,\|A\| \prod_{k=1}^n \sup_{r(\la_k')\leq r}\|\chi_{r(\la_k')}\|_2
=
\left(\frac{32}{\pi^2\,r}\right)^{n/2}\cdot\|A\|
\,.\label{bnd-4567}
\end{align}
This bound can be slightly improved as follows: By virtue of Cauchy's integral formula, $(A\Om)_n(\bte-i\bla)$ can be written as an $n$-fold contour integral over $\partial D_r(\zeta_1)\times ... \times \partial D_r(\zeta_n)$. Since $(A\Om)_n(\bte-i\bla)$ is bounded in $\bte\in\Rl^n$ for fixed $\bla\in\Tu_n$ \bref{bnd-4567}, the integration contour $\partial D_r(\zeta_k)$ can be deformed to $(\Rl-i\la_k-i r)\cup(\Rl-i\la_k+i r)$, yielding
\begin{align*}
(A\Om)_n(\bte-i\bla)
&=
(2\pi i)^{-n} \sum_{\sbeps} \eps_1\cdots \eps_n \int_{\Rl^n} d^n\bte'\,
\frac{(A\Om)_n(\bte'-i\bla-i r\beps)}{\prod_{k=1}^n(\te_k'-\te_k-ir\eps_k )}
\,.
\end{align*}
Here the sum $\sum_{\sbeps}$ runs over $\beps=(\eps_1,...,\eps_n)$, $\eps_k=\pm 1$. Using the bound \bref{bnd34} and the $L^2$-norm of the Cauchy kernel, $\|\te\mapsto(\te\pm i r)^{-1}\|_2=\pi^{1/2} r^{-1/2}$, we arrive at
\begin{align}\label{bnd789}
|(A\Om)_n(\bte-i\bla)|
&\leq
2^n\,\|A\| \cdot (\pi r)^{-n/2}\,.
\end{align}
Letting $r\to d(\bla)$ yields the claim.
%
%
\end{proof}
In order to mimic the proof of the nuclearity condition in the one particle case (Corollary \ref{xi1nuc}), we need to extend the domain of analyticity of $(A\Om)_n$ in such a way that it contains $(-\frac{i\pi}{2},...,-\frac{i\pi}{2})$, because continuation of $(A\Om)_n$ to $\Rl^n-(\frac{i\pi}{2},...,\frac{i\pi}{2})$ corresponds to the action of the modular operator $\Delta^{1/4}$ (cf. \bref{xi-n-ana}). To achieve an enlargement of the domain $\Tu_n$ of analyticity (which contains the point $(-\frac{i\pi}{2},...,-\frac{i\pi}{2})$ only in its boundary), more specific information on the underlying scattering function is needed. To this end, we exploit the fact that $S_2\in\SF_0$ can be continued to the enlarged strip $S(-\kappa(S_2),\pi+\kappa(S_2))$, where
\begin{align}
 \kappa(S_2)	&:=	\inf\{{\rm Im}\,\zeta\,:\, \zeta\in S(0,\tfrac{\pi}{2})\;,S_2(\zeta)=0\}>0\,.
\end{align}
For simplicity, let us consider the two-dimensional case first. In view of the $S_2$-symmetry \bref{eq:S2sym} of $(A\Om)_2\in\Hil_2$, we have
\begin{align}\label{refl2}
(A\Om)_2(\te_1,\te_2)		&=		S_2(\te_2-\te_1)\cdot(A\Om)_2(\te_2,\te_1)
\,,\qquad\te_1,\te_2\in\Rl\,.
\end{align}
We know from Corollary \ref{cor} that $(A\Om)_2$ is analytic in $\Tu_2=\Rl^2-i\La_2$, and hence $(\te_1,\te_2)\lmto(A\Om)_2(\te_2,\te_1)$ has an analytic continuation to the "flipped" tube $\Rl^2-i\La_2'$, with $\La_2'=\{(\la_1,\la_2)\,:\,0<\la_1<\la_2<\pi\}$. As $(\te_1,\te_2)\lmto S_2(\te_2-\te_1)$ is analytic in $-\kappa(S_2)<{\rm Im}\,\te_2-{\rm Im}\,\te_1<\pi+\kappa(S_2)$, it follows that also the right hand side of \bref{refl2} has an analytic continuation. Since the right and left hand sides agree on the real subspace, we may apply Epstein's generalization of the edge of the wedge theorem \cite{epstein-eotw} to enlarge the domain of holomorphy of $(A\Om)_2$. 
%
%
The enlarged region is depicted in figure \ref{fig:ana2}. Note in particular that it contains the cube with center $(-\frac{\pi}{2},-\frac{\pi}{2})$ and side length $\kappa(S_2)$.

The proof for the $n$-dimensional case is carried out in Proposition \ref{prop:ana} below. 
\\
\\
As a higher-dimensional analogue of the situation depicted in figure \ref{fig:ana2}, we consider for $\kappa>0$ the region
\begin{align}\label{bibobase}
      \Ba_n(\kappa)
      := 
      \big\{\bla\in \Rl^n
      \,:\, 
      0<\la_1,...,\la_n < \pi,\;\,
      \la_k-\la_l < \kappa,\,
      1\leq l< k\leq n
      \big\}
\end{align}
and the cube $\Cu_n(\kappa)+\bla_0 \subset \Ba_n(\kappa)$ defined by
\begin{align}\label{def:Tnk}
\bla_0:=(-\tfrac{\pi}{2},...,-\tfrac{\pi}{2})
\,,\qquad
\Cu_n(\kappa):=(-\tfrac{\kappa}{2},\tfrac{\kappa}{2})^{\times n}\,.
\end{align}
The tube based on this cube is denoted
\begin{align}\label{def:tkn}
\Tu_n(\kappa)
&:=
\Rl^n + i(\bla_0+\Cu_n(\kappa))
\,.
\end{align}


\begin{proposition}\label{prop:ana}
  Consider a model with scattering function $S_2\in\mathcal{S}_0$, and $A\in\A(W_R)$.
  \begin{enumerate}
  \item $(A\Omega)_n$ is analytic in the tube $\Rl^n-i\Ba_n(\kappa(S_2))$.
\item Let $0<\kappa<\kappa(S_2)$. There holds the bound, 
    \begin{align}\label{masterbound}
      |(A\Omega)_n(\bze)|
      &\leq
	\left(\frac{2\sqrt{2}\,\|S_2\|_\kappa}{\sqrt{\pi (\kappa(S_2)-\kappa)}}\right)^n\cdot\|A\|
      \,,\qquad\bze\in\Tu_n(\kappa)\,.
    \end{align}
  \end{enumerate}
\end{proposition}
\begin{proof}
{\em a)} Let $\frS_n$ denote the group of permutations of $n$ objects and consider the ``permuted wavefunctions'' 
\begin{align*}
  (A\Omega)_n^\rho(\bte) 
  &:=
  (A\Omega_n)(\rho^{-1}\bte)
  \;\,=\;\,
  (A\Omega)_n(\te_{\rho(1)},...,\te_{\rho(n)})
  \,,\qquad \rho\in\frS_n\,,
\end{align*}
 which by Corollary \ref{cor} are analytic in the permuted tubes $\Tu_n^\rho :=\Rl^n-i\Lambda_n^\rho$\,,
\begin{align*}
  \Lambda_n^\rho
  &:=
  \rho\Lambda_n
  =
  \big\{\bla\in\Rl^n\,:\,  \pi>\la_{\rho(1)}>...>\la_{\rho(n)}>0\big\}
  \;.
\end{align*}
Recall that $(A\Om)_n\in\Hil_n$ is invariant under the representation $D_n$ of $\frS_n$ \bref{def:Dn}, 
\begin{align}\label{psi-perm}
  (A\Omega)_n(\bte)
  =
  (D_n(\rho)(A\Om)_n)(\bte)
  &=
  S^\rho(\bte)\cdot(A\Omega)_n^\rho(\bte)
  \,,\\
  S^\rho(\bte)
  &=
  \prod_{1\leq l < k \leq n \atop \rho(l) > \rho(k)}
  S_2(\te_{\rho(l)}-\te_{\rho(k)})
  \;.
\end{align} 
As $S_2\in\SF_0$ is analytic in $S(-\kappa(S_2),\pi+\kappa(S_2))$, all the functions $S^\rho$, $\rho\in\frS_n$, are
analytic in the tube $\Rl^n + i \,\Ba'_n(\kappa(S_2))$ with base (see figure \ref{fig:ana3})
\begin{align*}
  \Ba'_n(\kappa(S_2))
  &:=
  \big\{\bla\in\Rl^n
  \,:\, 
  -\kappa(S_2) < \la_k-\la_l < \pi + \kappa(S_2),
  \quad 1\leq l<k\leq n
  \big\}\,.
\end{align*}
Hence the right hand side of \bref{psi-perm} can be analytically continued to the tube based on $\Ba'_n(\kappa(S_2))\cap(-\Lambda_n^\rho)$. But the left hand side of \bref{psi-perm} is analytic in $\Rl^n-i\Lambda_n$, and both sides converge in the sense of distributions to the same boundary values on $\Rl^n$. So we may apply Epstein's generalization of the Edge of the Wedge Theorem \cite{epstein-eotw} to conclude that $(A\Omega)_n$ has an analytic continuation to the tube whose base is the convex closure of 
\begin{align*}
  \bigcup_{\rho\in\frS_n} \Ba'_n(\kappa(S_2))\cap(-\Lambda_n^\rho)\,.
\end{align*}
Since the convex closure of $\bigcup_{\rho}\Lambda_n^\rho$ is the cube $(0,\pi)^{\times n}$, it follows that $(A\Omega)_n$ is
analytic in the tube based on $-(0,\pi)^{\times n}\cap\Ba'_n(\kappa(S_2))=-\Ba_n(\kappa(S_2))$.
\\
\begin{figure}[here]
    \noindent
    \psfrag{0}{\snull}
    \psfrag{im1}{\small{Im}$\,\zeta_1$}
    \psfrag{im2}{\small{Im}$\,\zeta_2$}
    \psfrag{k}{\minskappa}
    \psfrag{0p}{\minpi}
    \psfrag{BB}{\bnstrich}
    \psfrag{BBB}{\bbase}
    \psfrag{L1}{\lone}
    \psfrag{L2}{\ltwo}
    \centering\epsfig{file=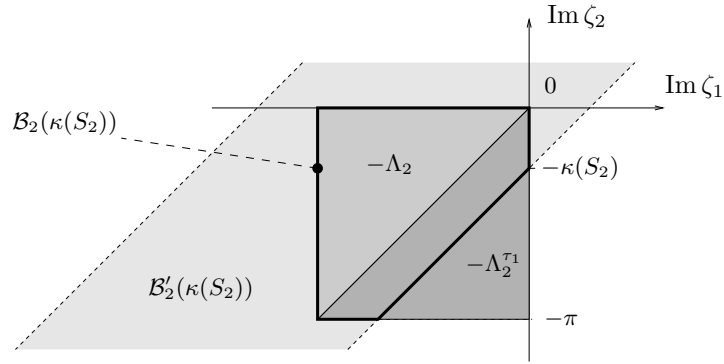,width=9cm}
     \caption{The bases of the tubes involved in the proof of Proposition \ref{prop:ana}}
    \label{fig:ana3}
\end{figure}
\\
{\em b)} To derive the desired bound, we proceed along the same lines as in the proof of Corollary \ref{cor}. Firstly, \bref{bnd34} needs to be generalized to $(A\Om)_n(\bte+i\bla)$, with $\bla\in\Cu_n(\kappa)+\bla_0$, $0<\kappa<\kappa(S_2)$. To this end, we need to estimate $S^\rho$ \bref{psi-perm}. Clearly, the functions $S^\rho$ are bounded on $\Rl^n +i\bla_0 + i \Cu_n(\kappa)$ for $0<\kappa<\kappa(S_2)$, because each factor $S_2(\zeta_k-\zeta_l)$ is bounded (Def. \ref{def:S0family}). By the multidimensional analogue of the three lines theorem \cite{bochner-martin}, the supremum of $S^\rho$ over this tube is attained on a subspace of the form $\Rl^n+i\bla_0+i\bxi$, where $\bxi$ is a vertex of $\Cu_n(\kappa)$, i.e. $\xi_k=\pm \frac{\kappa}{2}$ and hence $\xi_k-\xi_l\in\{0,\kappa,-\kappa\}$. We have $|S_2(0)|=1$, $|S_2(i\kappa)|\leq 1$, and $|S_2(-i\kappa)|\leq\|S_2\|_\kappa<\infty$. At most $(n-1)$ of the differences $\xi_k-\xi_l$ can equal $-\kappa$ simultaneously. So we arrive at $|S^\rho(\bze)| \leq \|S_2\|_\kappa^{\,n-1} \leq \|S_2\|_\kappa^n$, $\bze\in\Tu_n(\kappa)$, and conclude that the bound \bref{bnd34} holds for $\bla-\bla_0\in\Cu_n(\kappa)$ if the right hand side is multiplied with $\|S_2\|_\kappa^{n}$, i.e.
\begin{align}
|(A\Om)_n * (f_1\otimes ... \otimes f_n)(\bte+i\bla)|
&\leq
\left(2\|S_2\|_\kappa\right)^n\|A\|\prod_{k=1}^n\|f_k\|_2\,,\qquad f_k\in L^2(\Rl)\,.
\end{align}
As before, we can use the mean value property to represent $(A\Om)_n(\bte-i\bla)$ as an integral over a polydisc $D_r(\te_1-i\la_1)\times...\times D_r(\te_n-i\la_n)\subset \Tu_n(\kappa(S_2))$. But the radius $r$ can now be chosen larger, as only $\bla+(-r,r)^{\times n}\subset \Cu_n(\kappa(S_2))+\bla_0$ has to be satisfied. The maximal admissible radius is $R:=\min\limits_{k=1,...,n}\{\frac{\kappa(S_2)}{2}\pm(\frac{\pi}{2}+\la_k)\}$. Note that for $\bla-\bla_0\in\Cu_n(\kappa)$, this entails $R\geq \frac{1}{2}(\kappa(S_2)-\kappa)$.

Now we can repeat the arguments leading to the bounds \bref{bnd789} for $\bla\in\bla_0+\Cu_n(\kappa)$, the only differences being the different radius $R$ of the polydisc and the additional factor $\|S_2\|_\kappa^{\,n}$. So we end up with the claimed bound
\begin{align*}
|(A\Om)_n(\bze)|
&
\leq
2^n \|S_2\|_\kappa^{\,n}\|A\|\,(\pi R)^{-n/2}
\leq
	\left(
		\frac{2\sqrt{2}\|S_2\|_\kappa}{\sqrt{\pi\,(\kappa(S_2)-\kappa)}}
	\right)^n
	\cdot\|A\|
\,,\qquad \bze\in\Tu_n(\kappa)\,.
\end{align*}
\end{proof}

The following corollary provides the appropriate generalization of the Hardy space structure of the single particle wavefunctions to the higher-dimensional setting, and is the final result of this section.
\begin{corollary}\label{cor:hardy}
Consider a model with $S_2\in\SF_0$, and let $0<\kappa<\kappa(S_2)$. The restrictions of the functions $(A(\bs)\Om)_n$ to the tube $\Tu_n(\kappa)$ \bref{def:tkn} are elements of the Hardy space $H^2(\Tu_n(\kappa))$, and their Hardy norms are bounded by
\begin{align}
\bno{(A(\bs)\Om)_n}
&\leq
\sigma(2s,\kappa)^n\cdot\|A\|\,.
\end{align}
The constants $\sigma(2s,\kappa)$ can be chosen as
\begin{align}
\sigma(2s,\kappa)
&:=
\frac{2\sqrt{2}\,e^{-2ms\cos\kappa}\|S_2\|_\kappa}{\sqrt{ms\cos\kappa\cdot(\kappa(S_2)-\kappa)}}
\,.\label{an-hardy-bound}
\end{align}
\end{corollary}
\begin{proof}
The analyticity of $(A\Om)_n$ in $\Tu_n(\kappa)$ has been shown before, and as $U(\bs)$ acts by multiplication with the entire function $u_{n,s}(\bze):=\prod_{k=1}^n e^{-ims\sinh\zeta_k}$, also $(A(\bs)\Om)_n(\bze)=u_{n,s}(\bze)\cdot(A\Om)_n(\bze)$ is analytic in this tube. Since $|u_{n,s}(\bze+i\bla_0)|=\prod_{k=1}^n \exp(-ms\cos({\rm Im}\zeta_k)\cosh({\rm Re}\zeta_k))$ and $|{\rm Im}\zeta_k|\leq \kappa<\frac{\pi}{2}$ for $\bze\in\Cu_n(\kappa)$, it follows that $u_{n,s}$ is an element of $H^2(\Tu_n(\kappa))$ for any $s>0$. Using $\cosh\te\geq 1+\frac{1}{2}\te^2$ yields
\begin{align}\label{un-hardy-bound}
\bno{u_{n,s}}
=
\left(\int_\Rl d\te\, e^{-2ms\cos\kappa\,\cosh\te}\right)^{n/2}
\leq
\left(
\frac{\pi^{1/2}e^{-ms\cos\kappa}}{(ms\cos\kappa)^{1/2}}\right)^n\,.
\end{align}
As $(A\Om)_n$ is uniformly bounded on $\Tu_n(\kappa)$, $\kappa<\kappa(S_2)$, this implies $(A(\bs)\Om)_n \in H^2(\Tu_n(\kappa)$, and the bound \bref{an-hardy-bound} is given by the product of \bref{masterbound} and \bref{un-hardy-bound}.
\end{proof}
\noindent The concrete structure of $\sigma(s,\kappa)$ specified in \bref{an-hardy-bound} is not essential in the following. It should only be noticed that for fixed $\kappa$, $\sigma(s,\kappa)$ is a monotonously decreasing function of $s>0$, with the $\sigma(s,\kappa)	\to0$ for $s\to\infty$ and $\sigma(s,\kappa)\to\infty$ for $s\to 0$.

\section{Proof of the Nuclearity Condition}

We now exploit the analytic structure of the wavefunctions $(A\Om)_n$ to prove the nuclearity of the maps $\Xi_n(s)$ \bref{def:Xin}, and later on, the nuclearity of the total map $\Xi(s)$.

As in the proof of the nuclearity of $\Xi_1(s)$, we also treat $\Xi_n(s)$, $n>1$, as the concatenation of two maps, as depicted in the following commutative diagram.
\\
\vspace*{-4mm}
\\
\begin{figure}[here]
  \noindent
  \psfrag{A}{$\A(W_R)$}
  \psfrag{H}{$H^2(\Tu_n(\kappa))$}
  \psfrag{Hi}{$\,\Hil_n$}
  \psfrag{X}{$\Xi_n(s)$}
  \psfrag{S}{$\Sigma_n(s,\kappa)$}
  \psfrag{D}{$\Delta_n(s,\kappa)$}
  \centering\epsfig{file=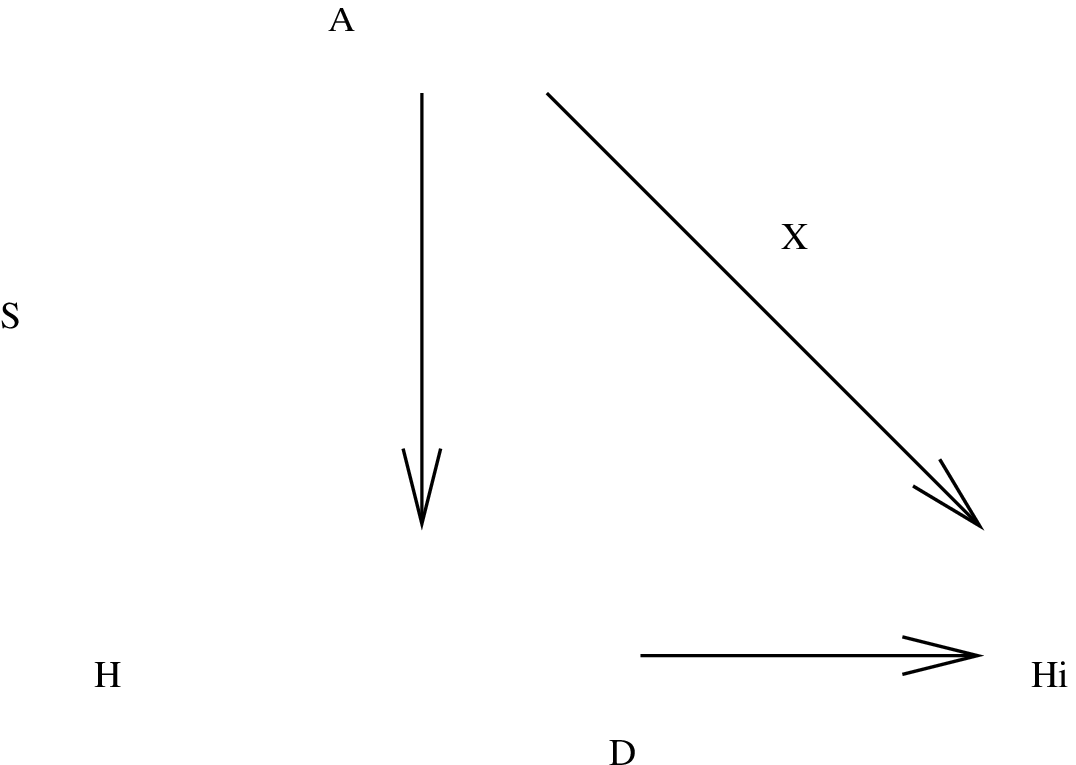,width=35mm}
\end{figure}
\\
Here $\kappa$ is chosen in the interval $(0,\kappa(S_2))$,  and the notations 
\begin{align*}
\bla_0=-\left(\tfrac{\pi}{2},...,\tfrac{\pi}{2}\right)\in\Rl^n
\,,\qquad
\Cu_n(\kappa)=\left(-\tfrac{\kappa}{2},\tfrac{\kappa}{2}\right)^{\times n}
\,,\qquad
\Tu_n(\kappa)=\Rl^n+i(\bla_0+\Cu_n(\kappa))
\end{align*}
are the same as before \bref{def:Tnk}. The maps $\Sigma_n(s,\kappa)$ and $\Delta_n(s,\kappa)$ appearing in the diagram are defined as
\begin{align}
\Sigma_n(s,\kappa):	\A(W_R) &\to H^2(\Tu_n(\kappa))
\,,\qquad
\Sigma_n(s,\kappa)A := (A(\tfrac{1}{2}\bs)\Om)_n\,,\label{def:Sigma_n}
\\
\Delta_n(s,\kappa):	H^2(\Tu_n(\kappa)) &\to \Hil_n
\,,\quad
(\Delta_n(s,\kappa)F)(\bte) := \prod_{k=1}^n e^{-\frac{ms}{2}\cosh\te_k}\cdot F(\bte+i\bla_0)\,,
\label{def:Delta_n}
\end{align}
and in view of \bref{xi-n-ana} we have
\begin{align}\label{xin-split}
\Xi_n(s)A	&=	\left(\Delta_n(s,\kappa)\circ\Sigma_n(s,\kappa)\right)A\,,\qquad A\in\A(W_R)\,.
\end{align}
Corollary \ref{cor:hardy} states that $\Sigma_n(s,\kappa)$ is a bounded linear map between the Banach spaces $(\A(W_R),\|\cdot\|_{\B(\Hil)})$ and $(H^2(\Tu_n(\kappa)),\bno{\cdot})$, and that its operator norm is bounded by $\|\Sigma_n(s,\kappa)\|\leq\sigma(s,\kappa)^n$ \bref{an-hardy-bound}. To show that $\Xi_n(s)$ is nuclear, it therefore suffices to establish the nuclearity of $\Delta_n(s,\kappa)$.
\begin{lemma}\label{lemma:delta}
  Let $s>0$, $\kappa > 0$, and $\Delta_n(s,\kappa)$ be defined as in \bref{def:Delta_n}.
  \begin{enumerate}
  \item $\Delta_n(s,\kappa)$ is a nuclear
    map between the Banach spaces $(H^2(\Tu_n(\kappa)),\bno{\cdot})$
    and $(\Hil_n,\|\cdot\|)$.
  \item Define $T_{s,\kappa}$ as the integral operator on $L^2(\Rl,d\te)$ with kernel
  \begin{eqnarray}\label{int-kernel2}
    T_{s,\kappa}(\te,\te')
    &=&
    \frac{e^{-\frac{ms}{2}\cosh\te}}
    {i\pi \,(\te'-\te - \frac{i\kappa}{2})}   
     \;.
   \end{eqnarray}
   $T_{s,\kappa}$ is of trace class, and there holds the bound
  \begin{eqnarray}\label{delta-bound}
    \|\Delta_n(s,\kappa)\|_1
    &\leq&
    \|T_{s,\kappa}\|_1^{\,n}
    < \infty
    \,.
  \end{eqnarray}
  \end{enumerate}
 \end{lemma}
\noindent In the proof, we will need the following two properties of functions $F\in H^2(\Tu_n(\kappa))$: Firstly, $F_\sbla(\te_1,...,\te_k,...,\te_n)\to 0$ for $|\te_k|\to\infty$ if $\te_1,...,\te_{k-1},\te_{k+1},...,\te_n\in\Rl$ are held fixed, and this limit is uniform in $\bla$ if $\bla$ varies over a compact subset of $\Cu_n(\kappa)$. Secondly, $F_\sbla$ converges in the norm topology of $L^2(\Rl^n)$ as $\bla-\bla_0$ approaches the boundary of $\Cu_n(\kappa)$. For the functions $F=(A(\frac{1}{2}\bs)\Om)_n$, these properties can be derived in a straightforward manner from Proposition \ref{prop:ana}, but they also hold for arbitrary $F\in H^2(\Tu_n(\kappa))$ since $\Cu_n(\kappa)$ is a polyhedron (see appendix \ref{app-hardy}).
\begin{proof}
Let $F\in H^2(\Tu_n(\kappa))$, and pick $\bte\in\Rl^n$ and a polydisc $\DD_n(\bte)\subset\Tu_n(\kappa)$ with center $\bte+i\bla_0$. By virtue of Cauchy's integral formula, we can represent $F(\bte+i\bla_0)$ as a contour integral over $\DD_n(\bte)$,
\begin{align*}
    F(\bte+i\bla_0)
    &=
    \frac{1}{(2\pi i)^n}
    \oint_{\DD_n(\sbte)} d^n\bze'\,
    \frac{F(\bze')}{\prod_{k=1}^n(\zeta'_k-\te_k + \frac{i\pi}{2})}
    \;\;.
\end{align*}
Taking advantage of the two properties of $F$ mentioned above, we can deform the contour of integration to the boundary of $\Tu_n(\kappa)$. After multiplication with the exponential factor \bref{def:Delta_n} we arrive at
\begin{align*}
    (\Delta_n(s,\kappa)F)(\bte)
    &=
    \frac{1}{(2\pi i)^n}\sum_\sbeps
    \int_{\Rl^n}d^n\bte'\,
\left(\,
    \prod_{k=1}^n \frac{\eps_k\;e^{-\frac{ms}{2}\cosh\te_k}}{(\te_k'-\te_k-\frac{i\eps_k\kappa}{2})}
\right)
    \cdot F_{\sbla_0-\frac{\kappa}{2} \sbeps}(\bte')
    \;,
\end{align*}
where the summation runs over $\beps=(\eps_1,...,\eps_n)$, $\eps_1,...,\eps_n=\pm 1$. Expressed in terms of the integral operator $T_{s,\kappa}$, this equation reads
\begin{align}\label{delta-eqn}
    \Delta_n(s,\kappa)F
    &=
    2^{-n}\sum_{\sbeps} \eps_1\cdots\eps_n (T_{s,\eps_1\kappa}\otimes ... \otimes T_{s,\eps_n\kappa})F_{\sbla_0-\frac{\kappa}{2}\sbeps}\;.
\end{align}
The integral operators $T_{s,\pm\kappa}$ are of trace class on $L^2(\Rl)$, as is shown in appendix \ref{app:intop} by a standard argument. Hence $T_{s,\eps_1\kappa}\otimes ... \otimes T_{s,\eps_n\kappa}$ is a trace class operator on $L^2(\Rl^n)$ for any $\eps_1,...,\eps_n=\pm 1$.  Note that as $T_{s,-\kappa}$ is unitary equivalent to $T_{s,\kappa}$ (the equivalence being implemented by $V$, $(Vf)(\te) := i\cdot f(-\te)$), there holds $\|T_{s,\eps_1\kappa}\otimes ... \otimes T_{s,\eps_n\kappa}\|_1=\|T_{s,\kappa}\|_1^n$.

Moreover, it follows from the $L^2$-convergence of $F$ to its boundary values that the maps $F\lmto F_{\sbla_0-\frac{\kappa}{2}\beps}$ are bounded as operators from $H^2(\Tu_n(\kappa))$ to $L^2(\Rl^n)$ for any $\beps$, with norm not exceeding one. Hence the nuclearity of $\Delta_n(s,\kappa)$ \bref{delta-eqn} follows, and since the sum in \bref{delta-eqn} runs over $2^n$ terms, we also obtain the claimed bound $\|\Delta_n(s,\kappa)\|_1\leq\|T_{s,\kappa}\|_1^{\,n}$.
\end{proof}

Lemma \ref{lemma:delta} implies our first nuclearity result for the maps $\Xi(s)=\sum_{n=0}^\infty\Xi_n(s)$ \bref{xi-series}.

\begin{theorem}\label{thm:distalsplit}
For each model theory with scattering function $S_2\in\mathcal{S}_0$, there exists a splitting distance $s_{\min}<\infty $ such that $\Xi(s)$ is nuclear for all $s>s_{\min}$.
\\
Hence in these models, for each double cone $\OO_{a,b}=(W_R+a)\cap(W_L+b)$ with $b-a\in W_R$ and $-(b-a)^2>s_{\min}^2$, the corresponding observable algebra $\A(\OO_{a,b})=\A(W_R+a)\cap\A(W_L-b)$ has $\Om$ as a cyclic vector.
\end{theorem}
\begin{proof}
  Let $\kappa\in(0,\kappa(S_2))$. 
  We have $\Xi_n(s) = \Delta_n(s,\kappa) \circ \Sigma_n(s,\kappa)$, and in view of the previously established bounds (\ref{an-hardy-bound}, \ref{delta-bound}),
\begin{align}
    \|\Xi_n(s)\|_1 
    \;\leq\;
    \|\Sigma_n(s,\kappa)\|\cdot\|\Delta_n(s,\kappa)\|_1
    \;\leq\;
    \left(\sigma(s,\kappa)\cdot\|T_{s,\kappa}\|_1\right)^n
    \;.
\end{align}
For $s\to\infty$, $\|T_{s,\kappa}\|_1$ and $\sigma(s,\kappa)$ converge strictly monotonously to zero (cf. \bref{an-hardy-bound} and \bref{int-kernel2}, see also appendix \ref{app:intop} for an explicit bound on $\|T_{s,\kappa}\|_1$). So there exists $s_{\min}<\infty$ such that $\sigma(s,\kappa)\|T_{s,\kappa}\|_1 < 1$ for all $s>s_{\min}$. But for these values of $s$, there holds
\begin{align}
\sum_{n=0}^\infty \|\Xi_n(s)\|_1 
\leq
\sum_{n=0}^\infty \left(\sigma(s,\kappa)\,\|T_{s,\kappa}\|_1\right)^n
<
\infty
\,,
\end{align}
and hence the series $\sum_{n=0}^\infty \Xi_n(s)$ of nuclear operators converges in nuclear norm to $\Xi(s)$. Since the set of nuclear operators between two Banach spaces is closed with respect to convergence in $\|\cdot\|_1$, the nuclearity of $\Xi(s)$ follows.

The Reeh-Schlieder property for the double cone algebras $\A(W_R\cap(W_L+(0,s)))$, $s>s_{\min}$, is a consequence of the nuclearity of $\Xi(s)$ (Theorem \ref{thm:Ch2summary}). As the double cone $\OO_{a,b}$, $b-a\in W_R$, $-(b-a)^2>s_{\min}^2$, can be transformed to $W_R\cap(W_L+(0,s))$, $s>s_{\min}$, by a translation and a boost, the Reeh-Schlieder property for $\A(\OO_{a,b})$ follows by covariance.
\end{proof}
Theorem \ref{thm:distalsplit} establishes the Reeh-Schlieder property (and all the other consequences of the modular nuclearity condition listed in Thm. \ref{thm:Ch2summary}) for double cones having a minimal "relativistic size". This size is measured by the length $s_{\min}$ and depends on the scattering function $S_2$ and the mass $m$. For example, if we consider the scattering function with a simple pole at $-\frac{i\pi}{4}$,
\begin{align}
S_2(\te)
&=
\frac{i -\sqrt{2}\sinh\te}{i+\sqrt{2}\sinh\te}
\,,
\end{align}
one can use the estimates on $\|T_{s,\kappa}\|_1$ calculated in Lemma \ref{intop-lemma} in appendix \ref{app:intop} to show that $s_{\min} < k/m$, where $k$ is a constant of the order of magnitude 1. In SI units, the bound reads $s_{\min}<\frac{k\,\hbar}{m\,c}$, and is in good agreement with the Compton wavelength $\la_C=\frac{2\pi\hbar}{mc}$ corresponding to the mass $m$. For example, if $m$ is taken to be the mass of the electron, this implies that $s_{\min}$ is of the order of magnitude $10^{-12}$m. But as $\|T_{s,\kappa}\|_1$ and $\sigma(s,\kappa)$ diverge for $s\to 0$, we cannot establish the nuclearity of $\Xi(s)$ for arbitrary small $s$ with these methods.

Whereas the occurrence of a minimal localization length (the Planck length $\la_P\approx 10^{-35}$m) in theories describing quantum effects of gravity is expected for physical reasons, we strongly believe that the minimal length $s_{\min}$ appearing here is an artifact of our estimates, without any physical content. This conjecture is supported by a second theorem, stated below, which improves the previous one under an additional assumption on the underlying scattering function. Namely, we establish the nuclearity of $\Xi(s)$ without restriction on the splitting distance $s$ if $S_2(0)=-1$.
\\
\\
The estimate $\|\Xi_n(s)\|_1 \leq \left(\sigma(s,\kappa)\|T_{s,\kappa}\|_1\right)^n$ used above is rather crude because it does not fully take into account the effects of the ``$S_2$-statistics'', i.e. the symmetry structure \bref{eq:S2sym} of the functions in $\Hil_n$. In principle, it is an estimate on the unsymmetrized Fock space $\F_{\Hil_1}=\bigoplus_n\Hil_1\tp{n}$ describing states of distinguishable particles, and can be improved by working on the much smaller subspace $\Hil\subset\F_{\Hil_1}$.

The symmetry properties of $\Hil$ can most easily be used for an enhanced estimate on $\|\Xi_n(s)\|_1$ in the two special cases of the constant scattering functions $S_2=\pm 1$, where the Hilbert space $\Hil$ of the model coincides with the Bosonic or Fermionic Fock space over $\Hil_1$. Here the combinatorial problems appearing in estimating the nuclear norms of $\Xi(s)$ have been settled in \cite{BuWi} and \cite{GL-1}, respectively. The proofs of the modular nuclearity condition in these two models (without restriction on the splitting distance $s$) can be found in appendix \ref{chapter:mnc+-1}. 

In the generic case of a non-constant scattering function, we map the model formulated on the $S_2$-symmetric Hilbert space $\Hil$ to the Bose or Fermi Fock space with a certain unitary operator to be constructed below. The Bose/Fermi alternative corresponds here to the sign of the scattering function at the origin, where it can take only the values $S_2(0)=\pm 1$. We thus subdivide $\SF$ and $\SF_0$ into a "Bosonic" and a "Fermionic" class according to
\begin{align}\label{def:SF0pm}
\SF^\pm	&:=	\{S_2\in\SF\,:\,S_2(0)=\pm1\}\,,\quad& \SF&=\SF^+\cup\SF^-\,,
\\
\SF_0^\pm	&:=	\{S_2\in\SF_0\,:\,S_2(0)=\pm1\}\,,\quad& \SF_0&=\SF_0^+\cup\SF_0^-\,.
\end{align}
It has to be kept in mind, however, that independently of the scattering function, all the models under consideration describe Bosons, as follows from the symmetry properties of their two-particle scattering states (section \ref{sect:wedgescattering}). But the choice of sign in $S_2(0)=\pm 1$ implies certain similarities between the wavefunctions $(\te_1,...,\te_n)\lmto\Psi_n(\te_1,...,\te_n)\in\Hil_n$ and Bose/Fermi wavefunctions for small rapidity differences $|\te_l-\te_k|<\eps$, which will be exploited in the following.

In order to distinguish between the different scattering functions involved, we adopt the convention that the usual notations $z,\zd,D_n,P_n,\Hil_n,\Hil$ refer to the generic scattering function under consideration. All objects corresponding to the special functions $S_2=\pm 1$ are tagged with an index "$\pm$", i.e. we write $z_\pm,\zd_\pm,D_n^\pm,P_n^\pm,\Hil_n^\pm,\Hil^\pm$.
\\
\\
In preparation for the construction of the mentioned unitaries, recall that each scattering function $S_2\in\mathcal{S}_0$ is analytic and nonvanishing in the strip $S(-\kappa(S_2),\kappa(S_2))$. So there is an analytic function $\delta:S(-\kappa(S_2),\kappa(S_2))\to\Cl$ (the phase shift) such that
\begin{align}
  S_2(\zeta)
  &=
  S_2(0)e^{2i\delta(\zeta)},
  \qquad
  \zeta\in S(-\kappa(S_2),\kappa(S_2))\,.
\end{align}
Since $S_2$ has modulus one on the real line, $\delta$ takes real values on $\Rl$, and we fix it uniquely by the choice $\delta(0)=0$. Note that in view of $S_2(-\te)=\overline{S_2(\te)}$, $\te\in\Rl$, $\delta$ is odd.
\begin{lemma}\label{lemma:y}
  Let $S_2\in\SF_0^\pm$ and $\delta:S(-\kappa(S_2),\kappa(S_2))\lto\Cl$ be defined as above. Consider the functions
  \begin{align}\label{def-yn}
    Y_n^\pm (\bze)
    &:=
    \prod_{1\leq k < l \leq n} 
    \left(
      \pm
      e^{i\delta(\zeta_k-\zeta_l)}\right)
    \,,\;\;n\geq 2\,,
  \qquad
 Y_0=1,\;\;Y_1(\zeta)=1\,,
  \end{align}
  and the corresponding multiplication operators (denoted by the same
  symbol $Y_n^\pm$).
  \begin{enumerate}
  \item Let $\kappa\in(0,\kappa(S_2))$. Viewed as an operator on $H^2(\Tu_n(\kappa))$, $Y_n^\pm$ is a bounded map with $\|Y_n^\pm\|_{\B(H^2(\Tu_n(\kappa)))}\leq\|S_2\|_\kappa^{\,n/2}$.
  \item Viewed as an operator on $L^2(\Rl^n)$, $Y_n^\pm$ is a unitary intertwining the representations $D_n$ and $D_n^\pm$ of the symmetric group $\frS_n$, and hence mapping the subspace $\Hil_n\subset L^2(\Rl^n)$ onto the subspace $\Hil_n^\pm\subset L^2(\Rl^n)$.
 \end{enumerate}
\end{lemma}
\begin{proof}
{\em a)} Since $\delta$ is analytic in $S(-\kappa,\kappa)$, so is the function $Y_n^\pm$ in $S(-\frac{\kappa}{2},\frac{\kappa}{2})^{\times n}$. Depending only on differences of rapidities, $Y_n^\pm$ is also analytic in $S(-\frac{\kappa}{2},\frac{\kappa}{2})^{\times n} + i\bla_0 = \Tu_n(\kappa)$. By application of the same argument as in the proof of the bounds in Prop. \ref{prop:ana} b), it follows that
  \begin{align}\label{y-bound}
    \left|
      Y_n^\pm(\bze)
    \right|
  &\leq
  \|S_2\|_\kappa^{n/2}
  \,,\qquad \bze\in\Tu_n(\kappa)\,.
\end{align}
Hence $\bno{Y_n^\pm \cdot{F}} \leq \|S_2\|_\kappa^{\,n/2}\cdot\bno{F}$, $F\in H^2(\Tu_n(\kappa))$, which proves {\em a)}.

{\em b)} Considered as a multiplication operator on $L^2(\Rl^n)$, $Y_n^\pm$ multiplies with a phase and is hence unitary. Let $\tau_j\in\frS_n$ denote the transposition exchanging $j$ and $j+1$, $j\in\{1,...,n\}$, and pick arbitrary $\Psi_n \in L^2(\Rl^n)$, $\bte\in\Rl^n$. 
 \begin{align*}
   (D_n^\pm(\tau_j)Y_n^\pm\Psi_n)(\bte)
   &=
   \pm\prod_{1\leq k < l \leq n\atop{(k,l)\neq (j,j+1)}}
   \left(\pm e^{i\delta(\te_k-\te_l)}\right)
   \cdot
   \left(\pm e^{i\delta(\te_{j+1}-\te_j)}\right)
   \Psi_n(\te_1,...,\te_{j+1},\te_j,...,\te_n)\\
   &=
   \prod_{1\leq k < l \leq n}
   \left(\pm e^{i\delta(\te_k-\te_l)}\right)
   \cdot
   S_2(\te_{j+1}-\te_j)
   \cdot
   \Psi_n(\te_1,...,\te_{j+1},\te_j,...,\te_n)\\
   &=
   (Y_n^\pm D_n(\tau_j)\Psi_n)(\bte)
  \end{align*}
As the transpositions $\tau_j$ generate $\frS_n$, this calculation shows that $Y_n^\pm $ intertwines $D_n^\pm$ and $D_n$. In particular, $Y_n^\pm$ restricts to a unitary mapping $\Hil_n$ onto $\Hil_n^\pm$.
\end{proof}

The operator
\begin{align}
Y^\pm := \bigoplus_{n=0}^\infty Y_n^\pm : \Hil\lto\Hil^\pm
\end{align}
will be used to improve the estimate on $\|\Xi_n(s)\|_1$ underlying Theorem \ref{thm:distalsplit}. A similar construction of distinguished isomorphisms between Fock spaces with different "statistics" has been carried out by Liguori and Mintchev \cite{LiMi}. But whereas in that work, the essential quality of $Y_n^\pm$ was property {\em b)} of the preceding Lemma, here also the preservation of the Hardy space structure, as stated in part {\em a)} of Lemma \ref{lemma:y}, is important. For it allows us to use a splitting of $\Xi_n(s)$ into a bounded and a nuclear operator as before \bref{xin-split}, and to work at the same time with a simpler symmetry structure. 

Explicitly, we consider in a model theory with scattering function $S_2\in\SF_0^\pm$ the maps 
\begin{align*}
  \Xi_n^\pm(s)
	&:=
	Y_n^\pm\Xi_n(s) \,:\,  \A(W_R) \longrightarrow \Hil_n^\pm
  \,,\qquad \Xi^\pm(s):=Y^\pm\Xi(s)\,.
\end{align*}
Since $Y^\pm : \Hil\to\Hil^\pm$ is unitary, $\Xi(s)$ is nuclear if and only if $\Xi^\pm(s)$ is, and in this case $\|\Xi(s)\|_1=\|\Xi^\pm(s)\|_1$. Moreover, as $Y_n^\pm$ acts by multiplication with a function depending only on differences of rapidities, we see that this operator commutes with the translation $U(\bs)$ and the modular operator, i.e. we have
\begin{align*}
  \Xi_n^\pm(s)A 
  &=
  \Delta^{1/4}U(\tfrac{1}{2}\bs)Y_n^\pm
  (A(\tfrac{1}{2}\bs)\Omega)_n
  \;=:\;
  \left(\Delta_n^\pm(s,\kappa) \circ Y_n^\pm \Sigma_n(s,\kappa)\right)A
  \;,\qquad A\in\A(W_R).
\end{align*}
Here $\Sigma_n(s,\kappa)$ is defined as in \bref{def:Sigma_n} and $\Delta^\pm_n(s,\kappa)$ acts as $\Delta_n(s,\kappa)$
\bref{def:Delta_n}, but is now considered as a map from the subspace $H^2_\pm(\Tu_n(\kappa))\subset H^2(\Tu_n(\kappa))$, consisting of the totally (anti-) symmetric functions in $H^2(\Tu_n(\kappa))$, to $\Hil_n^\pm$.

By Proposition \ref{prop:ana} and Lemma \ref{lemma:y} {\em a)}, $Y_n^\pm\Sigma_n(s,\kappa)$ is a bounded linear map from $\A(W_R)$ to $H^2_\pm(\Tu_n(\kappa))$, $\kappa\in(0,\kappa(S_2))$. Its norm is bounded by
\begin{align}\label{sigmaprime}
  \|Y_n^\pm \Sigma_n(s,\kappa)\|
  &\leq
	\left(  \|S_2\|_\kappa^{1/2}\cdot\sigma(s,\kappa)\right)^n
  \,.
\end{align}
The worsening of this bound in comparison to $\|\Sigma_n(s,\kappa)\|\leq\sigma(s,\kappa)^n$ is more than balanced by the improvement we get for the bound on $\|\Delta^-_n(s,\kappa)\|_1$ in the case $S_2\in\SF_0^-$, where the Pauli principle becomes effective.
\begin{theorem}\label{thm:-1nuclearity}
In a model theory with scattering function $S_2\in\SF_0^-$, the maps $\Xi(s)$ are nuclear for every splitting distance $s>0$, and there holds the bound, $\kappa\in(0,\kappa(S_2))$,
  \begin{align}\label{xi-bound--}
    \|\Xi(s)\|_1
    &\leq
    \sum_{n=0}^\infty
    \frac{\left(\sigma(s,\kappa)\,\|S_2\|_\kappa^{1/2}\,\|T_{s,\kappa}\|_1\right)^n}{\sqrt{n!}}
    \;<\; 
    \infty
    \;.
  \end{align}
In particular, in these models $\Om$ is a cyclic vector for the observable algebras $\A(\OO)$ localized in arbitrarily small open regions $\OO\subset\Rl^2$.
\end{theorem}
\begin{proof}
Proceeding along the same lines as in the proof of Lemma \ref{lemma:delta}, we infer that $\Delta_n^-(s,\kappa)$ is nuclear and can be represented as in \bref{delta-eqn}. With the notations $\beps=(\eps_1,...,\eps_n)$, $\eps_k=\pm 1$, there holds for $F^-\in H^2_-(\Tu_n(\kappa))$
\begin{align}\label{xi-t}
 \Delta_n^-(s,\kappa)F^-
 &=
 2^{-n}\sum_{\sbeps} \eps_1\cdots\eps_n (T_{s,\eps_1\kappa}\otimes ... \otimes T_{s,\eps_n\kappa}) 
    F^-_{\sbla_0-\frac{\kappa}{2}\sbeps}\;.
\end{align}
Consider the positive operator $\hat{T}_{s,\kappa} := (|T_{s,\kappa}^*|^2 + |T_{s,-\kappa}^*|^2)^{1/2}$, which is of trace class on $L^2(\Rl)$ and satisfies $\|\hat{T}_{s,\kappa}\|_1 \leq \|T_{s,\kappa}\|_1 + \|T_{s,-\kappa}\|_1 = 2\,\|T_{s,\kappa}\|_1$ \cite{kosaki}. We choose an orthonormal basis $\{\psi_k\}_k$ of $L^2(\Rl)$, consisting of eigenvectors $\psi_k$ of $\hat{T}_{s,\kappa}$, with eigenvalues $t_k\geq 0$. So $\hat{T}_{s,\kappa}$ acts as $\hat{T}_{s,\kappa}\xi = \sum_{k=1}^\infty t_k \langle \psi_k,\xi\rangle \psi_k$, $\xi\in L^2(\Rl)$, and its trace norm is $\|\hat{T}_{s,\kappa}\|_1 = \sum_{k=1}^\infty t_k <\infty$.

As a consequence of the Pauli principle, the vectors
\begin{align}\label{ONB}
  \Psi_\sbk^-
  &:=
  \zd_-(\psi_{k_1})\cdots\zd_-(\psi_{k_n})\Omega
  \;=\;
  \sqrt{n!}\,P_n^-(\psi_{k_1}\otimes ... \otimes \psi_{k_n})
	\nonumber
	\\
  &=
	\frac{1}{\sqrt{n!}}\sum_{\rho\in\frS_n}{\rm sign}(\rho)\,\psi_{\rho(k_1)}\otimes...\otimes\psi_{\rho(k_n)}
\end{align}
form an orthonormal basis of $\Hil^-_n$ if $\bk=(k_1,...,k_n)$ varies over $k_1<k_2<...<k_n$, $k_1,...,k_n\in\N$.

Expanding the right hand side of \bref{xi-t} in this basis, we find
\begin{align*}
\Delta_n^-(s,\kappa)F^-
&=
2^{-n}\sum_{\sbeps}\eps_1\cdots\eps_n
\sum_{k_1<...<k_n}
\langle \Psi_\sbk^-,\, (T_{s,\eps_1\kappa}\otimes ... \otimes T_{s,\eps_n\kappa})F^-_{\sbla_0-\frac{\kappa}{2}\sbeps}\rangle
\cdot \Psi_\sbk^-
\\
&=
\frac{2^{-n}}{\sqrt{n!}}\sum_{\sbeps,\rho}\eps_1\cdots\eps_n\,{\rm sign}\rho
\!\!\!\!
\sum_{k_1<...<k_n}
\langle T_{s,\eps_1\kappa}^*\psi_{\rho(k_1)}\otimes .. \otimes T_{s,\eps_n\kappa}^*\psi_{\rho(k_n)},\,
F^-_{\sbla_0-\frac{\kappa}{2}\sbeps}\rangle
\cdot \Psi_\sbk^-
\,.
\end{align*}
Taking into account $\|T_{s,\pm\kappa}^*\psi_{k_j}\|\leq \|\hat{T}_{s,\pm\kappa}\psi_{k_j}\|=t_{k_j}$ and $\|F^-_{\sbla_0-\frac{\kappa}{2}\sbeps}\|\leq\bno{F^-}$ as well as $\sum_{\sbeps,\rho}1=2^n\cdot n!$, this expansion leads to the estimate
\begin{align}
\|\Delta_n^-(s,\kappa)\|_1
\leq
\sqrt{n!}
\sum_{k_1<...<k_n}t_{k_1}\cdots t_{k_n}
\leq
\frac{1}{\sqrt{n!}}
\sum_{k_1,...,k_n}t_{k_1}\cdots t_{k_n}
=
\frac{\|T_{s,\kappa}\|_1^{\,n}}{\sqrt{n!}}
\;.
\end{align}
In view of the bound \bref{sigmaprime} on $Y_n^-\Sigma_n(s,\kappa)$, we arrive at the following estimate for the nuclear norm of $\Xi^-(s)=\sum_{n=0}^\infty \Delta^-_n(s,\kappa)\circ Y_n^-\Sigma_n(s,\kappa)$,
\begin{align}
\|\Xi^-(s)\|_1
&\leq
\sum_{n=0}^\infty \frac{\left(\sigma(s,\kappa)\,\|S_2\|_\kappa^{1/2}\,\|T_{s,\kappa}\|_1\right)^n}{\sqrt{n!}}\,.
\end{align}
This series converges for each
$\sigma(s,\kappa)\|S_2\|_\kappa^{1/2}\,\|T_{s,\kappa}\|_1$, i.e. for each $s>0$, and yields the claimed bound \bref{xi-bound--}.
\end{proof}

Theorem \ref{thm:-1nuclearity} provides the full proof of the modular nuclearity condition for wedge algebras in the class of theories with scattering functions $S_2\in\SF_0^-$. Consequently, the models with such scattering functions enjoy all the properties derived from the modular nuclearity condition in chapter \ref{chapter:netsin2d}. In particular, these models are hereby rigorously established as examples of interacting quantum field theories in the sense of local quantum physics. 

For the sake of clarity, we restate the results of chapter \ref{chapter:netsin2d} (Theorem \ref{thm:Ch2summary}) for the present more concrete situation.
\begin{theorem}
Consider a scattering function $S_2 \in \SF_0^-$ and the associated model theory, defined in terms of the local net $\A$ \bref{def:AWnet}, \bref{def:AOconc}.
\begin{enumerate}
	\item The net $\A$ has the split property for inclusions of wedges (and hence, also for inclusions of double cones).
	\item The wedge algebras and the double cone algebras are all isomorphic to the hyperfinite type III$_1$ factor.
	\item Haag duality holds, i.e. $\A(\OO)'=\A(\OO')$ for any region $\OO\subset\Rl^2$.
	\item Strong additivity as expressed by Lemma \ref{lemma-loc} and Lemma \ref{lemma-mue} holds.
	\item The Reeh-Schlieder property holds, i.e. the vacuum vector $\Omega$ is cyclic and separating for any algebra $\A(\OO)$ of observables localized in a non-empty open bounded region $\OO$ with non-empty causal complement.
\end{enumerate}
For models with scattering function $S_2$ from the larger class $\SF_0\supset\SF_0^-$, the above statements hold for double cones above a minimal size (cf. Thm. \ref{thm:distalsplit}).
\end{theorem}

To conclude the chapter, we discuss the quality of the bound \bref{xi-bound--}, and mention some related conjectures.

The proof of Theorem \ref{thm:-1nuclearity} relies on the Pauli principle, and does not carry over to the family $\SF_0^+$. In particular, the free field theory, corresponding to the scattering function $S_2=+1\in\SF_0^+$, cannot be treated by these methods. However, the underlying Zamolodchikov algebra simplifies to the CCR algebra in the free case, and allows for an alternative estimate of the nuclear norm of $\Xi(s)$ in this particular model. This alternative proof heavily relies on the analysis of nuclear maps on the Bose Fock space developed by Buchholz and Wichmann \cite{BuWi}, and it is demonstrated in appendix \ref{chapter:mnc+-1}. It is shown there that for scattering function $S_2=+1$, the maps $\Xi(s)$ are nuclear for any splitting distance $s>0$.

Furthermore, appendix \ref{chapter:mnc+-1} contains an analysis of the nuclearity properties of the model with the other constant scattering function, $S_2=-1\in\SF_0^-$. This model is also more easily manageable by algebraic methods, since the Zamolodchikov algebra coincides here with the CAR algebra. It can be shown that for $S_2=-1$, there holds a bound of the form
\begin{align}\label{-1bndbesser}
\|\Xi(s)\|_1
\leq
e^{2(\|T_\varphi(s)\|_1+\|T_\pi(s)\|_1)}
\,,
\end{align}
where $T_\varphi(s)$, $T_\pi(s)$ are trace class integral operators on $L^2(\Rl)$, similar to $T_{s,\kappa}$ \bref{int-kernel2}. A comparison of \bref{-1bndbesser} with \bref{xi-bound--} shows that the bound obtained in Thm. \ref{thm:-1nuclearity} is not optimal.  Rather, the factor $\sqrt{n!}$ appearing there in the denominator can be replaced by $n!$.
We conjecture that by a more refined analysis, the sharpened bound given by this replacement is true for arbitrary $S_2\in\SF_0^-$, i.e. that the series in \bref{xi-bound--} can be replaced by an exponential function. Although the actual value of $\|\Xi(s)\|_1$ does not matter for the nuclearity condition, which only requires $\|\Xi(s)\|_1<\infty$, it is useful to have a sharp bound on this quantity if one is interested in estimating the thermodynamical partition function of such models. This point will be discussed in section \ref{sec:thermo}.
\\
\\
In conclusion, we mention that already the result of Thm. \ref{thm:distalsplit}, namely the Reeh-Schlieder property for double cones above a minimal size, is completely sufficient for the investigation of the interaction by doing scattering theory, in any model with $S_2\in\SF_0$. This opens up the possibility to calculate the S-matrix of such models, which is the topic of the following chapter.

\chapter{Physical Properties of the Constructed Models}\label{chapter:reconstructS}

The results obtained in the previous chapters show that the model theories defined here comply with all principles of quantum field theory if the scattering function is chosen from an appropriate class. In this chapter, we discuss two aspects of the interaction in these models: In the first section, we do collision theory and obtain formulae for multiparticle scattering states. We find that the constructed theories are asymptotically complete, and that their S-matrices coincide with the ones associated to the scattering functions $S_2$ used in their definition.

In the second section, we briefly consider the thermodynamical behavior of the models and show that their partition function is finite for any bounded region and any temperature.

 \section{The Reconstruction of the S-Matrix} \label{sec:reconstructS}

To compute $n$-particle collision states, it is sufficient to restrict to the family $\SF_0$ of scattering functions (Definition \ref{def:S0family}), as Theorem \ref{thm:distalsplit} ensures that in this case there exist compactly localized observables, at least in double cones above some minimal size. Since arbitrarily many double cones of {\em any} size can be spacelike separated by translation, it is possible to apply the usual methods of collision theory in this class of theories -- localization in arbitrarily small regions is not needed.
\\
\\
In the following, we employ the Haag-Ruelle scattering theory \cite{araki, bs-scatter} in the same form as in \cite{BBS}, where scattering properties of polarization-free generators have been analyzed. As usual in this approach, we consider quasilocal operators of the form
\begin{align}\label{a-quasilocal}
A(f_t)
&=
\int d^2x\,
f_t(x)A(x)\,,\qquad A(x)=U(x)A\,U(x)^{-1}\,,
\end{align}
where $A\in\A(\OO)$ is localized in a double cone $\OO$ and the functions $f_t$, $t\in \Rl$, are defined in terms of momentum space wavefunctions $\fti$ by
\begin{align}
f_t(x)
&:=
\frac{1}{2\pi}
\int d^2p\,
\fti(p_0,p_1)\,e^{i(p_0-\omega_p)t}\,e^{-ip\cdot x}\;,\qquad \omega_p := \big(m^2+p_1^2\big)^{1/2}\,.
\end{align}
The functions $\fti$ are taken to be Schwartz test functions, such that the integral \bref{a-quasilocal} converges in operator norm.

\begin{align}
\VV(f)
&:=
\left\{(1,p_1/\omega_p)\,:\,(p_0,p_1)\in\supp\fti\,\right\}\,.
\label{velsup}
\end{align}
Recall that the support of $f_t$ is essentially contained in $t\,\VV(f)$ for asymptotic times $t$ \cite{hepp}. More precisely, let $\chi$ be a smooth function which is equal to 1 on $\VV(f)$ and vanishes in the complement of a slightly larger region. Then $\fhat_t(x):=\chi(x/t)f_t(x)$ is the asymptotically dominant part of $f_t$, i.e. the difference $f_t-\fhat_t$ converges to zero in the topology of $\Ss(\Rl^2)$ as $t\to\pm\infty$ \cite{BBS}.

As in section \ref{sect:wedgescattering}, we adopt the notation to write $f\prec g$ if $\VV(g)-\VV(f)\subset \{0\}\times(0,\infty)$.
\\
\\
If the support of $\fti$ is concentrated around a point $(\omega_p,p_1)$ on the upper mass shell and does not intersect the energy momentum spectrum elsewhere, $A(f_t)\Om$ is a single particle state which does not depend on the time parameter $t$. Moreover, we have the strong limits
\begin{align}
\lim_{t\to\infty}A(f_t)\Psi	=	A(f)\oout\Psi
\,,\qquad
\lim_{t\to-\infty}A(f_t)\Psi	=	A(f)\iin\Psi\,,
\end{align}
to the asymptotic creation operators $A(f)\oout$ and $A(f)\iin$, creating the single particle state $A(f)\Om$ from the vacuum. The adjoint operators converge to the corresponding annihilation operators,
\begin{align}
\lim_{t\to\infty}A(f_t)^*\Psi	=	{A(f)\oout}^*\Psi
\,,\qquad
\lim_{t\to-\infty}A(f_t)^*\Psi	=	{A(f)\iin}^*\Psi\,,
\end{align}
These limits are known to hold for a certain dense set of collision states $\Psi$ \cite{hepp, araki}. But by a result of Buchholz \cite{DB-harmonic}, it follows that they are also valid for all scattering states $\Psi$ of finite energy, in particular, for all single particle states of the form $\phi(f)\Om=f^+$, where $f^+$ has compact support \cite{BBS}.

These operators are related to the Bose creation and annihilation operators through the M{\o}ller operators $V_{\rm in/out} : \Hil^+\lto\Hil_{\rm in/out}\subset\Hil$. These creators and annihilators coincide with the Zamolodchikov operators $\zd_+$, $z_+$ with scattering function $S_2=1$, and act on the subspace of finite particle number of the totally symmetric Bose Fock $\Hil^+$ space over $\Hil$. In the terminology of chapter \ref{chapter:nuclearity}, 
\begin{align}
A(f)_{\rm in/out}		&=	V_{\rm in/out}\zd_+(A(f)\Om)V_{\rm in/out}^*,\quad 
{A(f)_{\rm in/out}}^*	=	V_{\rm in/out} z_+(\overline{A(f)\Om}) V_{\rm in/out}^*.
\end{align}
Having recalled these basic facts of scattering theory, we now compute $n$-particle collision states in the model theory based on a scattering function $S_2\in\SF_0$. To this end, we use the field $\phi$ and follow the analysis in \cite{BBS}. Recall that $\phi(f)$ is affiliated to the wedge algebra $\A(W_L+\supp f)$, $f\in\Ss(\Rl^2)$.
\begin{lemma}
Consider testfunctions $\fti_1,...,\fti_n\in\Ss(\Rl^2)$ having ordered, pairwise disjoint, compact supports concentrated around points on the upper mass shell such that $f_1\prec...\prec f_n$. Then
\begin{align}
\phi(f_1)\cdots\phi(f_n)\Om
&=
(f_1^+ \times ... \times f_n^+)\oout
\,,\label{n-out}\\
\phi(f_n)\cdots\phi(f_1)\Om
&=
(f_1^+\times ... \times f_n^+)\iin\,.\label{n-in}
\end{align}
\end{lemma}
\begin{proof}
The proof is based on induction in the particle number $n$. For $n=1$, we have
\begin{align}
\phi(f_1)\Om	=	f_1^+	=	(f_1^+)\oout	=	(f_1^+)\iin
\,,
\end{align}
since $f_1^+$ is a single particle state. For the step from $n$ to $(n+1)$, let $A_1,...,A_n\in\A(\OO)$ be operators localized in an arbitrary double cone $\OO$. We want to establish commutation relations between $\phi(f)$ and the creation operators $A_k(g_k)\oout$, where $f\prec g_1\prec ... \prec g_n$ and the test functions $f, g_1,...,g_n$ have the same support properties as the $f_1,...,f_n$. We first note that
\begin{align}\label{phi-limit}
\phi(f)\Psi	=	\phi(f_t)\Psi	=	\lim_{t\to\infty}\phi(\fhat_t)\Psi
\,,\qquad
\Psi\in\DD\,.
\end{align}
The first equality follows from $f_t^+=f^+$, $f_t^-=0$, since the support of $\fti$ does not intersect the lower mass shell. The second equality in \bref{phi-limit} follows because we have the limit $f_t-\fhat_t\to 0$ in $\Ss(\Rl^2)$ and $f\lmto\phi(f)\Psi$ is a vector valued tempered distribution\footnote{This can for example be inferred from the estimate in Proposition \ref{prop:phi} a).}. Taking into account the strong convergence $A_k(\hat{g}_{k,t})\to A_k(g_k)\oout$ for $t\to\infty$, we obtain
\begin{align}
\langle\phi(\fbar)\Psi,(A_1(g_1)\Om\times ... \times A_n(g_n)\Om)\oout\rangle
&=
\lim_{t\to\infty}\langle\phi(\hat{\fbar}_t)\Psi\,,A_1(\ghat_{1,t})\cdots A_n(\ghat_{n,t})\Om\rangle
\,.
\end{align}
But $\hat{f}_t$ and $\hat{g}_{k,t}$ have supports in small neighborhoods of $t\,\VV(f)$ and $t\,\VV(g_k)$, respectively. Hence $\phi(\hat{\fbar}_t)$ is localized in a wedge $W_L(f_t)$ slightly larger than $W_L+t\,\VV(f)$, and $A_k(\hat{g}_{k,t})$ is localized in a neighborhood of $\OO+t\,\VV(g_k)$. For large enough $t>0$, these regions are spacelike separated, and their distance increases linearly with $t$. As $\phi(\hat{\fbar}_t)$ is affiliated with $\A(W_L(f_t))$, it follows that this operator commutes with $A_k(\ghat_{k,t})$, $k=1,...,n$. Hence
\begin{align*}
\langle\phi(\fbar)\Psi,(A_1(g_1)\Om\times ... \times A_n(g_n)\Om)\oout\rangle
&=
\lim_{t\to\infty}\langle\Psi\,,A_1(\ghat_{1,t})\cdots A_n(\ghat_{n,t})\phi(\fhat_t)\Om\rangle
\\
&=
\lim_{t\to\infty}\langle\Psi\,,A_1(\ghat_{1,t})\cdots A_n(\ghat_{n,t})\, \fhat_t^+\rangle
\,.
\end{align*}
Applying stationary phase methods, one can show that $\fhat_t^+$ converges rapidly in $L^2$-norm to $f^+$ as $t\to\infty$. Namely, for any $N\in\N$ there exists a constant $C_N$ such that for large enough $t$ there holds $\|\fhat_t^+-f^+\|_2\leq C_N\cdot t^{-N}$ \cite[Cor. to Thm. XI.14]{SimonReed3}. On the other hand, a straightforward estimate shows $\|A_k(\ghat_{k,t})\|\leq c_{g_k,A_k}\cdot t^2$ with constants $c_{g_k,A_k}>0$. As the operators $A_k(\ghat_{k,t})$ converge strongly to the asymptotic creation operators $A_k(g_k)\oout$ on the one particle state $f^+$, we find
\begin{align*}
\langle\phi(\fbar)\Psi,(A_1(g_1)\Om\times ... \times A_n(g_n)\Om)\oout\rangle
&=
\langle \Psi, A_1(g_1)\oout \cdots A_n(g_n)\oout f^+\rangle
\\
&=
\langle \Psi, (A_1(g_1)\Om\times ... \times A_n(g_n)\Om\times f^+)\oout\rangle
\\
&=
\langle \Psi, (f^+\times A_1(g_1)\Om\times ... \times A_n(g_n)\Om)\oout\rangle
\,,
\end{align*}
where in the last step we used the Bose symmetry of the scattering states. In view of the Reeh-Schlieder property of $\A(\OO)$, we can approximate $f_k^+$ with $A_k(g_k)\Om$. Given any $\eps>0$, there exist operators $A_1,...,A_n\in\A(\OO)$ and testfunctions $g_1,...,g_n$, with $g_k$ having support in an arbitrarily small neighborhood of the support of $f_k$, such that $\|f_k^+-A_k(g_k)\Om\|<\eps$. As the left and right hand side of the above equation are continuous in the single particle states $A_k(g_k)\Om$, we conclude
\begin{align}
\langle\phi(\fbar)\Psi,(f_1^+\times ... \times f_n^+)\oout\rangle
&=
\langle\Psi\,,(f^+\times f_1^+\times ... \times f_n^+)\oout\rangle
\,.
\end{align}
Taking into account that $\Psi\in\DD$ was arbitrary and $\DD\subset\Hil$ is dense, this implies via the induction hypothesis that for $f\prec f_1\prec...\prec f_n$,
\begin{align}
\phi(f)\phi(f_1)\cdots \phi(f_n)\Om
&=
\phi(f)(f_1^+\times ... \times f_n^+)\oout
=
(f^+\times f_1^+\times ... \times f_n^+)\oout
\,,
\end{align}
proving \bref{n-out}.

For incoming $n$-particle states, the order of the velocity supports of $f_1,...,f_n$ has to be reversed, since $W_L+t\,\VV(f_1)$ becomes spacelike separated from $\OO+t\,\VV(f_k)$ for $t\to-\infty$ if $f\succ f_k$. Apart from this modification, the same argument can be used to derive formula \bref{n-in}.
\end{proof}
\noindent{\em Remark:} In principle, our proof can also be applied to the more general situation of temperate polarization-free generators $G$ affiliated to $\A(W_L)$ in a local net $\A$ (Def. \ref{def:tPFG}). In this setting, we have to assume that the generators leave their domain of temperateness $\DT$ invariant, i.e. $G\DT\subset\DT$. Given this invariance property, we can apply the same argument as above and obtain, $f_1\prec ... \prec f_n$, 
\begin{align}
G(f_1)\cdots G(f_n)\Om	&= 	(G(f_1)\Om \times ... \times G(f_n)\Om)\oout\,,\\
G(f_n)\cdots G(f_1)\Om	&= 	(G(f_1)\Om \times ... \times G(f_n)\Om)\iin\,.
\end{align}
\\
\\
We now derive the explicit form of scattering states as vectors in the $S_2$-symmetric Fock space $\Hil$. Note that in \bref{n-out}, \bref{n-in}, the annihilation part of $\phi(f_k)$ vanishes: Since the support of $\fti$ does not intersect the lower mass shell, we have $f_k^-=0$, and $\phi(f_k)=\zd(f^+_k)$.

The relation $f\prec g$ implies in particular $\supp(g^+) -\supp(f^+) \subset (0,\infty)$, as can be seen from the definition of the velocity support \bref{velsup}. On the other hand, for single particle functions $\psi_1,\psi_2\in C_0^\infty(\Rl)$ with $\supp\psi_2-\supp\psi_1\subset(0,\infty)$ we can find $f_1,f_2\in\Ss(\Rl^2)$ such that $f_1^+=\psi_1$, $f_2^+=\psi_2$ and $f_1\prec f_2$. We therefore also write $\psi_1\prec\psi_2$ if $\supp\psi_2-\supp\psi_1\subset(0,\infty)$. By continuity of \bref{n-out},\bref{n-in} in $f_1^+$,...,$f_n^+$, we arrive at the following form of $n$-particle collision states.
\begin{align*}
(\psi_1\times ... \times \psi_n)\oout
&=
\zd(\psi_1)\cdots\zd(\psi_n)\Om
=
\sqrt{n!}\,P_n(\psi_1\otimes ... \otimes \psi_n)
\,,\qquad \psi_1 \prec ... \prec \psi_n\,,
\\
(\psi_1\times ... \times \psi_n)\iin
&=
\zd(\psi_n)\cdots\zd(\psi_1)\Om
=
\sqrt{n!}\,P_n(\psi_n\otimes ... \otimes \psi_1)
\,,\qquad \psi_1 \prec ... \prec \psi_n
\,.
\end{align*}
Both, the incoming and outgoing scattering states, form total sets in $\Hil$. To prove this, note that the functions $\psi_1\otimes...\otimes\psi_n$, $\psi_1\prec...\prec\psi_n$, form a total set in $L^2(E_n,d^n\bte)$, $E_n=\{(\te_1,...,\te_n)\in\Rl^n\,:\,\te_1\leq...\leq\te_n\}$. But $P_n:L^2(E_n,d^n\bte)\lto\Hil_n$ is linear, continuous and onto, and hence the totality of the constructed outgoing collision states in $\Hil_n$ follows. Analogously, one can also show that the incoming $n$-particle states form a total set in $\Hil_n$. Since the space $\DD$ of vectors of finite particle number is dense in $\Hil$, any vector $\Psi\in\Hil$ can be approximated by linear combinations of incoming or outgoing scattering states, i.e. the theory is asymptotically complete.
\\
\\
Having derived explicit formulae for the scattering states, we can now compute the M{\o}ller operators $V\iin$, $V\oout$ and the S-matrix $S$ of the model.

The asymptotic states span the Bosonic Fock space $\Hil^+=\bigoplus_{n=0}^\infty \Hil_n^+$ over $\Hil_1=L^2(\Rl)$, whose $n$-particle spaces $\Hil_n^+$ contains all totally symmetric functions in $L^2(\Rl^n)$. Denoting the orthogonal projection onto $\Hil_n^+$ by $P_n^+$, we infer from the above derived form of the collision states that the M{\o}ller operators are given by
\begin{align}
V\oout P_n^+(\psi_1\otimes ...\otimes \psi_n)	&=	P_n(\psi_1\otimes...\otimes\psi_n)
\,,\qquad \psi_1\prec...\prec\psi_n\,,\label{vnout}
\\
V\iin P_n^+(\psi_n\otimes ...\otimes \psi_1)	&=	P_n(\psi_n\otimes...\otimes\psi_1)
\,,\qquad \psi_1\prec...\prec\psi_n\,.
\end{align}
Note that as $\|P_n^+(\psi_1\otimes...\otimes\psi_n)\|=\|P_n(\psi_1\otimes...\otimes\psi_n)\|=(n!)^{-1/2}\|\psi_1\|\cdots\|\psi_n\|$, these equations uniquely determine $V\iin$ and $V\oout$ as isometries mapping $\Hil^+$ onto $\Hil$.

Analogously to the reasoning in section \ref{sect:wedgescattering}, the $n$-particle M{\o}ller operators can be seen to be multiplication operators. 
Recall that the $S_2$-symmetrization operator $P_n$ has the explicit form (Lemma \ref{lem:Dn})
\begin{align}
(P_n \Psi_n)(\te_1,...,\te_n)
&=
\frac{1}{n!}\sum_{\pi\in\frS_n}
S_n^\pi(\te_1,...,\te_n)\cdot\Psi_n(\te_{\pi(1)},...,\te_{\pi(n)})\,,\\
S_n^\pi(\te_1,...,\te_n)
&=
\prod_{1\leq l < k \leq n \atop \pi(l) > \pi(k)} S_2(\te_{\pi(l)}-\te_{\pi(k)})\,.\label{snexp}
\end{align}
Inserting this decomposition into \bref{vnout} yields
\begin{align}\label{s-sum}
{V\oout}^*\sum_{\pi\in\frS_n}  S_n^\pi \cdot \psi_{\pi^{-1}(1)}\otimes...\otimes\psi_{\pi^{-1}(n)}
&=
\sum_{\pi\in\frS_n} \psi_{\pi^{-1}(1)}\otimes...\otimes\psi_{\pi{-1}(n)}
\,.
\end{align}
Taking into account the support properties of $\psi_1 \prec ... \prec \psi_n$, we see that ${V\oout}^*$ acts on $n$-particle states by multiplication with the function 
\begin{align}\label{f1---}
V^{(n)*}\oout(\te_1,...,\te_n)		=	S^\pi(\te_1,...,\te_n)^{-1}\,,\qquad \te_{\pi(1)}\leq...\leq\te_{\pi(n)}\,.
\end{align}
Analogously, $V\iin$ is seen to multiply $n$-particle states with
\begin{align}\label{f2---}
V^{(n)}\iin(\te_1,...,\te_n)		=	S^\pi(\te_1,...,\te_n)\,,\qquad \te_{\pi(1)}\geq...\geq\te_{\pi(n)}\,.
\end{align}
So the $S$-matrix, defined as $\hat{S}= {V\oout}^*V\iin$, also acts as a multiplication operator on $\Hil_n^+$,
\begin{align}
(\hat{S}\Psi^+)_n(\te_1,...,\te_n) = S_n(\te_1,...,\te_n)\cdot\Psi_n^+(\te_1,...,\te_n)\,,\qquad\Psi^+\in\Hil^+\,.
\end{align}
The functions $S_n(\te_1,...,\te_n)$ are according to \bref{f1---} and \bref{f2---} given by
\begin{align}
S_n(\te_1,...,\te_n)
&=
S_n^\pi(\te_1,...,\te_n)^{-1}\cdot 
S_n^{\pi\circ \iota_n}(\te_1,...,\te_n)\,,\qquad
\te_{\pi(1)}\leq...\leq\te_{\pi(n)}\,,
\end{align}
where $\iota_n\in\frS_n$ is the total inversion permutation, $\iota_n(k):=n-k+1$. In view of the form \bref{snexp}, we can compute $S_n(\te_1,...,\te_n)$ in the region $\te_{\pi(1)}\leq...\leq\te_{\pi(n)}$ as
\begin{align*}
S_n(\te_1,...,\te_n)
&=
\prod_{1\leq l<k\leq n \atop \pi(l)>\pi(k)} S_2(\te_{\pi(l)}-\te_{\pi(k)})^{-1}
\cdot
\prod_{1\leq l<k\leq n \atop \pi(n-l+1)>\pi(n-k+1)} S_2(\te_{\pi(n-l+1)}-\te_{\pi(n-k+1)})
\\
&=
\prod_{1\leq l<k\leq n \atop \pi(l)>\pi(k)} S_2(\te_{\pi(k)}-\te_{\pi(l)})
\cdot
\prod_{1\leq l'<k'\leq n \atop \pi(l')<\pi(k')} S_2(\te_{\pi(k')}-\te_{\pi(l')})\\
&=
\prod_{1\leq l<k\leq n} S_2(\te_{\pi(k)}-\te_{\pi(l)})\,.
\end{align*}
(In the computation, the primed indices $l':=n-k+1$, $k':=n-l+1$ have been used.) Note that the rapidity differences appearing in the last line, $\te_{\pi(k)}-\te_{\pi(l)}$, $l<k$, are all positive as a consequence of the ordering $\te_{\pi(1)}\leq ... \leq \te_{\pi(n)}$. Hence we arrive at
\begin{align}
S_n(\te_1,...,\te_n)
=
\prod_{1\leq l<k\leq n} S_2(|\te_{\pi(k)}-\te_{\pi(l)}|)
=
\prod_{1\leq l<k\leq n} S_2(|\te_k-\te_l|)
\,.
\end{align}
This formula is valid for arbitrary $(\te_1,...,\te_n)\in\Rl^n$ since all reference to the permutation $\pi$ has been eliminated. We have thus proven that $\hat{S}$ indeed coincides with the S-matrix corresponding to $S_2$ \bref{snfacts}. As the M{\o}ller operators and the S-matrix $\hat{S}$ are multiplication operators in these models, it also follows that the S-matrix on $\Hil$, $S:=V\iin{V\oout}^*$, agrees with $\hat{S}$.


\begin{theorem}\label{thm:S-matrix}
Consider a model with scattering function $S_2\in\SF_0$. This theory is asymptotically complete, a total set of $n$-particle scattering states being given by
\begin{align}
(\psi_1\times ... \times \psi_n)\oout
&=
\zd(\psi_1)\cdots\zd(\psi_n)\Om
\,,\qquad  \psi_1\prec...\prec\psi_n\,,
\\
(\psi_1\times ... \times \psi_n)\iin
&=
\zd(\psi_1)\cdots\zd(\psi_n)\Om
\,,\qquad  \psi_1\succ...\succ\psi_n\,.
\end{align}
The S-matrix is $S:\Hil\lto\Hil$,
\begin{align}
(S\Psi)_n(\te_1,...,\te_n)
=
\prod_{1\leq k<l\leq n}S_2(|\te_k-\te_l|)
\cdot \Psi_n(\te_1,...,\te_n)
\,,
\end{align}
and coincides with the S-matrix corresponding to $S_2$.
{\hfill  $\square$}
\end{theorem}

\noindent Theorem \ref{thm:S-matrix} shows that the construction presented here is the solution of the inverse scattering problem for factorizing S-matrices with scattering functions in $\SF_0$. Altough the property of asymptotic completeness is expected to hold in most quantum field theories, the above result is, to the best of our knowledge, the first proof of it in an interacting model.
\\
\\
In terms of improper $n$-particle states with sharp rapidities, we have shown that
\begin{subequations}\label{z-in-out}
\begin{align}
\zd(\te_1)\cdots\zd(\te_n)\Om	&=	|\,\te_1,...,\te_n\rangle\oout
\,,&\quad
&\te_1<...<\te_n
\,,\\
\zd(\te_1)\cdots\zd(\te_n)\Om	&=	|\,\te_1,...,\te_n\rangle\iin
\,,&\quad
&\te_1>...>\te_n
\,,
\end{align}
\end{subequations}
are asymptotic collision states in the sense of the Haag-Ruelle scattering theory.

The identification of incoming and outgoing $n$-particle states with $n$-fold products of Zamolodchikov creation operators acting on the vacuum, arranged in order of decreasing, respectively increasing, rapidities, is one of the basic assumptions in the framework of the form factor program. In fact, it has motivated the very definition of Zamolodchikov's algebra (cf. section \ref{sec:ffp}).

In the usual approach to the construction of quantum field theories with a factorizing S-matrix, local quantum fields are characterized in terms of their form factors, which are the primary objects of interest. However, due to the complicated structure of the form factors of local operators, this strategy has not led to a construction of model theories in most cases. In particular, no manageable one particle generators are available and the interpretation of expressions like \bref{z-in-out} cannot be checked in collision theory; but rather has to be taken  for granted. 

It is therefore gratifying that with the help of the approach presented here, the heuristic picture motivating Zamolodchikov's algebra can be rigorously justified. As we have seen, this point of view allows for a complete construction of models with factorizing S-matrices in the algebraic framework of quantum field theory. By giving the Zamolodchikov operators $\zd(\te)$ a spacetime interpretation in terms of the associated wedge-local field $\phi$, we arrived at a family of asymptotically complete models in which the idea of factorized scattering underlying the equations \bref{z-in-out} can be proven in Haag-Ruelle scattering theory.


\newpage
\section{The Thermodynamical Partition Function}\label{sec:thermo}

\begin{figure}[here]
    \noindent
    \psfrag{x0}{$x_0$}
    \psfrag{x1}{$x_1$}
    \psfrag{s}{$s$}
    \psfrag{-s}{$-s$}
    \psfrag{r}{$r$}
    \psfrag{-r}{$-r$}
    \psfrag{Os}{$\OO_s$}
    \psfrag{Or}{$\OO_r$}
    \centering\epsfig{file=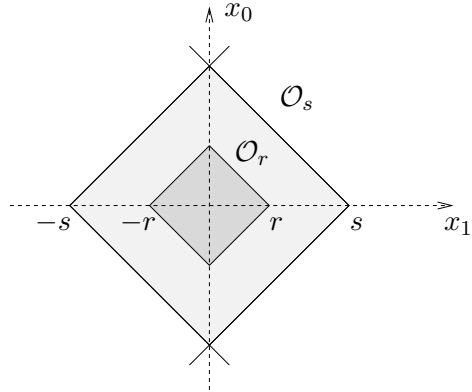,width=6cm}
     \caption{The geometrical situation considered for the proof of the nuclearity of $\Theta_{\beta,r}$.}
\label{fig:or-os}
\end{figure}

In addition to the results concerning scattering theory, also gross thermodynamical properties of the constructed models can be derived from the results of chapter \ref{chapter:nuclearity}. In this section, we will briefly discuss within the framework of the theories with $S_2\in\SF_0^-$ the thermodynamical partition function $Z(\beta,\OO)$ at inverse temperature $\beta>0$ and in restriction to a bounded spacetime volume $\OO$. The results found in this context do not give very accurate estimates on $Z(\beta,\OO)$, but rather serve to illustrate the fact that with the help of the nuclearity theorems on the modular structure of wedge algebras, also interesting local information can be obtained.
\\
\\
The mechanism to be used is the well-known connection between the modular nuclearity condition and the energy nuclearity condition, related to the Hamiltonian $H=P_0$ (see section \ref{sec:mnc}). As usual, the nuclear norm of
\begin{align}\label{def:thetabr}
\Theta_{\beta,r}	: 	\A(\OO_r) \lto \Hil\,,\qquad \Theta_{\beta,r}(A) :=	e^{-\beta H}A\Om\,,\qquad \OO_r:=(\{0\}\times(-r,r))''\,,
\end{align}
is interpreted as the thermodynamical partition function of the system at temperature $\beta^{-1}$, confined to the spacetime volume $\OO_r$ (see figure \ref{fig:or-os}) \cite{BuWi}.
 
Following closely the analysis in \cite{nuclearmaps2}, we show how estimates on $\|\Theta_{\beta,r}\|_1$ can be inferred from the modular nuclearity condition. We consider two concentric double cones
\begin{align}\label{def:Os}
	\OO_s &:= (W_R-\bs) \cap (W_L+\bs),\qquad \bs=(0,s),\;\;s>0\,,\\
	\OO_r &:= (W_R-\br) \cap (W_L+\br),\qquad \br=(0,r),\;\;r>0\,,
\end{align}
as depicted in figure \ref{fig:or-os} below. (The ratio $s/r$ will be fixed in the course of the argument.) 
\\
\\
Using the geometric action of the modular operator $\Delta$ of $(\A(W_R),\Om)$, it has been shown in \cite{nuclearmaps2} that there exist two bounded operators $Q_\pm\in\B(\Hil)$, $\|Q_\pm\|\leq 1$, such that
\begin{align}\label{cooleqn}
e^{-\sin(2\pi\mu)sH}A\Om	&=	Q_+\Delta^\mu U(\cos(2\pi\mu)\bs)A\Om + Q_-\Delta^{-\mu}U(-\cos(2\pi\mu)\bs)A\Om
\,.
\end{align}
This equation is valid for $0<\mu<\frac{1}{4}$, and for all $A\in\A(\OO_r)$, with $r\leq s\cos(2\pi\mu)$. The left hand side of \bref{cooleqn} coincides with $\Theta_{s\sin(2\pi\mu),r}(A)$, and the right hand side bears some similarity with the maps $\Xi(s)$ appearing in the modular nuclearity condition.

We define the maps 
\begin{align}\label{ximus}
\Xi(\mu; s)	&: \A(W_R)\lto\Hil\,,\quad & \Xi(\mu;s)A	&:= \Delta^\mu U(\bs)A\Om\,,
\\
\Xi'(\mu; s)	&: \A(W_L)\lto\Hil\,,\quad & \Xi'(\mu;s)A'	&:= \Delta^{-\mu} U(-\bs)A'\Om\,.
\end{align}
The map $\Xi(s)$ \bref{Xis} considered in the modular nuclearity condition is a special member of this family, given by $\Xi(s)=\Xi(\frac{1}{4};s)$. Note that $\Xi'(\mu;s)$ is nuclear if and only if $\Xi(\mu;s)$ is, with the same nuclear norm. 

We will now use \bref{cooleqn} to estimate $\|\Theta_{\beta,r}\|_1$ in terms of $\|\Xi(\mu;s\cos 2\pi\mu)\|_1$. Putting $r:=\frac{1}{2}\cos(2\pi\mu)s$, $\beta:=2r\tan(2\pi\mu)$ yields
\begin{align}
e^{-\beta H}A\Om	&=	Q_+\Delta^\mu U(2\br)A\Om + Q_-\Delta^{-\mu}U(-2\br)A\Om
\,,\qquad A\in\A(\OO_r)\,,
\end{align}
where $\mu=\frac{1}{2\pi}\arctan\frac{\beta}{2r}$ depends on $\beta>0$. In view of $U(\pm\br)\A(\OO_r)U(\mp\br)\in\A(W_{R/L})$ and $\|Q_\pm\|\leq 1$, this implies the estimate
\begin{align}\label{thermoest1}
\|\Theta_{\beta,r}\|_1		&\leq		2\,\|\Xi(\tfrac{1}{2\pi}\arctan\tfrac{\beta}{2r};r)\|_1 \,.
\end{align}
This bound will be used to estimate the thermodynamical partition function.\footnote{It has been noticed only recently that this bound can be slightly improved \cite{nuclearmaps3}: Considering the tensor product nets $\A(\OO)^{\otimes n}$ on $\Hil^{\otimes n}$, $n\in\N$, also the maps \bref{def:thetabr} and \bref{ximus} are given by $n$-fold tensor products $\Theta_{\beta,r}^{\otimes n}$ and $\Xi(\mu;r)^{\otimes n}$, respectively. Hence \bref{thermoest1} also holds if the nuclear norms on both sides are replaced by arbitrary powers $\|\Theta_{\beta,r}\|_1^n$ and $\|\Xi(\tfrac{1}{2\pi}\arctan\tfrac{\beta}{2r};r)\|_1^n$. This implies that actually there holds the bound $\|\Theta_{\beta,r}\|_1\leq		\|\Xi(\tfrac{1}{2\pi}\arctan\tfrac{\beta}{2r};r)\|_1$.}
This bound will be used to estimate the thermodynamical partition function.
\\
\\
The nuclearity of $\Xi(r)=\Xi(\frac{1}{4};r)$ for any $r>0$ is not quite sufficient to derive the nuclearity of $\Theta_{\beta,r}$, since $\mu=\frac{1}{4}$ is only realized in the limits $\beta\to\infty$ and $r\to 0$. In order to prove the energy nuclearity condition, we need to show that in the considered models, $\Xi(\mu;r)$ is also nuclear for $\mu<\frac{1}{4}$.

Recall how the nuclearity of $\Xi(r)$ has been proven in chapter \ref{chapter:nuclearity}: Exploiting the analyticity and boundedness properties of the wavefunctions $(A\Om)_n$, $A\in\A(W_R)$, we showed that $A\lmto(A(\frac{1}{2}\br)\Om)_n$ is a bounded map from $\A(W_R)$ into the Hardy space on a certain tube domain. The nuclearity of $\Xi(s)$ then followed by expressing
\begin{align}
(A(\br)\Om)_n(\bte)=\prod_{k=1}^n e^{-mr\cosh\te_k}\cdot (A\Om)_n(\te_1-\tfrac{i\pi}{2},...,\te_n-\tfrac{i\pi}{2})
\end{align}
as a Cauchy integral over the boundary of this tube.

The maps $\Xi(\mu;r)$ act on $A\in\A(W_R)$ explicitly as
\begin{align}\label{xinun}
(\Xi(\mu;r)A)_n(\te_1,...,\te_n)
=
\prod_{k=1}^n e^{-irm\sinh(\te_k-2\pi i\mu)} \cdot (A\Om)_n(\te_1-2\pi i\mu,...,\te_n-2\pi i \mu)
\,,
\end{align}
and thus have a structure very similar to $\Xi(s)$ \bref{xi-n-ana}. Taking into account the slightly weaker decay of the exponential function in \bref{xinun}, and the possibly shorter distance of $(\te_1-2\pi i\mu,...,\te_n-2\pi i \mu)$ to the boundary of the tube in which $(A\Om)_n$ is analytic, one can apply the arguments of chapter \ref{chapter:nuclearity} also to this map. It follows that $\Xi(\mu,r)$ is nuclear for any $0<\mu<\frac{1}{4}$ and any $r>0$ if the underlying scattering function $S_2$ is in the class $\SF_0^-$. 
\begin{proposition}
Consider a model theory with scattering function $S_2\in\SF_0$. Then the energy nuclearity condition holds, i.e. the maps 
\begin{align}
\Theta_{\beta,\OO}	:	\A(\OO)	\lto	\Hil\,,\qquad A\lmto e^{-\beta H}A\Om
\end{align}
are nuclear for any bounded region $\OO$ and any inverse temperature $\beta$.{\hfill $\square$}
\end{proposition}
Repeating the analysis of chapter \ref{chapter:nuclearity}, one can also derive bounds on $\|\Xi(\mu;r)\|_1$, and hence on the thermodynamical partition function $\|\Theta_{\beta,r}\|_1$. However, these bounds are rather crude and not sharp enough to  to give meaningful estimates on the partition function. In particular, the thermodynamically motivated bound $\|\Theta_{\beta,r}\|_1\leq e^{(\beta_0/\beta)^n}$ \bref{theta-bound} cannot be established without improving these estimates.
\\
\\
As explained in chapter \ref{chapter:nuclearity}, we surmise that the bounds given in Thm. \ref{thm:-1nuclearity} can be improved to
\begin{align}\label{betterbound}
 \|\Xi(s)\|_1
    &\leq
    \exp(\sigma(s,\kappa)\,\|S_2\|_\kappa^{1/2}\,\|T_{s,\kappa}\|_1)
    \;,\qquad S_2\in\SF_0^-\,.
\end{align}
The symbols used in this formula are the same as in chapter \ref{chapter:nuclearity}.

If this bound was established, similar bounds on $\|\Xi(\mu;r)\|_1$ could be derived in a straightforward manner, and these could then be used to compute meaningful estimates on the partition function with the help of \bref{thermoest1}. Employing the estimates on the trace norms of the integral operators $T_{s,\kappa}$ derived in appendix \ref{app:intop}, it would in particular follow that the partition function behaves like $e^{\beta_0/\beta}$ for $\beta\to 0$.

Such a behavior of the partition function can be expected to hold in particle theories. In view of the results obtained in the preceding section, the models considered here have a complete particle interpretation. Therefore the mentioned form of the partition function seems likely to be valid, and further supports the conjecture of the improved bound \bref{betterbound}.



\chapter{Conclusion and Outlook}\label{chapter:conclusions}

In the present work, a new construction method for two-dimensional quantum field theories has been presented. The main idea of this approach is to base the construction of models not on pointlike localized quantum fields, but rather on wedge-localized polarization-free generators. It has been shown that these generators are much easier to construct, and that they can be used for the definition of interacting theories.

Such a construction was carried out in full detail for a family of models with factorizing S-matrices. Employing an inverse scattering approach, we started from a prescribed factorizing S-matrix, specified by its scattering function $S_2$. We then considered the Zamolodchikov-Faddeev algebra associated with $S_2$ and investigated the corresponding quantum fields invented by Schroer \cite{Schroer:1997cx, Schroer:1999xi,Schroer-Wiesbrock}. These fields were shown to be localizable in wedges, and were used to generate a covariant net of wedge algebras on two-dimensional Minkowski space. Algebras of observables localized in bounded spacetime regions could then be defined as appropriate intersections of wedge algebras.

As a crucial step in the construction, we verified the modular nuclearity condition of Buchholz, D'Antoni and Longo \cite{nuclearmaps1,nuclearmaps2} for wedge algebras for a large class of scattering functions. This condition ensures that the local observables of the theory have all basic properties needed in relativistic quantum physics. In particular, it implies that the considered S-matrices are related to meaningful quantum field theories.

Furthermore, we assured ourselves of the fact that the models constructed in this way solve the inverse scattering problem for the considered family of S-matrices. Namely, the S-matrix corresponding to a particular scattering function $S_2$ can be recovered from the collision states from the model defined in terms of $S_2$.

The approach presented here provides an alternative point of view on the form factor program, which aims at the construction of local quantum fields and their $n$-point functions associated with a given factorizing S-matrix. The question whether a particular S-matrix is realized as the scattering operator of a quantum field theory and the rigorous construction of the corresponding models are problems which are very difficult to solve in the setup of the form factor program. As we have seen, the algebraic framework is better suited for the investigation of such problems. Although no formulae for interacting local fields are obtained in this approach, it is possible to compute all collision states and to derive estimates on the thermodynamical partition function of the theory.

For more detailed local information on these models, like the determination of correlation functions or the field content, the approach of the form factor program seems to be better suited. In fact, the bootstrap form factor program and the algebraic construction complement each other, and the combination of both approaches is likely to lead to a thorough understanding of the family of models with factorizing S-matrices.
\\
\\
In this thesis, we considered theories describing a single species of scalar massive particles, but a generalization to theories with a richer particle spectrum seems possible. To this end, the Zamolodchikov operators $z(\te)$, $\zd(\te)$ have to be replaced by vector-valued objects $z_k(\te)$, $\zd_k(\te)$, where the index $k$ denotes the species of particles in the model. The scattering functions of such models are matrix-valued, and the determination of the possible two-particle S-matrices becomes more complicated, because also the Yang-Baxter equations have to be solved. Such scattering operators have been analyzed extensively in the form factor program, and we surmise that the results obtained there can be used for the construction of wedge-local fields similar to the case discussed here.
\\
\\
An intriguing and in my point of view more important extension of the approach presented here would be a generalization to higher-dimensional spacetime. Such a generalization faces two major problems: Firstly, in more than two dimensions, factorizing S-matrices do not exist, and hence the input for the construction has to be chosen differently. In particular, as no non-trivial S-matrix is known in higher dimensions, a construction in an inverse scattering approach is presently out of sight. But it might well be possible that wedge algebras can also be constructed in physical spacetime by other methods.

Such wedge algebras can then be used analogously to the construction in chapter \ref{chapter:netsin2d} to generate a net of local algebras, only the representation $U$ of the translation group has to be replaced by a representation of the identity component of the Poincar\'e group. Although the characterization of the interaction in these theories will be more difficult, it should be possible to analyze two-particle scattering processes, which can be treated with the help of wedge-localized observables.

The second major problem for the generalization to higher dimensions is the fact that inclusions of wedge algebras cannot be split in more than two dimensions. Hence the modular nuclearity condition for wedges, which was a tool of crucial importance here, cannot be applied.

On the other hand, it is well-known that the split property for wedges is not necessary for relative commutants of wedge algebras to be nontrivial, and in fact, this condition is much too strong if one is merely interested in the existence of local observables. What is needed for the construction of higher-dimensional quantum field theories is a new effective criterion to control the structure of the local observable algebras in models defined in terms of wedge-localized objects.

The generalization of the construction demonstrated here to four dimensions would be a major advance in rigorous quantum field theory, for it might provide a way to finally settle the long-standing existence problem for interacting quantum field theories in physical spacetime.

\appendix
\chapter[The Nuclearity Condition for Constant Scattering Functions]{The Modular Nuclearity Condition for Constant Scattering Functions}\label{chapter:mnc+-1}

In this appendix we consider the modular nuclearity condition in two special models within the family of theories constructed in chapter \ref{chapter:wedgenet}. These are the models given by the two constant scattering functions $S_2=\pm 1$ in which the underlying Zamolodchikov algebra $\Z(S_2)$ simplifies to the CCR ($S_2=1$) and CAR ($S_2=-1$) algebra, respectively. This simplified algebraic structure allows for an alternative analysis of the modular nuclearity condition.

\section[$S_2=1$ : The Free Bose Field]{\Bose : The Free Bose Field}\label{sect:+1}

In this section we consider the free scalar Bose field of mass $m>0$, which in the setup of chapter \ref{chapter:wedgenet} is given by the constant scattering function $S_2=1$. The free field is the best studied system in quantum field theory, and most of the properties that we want to derive from the modular nuclearity condition are well-known in this context and have been proven by other methods already (cf., for example, \cite{Jost,Araki:free}). In fact, only for the split property for wedges, there seemed to exist no proof in the literature until \cite{BuLe}, although also this property was known to hold before \cite{Mue-SPW}.

So the proof of the modular nuclearity condition for wedges in the case $S_2=1$ provides little new information about the free field. But as the methods used in chapter \ref{chapter:nuclearity} did not suffice to establish this condition for the free field, because of the lacking Pauli principle, we feel obliged to provide an alternative argument that this simplest system complies with the condition of modular nuclearity, too.

The proof which we will give heavily relies on the analysis of nuclear subsets of the Bose Fock space, which was carried out by Buchholz and Wichmann in the context of the energy nuclearity condition in \cite{BuWi}, and generalized to the modular setting by Buchholz and Jacobi \cite{BuJa}.

The results of the present section have been published in \cite{BuLe}.
\\
\\
The construction of the net $\OO\lmto\A(\OO)$ carried out in chapter \ref{chapter:wedgenet} simplifies considerably if $S_2=1$, since the fields $\phi$ and $\phi'$ coincide in this case, and $\phi$ is a {\em local} Wightman field. In contrast to the situation for the other scattering functions, the local algebras of this model are explicitly known: For any open region $\OO\subset\Rl^2$, the algebra $\A(\OO)$ of observables localized in $\OO$ is generated by the Weyl operators $e^{i\phi(f)}$, $f\in C_0^\infty(\OO)$ real.

There exist several formulations of the theory of free fields \cite{Jost,Araki:free, BraRob2, BGL}. In order to use the results of \cite{BuWi,BuJa}, we need to formulate the theory in terms of the time-zero fields $\varphi(x_1)=\phi(0,x_1)$, $\pi(x_1)=(\partial_0\phi)(0,x_1)$, $x_1\in\Rl$, of the free field $\phi$. This can be done as follows.

In view of the form \bref{def:phipi} of these field operators, the one particle states localized in $W_R$ that are created from the vacuum by $\varphi$ and $\pi$, respectively,. span the spaces
\begin{subequations}\label{def:Lphipi}
\begin{align}
\LL_\varphi(W_R)	&:=	\{\fhat \,:\, f\in\Ss(\Rl_+)\}^-,\\
\LL_\pi(W_R)	&:=	\{\omega\fhat \,:\, f\in\Ss(\Rl_+)\}^-\,,
\end{align}
\end{subequations}
where the bar denotes closure in $L^2(\Rl)$, and the notations $\fhat(\te):=\fti(m\sinh\te)$ and $(\omega\fhat)(\te)=m\cosh(\te)\fhat(\te)$ have been used..

These subspaces are related to the algebra of observables localized in $W_R$ as follows: With the help of the antiunitary involution $\Gamma$, which was defined in Lemma \ref{lem:extendU} and represents the time reflection, $(\Gamma\psi)(\te)=\overline{\psi(-\te)}$, we define a real linear subspace of $\Hil_1$ as
\begin{align}\label{def:KWL}
\K(W_R)
&:=
(1+\Gamma)\LL_\varphi(W_R) + (1-\Gamma)\LL_\pi(W_R)\;.
\end{align}
The algebra of the right wedge can be expressed as
\begin{align}\label{AdurchK}
	\A(W_R)
	&=
	\big\{ e^{i\phi(f)}\,:\, \phi(f)\Om\in\K(W_R)\big\}''\,.
\end{align}
We will not prove this relation here, but only indicate how it can be understood in terms of the modular structure of the wedge algebras.

It is known from the work of Brunetti, Guido and Longo \cite{BGL} that $\A(W_R)$ is given by a formula like \bref{AdurchK}, but with the space $\K(W_R)$ replaced by the eigenspace $\hat{\K}(W_R)$ of the Tomita operator of $(\A(W_R),\Om)$, corresponding to eigenvalue one,
\begin{align}
\hat{\K}(W_R)
&:=
\{\psi\in\dom\Delta^{1/2}\cap\Hil_1\,:\,J\Delta^{1/2}\psi=\psi\}\,.
\end{align}
In fact, $\K(W_R)=\hat{\K}(W_R)$, and we will show the inclusion $\K(W_R)\subset\hat{\K}(W_R)$ here.
Note that in view of the support restriction $\supp f\subset\Rl_+$, the functions $\fhat$ and $\omega\fhat$ have an analytic continuation to the strip $S(-\pi,0)=\{\zeta\in\Cl\,:\,-\pi<{\rm Im}\zeta<0\}$. Moreover, as $f$ is a Schwartz function, $\te\mapsto\fhat(\te-i\la)$ and $\te\mapsto(\omega\fhat)(\te-i\la)$ are square integrable for any $\la\in[0,\pi]$. This implies that  $\fhat$ and $\omega\fhat$ lie in the domain of the modular operator $\Delta^{1/2}$, which acts as $(\Delta^{1/2}\psi)(\te)=\psi(\te-i\pi)$, $\psi\in\dom\Delta^{1/2}$.

Taking into account $\sinh(\te-i\pi)=\sinh(-\te)$ and $\cosh(\te-i\pi)=-\cosh(-\te)$, it also follows that $\fhat(\te-i\pi)=\fhat(-\te)$, $(\omega\fhat)(\te-i\pi)=-(\omega\fhat)(-\te)$. Hence the corresponding eigenfunctions of $\Gamma$, $\fhat_\pm := (1\pm\Gamma)\fhat$, $\Gamma\fhat=\pm\fhat$, satisfy
\begin{align}
\fhat_\pm(\te-i\pi)	&=	\fhat_\pm(-\te)	=	\pm (\Gamma\fhat_\pm)(-\te)	 = \pm \overline{\fhat_\pm(\te)}\,.
\end{align}
As $\Gamma$ and $\omega$ commute, we furthermore obtain 
\begin{align}
(\omega\fhat_-)(\te-i\pi)	&=	-(\omega\fhat_-)(-\te)	=	(\Gamma\omega\fhat_-)(-\te)		=	
\overline{(\omega\fhat_-)(\te)}\,.
\end{align}
But the action of the modular conjugation $J$ on $\Hil_1$ is just complex conjugation, and so we conclude
\begin{align}
J\Delta^{1/2}\fhat_+	=	\fhat_+
\,,\qquad
J\Delta^{1/2}\omega\fhat_-	= \omega\fhat_-\,,
\end{align}
i.e. $\K(W_R)\subset\hat{\K}(W_R)$. The converse direction can be proven in a similar way.
\\
\\
By systematically exploiting the second quantization structure and Bose symmetry of the free theory, Buchholz and Wichmann managed to simplify the energy nuclearity condition to a problem on the single particle space \cite{BuWi}. In \cite{BuJa}, this analysis was generalized to cover the modular nuclearity condition as well. The result which we will use is formulated in terms of the orthogonal projections $E_\varphi(W_R)$ and $E_\pi(W_R)$, projecting onto $\LL_\varphi(W_R)$ and $\LL_\pi(W_R)$, respectively. As before, we write $\bs=(0,s)$, $s>0$.
\begin{lemma}{\bf \cite[Thm. 2.1]{BuJa}}\label{lemma:BuJa}\\
Consider the free theory, given by the scattering function $S_2=1$, and assume that both, $E_\varphi(W_R)U(-\bs)\Delta^{1/4}$ and $E_\pi(W_R)U(-\bs)\Delta^{1/4}$, are trace class operators on $\Hil_1$, with norms less than one.
\\
Then the map $\Xi(s)$ is nuclear, and its nuclear norm is bounded by
\begin{align}
	\|\Xi(s)\|_1
	&\leq
	\det(1-|E_\varphi(W_R)U(-\bs)\Delta^{1/4}|)^{-2}
	\cdot
	\det(1-|E_\pi(W_R)U(-\bs)\Delta^{1/4}|)^{-2}\,.
\end{align}
\end{lemma}
\noindent We will not repeat the proof of this statement here, but refer the reader to section \ref{sect:-1}, where a similar result is demonstrated for the model given by the scattering function $S_2=-1$.
\\
\\
To be able to apply Lemma \ref{lemma:BuJa}, we need to check if its assumptions are valid here. We first establish the estimates on the norms of the operators appearing in the Lemma. 

Let $\psi_\varphi\in\LL_\varphi(W_R)\subset\dom\Delta^{1/2}$. As $\psi_\varphi(\te-i\pi)=\psi_\varphi(-\te)$, we have $\|\Delta^{1/2}\psi_\varphi\|=\|\psi_\varphi\|$, and hence $\|\Delta^{1/4}\psi_\varphi\|=\langle\psi_\varphi,\Delta^{1/2}\psi_\varphi\rangle^{1/2}\leq\|\psi_\varphi\|$. As $U(\bs)$ acts by multiplication with $e^{-ims\sinh\te}$, we obtain after analytic continuation to $\te-i\pi/2$
\begin{align}
(\Delta^{1/4}U(\bs)\psi)(\te)	
=
e^{-sm\cosh\te}\cdot\psi(\te-\tfrac{i\pi}{2})\,,
\end{align}
and consequently
\begin{align}
\|\Delta^{1/4}U(\bs)\psi_\varphi\|
	\leq
e^{-sm}\|\psi_\varphi\|
<
\|\psi_\varphi\|
\,,\qquad s>0\,.
\end{align}
Since $\psi_\varphi\in\LL_\varphi(W_R)$ was arbitrary, this estimate implies $\|\Delta^{1/4}U(\bs)E_\varphi(W_R)\|<1$, $s>0$. But the adjoint operator $E_\varphi(W_R)U(-\bs)  \Delta^{1/4}$ has the same norm, and so the desired bound follows. In the same manner, one shows that also $E_\pi(W_R)U(-\bs)\Delta^{1/4}$ has norm less than one.

It remains to establish the trace class property of these operators. To this end, we use the analyticity and boundedness properties of $\psi_\varphi\in\LL_\varphi(W_R)$ in the strip $S(-\pi,0)$, as well as the relation $\psi_\varphi(\te-i\pi)=\psi_\varphi(-\te)$, to express $(\Delta^{1/4}U(\bs)\psi_\varphi)(\te)=e^{-ms\cosh\te}\psi_\varphi(\te-\frac{i\pi}{2})$ as a Cauchy integral,
\begin{align}
(\Delta^{1/4}U(\bs)\psi_\varphi)(\te)
=
\frac{1}{2\pi i}\int_\Rl d\te'\,
\left(
\frac{e^{-sm\cosh\te}}{-\te'-\te-\frac{i\pi}{2}}
-
\frac{e^{-sm\cosh\te}}{\te'-\te+\frac{i\pi}{2}}
\right)
\psi_\varphi(\te')\,.
\end{align}
Hence $\Delta^{1/4}U(\bs)E_\varphi(W_R) = T_\varphi(s)E_\varphi(W_R)$, where $T_\varphi(s)$ is the integral operator with kernel
\begin{align}\label{Tphikernel}
T_\varphi(s)(\te,\te')
&=
\frac{1}{2\pi i}
\left(
\frac{e^{-sm\cosh\te}}{-\te'-\te-\frac{i\pi}{2}}
-
\frac{e^{-sm\cosh\te}}{\te'-\te+\frac{i\pi}{2}}
\right)
\,.
\end{align}
This integral operator is of trace class on $\Hil_1=L^2(\Rl,d\te)$ for any $s>0$, cf. appendix \ref{app:intop}.
Given the nuclearity of $T_\varphi(s)$, the nuclearity of $E_\varphi(W_R)T_\varphi(s)^*=E_\varphi(W_R)U(-\bs)\Delta^{1/4}$ follows since the trace class is a $*$-ideal in $\B(\Hil_1)$.

The operator $E_\pi(W_R)U(-\bs)\Delta^{1/4}$ can be treated analogously, the only difference being that the relation $\psi_\pi(\te-i\pi)=-\psi_\pi(-\te)$ results in a minus sign in front of the first term in \bref{Tphikernel}. Hence $\Delta^{1/4}U(\bs)E_\pi(W_R) = T_\pi(s)E_\pi(W_R)$, where $T_\pi(s)$ has the integral kernel
\begin{align}\label{Tpikernel}
T_\pi(s)(\te,\te')
&=
\frac{1}{2\pi i}
\left(
\frac{e^{-sm\cosh\te}}{\te'+\te+\frac{i\pi}{2}}
-
\frac{e^{-sm\cosh\te}}{\te'-\te+\frac{i\pi}{2}}
\right)
\,,
\end{align}
and is of trace class, too (cf. Lemma \ref{intop-lemma}).

To summarize, we have shown that the model theory given by $S_2=1$ complies with the assumptions of Lemma \ref{lemma:BuJa}. Consequently, the following Proposition holds.
\begin{proposition}\label{prop:S2=1nuc}{\bf \cite{BuLe}}\\
Consider the free theory, given by the scattering function $S_2=1$.
\\
The maps $\Xi(s)$ \bref{Xis} are nuclear for any $s>0$, and there holds the bound
\begin{align}
	\|\Xi(s)\|_1
	&\leq
	\det(1-|E_\varphi(W_R)T_\varphi(s)^*|)^{-2}
	\cdot
	\det(1-|E_\pi(W_R)T_\pi(s)^*|)^{-2}\,,
\end{align}
where $T_\varphi(s)$ and $T_\pi(s)$ are the trace class integral operators on $L^2(\Rl)$, given by the kernels \bref{Tphikernel} and \bref{Tpikernel}.{\hfill $\square$}
\end{proposition}


\section[The scattering function $S_2=-1$]{The scattering function \Fermi}\label{sect:-1}

In this section we study the nuclearity properties of the model theory given by the scattering function $S_2=-1$. Although the two-particle scattering states of this model, as constructed in section \ref{sect:wedgescattering}, are of Bose type, the $S_2$-symmetric Hilbert space $\Hil$ \bref{def:Hil} coincides with the antisymmetric Fermi Fock space over $\Hil_1$ since $S_2=-1$. This fact implies a number of algebraic similarities between this model and models of a free (local) Fermionic field on $d$-dimensional Minkowski space, $d\geq 2$. Our main interest lies in proving the modular nuclearity condition for the former theory. But in view of the formal analogy to systems of free Fermions, it is possible to use a formulation wide enough to be applicable to nuclearity properties of such theories as well. This has the advantage that, as a byproduct of our investigation of models with a factorizing S-matrix in two dimensions, we also obtain a proof of the energy nuclearity condition of theories describing free Fermions in $d\geq 2$ dimensions. 

In the following section, we study nuclear maps in Fermi Fock space in a slightly more abstract setting. Subsequently, the results obtained there are applied to prove the modular nuclearity condition for wedges in the model with scattering function $S_2=-1$ and the energy nuclearity condition (for bounded regions) in theories of free Fermions.

The contents of this section can be found in the article \cite{GL-1}.

\subsection{Nuclear Maps in Fermionic Fock Space}\label{-1nuci-general}

The mathematical structure which is common to the two kinds of models mentioned above is the following: We consider a separable Hilbert space $\Hil_1$ (the one particle space) with scalar product $\langle\,.\,,\,.\,\rangle$, and the CAR algebra generated by the symbols $a(\psi)$ and $a^*(\psi)=a(\psi)^*$, $\psi\in\Hil_1$, and a identity $1$. We adopt the convention that $a^*(\psi)$ depends complex linearly on $\psi$. The structure of the CAR algebra is fixed by the canonical anticommutation relations
 \begin{align}
  [a(\psi_1),a(\psi_2)]_+ 			&= 0,\label{CAR1}\\
  \mbox{[}a(\psi_1),a^*(\psi_2)]_+ 	&= \langle \psi_1,\psi_2 \rangle \cdot 1\,.\label{CAR2}
\end{align}
Here $[A,B]_\pm = AB \pm BA$ denotes the (anti-) commutator. By second quantization, we obtain the Fock space $\Hil$ over $\Hil_1$, which is acted upon by the CAR algebra in the standard representation \cite{BraRob2}, and the Fock vacuum $\Om\in\Hil$. Recall that $a(\psi)$, $a^*(\psi)$ are bounded operators, $\|a(\psi)\|=\|a^*(\psi)\|=\|\psi\|$ \cite{BraRob2}.

On $\Hil_1$, we consider an antiunitary involution $\Gamma$, and two closed, complex subspaces $\LL_\varphi$ and $\LL_\pi$ of $\Hil_1$, which are invariant under $\Gamma$. These are used to define a real linear subspace $\K$ of $\Hil_1$ as
\begin{equation}\label{def:K}
  \K := (1+\Gamma)\LL_\varphi + (1-\Gamma)\LL_\pi \;.
\end{equation}
Furthermore, we introduce a Fermionic field operator
\begin{align}\label{def:phi-eckig}
	\phi[\psi]		&:= a^*(\psi) + a(\psi)\,,\qquad\psi\in\Hil_1\,.
\end{align}
The square brackets are used in order to distinguish $\phi[\psi]$ from the field $\phi(f)$ \bref{def:phi}. In application to the model with scattering function $S_2=-1$, we will see that $\phi[\,\cdot\,]$ and $\phi(\cdot)$ are closely related. In the present abstract formulation, we consider the von Neumann algebra generated by the field \bref{def:phi-eckig},
\begin{align}\label{def:AK}
\A_\K	&:=	\big\{\phi[\psi]\,:\,\psi\in\K\}'' \subset \B(\Hil)\,,
\end{align}
and assume that the vacuum vector $\Om$ is separating for this algebra.

To formulate the maps \bref{Xis} in the present setting, we introduce a densely defined, strictly positive operator $X$ on $\Hil_1$, which commutes with the involution $\Gamma$. So $X$ is in particular assumed to be invertible, but it need not be bounded. We use the same symbol $X$ to denote its second quantization $\bigoplus_{n=0}^\infty X\tp{n}$ and assume that $\A_\K\Om$ is contained in its domain. It is our aim to find sufficient conditions on the real subspace $\K$ and the operator $X$ that imply the nuclearity of the map 
\begin{eqnarray}\label{def:xi-abstract}
  \Xi : \A_\K \lto \Hil\,,\qquad\qquad\Xi(A) := XA\Omega\,.
\end{eqnarray}
Denoting by $E_{\varphi}, E_\pi \in \B(\Hil)$ the orthogonal projections onto $\LL_{\varphi}$, $\LL_\pi$, respectively, the nuclearity properties of \bref{def:xi-abstract} are characterized in the following Proposition.
\begin{proposition}{\bf \cite{GL-1}}\label{prop:-1nuc}\\
  Assume that $E_\varphi X$ and $E_\pi X$ extend to trace class operators on $\Hil_1$. Then $\Xi$ is a nuclear map, and its nuclear norm is bounded by
  \begin{equation}\label{xi-bound}
  \|\Xi\|_1 \leq
  e^{2\|E_\varphi X\|_1}\cdot e^{2\|E_\pi X\|_1}\,.
  \end{equation}
\end{proposition}
This Proposition has to be seen in analogy to Lemma \ref{lemma:BuJa} from the discussion of the free Bose field in the preceding section. It simplifies the study of nuclear maps on the full Hilbert space to a problem on the one particle space, which can be solved in concrete applications, where the operator $X$ is given by $\Delta^{1/4}U(\bs)$ or $e^{-\beta H}$.

In comparison with the analogous result for free Bosons one notices two differences: Firstly, the conditions on $E_\varphi X$, $E_\pi X$ are relaxed since the bounds $\|E_\varphi X\|<1$, $\|E_\pi X\|<1$ on their operator norms are not required here. Secondly, the bound on the nuclearity index is smaller than the corresponding one for Bosons, $\det(1-|E_\varphi X|)^{-2}\cdot \det(1-|E_\pi X|)^{-2}$, obtained in \cite{BuJa}. This can be seen from the following simple inequality, valid for any non-zero trace class operator $T$ with norm $\|T\| < 1$. The singular values of $T$ are denoted by $t_n$, repeated according to multiplicity.
\begin{equation*}
  e^{2\|T\|_1} = e^{2\sum_{n=1}^\infty |t_n|}
  =
    \prod_{n=1}^\infty \left(e^{-|t_n|}\right)^{-2}
    < \prod_{n=1}^\infty (1-|t_n|)^{-2}
    = \det(1-|T|)^{-2}\,.
\end{equation*}
This result is due to the Pauli principle; it may be understood in analogy to the difference between the partition functions of the
non-interacting Bose and Fermi gases in the grand canonical ensemble.
\\\\
The rest of this section is devoted to the proof of Proposition \ref{prop:-1nuc}. To begin with, we consider the polynomial algebra generated by the field,
\begin{equation}\label{pol}
  \pol_\K := \mbox{span}\{\phi[\psi_1]\cdots\phi[\psi_n]\,:\, n\in\N\,,\psi_1,...,\psi_n\in\K\}\,.
\end{equation}
As $\phi[\psi]$ is bounded, $\pol_\K$ is a weakly dense subalgebra of $\A_\K$. 
The polynomial algebra has the structure of a $\Zl_2$-graded $*$-algebra, with the even and odd parts $\pol^+_\K$
and $\pol^-_\K$ given by the even and odd field polynomials, respectively. On $\pol_\K$ acts the
grading automorphism
\begin{align}\label{gammagrader}
  \gamma(A^++A^-) := A^+-A^-\,,\qquad A^\pm \in \pol_\K^\pm\,.
\end{align}
As $\|\gamma\|=1$ and $A^\pm = \frac{1}{2}(A\pm\gamma(A))$, we
conclude $\|A^\pm\|\leq\|A\|$.\\

The following Lemma about the interplay of the CAR algebra and $\pol_\K$ in connection with the real linear structure of $\K$
is the main technical tool in the proof of Proposition \ref{prop:-1nuc}. We will use the analogues of the time-zero fields \bref{def:phipi}, which are here defined as
\begin{align}
\varphi(\psi)	&:=	a^*(\psi)	+	a(\Gamma\psi)\,,\\
\pi(\psi)		&:=	i\big(a^*(\psi)	-	a(\Gamma\psi)\big)\,.
\end{align}
Note that $\varphi(\psi)^*=\varphi(\Gamma\psi)$, $\pi(\psi)^*=\pi(\Gamma\psi)$. Furthermore, these operators anticommute: Using the CAR relations and the fact that $\Gamma$ is an antiunitary involution, we find for arbitrary $\psi_1,\psi_2\in\Hil_1$:
\begin{align*}
	[\varphi(\psi_1),\,\pi(\psi_2)]_+	&=	[a^*(\psi_1),\,-i\,a(\Gamma\psi_2)]_+ + [a(\Gamma\psi_1),\,i\,a^*(\psi_2)]_+\\
								&=-i\,\langle\Gamma\psi_2,\,\psi_1\rangle +i\,\langle\Gamma\psi_1,\,\psi_2\rangle = 0\,.
\end{align*}
For later use we also state 
\begin{align}
  \label{a}
  a(\Gamma\psi) = \frac{1}{2}\big(\varphi(\psi)+i\pi(\psi)\big)\,.
\end{align}
In preparation for the Lemma stated below, recall that the symplectic complement of $\K$ is 
\begin{equation}
\K'=\{\psi\in\Hil_1\,:\,\langle\psi,\xi\rangle=\langle\xi,\psi\rangle\quad\forall\xi\in\K\}\,,
\end{equation}
and that an odd derivation on a $\Zl_2$-graded algebra $\pol$ is a linear map $\delta:\pol\to\pol$ which satisfies $\delta(\pol^\pm) \subset \pol^\mp$ and obeys the graded Leibniz rule
\begin{align}\label{leibniz}
  \delta(A B) = \delta(A)B + \gamma(A)\delta(B),\qquad\quad
  A, B\in\pol\,,
\end{align}
where $\gamma$ is the grading automorphism \bref{gammagrader}.
\begin{lemma}\label{lemma1}
For arbitrary $\psi\in\Hil_1$, the assignments
\begin{align}\label{def-delta}
	\delta^\pm_\psi(A) 
    	:=\;
    	&\tfrac{1}{2}\left[\varphi((1\mp\Gamma)\psi)+i\pi((1\pm\Gamma)\psi),A^+\right]_-
    	\\
     	+ &\tfrac{1}{2}\left[\varphi((1\mp\Gamma)\psi)+i\pi((1\pm\Gamma)\psi),A^-\right]_+\nonumber
 \end{align}
 define two odd derivations on $\pol_\K$ which are real linear in $\psi$. These maps satisfy the bounds
 \begin{align}
\|\delta^+_\psi(A^\pm)\| &\leq \left(\|(1 - \Gamma)E_\varphi\psi\|^2
+
\|(1 + \Gamma)E_\pi\psi\|^2\right)^{1/2}\cdot\|A^\pm\|
\,,\label{d-norm}\\
\|\delta^-_\psi(A^\pm)\| &\leq \left(\|(1 + \Gamma)E_\varphi\psi\|^2 
+ \|(1 - \Gamma)E_\pi\psi\|^2\right)^{1/2}\cdot\|A^\pm\|
\,.\label{d-norm2}
\end{align}
Moreover, if $\psi\in\K'$,
\begin{align}\label{kill}
	\delta^+_\psi=0\,,\qquad\qquad \delta^-_{i\psi}=0\,.
\end{align}
\end{lemma}
\begin{proof}
The real linearity of $\psi\longmapsto\delta^\pm_\psi$ follows directly from the definition \bref{def-delta} and the real linearity of $\varphi,\pi$ and $\Gamma$.

As $\delta^\pm_\psi$ are complex linear maps on $\pol_\K$, it suffices to consider their action on field monomials $\phi[\xi_1]\cdots\phi[\xi_n]$, $\xi_1,...,\xi_n\in \K$ to prove the assertion about the derivation property and \bref{kill}. We also write $\phi_k := \phi[\xi_k]$ and carry out a proof based on induction in the field number $n$. For $n=0$, we have $\delta^\pm_\psi(1)=0 \in \pol_\K^-$ for any $\psi\in\Hil_1$, and \bref{kill} holds trivially. For $n=1$, the canonical anticommutation relations (\ref{CAR1},\ref{CAR2}) imply
\begin{align}
\delta^\pm_\psi(\phi[\xi])
	&=
\tfrac{1}{2} \left[\varphi((1\mp\Gamma)\psi)+ i\pi((1\pm\Gamma)\psi),\phi[\xi]\right]_+
	\nonumber\\
    	&=
\tfrac{1}{2} \left(\langle \xi,(1\mp\Gamma)\psi\rangle\mp \langle (1\mp\Gamma)\psi,\xi\rangle
  	  -
\langle\xi,(1\pm\Gamma)\psi\rangle \pm \langle(1\pm\Gamma)\psi,\xi\rangle\right)\cdot 1
\nonumber\\
	&=
\left( \langle\Gamma\psi,\xi\rangle \mp \langle\xi,\Gamma\psi\rangle\right)\cdot 1
\,.\label{calc1}
\end{align}
  As $\K$ is $\Gamma$-invariant, so is $\K'$, and hence $\psi\in\K'$ implies $\delta^+_\psi(\phi[\xi])=0$, $\delta^-_{i\psi}(\phi[\xi])=0$. Being a multiple of the identity, $\delta^\pm_\psi(\phi[\xi])$ is contained in $\pol^+_\K$ for arbitrary $\psi\in\Hil_1$. The step from $n$ to $n+1$ fields is achieved by considering
  \begin{align}\label{recurse}
    \begin{array}{rcl}
    [F,\phi_1\cdots\phi_{2n}]_-
    &=& 
    [F,\phi_1\cdots\phi_{2n-1}]_+\cdot\phi_{2n}
    -
    \phi_1\cdots\phi_{2n-1}\cdot[F,\phi_{2n}]_+\,,\\\vspace*{-3mm}&&\\
    \mbox{[}F,\phi_1\cdots\phi_{2n+1}]_+
    &=& 
    [F,\phi_1\cdots\phi_{2n}]_-\cdot\phi_{2n+1}
    +
    \phi_1\cdots\phi_{2n}\cdot[F,\phi_{2n+1}]_+\,,
  \end{array}
\end{align}
with $F=\frac{1}{2}(\varphi((1\mp\Gamma)\psi)+i\pi((1\pm\Gamma)\psi))$. It follows from these formulae inductively that $\delta^\pm_\psi$ turn even elements of $\pol_\K$ into odd ones and vice versa. Moreover, $\delta^+_\psi=0$, $\delta^-_{i\psi}=0$ for $\psi\in\K'$ because of  the corresponding result for $n=1$. By direct calculation, one can also verify the Leibniz rule \bref{leibniz}.
We have  thus shown that $\delta^\pm_\psi$ are odd derivations of $\pol_\K$ satisfying \bref{kill}.

To prove the norm estimate \bref{d-norm}, we first note that
\begin{align}
    \psi':=\left(\tfrac{1}{2}(1+\Gamma)(1-E_\pi)+\tfrac{1}{2}(1-\Gamma)(1-E_\varphi)\right)\psi
\end{align}
is an element of the symplectic complement $\K'$ for arbitrary $\psi\in\Hil_1$, as can be easily verified using \bref{def:K}. Since $\delta^+_{\psi'}=0$ and $\delta^+_\psi$ is real linear in $\psi$, we have
\begin{align}
    \|\delta^+_\psi(A^\pm)\|
    &=
    \|\delta^+_{\psi-\psi'}(A^\pm)\|
    \nonumber\\
    &=
    \tfrac{1}{2} \|
    [\varphi((1-\Gamma)E_\varphi\psi)+i\,\pi((1+\Gamma)E_\pi\psi),A^\pm]_\mp\|\nonumber\\
    &\leq
    \|\varphi((1-\Gamma)E_\varphi\psi)+i\,\pi((1+\Gamma)E_\pi\psi)\|\cdot\|A^\pm\|\,.\label{d-norm-1}
  \end{align}
  To proceed to the estimate \bref{d-norm}, let $\chi_-:=(1-\Gamma)E_\varphi\psi$, $\chi_+ := (1+\Gamma)E_\pi\psi$. As $(\varphi(\chi_-)+i\pi(\chi_+))^* = -(\varphi(\chi_-)+i\pi(\chi_+))$ and $\varphi(\chi_-)$ anticommutes with $\pi(\chi_+)$,
  \begin{align*}
    \|\varphi(\chi_-)+i\pi(\chi_+)\| = \|\varphi(\chi_-)^2 -
    \pi(\chi_+)^2\|^{1/2}
    = \left(\|\chi_-\|^2 + \|\chi_+\|^2\right)^{1/2}.
  \end{align*}
Together with \bref{d-norm-1} this implies the claimed norm bound \bref{d-norm} for $\delta^+_\psi$. To establish the corresponding inequality \bref{d-norm2} for $\delta^-_\psi$, consider the vector
\begin{align}
    \psi'':=\left(\tfrac{1}{2}(1-\Gamma)(1-E_\pi)+\tfrac{1}{2}(1+\Gamma)(1-E_\varphi)\right)\psi\,,
 \end{align}
  which is contained in $i\K'$ for any $\psi\in\Hil_1$. The norms of $\delta^-_\psi(A^\pm) = \delta^-_{\psi-\psi''}(A^\pm)$
  can then be estimated along the same lines as before.
\end{proof}

After these preparations, we now turn to the proof of the nuclearity of $\Xi$ by estimating the size of its image in $\Hil$. Let $\psi_1,...,\psi_n\in\Hil_1 \cap \dom(X)$ and $A\in\pol_\K$. In view of the second quantization structure of $X$ and the annihilation property of $a(\psi_j)$, we have
\begin{align}\label{f1}
\langle a^*(\Gamma\psi_1)\cdots a^*(\Gamma\psi_n)\Omega,&XA^\pm\Omega\rangle =
  \langle \Omega,\, a(X\Gamma\psi_n)\cdots a(X\Gamma\psi_1)A^\pm\Omega\rangle
  \\
  & = 
  \langle \Om,\,[a(X\Gamma\psi_n),[\,...\,[a(X\Gamma\psi_2),[a(X\Gamma\psi_1),A^\pm]_\mp]_\pm\,...\,]_\pm]_\mp\Om
  \rangle
  \,.\nonumber
\end{align}
From the inside to the outside, commutators and anticommutators are applied alternatingly. We start with a commutator
$[a(X\Gamma\psi_1),A^+]_-$ if $A=A^+$ is even and with an anticommutator $[a(X\Gamma\psi_1),A^-]_+$ if $A=A^-$ is odd. Writing the annihilation operator as a linear combination of the auxiliary fields \bref{a} and recalling that $X$ commutes with $\Gamma$, one notes that the innermost (anti-) commutator is
\begin{align}
[a(X\Gamma\psi_1),A^\pm]_\mp = \frac{1}{2}(\delta_{X\psi_1}^+ + \delta_{X\psi_1}^-)(A^\pm)\,.
\end{align}
Making use of this equality for all of the $n$ (anti-) commutators, it becomes apparent
that \bref{f1} can be rewritten as
\begin{align}\label{d-eqn}
\langle a^*(\Gamma\psi_1)\cdots
a^*(\Gamma\psi_n)\Omega,XA^\pm\Omega\rangle =
2^{-n}\,\langle\Omega, ((\delta^+_{X\psi_n}+\delta^-_{X\psi_n})\cdots(\delta^+_{X\psi_1}+\delta^-_{X\psi_1})(A^\pm))\Omega\rangle\,.
\end{align}
We now turn to the operators
\begin{align}
  T_\varphi := E_\varphi X\,, \qquad\qquad T_\pi := E_\pi X\,,
\end{align}
which are of  trace class on $\Hil_1$ according to the assumptions of Proposition \ref{prop:-1nuc}. Taking into account that
$\delta^\pm_{X\psi_j}$ are odd derivations on $\pol_\K$, an
application of the bounds (\ref{d-norm}, \ref{d-norm2}) to \bref{d-eqn} yields
\begin{align*}
  |\langle a^*(\Gamma\psi_1)\cdots
  a^*(\Gamma\psi_n)\Omega,XA^\pm\Omega\rangle|
  \leq&
  2^{-n}\prod_{j=1}^n 
  \big(\!
    \left(
      \|(1 - \Gamma)T_\varphi\psi_j\|^2 + \|(1 +\Gamma)T_\pi \psi_j\|^2
    \right)^\frac{1}{2}\\
    &+
    \left(
      \|(1+ \Gamma)T_\varphi \psi_j\|^2 + \|(1 -\Gamma)T_\pi\psi_j\|^2
    \right)^\frac{1}{2}
  \big)\cdot\|A^\pm\|\,.
\end{align*}
Following \cite{BuWi,BuJa} we now consider the positive operator
\begin{align}
  T := \left(|T_\varphi|^2 + |T_\pi|^2 \right)^{1/2}
\end{align}
which is in the trace class, too, satisfies $\|T\|_1 \leq \|T_\varphi\|_1 + \|T_\pi\|_1$ \cite{kosaki} and commutes with $\Gamma$ since $T_\varphi$ and $T_\pi$ do. As $T^2 \geq |T_\varphi|^2$, $T^2\geq |T_\pi|^2$,
\begin{align*}
  \|\tfrac{1}{2}(1 \mp \Gamma)T_\varphi\psi_j\|^2 + \|\tfrac{1}{2}(1
  \pm\Gamma)T_\pi \psi_j\|^2 \leq  \|\tfrac{1}{2}(1 \mp
  \Gamma)T\psi_j\|^2 + \|\tfrac{1}{2}(1  \pm\Gamma)T \psi_j\|^2 = \|T\psi_j\|^2\,.
\end{align*}
In terms of $T$, we thus arrive at the estimate
\begin{align}\label{est-T}
  |\langle a^*(\Gamma\psi_1)\cdots a^*(\Gamma\psi_n)\Omega, XA^\pm\Omega\rangle|
  \leq 2^n\,\|A^\pm\| \cdot\prod_{j=1}^n \|T\psi_j\|\,.
\end{align}
Although this bound was derived for $\psi_1,...,\psi_n\in\Hil_1\cap\dom(X)$ only, it holds for arbitrary $\psi_1,...,\psi_n\in\Hil_1$ since $\K\cap\dom(X)\subset\K$ is dense and the left- and right hand sides of \bref{est-T} are continuous in the $\psi_j$. Moreover, it can be extended to $A\in\A_\K$ as follows.

The polynomial algebra $\pol_\K$ is a weakly dense subalgebra of $\A_\K$, and the grading automorphism $\gamma$ extends to $\A_\K$. In view of Kaplansky's density theorem \cite{Takesaki1}, given $A^\pm=\frac{1}{2}(1\pm\gamma(A))\in\A_\K$, we can find a sequence $\{A_k^\pm\}\subset\pol_\K$ such that $A_k^\pm\Om\to A^\pm\Om$ weakly and $\|A_k^\pm\|\leq\|A^\pm\|$. This implies that \bref{est-T} holds for $A\in\A_\K$ also, and we are now able to give a bound on the nuclear norm of $\Xi$ \bref{def:xi-abstract}.
 
The positive trace class operator $T$ acts on $\xi\in\Hil_1$ as $T\xi = \sum_{k=1}^\infty t_k\langle \psi_k,\xi\rangle \psi_k$, where $\psi_k\,,k\in\N$, is an orthonormal basis of $\Hil_1$ and $t_k$ the (positive) eigenvalues of $T$, repeated according
to multiplicity, i.e. $\sum_{k=1}^\infty t_k = \|T\|_1 < \infty$. Moreover, since $\Gamma$ and $T$ commute, we may choose the
basis vectors $\psi_k$ to be eigenvectors of $\Gamma$ as well. As a consequence of the Pauli principle, the vectors
\begin{align}\label{basis-n}
  \Psi_\sbk := a^*(\Gamma \psi_{k_1})\cdots a^*(\Gamma \psi_{k_n})\Omega = \pm\, a^*(\psi_{k_1})\cdots a^*(\psi_{k_n})\Omega\,,
\end{align}
form an orthonormal basis of the totally antisymmetric subspace of $\Hil_1\tp{n}$ (the fermi\-onic $n$-particle space $\Hil_n$ in the terminology of chapter \ref{chapter:wedgenet}) if the multi-index $\bk:=(k_1,...,k_n)$ varies over $k_1<k_2<...<k_n$, $k_1,...,k_n\in\N$.

Note that $XA\Om$ has even (odd) particle number if $A\in\A_\K$ is even (odd). By the Fock structure of $\Hil$, we have for each $\Xi(A) = XA\Om$, $A\in\A_\K$, the decomposition  
\begin{align}
  \Xi(A) = \sum_{n=0}^\infty \sum_{k_1<...<k_{2n}} \langle
    \Psi_\sbk,XA^+\Omega\rangle\cdot \Psi_\sbk
     + \sum_{n=0}^\infty \sum_{k_1<...<k_{2n+1}} 
     \langle \Psi_\sbk,XA^-\Omega\rangle\cdot \Psi_\sbk\,,
\end{align}
as an example for a representation of the type \bref{nuc-decc} of $\Xi$. As $\|\Psi_\sbk\|=1$ for all $k_1,...,k_n\in\N$, and $\|A^\pm\|\leq\|A\|$, the sum of the expansion coefficients can be estimated with the help of \bref{est-T} as follows:
\begin{align}
\sum_{n=0}^\infty& \bigg(
  \sum_{1\leq k_1<...<k_{2n}} |\langle \Psi_\sbk,XA^+\Omega\rangle| 
  +
  \sum_{1\leq k_1<...<k_{2n+1}} |\langle \Psi_\sbk,\,XA^-\Omega\rangle|\bigg)
    \nonumber\\
&\leq \|A^+\|\sum_{n=0}^\infty 2^{2n}\!\!
  \sum_{1\leq k_1<...<k_{2n}}\prod_{j=1}^{2n} \|T \psi_{k_j}\|
  +
  \|A^-\|\sum_{n=0}^\infty 2^{2n+1}\!\! \sum_{1\leq k_1<...<k_{2n+1}} \prod_{j=1}^{2n+1}\|T \psi_{k_j}\|\nonumber\\
&\leq \|A\|\cdot \sum_{n=0}^\infty\, \sum_{1\leq k_1<...<k_n}
\prod_{j=1}^n \,2 t_{k_j}\label{part}\;.
\end{align}
According to \bref{def:nuclearnorm}, the sum \bref{part} provides an upper bound for the nuclear norm of $\Xi$. To compute this sum, note that \bref{part} is nothing else but the partition function of the ideal Fermi gas with Hamiltonian $e^{-\beta H}=2T$ and zero chemical potential in the grand canonical ensemble. This leads to the estimate (cf., for example, \cite{BraRob2})
\begin{align}
  \|\Xi\|_1 \leq \sum_{n=0}^\infty\, \sum_{1\leq k_1<...<k_n}
\prod_{j=1}^n \,2 t_{k_j} = \prod_{j=1}^\infty (1+2t_j) = \det(1+2T)\,.
\end{align}
As $\det(1+2T)\leq \exp(2\|T\|_1) < \infty$, the nuclearity of $\Xi$ follows. Taking into account 
\begin{align}
  \|T\|_1 \leq \|T_\varphi\|_1 + \|T_\pi\|_1 = \|E_\varphi X\|_1 +
  \|E_\pi X\|_1,
\end{align}
we also obtain the bound \bref{xi-bound}, finishing the proof of Proposition \ref{prop:-1nuc}. {\hfill $\square$ \\[2mm] \indent}

\subsection{Application to the Scaling Ising Model}
Our main interest in Proposition \ref{prop:-1nuc} derives from the fact that this result can be used to prove the modular nuclearity condition for wedges in the model given by the scattering function $S_2=-1$. This quantum field theory is related to the two-dimensional Ising model, which is a model of $\Zl_2$-spins on a two-dimensional lattice with nearest neighbor interaction. For detailed information on the Ising model and a guide to the literature, see the textbook \cite{ising2d}. We only note here that the Ising model is known to undergo a second order phase transition at some critical temperature.

Contact to quantum field theory is made by interpreting the $n$-spin correlation functions, evaluated in a suitable scaling limit, as $n$-point Schwinger functions of a field theory. Explicit expressions for these functions have been found in \cite{CTW, BCTW}. The field theory obtained in the scaling limit can be described by a Majorana Fermion, and is known to have a factorizing S-matrix, which is given by the scattering function $S_2=-1$ \cite{BKW, SMJ}.

This field theory has also been studied in the framework of the form factor program \cite{BKW, YuroZam, Babujian:2003sc}. Starting from the scattering function $S_2=-1$, the form factors and Greens functions of this model have been calculated in \cite{BKW}, and the results are in agreement with the findings in the Ising model. Thus this model is one of the few examples where the form factor program can be carried out to the end, yielding expressions for $n$-point functions \cite{Babujian:2003sc}.

So as the scattering function $S_2=+1$, also $S_2=-1$ gives rise to a well-studied model theory. But despite the detailed information one has about the structure of this theory, there seems to exist no proof of the Wightman axioms for the emerging family of $n$-point functions. Thus an inspection from the algebraic point of view taken in our approach seems warranted, as it establishes this model in a rigorous manner.

It is interesting to note that the model with scattering function $S_2=-1$ can also be formulated in higher dimensions, in contrast to the models with generic $S_2\in\SF$. To do so, one considers a Klein-Gordon field on $d$-dimensional Minkowski space, $d\geq 2$, whose positive and negative frequency parts obey canonical equal time anticommutation relations. This model was studied by Jost in \cite{Jost} as an example of a weakly local but nonlocal quantum field. More recently, Buchholz and Summers reexamined the model in an algebraic setting \cite{BuSu-1}. Defining the net of wedge algebras and double cone algebras as in chapter \ref{chapter:netsin2d}, they proved that the vacuum vector is cyclic for intersection of opposite wedge algebras by different methods than the ones presented in chapter \ref{chapter:netsin2d}.
\\
\\
We now turn to the proof of the modular nuclearity condition for wedge algebras in this model. For this proof, we use Proposition \ref{prop:-1nuc}, and are therefore obliged to explain how the mathematical framework used in the preceding section translates to the more concrete setting of the model theory.

The one particle space is $\Hil_1=L^2(\Rl,d\te)$ and the full Hilbert space $\Hil$ is the Fermionic Fock space over $\Hil_1$, in agreement with the general construction \bref{def:Hil}. The representation of the CAR algebra is related to the Zamolodchikov-Faddeev algebra with scattering function $S_2=-1$ by
\begin{align}
\zd(\psi)=a^*(\psi)\,,\quad z(\psi)=a(\overline{\psi})\,,\qquad\psi\in\Hil_1\,.
\end{align}
The involution $\Gamma$ on $\Hil_1$ is given by the antiunitary representing time reflection, $(\Gamma\psi)(\te)=\overline{\psi(-\te)}$, and the two closed subspaces $\LL_\varphi, \LL_\pi \subset\Hil_1$ are defined similarly to \bref{def:Lphipi}. But in order to use the field $\phi$ instead of $\phi'$, we here consider observables localized in the left wedge and thus put
\begin{align}
\LL_\varphi := \LL_\varphi(W_L)  &:= \{\fhat\,:\,f\in\Ss(\Rl_-)\}^-
\,,\\
\LL_\pi := \LL_\pi(W_L)  &:= \{\omega\fhat\,:\,f\in\Ss(\Rl_-)\}^-
\,,
\end{align}
and write $E_\varphi(W_L)=E_\varphi$, $E_\pi(W_L)=E_\pi$ for the associated orthogonal projections.
With these assignments, $\K$ \bref{def:K} is the eigenspace of the Tomita operator $J\Delta^{-1/2}$ of $(\A(W_L),\Om)$, with eigenvalue one.

In the context of the family of models with a factorizing S-matrix, the field operator $\phi(f)$ \bref{def:phi} is defined as an operator-valued distribution on two-dimensional Minkowski space, whereas here $\phi[\psi]$ \bref{def:phi-eckig} takes one-particle vectors $\psi\in\Hil_1$ as its arguments. As the vacuum is separating for the field $\phi(f)$, it can be uniquely described by the one-particle vector $\phi(f)\Om=f^+$, and the correspondence between the two formulations of $\phi$ is given by $\phi(f)=\phi[f^+]$, $f\in\Ss(W_L)$ real. Namely, for real $f\in\Ss(W_L)$ we have $f^+\in\K$ and thus
\begin{align}
	f^+(\te)	= 	(J\Delta^{-1/2}f^+)(\te)	&=	\overline{f^+(\te+i\pi)}	=	\overline{f^-(\te)}\,.
\end{align}
Hence
\begin{align*}
	\phi(f)	&=	\zd(f^+)	+	z(f^-)
			=	a^*(f^+)	+	a(\overline{f^-})
			=	a^*(f^+)	+	a(f^+)
			=	\phi[f^+]\,.
\end{align*}
Therefore the algebra $\A_\K$ \bref{def:AK} is 
\begin{align*}
	\A_\K		&=	\{\phi[\psi] \,:\, \psi\in\K\}''
			=	\{\phi(f) \,:\, f\in\Ss(W_L)\, {\rm real}\}''\,.
\end{align*}
But the field $\phi(f)$ is bounded, and so $\A_\K$ can equivalently be written as \cite{KadRin1}
\begin{align}
	\A_\K		&=	\{e^{i\phi(f)} \,:\, f\in\Ss(W_L)\, {\rm real}\}''
			=	\A(W_L)\,,
\end{align}
coinciding with the definition of the algebra of the left wedge made in chapter \ref{chapter:wedgenet} \bref{def:AWL}.

Finally, $X:=U(\bs)\Delta^{-1/4}U(-\bs)$ is strictly positive, contains $\A(W_L)\Om$ in its domain for $s>0$, and has the required second quantization structure. Moreover, the involution $\Gamma$ commutes with $U(\pm \bs)$ and $\Delta^{-1/4}$, and hence with $X$.

The map \bref{def:xi-abstract} is thus
\begin{align}
	\Xi:\A(W_L)\longrightarrow\Hil,\qquad A\longmapsto U(\bs)\Delta^{-1/4}U(-\bs)A\Om\,,
\end{align}
and is nuclear if and only if $\Xi(s)$ \bref{Xis} is, with the same nuclear norm. ($\Xi$ and $\Xi(s)$ are related by the antiunitary modular involution $J$.) So the result of Proposition \ref{prop:-1nuc} is the desired modular nuclearity condition for wedges in the model with scattering function $S_2=-1$.

To conclude the proof of this property, we only need to establish that $E_\varphi(W_L) U(\bs)\Delta^{-1/4}U(-\bs)$ and $E_\pi(W_L) U(\bs)\Delta^{-1/4}U(-\bs)$ are trace class operators on $\Hil_1$. But the projections $E_\varphi(W_L)$, $E_\pi(W_L)$ are related to $E_\varphi(W_R)$, $E_\pi(W_R)$ by the adjoint action of $J$, and so we can apply the results obtained in the analysis of the free field in the previous section to arrive at
\begin{align}
 E_\varphi(W_L) U(\bs)\Delta^{-1/4}U(-\bs)	&=	JE_\varphi(W_R)U(-\bs)\Delta^{1/4}U(\bs)J	=J T_\varphi(s)^* U(\bs)J\,,\\
 E_\pi(W_L) U(\bs)\Delta^{-1/4}U(-\bs)	&=	JE_\pi(W_R)U(-\bs)\Delta^{1/4}U(\bs)J	=J T_\pi(s)^* U(\bs)J\,,
\end{align}
where $T_\varphi(s)$ and $T_\pi(s)$ are the integral operators with kernels \bref{Tphikernel} and \bref{Tpikernel}. As these have been shown to be nuclear, and $U(\bs)$ and $J$ are (anti-) unitary, we obtain
\begin{align}
 \|E_\varphi(W_L) U(\bs)\Delta^{-1/4}U(-\bs)\|_1=\|T_\varphi(s)\|_1<\infty
\,,\\
\|E_\pi(W_L) U(\bs)\Delta^{-1/4}U(-\bs)\|_1=\|T_\pi(s)\|_1<\infty
\,,
\end{align}
 and see that the assumptions of Proposition \ref{prop:-1nuc} are satisfied in this model.
\begin{proposition}
Consider the model theory with scattering function $S_2=-1$.\\
The maps $\Xi(s)$ \bref{Xis} are nuclear for any $s>0$, and there holds the bound
\begin{align} \label{xi-1bound}
\|\Xi(s)\|_1	&\leq		e^{2\|T_\varphi(s)\|_1}\cdot e^{2\|T_\pi(s)\|_1}\,,
\end{align}
where $T_\varphi(s)$ and $T_\pi(s)$ are the trace class integral operators on $L^2(\Rl)$ given by the kernels \bref{Tphikernel} and \bref{Tpikernel}.{\hfill $\square$}
\end{proposition}
According to Theorem \ref{thm:Ch2summary}, this implies in particular that all double cone algebras are isomorphic to the hyperfinite type III$_1$ factor, and that the Reeh-Schlieder property holds in this model.

In comparison to the estimate \bref{xi-bound--} obtained in chapter \ref{chapter:nuclearity}, we note that the bound \bref{xi-1bound} is much smaller. As mentioned before, we conjecture that by a more refined analysis, an exponential bound similar to \bref{xi-1bound} can be shown to hold for generic scattering functions, but this remains to be proven.

\subsection{Application to the Thermodynamics of Relativistic Free Fermions}
The original work of Buchholz and Wichmann \cite{BuWi} applies to Bose fields only, but in view of the thermodynamical interpretation of the energy nuclearity condition, it can be expected to hold for Fermi fields as well. For the case of a free Dirac field on a globally hyperbolic spacetime, such a proof has been given by D'Antoni and Hollands in \cite{stephan-nuclear}.
\\
\\
Without going into any details, we briefly mention how Proposition \ref{prop:-1nuc} can be used to prove the energy nuclearity condition for free Fermi fields on $d$-dimensional Minkowski space ($d\geq 2)$.

Consider a net $\OO\lmto\A(\OO)$ on the antisymmetric Fock space $\Hil$ over $\Hil_1$ which is generated by a free quantum field satisfying canonical equal time anticommutation relations. Fixing a double cone $\OO_r$ of radius $r$, the corresponding algebra $\A(\OO_r)$ can be formulated in terms of subspaces $\LL_\varphi(r), \LL_\pi(r)\subset\Hil_1$ as in \bref{def:K}. Putting $X:=e^{-\beta H}$, the map $\Xi$ \bref{def:xi-abstract} coincides with
\begin{align}
\Theta_{\beta,r}	:	\A(\OO_r) \lto	\Hil,\qquad A\lmto e^{-\beta H}A\Om\,.
\end{align}
By Prop. \ref{prop:-1nuc}, the nuclearity of $\Theta_{\beta,r}$ can be derived from the nuclearity of $e^{-\beta H}E_{\varphi\slash\pi}(r)$, where $E_{\varphi\slash\pi}(r)\in\B(\Hil_1)$ are the orthogonal projections onto the subspaces $\LL_{\varphi\slash\pi}(r)\subset\Hil_1$.

Realizing the single particle space as a direct sum of $L^2$-spaces (depending on the spin of the underlying field), the operators $e^{-\beta H}E_{\varphi\slash\pi}(r)$ can be shown to be of trace class. Hence $\Theta_{\beta,r}$ is nuclear, and as usual, the nuclear norm $\|\Theta_{\beta,r}\|_1$ is interpreted as the thermodynamical partition function of the system restricted to the region $\OO_r$, at temperature $\beta^{-1}$.

For explicit estimates on the trace norms $\|e^{-\beta H}E_{\varphi\slash\pi}(r)\|_1$in terms of $\beta$ and $r$, we refer to \cite{BuWi}.

\chapter{Technical Proofs}\label{app:zamo-calcs}
\section{The General Form of a Scattering Function}\label{sec:formofs2}

In this section we give an explicit characterization of the family $\SF$ of scattering functions (Definition \ref{def:s2}) in terms of their zeros by proving Proposition \ref{prop:s2}, which is repeated below. A similar computation has been carried out by Mitra \cite{mitra}.
\\\\
{\bf Proposition 3.2.2} {\em The set $\SF$ of scattering functions is
\begin{align}\label{s2-rep-app}
      \SF &=
      \left\{
        \zeta\lmto  \eps
      \cdot
      e^{ia\sinh\zeta}
      \cdot
      \prod_{k}
      \frac{\sinh\beta_k-\sinh\zeta}{\sinh\beta_k+\sinh\zeta}
      \;:\;
      \eps=\pm1,\,a\geq 0,\;\{\beta_k\}\in\ZZZ
      \right\}\,,
    \end{align}
    where the family $\ZZZ$ consists of the finite or infinite sequences $\{\beta_k\}\subset\Cl$ satisfying the following conditions:
\begin{itemize}
\item[i)] $0<\mathrm{Im}\beta_k\leq\frac{\pi}{2}$,
\item[ii)] $\beta_k$ and $-\overline{\beta_k}$ appear the same finite number of times in the sequence $\{\beta_k\}$,
\item[iii)] $\{\beta_k\}$ has no finite limit point,
\item[iv)] $\sum_k \mathrm{Im}\frac{1}{\sinh\beta_k}<\infty$.
\end{itemize}
    The product in \bref{s2-rep-app} converges absolutely and uniformly in $\zeta$ on compact
    subsets of the strip $S(0,\pi)$.
}
\begin{proof}
We first show that any $S_2\in\SF$ has a representation of the form \bref{s2-rep}. To this end, let $\eps:=S_2(0)$, and define $\{\beta_k\}$ as the sequence of zeros of $S_2$ in $\overline{S(0,\frac{\pi}{2})}$, repeated according to multiplicity. The constraining equations for $S_2$ \bref{s2-rel},
\begin{align}\label{s2-rel-app}
\overline{S_2(\te)}
    \;=\;
    S_2(\te)^{-1}
    \;=\;
    S_2(-\te)
    \;=\;
    S_2(\te+i\pi)
    \,,\qquad
    \te\in\Rl\,,
\end{align}
imply $\eps=\pm 1$. Since $S_2$ has modulus unity on the real line, and $\overline{S_2(-\overline{\zeta})}=S_2(\zeta)$ holds for $\zeta\in \overline{S(0,\pi)}$ by analytic continuation, the sequence $\{\beta_k\}$ has the properties $i)$ and $ii)$.

In the derivation of \bref{s2-rep-app}, we may restrict to scattering functions without purely imaginary zeros. For if $S_2$ vanishes at $i\alpha_1,...,i\alpha_K$, $0<\alpha_k\leq\frac{\pi}{2}$  (There can be only finitely many such points in view of the analyticity and continuity of $S_2$ and the fact that $|S_2(0)|=1$), consider
\begin{align}
  R(\zeta)
  &:=
  \prod_{k=1}^K
  \frac{\sinh(i\alpha_k) - \sinh\zeta}{\sinh(i\alpha_k)+\sinh\zeta}
  \;.
\end{align}
It is readily verified that $R\in\SF$, and its only zeros in $\overline{S(0,\frac{\pi}{2})}$ are $i\alpha_1,...,i\alpha_K$. But as the equations \bref{s2-rel-app} are stable under taking products and reciprocals, it follows that $S_2(\zeta)/R(\zeta)$ is a scattering function as well, without imaginary zeros. This implies that formula \bref{s2-rep-app} holds for scattering functions with imaginary zeros if it can be proven for those without.

The not purely imaginary zeros a scattering function has in $S(0,\frac{\pi}{2})$ occur in pairs of the form $(\beta_k,-\overline{\beta_k})$. We denote the zeros with positive real part by $\gamma_k$, i.e. $\{\beta_k\,:\,k\in\N\}=\{\gamma_k,-\overline{\gamma_k}\,:\,k\in\N\}$, and assume that there are infinitely many of them. (The case of finitely many zeros leads only to simplifications in the proof.)
\\
\\
The hyperbolic sine is a biholomorphic map between the strip $S(0,\frac{\pi}{2})$ and the upper half plane with a cut along $i\,[1,\infty)$. The points $\frac{i\pi}{2}\pm\te$, $\te\in\Rl$, are mapped onto opposite sides of the cut. Since \bref{s2-rel-app} implies $S_2(\frac{i\pi}{2}+\te)=S_2(\frac{i\pi}{2}-\te)$, the function $\hat{S}_2(z):=S_2(\mathrm{arsinh}(z))$ is analytic in the upper half plane and bounded and continuous on its closure. The zeros of $\hat{S}_2$ are precisely $g_k := \sinh\gamma_k$ and $-\overline{g_k}=\sinh(-\overline{\gamma_k})$, $k\in\N$. Taking into account $|S_2(\te)|=|S_2(\te+i\pi)|=1$, $\te\in\Rl$, we conclude $|\hat{S}_2(z)|\leq 1$ for Im$\,z\geq 0$ from the three lines theorem \cite[Thm. 3.7]{conway}.

In the context of Hardy spaces, it is well-known (cf., for example \cite[Thm. 11.3]{duren}) that a function like $\hat{S}_2$ admits a factorization of the form $\hat{S}_2(z) = H(z)B(z)$, where $H$ is an analytic and nonvanishing function bounded by unity on the upper half plane, and $B$ is the Blaschke product with zeros $\{g_k,-\overline{g_k}\}$, 
\begin{align}\label{blaschke}
  B(z)
  =
  \prod_{k=1}^\infty
  \left(
    \frac{|g_k^2+1|}{g_k^2+1}\frac{z-g_k}{z-\overline{g_k}}
    \cdot
    \frac{|\overline{g_k}^2+1|}{\overline{g_k}^2+1}\frac{z+\overline{g_k}}{z+g_k}
  \right)
  =
  \prod_{k=1}^\infty
  \left(
    \frac{g_k-z}{g_k+z}\cdot\frac{-\overline{g_k}-z}{-\overline{g_k}+z}
  \right)
  \,.
\end{align}
The boundedness of $\hat{S}_2$ implies the Blaschke condition for the upper half plane, $\sum_k\mathrm{Im}\sinh\beta_k/(1+|\sinh\beta_k|^2)<\infty$ \cite{duren}. This condition is equivalent to the convergence (absolute and uniform as $z$ varies over compact subsets of the upper halfplane) of the product \bref{blaschke}. In view of the continuity of $\hat{S}_2$ on the real axis, and $|S_2(\te)|=1$, $\te\in\Rl$, it also follows that $\{\beta_k\}$ has no finite limit point (that is, property $iii)$ of $\{\beta_k\}$ holds), and the Blaschke condition simplifies to property $iv)$. In particular, $B$ extends continuously to the real axis \cite{garnett}.

As $H$ is nonvanishing, we find a function $h$ analytic in the upper halfplane such that $H(z)=\eps\, e^{ih(z)}$. Since $\hat{S}_2(0)=\eps$ and $B(0)=1$, we may choose $h(0)=0$ to fix $h$ uniquely. Moreover, using the factorization theorem for bounded analytic functions on the upper half plane and the fact that $\hat{S}_2$ has modulus one on the real line, we conclude that there exists a constant $a\geq 0$ such that Im$\,h(z)=a\cdot\mathrm{Im}\,z$ \cite[Thm. 6.5.4]{boas}. Taking into account the analyticity of $h$, we arrive at $h(z)=a\cdot z + h(0) = a\cdot z$.

Mapping back to the strip yields the claimed expression
\begin{align}\label{s-rep-1}
  S_2(\zeta) 
  =
H(\sinh\zeta) B(\sinh\zeta)
=
  \eps\,e^{ia\sinh\zeta}\cdot
  \prod_{k=1}^\infty
  \frac{\sinh\beta_k-\sinh\zeta}{\sinh\beta_k+\sinh\zeta}
  \;,\qquad \zeta\in S(0,\tfrac{\pi}{2})\,.
\end{align}
By reflection about $\Rl+\frac{i\pi}{2}$ (recall $S_2(i\pi-\zeta)=S_2(\zeta)$), this formula extends to $S(0,\pi)$.
\\
\\
To establish the converse direction, assume $a\geq 0$, $\eps=\pm 1$ and a sequence $\{\beta_k\}$ satisfying $i)$-$iv)$ are given. The product \bref{s-rep-1} converges to a bounded analytic function due to the Blaschke condition $iv)$, and the absence of finite limit points $iii)$ of $\{\beta_k\}$ implies continuous boundary values. Taking into account $a\geq 0$ and $ii)$, the boundedness and the relations \bref{s2-rel} can be checked by direct computation.
\end{proof}

\newpage

\section{The Family of Integral Operators $T_{s,\kappa}$}\label{app:intop}

In this section we consider the integral operators $T_{a,b}$ on $L^2(\Rl)$, given by the kernels
\begin{align}\label{def:Tab}
T_{a,b}(x,y)	&=	\frac{e^{-a\cosh x}}{x-y+ib}
\,,\qquad
a>0,\;b\in\Rl\backslash\{0\}\,,
\end{align}
which play a role in the proof of the modular nuclearity condition in chapter \ref{chapter:nuclearity}. The result we want to establish is
\begin{lemma}\label{intop-lemma}
The integral operators $T_{a,b}$ \bref{def:Tab} are trace class operators for any $a>0$, $b\in\Rl\backslash\{0\}$. Their trace norms are bounded by
\begin{align}
\|T_{a,b}\|_1
\leq
2^{1/4}\pi^{3/4}
\cdot
\frac{e^{-a}}{a^{1/4}}
\left(\sqrt{\frac{\pi}{2}}+\frac{1}{4a}\right)^{1/2}
\cdot
\left(\frac{b^4+4b^2+24}{b^5}\right)^{1/2}
\,.
\end{align}
\end{lemma}
\begin{proof}
By a Fourier transformation of the kernel $T_{a,b}(x,y)$ in $y$, one notes that $T_{a,b}$ can be expressed in terms of the quantum mechanical position and momentum operators $(X\psi)(x)=x\cdot\psi(x)$, $(P\psi)(x)=-\psi'(x)$, as
\begin{align}
T_{a,b}	&=	-2\pi \eps(b) i\, e^{-a\cosh X} \Theta(\eps(b)P)e^{-bP}\,,
\end{align}
where $\eps(b)=b/|b|$ is the sign of $b$ and $\Theta=\chi_{\Rl_+}$ the Heaviside step function. We write $f_a(x):=e^{-a\cosh x}$ and $g_b(p)= -2\pi \eps(b) i \, \Theta(\eps(b)p)e^{-bp}$, and decompose $T_{a,b}$ according to 
\begin{align}
T_{a,b}
&= R_a\cdot S_b\,,\\
R_a
&:= f_a(X)(i+P)^{-2}(i+X)\,,\qquad S_b:= (i+X)^{-1}(i+P)^2 g_b(P)\,.
\end{align}
Thus $R_a$ and $S_b$ are integral operators with the kernels
\begin{align}
R_a(x,y)		&=	\frac{1}{\sqrt{2\pi}}\,e^{-a\cosh x} \frac{i+y}{(i+(x-y))^2}
\\
S_b(x,y)		&=	-\sqrt{2\pi} \eps(b) i\,\frac{(i+x-y)^2}{i+x}  \Theta(\eps(b)(x-y))e^{-b(x-y)}
\end{align}
By inspection of these kernels, we find that $R_a$ and $S_b$ are Hilbert Schmidt operators, and hence $T_{a,b}$ is of trace class. Their Hilbert Schmidt norms can be estimated as
\begin{align}
\|R_a\|_2 		&\leq \left(\frac{\pi}{2a}\right)^{1/4}\,e^{-a}\,\left(\sqrt{\frac{\pi}{2}}+\frac{1}{4a}\right)^{1/2}\,,\\
\|S_b\|_2		&=\sqrt{2}\pi\sqrt{\frac{b^4+4b^2+24}{b^5}}\,,
\end{align}
and since $\|T_{a,b}\|_1 \leq \|R_a\|_2\|S_b\|_2$, this gives the claimed bound.
\end{proof}

%

\chapter{Mathematical Topics}\label{chapter:math}

\section{Nuclear Maps between Banach Spaces}\label{app:nuclearity}

The material in this subsection provides the mathematical background for the discussion of nuclearity criteria in quantum field theory, it is primarily used in chapter \ref{chapter:nuclearity}.  The basic notion is the following:
\begin{definition}\label{def:nucmap}
	Let $\X$ and $\Y$ be two Banach spaces. A linear map $T:\X\lto\Y$ is said to be nuclear if there exists a sequence of vectors $\{\Psi_n\}_n\subset\Y$ and a sequence of bounded linear functionals $\{\rho_n\}_n\subset\X_*$ such that
	\begin{align}\label{nuc-dec}
		T(\xi) &= \sum_{n=1}^\infty \rho_n(\xi)\,\Psi_n\,,\qquad \sum_{n=1}^\infty \|\rho_n\|_{\X_*} \|\Psi_n\|_\Y < \infty \,.
	\end{align}
	The nuclear norm of such a mapping is defined as
	\begin{align}
		\|T\|_1 &:= \inf_{\rho_n,\Psi_n} \sum_{n=1}^\infty \|\rho_n\|_{\X_*} \|\Psi_n\|_\Y \,,
	\end{align}
	where the infimum is taken over all sequences $\{\Psi_n\}_n\subset\Y$, $\{\rho_n\}_n\subset\X_*$  complying with the above conditions.
\end{definition}
The set of all nuclear maps between two Banach spaces $\X$, $\Y$ is denoted by $\NN(\X,\Y)$, the compact operators are $\K(\X,\Y)$, and the bounded operators $\B(\X,\Y)$. The operator norm of $\B(\X,\Y)$ is written as $\|\cdot\|$.

The following well-known Proposition summarizes some basic properties of nuclear maps. For the convenience of the reader, we give the proof here, see also \cite{pietsch, jarchow}.
\begin{proposition}\label{Lem:nuclearmaps--app}
	Let $\X,\X_1,\Y,\Y_1$ be Banach spaces.
	\begin{enumerate}
		\item $\|T\| \leq \|T\|_1$ for $T\in\NN(\X,\Y)$.
		\item $\NN(\X,\Y) \subset \K(\X,\Y)$.
		\item $(\NN(\X,\Y),\|\cdot\|_1)$ is a Banach space.
	\end{enumerate}
\end{proposition}
\begin{proof}
Let $T\in\NN(\X,\Y)$ be given by a representation of the form \bref{nuc-dec}, and $\xi\in\X$. Then
\begin{align*}
\|T\xi\|_\Y
\leq
\sum_{n=1}^\infty |\rho_n(\xi)|\,\|\Psi_n\|_\Y
\leq
\sum_{n=1}^\infty \|\rho_n\|_{\X^*}\,\|\Psi_n\|_\Y\cdot \|\xi\|_\X\,.
\end{align*}
Hence $T$ is bounded, with operator norm $\|T\|\leq\|T\|_1$, as claimed in a). Moreover, the operators $T_N:=\sum_{n=1}^N \rho_n(\cdot)\Psi_n$ have finite rank, and $\|T-T_N\|\leq\sum_{n=N+1}^\infty\|\rho_n\|_{\X^*}\|\Psi_n\|\to 0$ for $N\to\infty$. Thus $T$ is the $\|\cdot\|$-limit of a sequence of finite rank operators and hence compact, i.e. b) holds.

{\em c)} Let $T_1,T_2\in\NN(\X,\Y)$ be given by decompositions of the form \bref{nuc-dec}, with linear functionals $\rho_n^{(1)}$, $\rho_n^{(2)}\in\X^*$ and vectors $\Psi_n^{(1)}$, $\Psi_n^{(2)}\in\Y$. Then $T_1+T_2$ is, $\xi\in\X$,
\begin{align}\label{T-sum}
(T_1+T_2)\xi	&=	\sum_{n=1}^\infty \rho_n^{(1)}(\xi)\Psi_n^{(1)}
+
\sum_{n=1}^\infty \rho_n^{(2)}(\xi)\Psi_n^{(2)}, 
\end{align}
and thus also of the form \bref{nuc-dec}. As $\NN(\X,\Y)$ is clearly invariant under scalar multiplication, this shows that $\NN(\X,\Y)$ is a linear space. For the nuclear norm of $T_1+T_2$ we find 
\begin{align*}
\|T_1+T_2\|_1
\leq 
\sum_{n=1}^\infty \|\rho_n^{(1)}\|_{\X^*}\|\Psi_n^{(1)}\|_\X
+
\sum_{n=1}^\infty \|\rho_n^{(2)}\|_{\X^*}\|\Psi_n^{(2)}\|_\X \;, 
\end{align*}
and hence $\|T_1+T_2\|_1\leq\|T_1\|_1+\|T_2\|_1$. As $\|\la T\|_1 = |\la|\,\|T\|_1$, $\la\in\Cl$, holds trivially and $\|T\|_1=0 \Leftrightarrow T=0$, it follows that $\|\cdot\|_1$ is a norm on $\NN(\X,\Y)$.

To show completeness of $\NN(\X,\Y)$ with respect to this norm, we consider a $\|\cdot\|_1$-Cauchy sequence $T_k$ and pick numbers $\alpha_r\in\N$ such that
\begin{align*}
\|T_k-T_l\|_1 \leq 2^{-r}\,,\qquad k,l\geq \alpha_r\,.
\end{align*}
Since $T_{\alpha_{r+1}}-T_{\alpha_r} \in \NN(\X,\Y)$, there exist functionals $\{\rho_n^{(r)}\}_n\subset\X_*$ and vectors $\{\Psi_n^{(r)}\}_n\subset\Y$ such that
\begin{align*}
(T_{\alpha_{r+1}}-T_{\alpha_r})\xi
=
\sum_{n=1}^\infty \rho_n^{(r)}(\xi)\Psi_n^{(r)}\,,\qquad
\sum_{n=1}^\infty \|\rho_n^{(r)}\|_{\X^*}\|\Psi_n^{(r)}\|_\Y \leq 2^{-r}\,.
\end{align*}
Consequently, we have 
\begin{align*}
(T_{\alpha_{r+m}}-T_{\alpha_r})\xi
=
\sum_{s=r}^{r+m-1}\sum_{n=1}^\infty \rho_n^{(s)}(\xi)\Psi_n^{(s)}\,.
\end{align*}
Note that $T:=\lim\limits_{m\to\infty}T_{\alpha_{r+m}}$ exists as a bounded operator in view of {\em a)}. So by taking the limit $m\to\infty$ we obtain
\begin{align*}
(T-T_{\alpha_r})\xi
=
\sum_{s=r}^\infty\sum_{n=1}^\infty \rho_n^{(s)}(\xi)\Psi_n^{(s)}\,,
\end{align*}
and the nuclear norm of $T-T_{\alpha_r}$ can be estimated by 
\begin{align*}
\|T-T_{\alpha_r}\|_1
\leq
\sum_{s=r}^\infty 2^{-s}
=
2\cdot 2^{-r}\,.
\end{align*}
Hence $T$ is nuclear. Finally, 
\begin{align*}
\|T-T_k\|_1
\leq
\|T-T_{\alpha_r}\|_1+\|T_{\alpha_r}-T_k\|_1
\leq
2\cdot 2^{-r} + 2^{-r}\,,\qquad k\geq\alpha_r\,,
\end{align*}
shows $\|T-T_k\|_1\to 0$ for $k\to\infty$.
\end{proof}

Nuclear maps can be regarded as a generalization of the concept of trace class operators to Banach spaces. In the following Lemma, it is shown that in the case of Hilbert spaces, the nuclear operators form precisely the trace class. The familiar ideal properties of trace class operators are shown to hold also for nuclear maps.
\begin{lemma}
{\qquad}\\
\vspace*{-11mm}
\\
\begin{enumerate}
\item Let $T\in \NN(\X,\Y), A_1\in \B(\Y,\Y_1), A_2 \in\B(\X_1,\X)$. Then $A_1 T A_2 \in \NN(\X_1,\Y_1)$, and 
			\begin{align}
				\|A_1 T A_2\|_1 &\leq \|A_1\| \cdot \|T\|_1 \cdot \|A_2\|\,.
			\end{align}
		\item Let $\Hil$ be a separable Hilbert space. Then any trace class operator $T$ on $\Hil$ lies in $\NN(\Hil,\Hil)$ and satisfies $\|T\|_1\leq{\rm Tr}\,|T|$.
\end{enumerate}
\end{lemma}
\begin{proof}
{\em a)} Consider $T,A_1,A_2$ as above, where $T\in\NN(\X,\Y)$ is given by a decomposition of the form \bref{nuc-dec}. Then $A_1TA_2$ acts on $\xi_1\in\X_1$ according to
\begin{align*}
A_1TA_2\xi_1
&=
\sum_{n=1}^\infty \rho_n(A_2\xi_1) A_1\Psi_n\,,
\end{align*}
and its nuclear norm can be estimated by
\begin{align*}
\sum_{n=1}^\infty\|\rho_n\circ A_2\|_{\X_1^*} \|A_1\Psi_n\|_{\Y_1}
\leq
\|A_2\|_{\B(\X_1,\X)}\cdot \sum_{n=1}^\infty\|\rho_n\|_{\X^*} \|\Psi_n\|_{\Y}\cdot \|A_1\|_{\B(\Y,\Y_1)}<\infty\,.
\end{align*}
Hence $A_1TA_2\in\NN(\X_1,\Y_1)$, and varying the nuclear decomposition \bref{nuc-dec} of $T$ gives $\|A_1TA_2\|_1\leq\|A_1\|\|T\|_1\|A_2\|$.

{\em b)} Given a trace class operator $T\in\B(\Hil)$, there exist two orthonormal systems $\{\Psi_k\}_k$ and $\{\Phi_k\}_k$ in $\Hil$, and numbers $\la_k\geq 0$ (the singular values of $T$) such that
\begin{align}
T\xi=\sum_{k=1}^\infty \la_k\langle\Phi_k,\xi\rangle\Psi_k
\,,\qquad
{\rm Tr}|T|=\sum_{k=1}^\infty \la_k <\infty\,.
\end{align}
Clearly, this decomposition of $T$ is a decomposition of the form \bref{nuc-dec}, implying that $T$ is nuclear. As the vectors $\Psi_k$ and $\Phi_k$ have norm one, we also find $\|T\|_1\leq\sum_k\la_k={\rm Tr}|T|$.
\end{proof}


\newpage
\section{Hardy Spaces on Tube Domains}\label{app-hardy}

In the investigation of the analytic properties of wedge-local wavefunctions in section \ref{sec:analytic}, it turns out to be convenient to formulate some results in terms of Hardy spaces on certain tube domains. In this section, we collect the necessary definitions and results.
\begin{definition}
Let $\Cu\subset\Rl^n$ be an open, bounded, convex domain. The tube $\Tu_n(\Cu)$ over $\Cu$ is defined as 
\begin{align}
\Tu_n(\Cu)
&:=
\Rl^n + i\,\Cu
\;\subset\;
\Cl^n
\,.
\end{align}
\end{definition}
\noindent Note that as a convex tube, $\Tu_n(\Cu)$ is a domain of holomorphy in $\Cl^n$. The Hardy space $H^2(\Tu_n(\Cu))$ on the tube $\Tu_n(\Cu)$ is defined as follows.
\begin{definition}\label{def:hardy}
The Hardy space  $H^2(\Tu_n(\Cu))$ on the tube $\Tu_n(\Cu)$ consists of the functions $F:\Tu_n(\Cu)\lto\Cl$ having the following three properties.
\begin{itemize}
\item[i)] $F$ is analytic in $\Tu_n(\Cu)$.
\item[ii)] For each $\bla\in\Cu$, the function 
\begin{align}
F_\sbla	:	\Rl^n \lto \Cl\,,\qquad F_\sbla(\bte)	&:=	F(\bte+i\bla)
\end{align}
is an element of $L^2(\Rl^n,d^n\bte)$.
\item[iii)] Let $\|\cdot\|_2$ denote the usual $L^2$-norm. Then for $F\in H^2(\Tu_n(\Cu))$, 
\begin{align}
\bno{F}	&:= 	\sup_{\sbla\in\Cu}\|F_\sbla\|_2	< \infty\;.
\end{align}
\end{itemize}
\end{definition}
\noindent Clearly $H^2(\Tu_n(\Cu))$ is a linear space, and $\bno{\cdot}$ is a norm on it. Note that in view of the boundedness of $\Cu$, $H^2(\Tu_n(\Cu)) \subset L^2(\Tu_n(\Cu))$, where in $L^2(\Tu_n(\Cu))$ the tube $\Tu_n(\Cu)$ is regarded as a domain in $\Rl^{2n}$. Furthermore, $H^2(\Tu_n(\Cu))$ is complete with respect to the norm $\bno{\cdot}$, i.e. it is a Banach space. This and other results can be derived from the following inequality:
\begin{lemma}\label{lemma:h2-inequ}
Let $F\in H^2(\Tu_n(\Cu))$ and $\bte\in\Rl^n$, $\bla\in\Cu$. Then
\begin{align}
|F(\bte+i\bla)|
&\leq
\left(\frac{2}{\pi\,d_\infty(\bla)}\right)^{n/2}\cdot\bno{F}\,,
\end{align}
where $d_\infty(\bla)$ denotes the distance of $\bla$ to the boundary of $\Cu$, measured in maximum norm.
\end{lemma}
\begin{proof}
Let $\bte\in\Rl^n$, $\bla\in\Cu$ and consider the $n$-dimensional polydisc $\DD_n(\bze)$ with center $\bze:=\bte+i\bla$ and sufficiently small radius such that the tube $\Tu_n^\rho(\bze) := \bze+\Rl^n+i\,[-\rho,\rho]^{\times n}$ is contained in $\Tu_n(\Cu)$. (This is possible if and only if $\rho<d_\infty(\bla)$.) By the mean value property for analytic functions \cite{krantz} and the Cauchy-Schwarz inequality in $L^2(\Tu_n^\rho(\bze))$, we have
\begin{align*}
|F(\bze)|
&=
\frac{1}{(\pi\rho^2)^n}\left|\int_{\DD_n(\sbze)} d^n\bze' \, F(\bze') \right|
\\
&\leq
\frac{\sqrt{\vol(\DD_n(\bze))}}{(\pi\rho^2)^n} \left(\int_{\Tu_n^\rho(\sbze)} d^n\bze'\,  |F(\bze')|^2 \right)^{1/2}
\\
&=
(\pi\rho^2)^{-n/2} \left( \int_{[-\rho,\rho]^{\times n}} d^n\bla' \, \|F_{\sbla+\sbla'}\|_2^2 \right)^{1/2}
\\
&\leq
\left(\frac{2}{\pi \rho}\right)^{n/2}\cdot\bno{F}\,.
\end{align*}
In the limit $\rho\to d_\infty(\bla)$ we arrive at the claimed inequality.
\end{proof}
We now show the completeness of the Hardy space.
\begin{proposition}
$(H^2(\Tu_n(\Cu)),\bno{\cdot})$ is a Banach space.
\end{proposition}
\begin{proof}
Everything is clear except for completeness. So let $\{F_k\}_k\subset H^2(\Tu_n(\Cu))$ be a Cauchy sequence in the norm $\bno{\cdot}$. In view of $\|F_{k,\sbla}-F_{l,\sbla}\|_2 \leq \bno{F_k-F_l}\to 0$ for $k,l\to\infty$, $\bla\in\Cu$, we find that $F_{k,\sbla}$ converges in $L^2(\Rl^n)$ to a limit function $F_\sbla$. As this convergence is uniform in $\bla$, the function $F:\bte+i\bla\mapsto F_\sbla(\bte)$, $\bte\in\Rl^n$, satisfies $\bno{F}\leq\lim_k\bno{F_k}<\infty$, and we have $\lim_k\bno{F_k-F}=0$.

To show that $F$ is analytic, note that the previous Lemma implies that the convergence $F_k\to F$ is uniform on tubes of the form $\Rl^n+iK$, where $K\subset \Cu$ is compact. So $F$ is a normal limit \cite{krantz} of analytic functions and hence analytic, too.
\end{proof}
Besides its completeness, the following facts about $H^2(\Tu_n(\Cu))$ are needed in chapter \ref{chapter:nuclearity}.
\begin{proposition}
For $F\in H^2(\Tu_n(\Cu))$, the following holds.
\begin{itemize}
\item[i)]	Let $k\in\{1,...,n\}$, $\te_1,...,\te_{k-1},\te_{k+1},...,\te_n\in\Rl$, and $K\subset\Cu$ compact. Then
\begin{align}
\lim_{|\te_k|\to\infty} \sup_{\sbla\in K} |F(\bte+i\bla)|	&= 0\,.
\end{align}
\item[ii)] Assume $\Cu$ is a polyhedron, i.e. the convex closure of finitely many points in $\Rl^n$. Then $F$ can be extended to $\overline{\Tu_n(\Cu)}$ such that $F_\sbla\in L^2(\Rl^n)$ for $\bla\in\overline{\Cu}$, and $\overline{\Cu}\ni\bla \lmto F_\sbla \in L^2(\Rl^n)$ is continuous. These boundary values satisfy
\begin{align}\label{multi-three-lines}
\bno{F}	&=	\sup_{\sbla\in\partial\Cu}\|F_{\sbla}\|_2\,.
\end{align}
\end{itemize}
\end{proposition}
\begin{proof}
{\em i)} Consider $\bte\in\Rl^n$, $\bla\in\Cu$ and a polydisc $\DD_n(\bte+i\bla)\subset\Tu_n(\Cu)$ with center $\bte+i\bla$. As in Lemma \ref{lemma:h2-inequ}, we obtain
\begin{align}\label{tozeroint}
|F(\bte+i\bla)|
&\leq
(\pi\rho^2)^{-n/2}\left(\int_{\Tu_n(\Cu)} d^n\bze'\, |F(\bze')|^2\cdot \chi_{\DD_n(\sbte+i\sbla)}(\bze')\right)^{1/2}
\,,
\end{align}
where $\chi_{\DD_n(\sbte+i\sbla)}(\bze')$ is the characteristic function of the $\DD_n(\bte+i\bla)=\DD_n(0)+\bte+i\bla$. The functions  $G_\sbte(\bze'):= |F(\bze')|^2\cdot \chi_{\DD_n(\sbte+i\sbla)}(\bze')$ are integrable over $\Tu_n(\Cu)$ for any choice of $\bte\in\Rl^n$, and we have the integrable majorant $G_\sbte(\bze')\leq|F(\bze')|^2$. But as $|\te_k|\to\infty$, $\bte=(\te_1,...,\te_k,...,\te_n)$, the integrand $G_\sbte(\bze')$ converges to zero pointwise in $\bze'$. By dominated convergence, this implies that the above integral \bref{tozeroint} converges to zero for $|\te_k|\to\infty$. If $\bla$ is allowed to vary over a compact subset of $\Cu$, the right hand side of \bref{tozeroint} can be chosen independently of $\bla\in K$, and hence the claim follows.

{\em ii)} For the existence of $L^2$-boundary values for tubes based on polyhedrons, see \cite[Ch. III, Cor. 2.9]{steinweiss}. (This assumption on the shape of $\Cu$ is necessary.)  The second statement is a multidimensional analogue of the three-lines theorem \cite[Thm. 3.7]{conway}, which can be derived from the maximum principle: Let $F\in H^2(\Tu_n(\Cu))$ and $f\in L^2(\Rl^n)$. Then $F*f$ is analytic on $\Tu_n(\Cu)$, and bounded and continuous on its closure. By application of the maximum principle \cite[Cor. 1.3.5]{krantz}, 
\begin{align}
|(F*f)(i\bla)|
\leq
\sup_{\sbla_0\in\partial\Cu}
\sup_{\sbte\in\Rl^n}
|(F*f)(\bte+i\bla_0)|
\leq
\|f\|_2\sup_{\sbla_0\in\partial\Cu}\|F_{\sbla_0}\|_2
\,,\qquad
\bla\in\Cu\,.
\end{align}
Since $f\in L^2(\Rl^n)$ was arbitrary, we have $\|F_\sbla\|\leq \sup_{\sbla_0\in\partial\Cu}\|F_{\sbla_0}\|_2$. On the other hand, $\|F_{\sbla_0}\|_2\leq\bno{F}$ by the continuity of $\overline{\Cu}\ni\bla\lmto F_\sbla\in L^2(\Rl^n)$, and hence the claimed equality \bref{multi-three-lines} follows.
\end{proof}


\addcontentsline{toc}{chapter}{Bibliography}
\bibliographystyle{amsalpha}
\bibliography{PhD-bibliography}

\providecommand{\bysame}{\leavevmode\hbox to3em{\hrulefill}\thinspace}
\providecommand{\MR}{\relax\ifhmode\unskip\space\fi MR }
\providecommand{\MRhref}[2]{%
  \href{http://www.ams.org/mathscinet-getitem?mr=#1}{#2}
}
\providecommand{\href}[2]{#2}
\begin{thebibliography}{BMTW76}

\bibitem[AAR91]{abdalla}
E.~Abdalla\footnote{References to internet archives are abbreviated as follows:
  math-ph/yymmxxx or hep-th/yymmxxx refer to the preprint archive at {\tt
  http://www.arXiv.org}, tags like euclid.cmp/1104248958 refer to the Project
  Euclid page at {\tt http://projecteuclid.org} and LQP/yymmddnn refers to the
  LQP archive at {\tt http://www.uni-goettingen.de/papers.}}, M.~C.~B. Abdalla,
  and K.~D. Rothe, \emph{Non-perturbative methods in two-dimensional quantum
  field theory}, World Scientific, 1991.

\bibitem[AFZ79]{AFZ}
A.E. Arinshtein, V.A. Fateev, and A.B. Zamolodchikov, \emph{Quantum {S}-matrix
  of the $(1+1)$-dimensional {T}oda chain}, Phys. Lett. \textbf{B 87} (1979),
  389--392.

\bibitem[{\AA}ks65]{aks}
S.~{\AA}ks, \emph{Proof that scattering implies production in quantum field
  theory}, J. Math. Phys. \textbf{6} (1965), 516--532.

\bibitem[Ara63]{Araki:free}
H.~Araki, \emph{A lattice of von {N}eumann algebras associated with the quantum
  theory of a free {B}ose field}, J. Math. Phys. \textbf{4} (1963), 1343--1362.

\bibitem[Ara99]{araki}
\bysame, \emph{Mathematical theory of quantum fields}, Oxford Scientific
  Publications, 1999.

\bibitem[BB99]{BoBu-deSitter}
H.-J. Borchers and D.~Buchholz, \emph{Global properties of vacuum states in de
  {S}itter space}, Ann. H. Poincar\'e \textbf{A 70} (1999), 23--40,
  [\href{http://arxiv.org/abs/gr-qc/9803036}{gr-qc/9803036}].

\bibitem[BBS01]{BBS}
H.-J. Borchers, D.~Buchholz, and B.~Schroer, \emph{Polarization-free generators
  and the {S}-matrix}, Commun. Math. Phys. \textbf{219} (2001), 125--140,
  [\href{http://arxiv.org/abs/hep-th/0003243}{hep-th/0003243}].

\bibitem[BDF87]{Buchholz:1986bg}
D.~Buchholz, C.~D'Antoni, and K.~Fredenhagen, \emph{The universal structure of
  local algebras}, Commun. Math. Phys. \textbf{111} (1987), 123--135,
  [\href{http://projecteuclid.org/Dienst/UI/1.0/Summarize/euclid.cmp/110415947%
0}{euclid.cmp/1104159470}].

\bibitem[BDFS00]{4maenner}
D.~Buchholz, O.~Dreyer, M.~Florig, and S.~J. Summers, \emph{Geometric modular
  action and spacetime symmetry groups}, Rev. Math. Phys. \textbf{12} (2000),
  475--560, [\href{http://arxiv.org/abs/math-ph/9805026}{math-ph/9805026}].

\bibitem[BDL90a]{nuclearmaps1}
D.~Buchholz, C.~D'Antoni, and R.~Longo, \emph{Nuclear maps and modular
  structures. 1. {G}eneral properties}, J. Funct. Anal. \textbf{88} (1990),
  233--250.

\bibitem[BDL90b]{nuclearmaps2}
\bysame, \emph{Nuclear maps and modular structures. 2. {A}pplications to
  quantum field theory}, Commun. Math. Phys. \textbf{129} (1990), 115--138,
  [\href{http://projecteuclid.org/Dienst/UI/1.0/Summarize/euclid.cmp/110418064%
8}{euclid.cmp/1104180648}].

\bibitem[BDL06]{nuclearmaps3}
\bysame, \emph{Nuclearity and thermal states in conformal field theory},
  preprint (2006),
  [\href{http://arxiv.org/abs/math-ph/0603083}{math-ph/0603083}].

\bibitem[BEG65]{crossing}
J.~Bros, H.~Epstein, and V.~Glaser, \emph{A proof of the crossing property for
  two-particle amplitudes in general quantum field theory}, Commun. Math. Phys.
  \textbf{1} (1965), 240--264,
  [\href{http://projecteuclid.org/Dienst/UI/1.0/Summarize/euclid.cmp/110375877%
5}{euclid.cmp/1103758775}].

\bibitem[BF82]{BF}
D.~Buchholz and K.~Fredenhagen, \emph{Locality and the structure of particle
  states}, Commun. Math. Phys. \textbf{84} (1982), 1--54,
  [\href{http://projecteuclid.org/Dienst/UI/1.0/Summarize/euclid.cmp/110392104%
4}{euclid.cmp/1103921044}].

\bibitem[BFKZ99]{Babujian:1998uw}
H.~M. Babujian, A.~Fring, M.~Karowski, and A.~Zapletal, \emph{Exact form
  factors in integrable quantum field theories: The sine-{G}ordon model}, Nucl.
  Phys. \textbf{B538} (1999), 535--586,
  [\href{http://arxiv.org/abs/hep-th/9805185}{hep-th/9805185}].

\bibitem[BGL02]{BGL}
R.~Brunetti, D.~Guido, and R.~Longo, \emph{Modular localization and {W}igner
  particles}, Rev. Math. Phys. \textbf{14} (2002), 759--786,
  [\href{http://arxiv.org/abs/math-ph/0203021}{math-ph/0203021}].

\bibitem[BI83]{Bros:1983vf}
J.~Bros and D.~Iagolnitzer, \emph{Structure of scattering functions at m
  particle thresholds in a simplified theory and nonholonomic character of the
  {S}-matrix and {G}reen's functions}, Phys. Rev. \textbf{D27} (1983),
  811--824.

\bibitem[BJ86]{BuJu1}
D.~Buchholz and P.~Junglas, \emph{Local properties of equilibrium states and
  the particle spectrum in quantum field theory}, Lett. Math. Phys. \textbf{11}
  (1986), 51.

\bibitem[BJ87]{BuJa}
D.~Buchholz and P.~Jacobi, \emph{On the nuclearity condition for massless
  fields}, Lett. Math. Phys. \textbf{13} (1987), 313.

\bibitem[BJ89]{BuJu2}
D.~Buchholz and P.~Junglas, \emph{On the existence of equilibrium states in
  local quantum field theory}, Commun. Math. Phys. \textbf{121} (1989),
  255--270,
  [\href{http://projecteuclid.org/Dienst/UI/1.0/Summarize/euclid.cmp/110417806%
6}{euclid.cmp/1104178066}].

\bibitem[BK01]{Babujian:2001wp}
H.~Babujian and M.~Karowski, \emph{The 'bootstrap program' for integrable
  quantum field theories in 1+1 dimensions}, preprint (2001),
  [\href{http://arxiv.org/abs/hep-th/0110261}{hep-th/0110261}].

\bibitem[BK04]{Babujian:2003sc}
\bysame, \emph{Towards the construction of {W}ightman functions of integrable
  quantum field theories}, Int. J. Mod. Phys. \textbf{A 19S2} (2004), 34--49,
  [\href{http://arxiv.org/abs/hep-th/0301088}{hep-th/0301088}].

\bibitem[BKW79]{BKW}
B.~Berg, M.~Karowski, and P.~Weisz, \emph{Construction of {G}reen's functions
  from an exact {S}-matrix}, Phys. Rev. \textbf{D 19} (1979), 2477--2479.

\bibitem[BL75]{Bros:1974ad}
J.~Bros and M.~Lassalle, \emph{Analyticity properties and many particle
  structure in general quantum field theory. 2. {O}ne particle irreducible
  n-point functions}, Commun. Math. Phys. \textbf{43} (1975), 279--309,
  [\href{http://projecteuclid.org/Dienst/UI/1.0/Summarize/euclid.cmp/110389918%
6}{euclid.cmp/1103899186}].

\bibitem[BL04]{BuLe}
D.~Buchholz and G.~Lechner, \emph{Modular nuclearity and localization}, Ann. H.
  Poincar\'e \textbf{5} (2004), 1065--1080,
  [\href{http://arxiv.org/abs/math-ph/0402072}{math-ph/0402072}].

\bibitem[BLT75]{ovovov}
N.~N. Bogolubov, A.~A. Logunov, and I.~T. Todorov, \emph{Introduction to
  axiomatic quantum field theory}, Benjamin, 1975.

\bibitem[BLT90]{ovovov2}
\bysame, \emph{General principles of quantum field theory}, Kluwer, 1990.

\bibitem[BM48]{bochner-martin}
S.~Bochner and W.~T. Martin, \emph{Several complex variables}, 4$^{\rm th}$
  ed., Princeton University Press, 1948.

\bibitem[BMTW76]{BCTW}
E.~Barouch, B.~M. McCoy, C.~A. Tracy, and T.~T. Wu, \emph{Spin-spin correlation
  functions for the two-dimensional {I}sing model: Exact theory in the scaling
  region}, Phys. Rev. \textbf{B 13} (1976), 316--374.

\bibitem[Boa54]{boas}
R.~P. Boas, \emph{Entire functions}, Academic Press, 1954.

\bibitem[Bor60]{borchers:class}
H.-J. Borchers, \emph{{\"U}ber die {M}annigfaltigkeit der interpolierenden
  {F}elder zu einer kausalen {S}-{M}atrix}, Nuovo Cim. \textbf{15} (1960),
  784--794.

\bibitem[Bor92]{borchers-2d}
\bysame, \emph{The {CPT} theorem in two-dimensional theories of local
  observables}, Commun. Math. Phys. \textbf{143} (1992), 315--332,
  [\href{http://projecteuclid.org/Dienst/UI/1.0/Summarize/euclid.cmp/110424895%
8}{euclid.cmp/1104248958}].

\bibitem[Bor98]{borchers-halfsided}
\bysame, \emph{Half-sided translations and the type of von {N}eumann algebras},
  Lett. Math. Phys. \textbf{44} (1998), 283--290,
  [\href{http://www.lqp.uni-goettingen.de/papers/98/01/98012900.html}{LQP/9801%
2900}].

\bibitem[BP90]{BuPo}
D.~Buchholz and M.~Porrmann, \emph{How small is the phase space in quantum
  field theory?}, Ann. Poincar\'e \textbf{52} (1990), 237.

\bibitem[BR87]{BraRob1}
O.~Bratteli and D.~W. Robinson, \emph{Operator algebras and quantum statistical
  mechanics 1}, 2$^{\rm nd}$ ed., Springer, 1987,
  [\url{http://www.math.uio.no/~bratteli/}].

\bibitem[BR97]{BraRob2}
\bysame, \emph{Operator algebras and quantum statistical mechanics 2}, 2$^{\rm
  nd}$ ed., Springer, 1997, [\url{http://www.math.uio.no/~bratteli/}].

\bibitem[Bro86]{Bros:1985gy}
J.~Bros, \emph{Derivation of asymptotic crossing domains for multiparticle
  processes in axiomatic quantum field theory: A general approach and a
  complete proof for 2 $\to$ 3 particle processes}, Phys. Rept. \textbf{134}
  (1986), 325.

\bibitem[Bro03]{bros-compact}
J.~Bros, \emph{A proof of {H}aag-{S}wieca's compactness property for elastic
  scattering states}, Commun. Math. Phys. \textbf{237} (2003), 289--308.

\bibitem[BS91]{BraSa}
H.~W. Braden and R.~Sasaki, \emph{The {S}-matrix coupling dependence for a, d
  and e affine {T}oda field theory}, Phys. Lett. \textbf{B 255} (1991), no.~3,
  343--352.

\bibitem[BS05a]{bs-scatter}
D.~Buchholz and S.~J. Summers, \emph{Scattering in relativistic quantum field
  theory: Fundamental concepts and tools}, preprint (2005),
  [\href{http://arxiv.org/abs/math-ph/0509047}{math-ph/0509047}].

\bibitem[BS05b]{BuSu-1}
\bysame, \emph{String-localized fields in a strongly nonlocal model}, preprint
  (2005), [\href{http://arxiv.org/abs/math-ph/0512060}{math-ph/0512060}].

\bibitem[Buc74]{Buchholz:1973bk}
D.~Buchholz, \emph{Product states for local algebras}, Commun. Math. Phys.
  \textbf{36} (1974), 287--304,
  [\href{http://projecteuclid.org/Dienst/UI/1.0/Summarize/euclid.cmp/110385977%
3}{euclid.cmp/1103859773}].

\bibitem[Buc75]{buchholz-massless-fermis}
\bysame, \emph{Collision theory for massless {F}ermions}, Commun. Math. Phys.
  \textbf{42} (1975), 269--279,
  [\href{http://projecteuclid.org/Dienst/UI/1.0/Summarize/euclid.cmp/110389904%
9}{euclid.cmp/1103899049}].

\bibitem[Buc77]{buchholz-massless-bosons}
\bysame, \emph{Collision theory for massless {B}osons}, Commun. Math. Phys.
  \textbf{52} (1977), 147--173,
  [\href{http://projecteuclid.org/Dienst/UI/1.0/Summarize/euclid.cmp/110390049%
4}{euclid.cmp/1103900494}].

\bibitem[Buc90]{DB-harmonic}
\bysame, \emph{Harmonic analysis of local operators}, Commun. Math. Phys.
  \textbf{129} (1990), 631--641.

\bibitem[Buc06]{buchholz-pc}
\bysame, private communication (2006).

\bibitem[BW75]{BiWi1}
J.~J. Bisognano and E.~H. Wichmann, \emph{On the duality condition for a
  hermitian scalar field}, J. Math. Phys. \textbf{16} (1975), 985--1007.

\bibitem[BW76]{BiWi2}
\bysame, \emph{On the duality condition for quantum fields}, J. Math. Phys.
  \textbf{17} (1976), 303--321.

\bibitem[BW86]{BuWi}
D.~Buchholz and E.~H. Wichmann, \emph{Causal independence and the energy level
  density of states in local quantum field theory}, Commun. Math. Phys.
  \textbf{106} (1986), 321--344,
  [\href{http://projecteuclid.org/Dienst/UI/1.0/Summarize/euclid.cmp/110411570%
3}{euclid.cmp/1104115703}].

\bibitem[BW90]{baumwoll}
H.~Baumgärtel and M.~Wollenberg, \emph{Causal nets of operator algebras},
  Akademie Verlag, Berlin, 1990.

\bibitem[BY90]{BY90}
H.-J. Borchers and J.~Yngvason, \emph{Positivity of {W}ightman functionals and
  the existence of local nets}, Commun. Math. Phys. \textbf{127} (1990),
  607--615,
  [\href{http://projecteuclid.org/Dienst/UI/1.0/Summarize/euclid.cmp/110418022%
3}{euclid.cmp/1104180223}].

\bibitem[CA01]{castro}
O.~A. Castro-Alvaredo, \emph{Bootstrap methods in 1+1-dimensional quantum field
  theories: The homogeneous sine-{G}ordon models}, Ph.D. thesis, Universidad de
  Santiago de Compostela, 2001,
  [\href{http://arxiv.org/abs/hep-th/0109212}{hep-th/0109212}].

\bibitem[CAF03]{ffp-applications}
O.~A. Castro-Alvaredo and A.~Fring, \emph{From integrability to conductance,
  impurity systems}, Nucl. Phys. \textbf{B649} (2003), 449--490,
  [\href{http://arxiv.org/abs/hep-th/0205076}{hep-th/0205076}].

\bibitem[CM67]{CM}
R.~S. Coleman and J.~Mandula, \emph{All possible symmetries of the {S}-matrix},
  Phys. Rev. \textbf{159} (1967), 1251--1256.

\bibitem[Con73]{conway}
J.~B. Conway, \emph{Functions of one complex variable}, Springer, 1973.

\bibitem[DF77]{DrFr}
W.~Driessler and J.~Fr\"ohlich, \emph{The reconstruction of local observable
  algebras from the {E}uclidian {G}reen's functions of relativistic quantum
  field theory}, Ann. Inst. Henri Poincar\'e \textbf{27} (1977), no.~3,
  221--236.

\bibitem[DH06]{stephan-nuclear}
C.~D'Antoni and S.~Hollands, \emph{Nuclearity, local quasiequivalence and split
  property for {D}irac quantum fields in curved spacetime}, Commun. Math. Phys.
  \textbf{261} (2006), 133--159,
  [\href{http://arxiv.org/math-ph/0106028}{math-ph/0106028}].

\bibitem[DHR71]{DHR1}
S.~Doplicher, R.~Haag, and J.~E. Roberts, \emph{Local observables and particle
  statistics. 1}, Commun. Math. Phys. \textbf{23} (1971), 199--230,
  [\href{http://projecteuclid.org/Dienst/UI/1.0/Summarize/euclid.cmp/110385763%
0}{euclid.cmp/1103857630}].

\bibitem[DHR74]{DHR2}
\bysame, \emph{Local observables and particle statistics. 2}, Commun. Math.
  Phys. \textbf{35} (1974), 49--85,
  [\href{http://projecteuclid.org/Dienst/UI/1.0/Summarize/euclid.cmp/110385951%
8}{euclid.cmp/1103859518}].

\bibitem[DL83]{D'AntoniLongo:Flip}
C.~D'Antoni and R.~Longo, \emph{Interpolation by type {I} factors and the flip
  automorphism}, J. Funct. Anal. \textbf{51} (1983), 361--371.

\bibitem[DL84]{DoplicherLongo:StandardSplit}
S.~Doplicher and R.~Longo, \emph{Standard and split inclusions of von {N}eumann
  algebras}, Invent. Math. \textbf{75} (1984), 493--536.

\bibitem[DM71]{DixMar}
J.~Dixmier and O.~Marechal, \emph{Vecteurs totalisateurs d'une algèbre de von
  {N}eumann}, Commun. Math. Phys. \textbf{22} (1971), 44--50,
  [\href{http://projecteuclid.org/Dienst/UI/1.0/Summarize/euclid.cmp/110385741%
2}{euclid.cmp/1103857412}].

\bibitem[Dri75]{Driessler:1975cm}
W.~Driessler, \emph{Comments on lightlike translations and applications in
  relativistic quantum field theory}, Commun. Math. Phys. \textbf{44} (1975),
  133--141,
  [\href{http://projecteuclid.org/Dienst/UI/1.0/Summarize/euclid.cmp/110389929%
8}{euclid.cmp/1103899298}].

\bibitem[Dur00]{duren}
P.~L. Duren, \emph{Theory of ${H}^p$ spaces}, Dover publications, 2000.

\bibitem[Dyb05]{dybalski}
W.~Dybalski, \emph{{H}aag-{R}uelle scattering theory in presence of massless
  particles}, Lett. Math. Phys. \textbf{72} (2005), 27--38,
  [\href{http://arxiv.org/abs/hep-th/0412226}{hep-th/0412226}].

\bibitem[ELOP66]{eden}
R.~J. Eden, P.~V. Landshoff, D.~I. Olive, and J.~C. Polkinghorne, \emph{The
  analytic {S}-matrix}, Cambridge University Press, 1966.

\bibitem[Eps60]{epstein-eotw}
H.~Epstein, \emph{Generalization of the ``edge-of-the-wedge'' theorem}, J.
  Math. Phys. \textbf{1} (1960), 524--531.

\bibitem[Eps66]{Epstein66}
\bysame, \emph{Some analytic properties of scattering amplitudes in quantum
  field theory}, Axiomatic Field Theory (M.~Chretien and S.~Deser, eds.),
  Gordon \& Breach, New York, 1966.

\bibitem[Fad80]{Faddeev}
L.~D. Faddeev, \emph{Quantum completely integrable models of field theory},
  Sov. Sci. Rev. \textbf{C1} (1980), 107--155.

\bibitem[Fid01]{fid}
F.~Fidaleo, \emph{On the split property for inclusions of ${W}^*$-algebras},
  Proceedings of the American Mathematical Society \textbf{130} (2001),
  121--127.

\bibitem[Flo98]{florig}
M.~Florig, \emph{On {B}orchers' theorem}, Lett. Math. Phys. \textbf{46} (1998),
  289.

\bibitem[FMS93]{FMS}
A.~Fring, G.~Mussardo, and P.~Simonetti, \emph{Form-factors for integrable
  {L}agrangian field theories, the sinh-{G}ordon model}, Nucl. Phys.
  \textbf{B393} (1993), 413--441,
  [\href{http://arxiv.org/abs/hep-th/9211053}{hep-th/9211053}].

\bibitem[FOP05]{p-nuc}
C.~J. Fewster, I.~Ojima, and M.~Porrmann, \emph{{p}-{N}uclearity in a new
  perspective}, Lett. Math. Phys. \textbf{73} (2005), 1--15,
  [\href{http://arxiv.org/abs/math-ph/0412027}{math-ph/0412027}].

\bibitem[Fri06]{fringpr}
A.~Fring, private communication (2006).

\bibitem[FRS89]{FRS1}
K.~Fredenhagen, K.-H. Rehren, and B.~Schroer, \emph{Superselection sectors with
  braid group statistics and exchange algebras. 1. {G}eneral theory}, Commun.
  Math. Phys. \textbf{125} (1989), 201--226,
  [\href{http://projecteuclid.org/Dienst/UI/1.0/Summarize/euclid.cmp/110417946%
4}{euclid.cmp/1104179464}].

\bibitem[Gar81]{garnett}
J.~B. Garnett, \emph{Bounded analytic functions}, Academic Press, 1981.

\bibitem[GJ72]{GJ1}
J.~Glimm and A.~Jaffe, \emph{Boson quantum field models}, In *London 1971,
  Mathematics Of Contemporary Physics*, London 1972, 77-143.

\bibitem[GJ81]{glimmjaffe}
\bysame, \emph{Quantum physics - a functional integral point of view}, 2$^{\rm
  nd}$ ed., Springer, 1981.

\bibitem[Haa58]{haag-scatter}
R.~Haag, \emph{Quantum field theories with composite particles and asymptotic
  conditions}, Phys. Rev. \textbf{112} (1958), 669--673.

\bibitem[Haa87]{Hag}
U.~Haagerup, \emph{Connes' bicentralizer problem and uniqueness of the
  injective factor of type {III}$_1$}, Acta. Math. \textbf{158} (1987),
  95--148.

\bibitem[Haa92]{haag}
R.~Haag, \emph{Local quantum physics: Fields, particles, algebras}, Springer,
  1992.

\bibitem[Hep66]{hepp}
K.~Hepp, \emph{On the connection between {W}ightman and {LSZ} quantum field
  theory}, Brandeis University Summer Institute in Theoretical Physics 1965
  "Axiomatic Field Theory" (M.~Chretien and S.~Deser, eds.), vol.~1, Gordon and
  Breach, 1966, pp.~135--246.

\bibitem[Her71]{herbst}
I.~Herbst, \emph{One particle operators and local internal symmetries}, J.
  Math. Phys. \textbf{12} (1971), 2480--2490.

\bibitem[Hor90]{Horuzhy}
S.~S. Horuzhy, \emph{Introduction to algebraic quantum field theory}, Kluwer,
  1990.

\bibitem[HS62]{Haag-Schroer}
R.~Haag and B.~Schroer, \emph{Postulates of quantum field theory}, J. Math.
  Phys. \textbf{3} (1962), 248.

\bibitem[HS65]{Haag-Swieca}
R.~Haag and J.~A. Swieca, \emph{When does a quantum field theory describe
  particles?}, Commun. Math. Phys. \textbf{1} (1965), 308--320,
  [\href{http://projecteuclid.org/Dienst/UI/1.0/Summarize/euclid.cmp/110375894%
7}{euclid.cmp/1103758947}].

\bibitem[Iag78]{Iago-fact}
D.~Iagolnitzer, \emph{Factorization of the multiparticle {S}-matrix in
  two-dimensional space-time models}, Phys. Rev. \textbf{D18} (1978), 1275.

\bibitem[Iag93]{iago}
D.~Iagolnitzer, \emph{Scattering in quantum field theories}, Princeton
  University Press, 1993.

\bibitem[Jar81]{jarchow}
H.~Jarchow, \emph{Locally convex spaces}, B.G.Teubner, 1981.

\bibitem[Jos65]{Jost}
R.~Jost, \emph{General theory of quantized fields}, American Mathematical
  Society, 1965.

\bibitem[Kni61]{Knight}
J.~M. Knight, \emph{Strict localization in quantum field theory}, J. Math.
  Phys. \textbf{2} (1961), 459--471.

\bibitem[Kos84]{kosaki}
H.~Kosaki, \emph{On the continuity of the map $\varphi\mapsto|\varphi|$ from
  the predual of a $w^*$-algebra}, J. Funct. Anal. \textbf{59} (1984), 123.

\bibitem[KR83]{KadRin1}
R.~V. Kadison and J.~R. Ringrose, \emph{Fundamentals of the theory of operator
  algebras. vol. {I}: {E}lementary theory}, Academic Press, 1983.

\bibitem[KR86]{KadRin2}
\bysame, \emph{Fundamentals of the theory of operator algebras. vol. {II}:
  {A}dvanced theory}, Academic Press, 1986.

\bibitem[Kra92]{krantz}
S.~G. Krantz, \emph{Function theory of several complex variables}, 2$^{\rm nd}$
  ed., Wadsworth \& Brooks/Cole, Pacific Grove, California, 1992.

\bibitem[KTTW77]{karo}
M.~Karowksi, H.J. Thun, T.T. Truong, and P.H. Weisz, \emph{On the uniqueness of
  a purely elastic {S}-matrix in (1+1) dimensions}, Phys. Lett. \textbf{67 B}
  (1977), 321--322.

\bibitem[KW01]{Wiesbrock3+1}
R.~Kähler and H.-W. Wiesbrock, \emph{Modular theory and the reconstruction of
  four-dimensional quantum field theories}, J. Math. Phys. \textbf{42} (2001),
  74--86.

\bibitem[Lec02]{GLdipl}
G.~Lechner, \emph{Polarisationsfreie {G}eneratoren in zwei {D}imensionen},
  Diploma thesis, University of Göttingen (2002).

\bibitem[Lec03]{GL1}
\bysame, \emph{Polarization-free quantum fields and interaction}, Lett. Math.
  Phys. \textbf{64} (2003), 137--154,
  [\href{http://arxiv.org/abs/hep-th/0303062}{hep-th/0303062}].

\bibitem[Lec05a]{GL-1}
\bysame, \emph{On the existence of local observables in theories with a
  factorizing {S}-matrix}, J. Phys. \textbf{A38} (2005), 3045--3056,
  [\href{http://arxiv.org/abs/math-ph/0405062}{math-ph/0405062}].

\bibitem[Lec05b]{GL-bros}
\bysame, \emph{Towards the construction of quantum field theories from a
  factorizing {S}-matrix}, preprint, to appear in proceedings `Rigorous Quantum
  Field Theory` by Birkh\"auser (2005),
  [\href{http://arxiv.org/abs/hep-th/0502184}{hep-th/0502184}].

\bibitem[Lec06]{GL-nuci}
\bysame, \emph{An existence proof for interacting quantum field theories with a
  factorizing {S}-matrix}, preprint (2006),
  [\href{http://arxiv.org/abs/math-ph/0601022}{math-ph/0601022}].

\bibitem[Lic63]{Licht1}
A.~L. Licht, \emph{Strict localization}, J. Math. Phys. \textbf{4} (1963),
  1443--1447.

\bibitem[Lic66]{Licht2}
\bysame, \emph{Local states}, J. Math. Phys. \textbf{7} (1966), 1656--1669.

\bibitem[LM95]{LiMi}
A.~Liguori and M.~Mintchev, \emph{Fock representations of quantum fields with
  generalized statistics}, Commun. Math. Phys. \textbf{169} (1995), 635--652,
  [\href{http://arxiv.org/abs/hep-th/9403039}{hep-th/9403039}].

\bibitem[Lon79]{Lo79}
R.~Longo, \emph{Notes on algebraic invariants for noncommutative dynamical
  systems}, Commun. Math. Phys. \textbf{69} (1979), 195--207,
  [\href{http://projecteuclid.org/Dienst/UI/1.0/Summarize/euclid.cmp/110390548%
8}{euclid.cmp/1103905488}].

\bibitem[LR04]{ABC-QFT}
R.~Longo and K.-H. Rehren, \emph{Local fields in boundary conformal qft}, Rev.
  Math. Phys. \textbf{16} (2004), 909,
  [\href{http://arxiv.org/abs/math-ph/0405067}{math-ph/0405067}].

\bibitem[Mai68]{maison}
D.~Maison, \emph{Eine {B}emerkung zu {C}lustereigenschaften}, Commun. Math.
  Phys. \textbf{10} (1968), 48--51,
  [\href{http://projecteuclid.org/Dienst/UI/1.0/Summarize/euclid.cmp/110384098%
2}{euclid.cmp/1103840982}].

\bibitem[Mar69]{martin}
A.~Martin, \emph{Can one continue the scattering amplitude through the elastic
  cut?}, Problems in Theoretical Physics (Nauka, Moscow) (D.~I. Blokintseff,
  ed.), 1969, p.~113.

\bibitem[Mir99]{Miramontes}
J.~L. Miramontes, \emph{Hermitian analyticity versus real analyticity in
  two-dimensional factorized {S}-matrix theories}, Phys. Lett. \textbf{B455}
  (1999), 231--238,
  [\href{http://arxiv.org/abs/hep-th/9901145}{hep-th/9901145}].

\bibitem[Mit77]{mitra}
P.~Mitra, \emph{Elasticity, factorization and {S}-matrices in
  (1+1)-dimensions}, Phys. Lett. \textbf{72 B} (1977), 62--64.

\bibitem[MSY05]{Mund:2005cv}
J.~Mund, B.~Schroer, and J.~Yngvason, \emph{String-localized quantum fields and
  modular localization}, preprint (2005),
  [\href{http://arxiv.org/abs/math-ph/0511042}{math-ph/0511042}].

\bibitem[MTW77]{CTW}
B.~M. McCoy, C.~A. Tracy, and T.~T. Wu, \emph{Two-dimensional {I}sing model as
  an exactly solvable relativistic quantum field theory: Explicit formulas for
  $n$-point functions}, Phys. Rev. Lett. \textbf{38} (1977), 793--796.

\bibitem[M{\"u}g97]{Mue-PhD}
M.~M{\"u}ger, \emph{Superselection structure of quantum field theories in 1+1
  dimensions}, Ph.D. thesis, University of Hamburg, 1997, DESY-preprint 97-073.

\bibitem[M{\"u}g98]{Mue-SPW}
\bysame, \emph{Superselection structure of massive quantum field theory in 1+1
  dimensions}, Rev. Math. Phys. \textbf{10} (1998), 1147--1170,
  [\href{http://arxiv.org/abs/hep-th/9705019}{hep-th/9705019}].

\bibitem[Mun01]{MundBiWi}
J.~Mund, \emph{The {B}isognano-{W}ichmann theorem for massive theories}, Ann.
  H. Poincar\'e \textbf{2} (2001), 907--926,
  [\href{http://arxiv.org/abs/hep-th/0101227}{hep-th/0101227}].

\bibitem[Mun03]{Mund:2002yc}
\bysame, \emph{Modular localization of massive particles with any spin in d =
  2+1}, J. Math. Phys. \textbf{44} (2003), 2037--2057,
  [\href{http://arxiv.org/abs/hep-th/0208195}{hep-th/0208195}].

\bibitem[MW73]{ising2d}
B.~M. McCoy and T.~T. Wu, \emph{The two-dimensional {I}sing model}, Harvard
  University Press, 1973.

\bibitem[Nie98]{niedermaier}
M.~R. Niedermaier, \emph{A derivation of the cyclic form factor equation},
  Commun. Math. Phys. \textbf{196} (1998), 411--428,
  [\href{http://arxiv.org/abs/hep-th/9706172}{hep-th/9706172}].

\bibitem[Oli62]{olive}
D.~I. Olive, Nuovo Cimento \textbf{26} (1962), 73.

\bibitem[Pie72]{pietsch}
A.~Pietsch, \emph{Nuclear locally convex spaces}, Springer, 1972.

\bibitem[RS75]{SimonReed2}
M.~Reed and B.~Simon, \emph{Methods of modern mathematical physics vol {II}:
  {F}ourier analysis, self-adjointness}, Academic Press, 1975.

\bibitem[RS79]{SimonReed3}
\bysame, \emph{Methods of modern mathematical physics vol {III}: Scattering
  theory}, Academic Press, 1979.

\bibitem[RS80]{SimonReed1}
\bysame, \emph{Methods of modern mathematical physics vol {I}: Functional
  analysis}, Academic Press, 1980.

\bibitem[Rud87]{rudin}
W.~Rudin, \emph{Real and complex analysis}, 3$^{\rm rd}$ ed., McGraw-Hill,
  1987.

\bibitem[Rue62]{ruelle-scatter}
D.~Ruelle, \emph{On asymptotic condition in quantum field theory}, Helv. Phys.
  Acta \textbf{35} (1962), 147--163.

\bibitem[Sak71]{sakai}
S.~Sakai, \emph{${C}^*$- and ${W}^*$-algebras}, Springer, 1971.

\bibitem[Sch97]{Schroer:1997cq}
B.~Schroer, \emph{Modular localization and the bootstrap-formfactor program},
  Nucl. Phys. \textbf{B499} (1997), 547--568,
  [\href{http://arxiv.org/abs/hep-th/9702145}{hep-th/9702145}].

\bibitem[Sch99]{Schroer:1997cx}
\bysame, \emph{Modular wedge localization and the $d = $1+1 formfactor
  program}, Ann. Phys. \textbf{275} (1999), 190--223,
  [\href{http://arxiv.org/abs/hep-th/9712124}{hep-th/9712124}].

\bibitem[Sch00a]{Schroer:1999xi}
\bysame, \emph{New concepts in particle physics from solution of an old
  problem}, J. Phys. \textbf{A33} (2000), 5231,
  [\href{http://arxiv.org/abs/hep-th/9908021}{hep-th/9908021}].

\bibitem[Sch00b]{Schroer:2000dg}
\bysame, \emph{Particle physics and qft at the turn of the century: Old
  principles with new concepts. (an essay on local quantum physics)}, J. Math.
  Phys. \textbf{41} (2000), 3801--3831,
  [\href{http://arxiv.org/abs/hep-th/9810080}{hep-th/9810080}].

\bibitem[Sch04]{schroer-crossing}
\bysame, \emph{Constructive proposals for qft based on the crossing property
  and on lightfront holography}, preprint (2004),
  [\href{http://arxiv.org/abs/hep-th/0406016}{hep-th/0406016}].

\bibitem[Sch05]{tobias-phd}
T.~Schlegelmilch, \emph{Local scattering operators for ${P}(\phi)_2$ models and
  the time-dependent {S}chr\"odinger equation}, Ph.D. thesis, University of
  Hamburg, 2005,
  [\href{http://arxiv.org/abs/math-ph/0512091}{math-ph/0512091}].

\bibitem[Skl89]{Sklyanin:1989cf}
E.~K. Sklyanin, \emph{Exact quantization of the sinh-{G}ordon model}, Nucl.
  Phys. \textbf{B326} (1989), 719.

\bibitem[Smi92]{smirnov}
F.~A. Smirnov, \emph{Form factors in completely integrable models of quantum
  field theory}, World Scientific, 1992.

\bibitem[Smi06]{smirnov-pc}
F.~Smirnov, private communication, 2006.

\bibitem[SMJ77]{SMJ}
M.~Sato, T.~Miwa, and M.~Jimbo, report, Kyoto (unpublished) (1977).

\bibitem[SS05]{tobias}
T.~Schlegelmilch and R.~Schnaubelt, \emph{Wellposedness of hyperbolic evolution
  equations in {B}anach spaces}, preprint (2005),
  [\href{http://arxiv.org/abs/math-ph/0507013}{math-ph/0507013}].

\bibitem[Sum82]{Summers-split}
S.~J. Summers, \emph{Normal product states for fermions and twisted duality for
  {CCR} and {CAR} type algebras with application to the {Y}ukawa-2 quantum
  field model}, Commun. Math. Phys. \textbf{86} (1982), 111--141,
  [\href{http://projecteuclid.org/Dienst/UI/1.0/Summarize/euclid.cmp/110392161%
9}{euclid.cmp/1103921619}].

\bibitem[Sum90]{Summers:1990tp}
S.J. Summers, \emph{On the independence of local algebras in quantum field
  theory}, Rev. Math. Phys. \textbf{2} (1990), 201--247.

\bibitem[SW71]{steinweiss}
E.~Stein and G.~Weiss, \emph{Introduction to {F}ourier analysis on {E}uclidian
  spaces}, Princeton University Press, 1971.

\bibitem[SW78]{Shankar-Witten}
R.~Shankar and E.~Witten, \emph{The {S}-matrix of the supersymmetric nonlinear
  $\sigma$ model}, Phys. Rev. \textbf{D17} (1978), 2134.

\bibitem[SW80]{streater}
R.~F. Streater and A.~S. Wightman, \emph{P{C}{T}, spin and statistics, and all
  that}, 3$^{\rm rd}$ ed., Princeton University Press, 1980.

\bibitem[SW00]{Schroer-Wiesbrock}
B.~Schroer and H.-W. Wiesbrock, \emph{Modular constructions of quantum field
  theories with interactions}, Rev. Math. Phys. \textbf{12} (2000), 301--326,
  [\href{http://arxiv.org/abs/hep-th/9812251}{hep-th/9812251}].

\bibitem[Sym69]{symanzik}
K.~Symanzik, \emph{Euclidian quantum field theory}, Local quantum theory.
  Proceedings, 45th Course of the International School of Physics 'Enrico
  Fermi' (Varenna, Italy) (R.~Jost, ed.), Academic Press, 1969.

\bibitem[Tak79]{Takesaki1}
M.~Takesaki, \emph{Theory of operator algebras {I}}, Springer, 1979.

\bibitem[Wei95]{weinberg}
S.~Weinberg, \emph{The quantum theory of fields}, Cambridge University Press,
  1995.

\bibitem[Wie93]{Wiesbrock1+1}
H-W. Wiesbrock, \emph{Symmetries and half-sided modular inclusions of von
  {N}eumann algebras}, Lett. Math. Phys. \textbf{28} (1993), 107--114.

\bibitem[Wie98]{Wiesbrock2+1}
\bysame, \emph{Modular intersections of von {N}eumann algebras in quantum field
  theory}, Commun. Math. Phys. \textbf{193} (1998), 269--285.

\bibitem[Wig39]{Wigner}
E.~P. Wigner, \emph{On unitary representations of the inhomogeneous {L}orentz
  group}, Ann. Math. \textbf{40} (1939), 149--204.

\bibitem[YZ91]{YuroZam}
V.~P. Yurov and A.~B. Zamolodchikov, \emph{Correlation functions of integrable
  2d models of the relativistic field theory: {I}sing model}, Int. J. Mod.
  Phys. \textbf{A 6} (1991), 3419--3440.

\bibitem[Zam77]{Zamo-SG}
A.~B. Zamolodchikov, \emph{Exact two particle {S}-matrix of quantum
  sine-{G}ordon solitons}, Pisma Zh. Eksp. Teor. Fiz. \textbf{25} (1977),
  499--502.

\bibitem[ZZ79]{ZZ}
A.B. Zamolodchikov and A.B. Zamolodchikov, \emph{Factorized {S}-matrices in two
  dimensions as the exact solutions of certain relativistic quantum field
  models}, Ann. Phys. \textbf{120} (1979), 253--291.

\end{thebibliography}
\chapter*{Frequently Used Symbols}
\thispagestyle{empty}
\addcontentsline{toc}{chapter}{{Frequently Used Symbols}}

\begin{center}
\begin{longtable}[]{lp{8cm}l}
Symbol		&	Description				& Reference							\\	
\hline
																			\\
$\|\cdot\|_1$	&	nuclear norm				&	Def. \ref{def:nucmap}, page \pageref{def:nucmap}	\\
$\bno{\cdot}$	&	Hardy norm				&	Def. \ref{def:hardy}, page \pageref{def:hardy}		\\
$\Subset$	& 	inclusion relation			&	section \ref{Sec:Geo}, page \pageref{inclusionsymbol}			\\
$\A$		&	net of local algebras		&	section \ref{Sec:Axioms}, page \pageref{Sec:Axioms}		\\
$C_0^\infty$	&	smooth functions of compact support
										&									\\
$\DD$			&	subspace of $\Hil$ of finite particle number
										&	section \ref{sec:zf-algebra}, page \pageref{defineDD}		\\
$D_n$		&	representation of the symmetric group $\frS_n$
										& Lemma \ref{lem:Dn}, page \pageref{lem:Dn}	\\
$\Delta$		&	modular operator of $(\A(W_R),\Om)$
										&	Prop. \ref{prop:BiWi}, page \pageref{prop:BiWi}		\\
$\F_{\Hil_1}$	&	unsymmetrized Fock space over the one particle space $\Hil_1$
										&	section \ref{sec:zf-algebra}, page \pageref{defFH1}	\\
$\phi$		&	wedge-local quantum field		&	Def. \ref{def:phi}, page \pageref{def:phi}								\\
$\phi'$		&	wedge-local quantum field		&	\bref{def:phi'}, page \pageref{def:phi'}								\\
$\Gamma$		&	antiunitary implementing the time reflection
										&	\bref{def:Gamma}, page \pageref{def:Gamma}\\
$H$			& 	Hamiltonian, $H=P_0$, generator of time translations
										&									\\
$\Hil$			&	Hilbert space				&	\bref{def:Hil}, page \pageref{def:Hil}		\\
$\Hil_n$		&	$n$-particle spaces			&	\bref{def:Hil}, page \pageref{def:Hil}		\\
$\Hil^\pm$		&	Bose and Fermi Fock spaces over $\Hil_1$&									\\
$H^\pm_m$	&	upper and lower mass shell		&	section \ref{sec:SM}, page \pageref{uppermassshell}	\\
$H^2(\Tu)$		&	Hardy space over the tube $\Tu\subset\Cl^n$
										&	Def. \ref{def:hardy}, page \pageref{def:hardy}	\\
$J$			& 	modular conjugation of $(\A(W_R),\Om)$
										&	\bref{def:J}, page \pageref{def:J}			\\
$\kappa(S_2)$	&	distance of singularities of $S_2$ to the real line
										&	\bref{def:kappa}, page \pageref{def:kappa}	\\
$\La(\la)$		&	proper Lorentz boost with rapidity $\la$
										&	\bref	{def:boost}, page \pageref{def:boost}	\\
$\bla_0$		&	special vector in $\Cl^n$		&	\bref{def:Tnk}, page \pageref{def:Tnk}		\\	
$\M$			&	right wedge algebra			&	Def. \ref{def:srwa}, page \pageref{def:srwa}	\\
$\NN(\X,\Y)$	&	space of nuclear operators between two Banach spaces $\X$ and $\Y$
										&	section \ref{app-hardy}, page \pageref{app-hardy}\\
$\OO$		&	double cone				&	section \ref{Sec:Geo}, page \pageref{Sec:Geo}\\
$\OOO$		&	family of all double cones		&	section \ref{Sec:Geo}, page \pageref{Sec:Geo}\\
$\Om$		&	the vacuum vector			&									\\
$P,P_n$		&	projections onto $S_2$-symmetric spaces
										&	\bref{def:Pn}, page \pageref{def:Pn}		\\
$P_0,P_1$		&	energy and momentum operators	&									\\
$\PG$, $\PG_+$,	$\PGpo$
			&	Poincar\'e group and subgroups thereof
										&	section \ref{Sec:Geo}, page \pageref{Sec:Geo}\\
$p(\te)$		&	on-shell momentum vector with rapidity $\te$
										&	\bref{def:pte}, page \pageref{def:pte}		\\
$S$			&	the S-matrix on $\Hil$			&	section \ref{sec:SM}, page \pageref{sec:SM}	\\
$\hat{S}$		&	the S-matrix on $\Hil^+$		&	section \ref{sec:SM}, page \pageref{sec:SM}	\\
$S_{n,m}$		&	kernels of the S-matrix			&	section \ref{sec:SM}, page \pageref{sec:SM}	\\
$S_2$			&	scattering function			& 	Def. \ref{def:s2}, page \pageref{def:s2}		\\
$\|S_2\|_\kappa$	&	supremum norm of $S_2$		&	\bref{def:S2kappa}, page \pageref{def:S2kappa}\\
$S^\pi$		&	function related to the representation $D_n$
										&	Lemma \ref{lem:Dn}, page \pageref{lem:Dn}	\\
$\SF$			&	set of all scattering functions	& 	Def. \ref{def:s2}, page \pageref{def:s2}		\\
$\SF_0$		&	subfamilies of $\SF$			&	Def. \ref{def:S0family}, page \pageref{def:S0family}		\\
$\SF_0^\pm$	&	subfamilies of $\SF_0$			&	\bref{def:SF0pm}, page \pageref{def:SF0pm}	\\
$S(a,b)$		&	strip region in $\Cl$			&	\bref{strip-not}, page \pageref{strip-not}	\\
$\frS_n$		&	the group of permutations of $n$ objects
										&									\\
$\Ss$			&	Schwartz test functions		&									\\
$T$			&	time reflection				&	section \ref{Sec:Geo}, page \pageref{Sec:Geo}\\
$T_{s,\kappa}$	&	integral operators on $L^2(\Rl)$	&	section \ref{app:intop}	, page \pageref{app:intop}	\\
$\Theta_{\beta,\OO}$
			&	map appearing in the energy nuclearity condition 
										&	\bref{def:ThetaO}, page \pageref{def:ThetaO}	\\
$U, \hat{U}, \tilde{U}$
			&	representations of $\PG$, $\PG_+$, $\PGpo$ or $(\Rl^2,+)$
										&									\\
$V\iin$, $V\oout$&	M{\o}ller operators			&	section \ref{sec:SM}, page \pageref{sec:SM}	\\
$V^+$		&	forward light cone			&									\\
$W$			&	a wedge					&	section \ref{Sec:Geo}, page \pageref{Sec:Geo}\\
$\W$			& 	the set of all wedges			&	section \ref{Sec:Geo}, page \pageref{Sec:Geo}\\
$W_L,W_R$		&	the left and right wedge		&	section \ref{Sec:Geo}, page \pageref{Sec:Geo}\\
$W_L^\OO$, $W_R^\OO$
			&	left and right wedges of double cone $\OO$
										&	section \ref{Sec:Geo}, page \pageref{Sec:Geo}\\
$\Xi(s),\Xi_n(s)$	& 	maps needed for the modular nuclearity condition 
										&	\bref{Xis}, page \pageref{Xis}	\\
$(x,\la)$		&	proper orthochronous Poincar\'e transformation
										&	section \ref{Sec:Geo}, page \pageref{Sec:Geo}\\
$Y^\pm$		& 	unitaries mapping $\Hil$ onto $\Hil^\pm$
										&	Lemma \ref{lemma:y}, page \pageref{lemma:y}\\
$Z,\Zd$		& 	abstract Zamolodchikov creation and annihilation operators
										& section \ref{sec:zf-algebra}, page \pageref{sec:zf-algebra}	\\
$z,\zd$		& 	Zamolodchikov creation and annihilation operators on $\Hil$
										& Lemma \ref{z-lemma}, page \pageref{z-lemma}	\\
$\Z(S_2)$		&	Zamolodchikov algebra with scattering function $S_2$
										& section \ref{sec:zf-algebra}, page \pageref{sec:zf-algebra}	\\
$\ZZZ$		&	set of zeros of a scattering function
										& Prop. \ref{prop:s2}, page \pageref{prop:s2}		
\end{longtable}	
\end{center}
\thispagestyle{empty}
\addcontentsline{toc}{chapter}{Acknowledgements}
\thispagestyle{plain}
\vspace*{6em}
\noindent 
{\Huge \bf Acknowledgements}
\vspace{3.5em}\\

\noindent I would like to thank Prof.~D.~Buchholz for giving me the opportunity to work in this interesting field of mathematical physics. His ideas constituted the basis for my contributions presented in this thesis, and without his constant support and advice, this work would not have been possible.

I would also like to thank Prof.~K.~Fredenhagen, who immediately accepted to write the additional report.

During the years I worked on this thesis, I profited from discussions with many persons. Special thanks are due to B. Schroer, whose ideas form an essential part of the construction of the models presented here, and with whom I had the opportunity to discuss the topic several times. Regarding the theory of complex analysis and its applications to quantum field theory, conversations with J. Bros and H.-J. Borchers have been very helpful. In particular, I would like to thank J. Bros for repeatedly explaining analytic properties of the S-matrix to me. I am also grateful for useful comments and interesting discussions with K. Fredenhagen, J. Mund, M. M\"uger, D. Iagolnitzer, K.-H. Rehren, S.J. Summers and R. Verch at various stages of the work.

Many thanks are also due to G. Garrigos, who helped me finding a much sought after reference on Hardy spaces, and to A. Fring, who informed me about the status of the form factor program.

Highly appreciated help in the proofreading of the manuscript came from several directions. First and foremost, I would like to thank Saskia for unrelentingly working her way through incomprehensible physicists' writing, and for eliminating lots of mistakes. Furthermore, Claus Ropers, Helmut H\"olzler, Mathias Puhlmann, Philipp Knake, Stefan Hollands and Wojciech Dybalski helped correcting the manuscript.

Finally, financial support from the Lucia-Pfohe-Stiftung and the DFG is gratefully acknowledged, as well as travel grants from the DPG, the Centre d'\'Etudes de Saclay, the Universities of Hamburg and Leipzig, the FU Berlin, the University of Gainesville, Florida, the Max Planck Institute in Golm and the Research Institute in Oberwolfach.

\end{document}